\documentclass[]{aastex631}

\renewcommand{\thefootnote}{}
\shorttitle{A $\lambda$ 3\,mm line survey towards IRC\,+10216 (CW Leo)}
\shortauthors{Tuo et al.}
\graphicspath{{./}{figures/}}

\begin{document}

\title{A $\lambda$ 3\,mm  line survey towards the circumstellar envelope of the carbon-rich AGB star IRC\,+10216 (CW Leo)}
%
%\correspondingauthor{Xiaohu Li閳{*}
%}
%\email{xiaohu.li@xao.ac.cn}

\footnote{$^{*}$Corresponding author: X. Li,  E-mail address: xiaohu.li@xao.ac.cn
}

\author{Juan Tuo}

\affiliation{Xinjiang Astronomical Observatory, Chinese Academy of Sciences, 150 Science 1-Street, Urumqi, Xinjiang 830011, China}
\affiliation{ University of Chinese Academy of Sciences, Beijing 100049, China}
%\collaboration{16}{ (AAS Journals Data Editors)  }

\author[0000-0003-2090-5416]{Xiaohu Li$^{*}$}
\affiliation{Xinjiang Astronomical Observatory, Chinese Academy of Sciences, 150 Science 1-Street, Urumqi, Xinjiang 830011, China}
\affiliation{Xinjiang Key Laboratory of Radio Astrophysics, 150 Science1-Street, Urumqi, Xinjiang 830011, China}

\author{Jixian Sun}
\affiliation{Purple Mountain Observatory, Chinese Academy of Sciences, Nanjing 210034, China}
\affiliation{Key Laboratory of Radio Astronomy, Chinese Academy of Sciences, Nanjing 210034, China}

\author{Tom J. Millar}
\affiliation{Astrophysics Research Centre, School of Mathematics and Physics, Queen's University Belfast, Belfast, BT7 1NN, UK}

\author[0000-0002-1086-7922]{Yong Zhang}
\affiliation{Xinjiang Astronomical Observatory, Chinese Academy of Sciences, 150 Science 1-Street, Urumqi, Xinjiang 830011, China}
\affiliation{School of Physics and Astronomy, Sun Yat-sen University, Guangzhou 510275, China}
\affiliation{Laboratory for Space Research, Faculty of Science, The University of Hong Kong, Hong Kong, China}

\author[0000-0002-9829-8655]{Jianjie Qiu}
\affiliation{School of Physics and Astronomy, Sun Yat-sen University, Guangzhou 510275, China}

\author{Donghui Quan}
\affiliation{Xinjiang Astronomical Observatory, Chinese Academy of Sciences, 150 Science 1-Street, Urumqi, Xinjiang 830011, China}
\affiliation{Astronomical Computing Research Center, Zhejiang Laboratory, Hangzhou 311100, China}

\author{Jarken Esimbek}
\affiliation{Xinjiang Astronomical Observatory, Chinese Academy of Sciences, 150 Science 1-Street, Urumqi, Xinjiang 830011, China}
\affiliation{Xinjiang Key Laboratory of Radio Astrophysics, 150 Science1-Street, Urumqi, Xinjiang 830011, China}

\author[0000-0003-0356-818X]{Jianjun Zhou}
\affiliation{Xinjiang Astronomical Observatory, Chinese Academy of Sciences, 150 Science 1-Street, Urumqi, Xinjiang 830011, China}
\affiliation{Xinjiang Key Laboratory of Radio Astrophysics, 150 Science1-Street, Urumqi, Xinjiang 830011, China}

\author{Yu Gao$^{\ddag}$}
\footnote{$^{\ddag}$Prof. Yu Gao passed away on 21 May 2022}
\affiliation{Department of Astronomy, Xiamen University, Xiamen, Fujian 361005, China}
\affiliation{Purple Mountain Observatory, Chinese Academy of Sciences, Nanjing 210034, China}
\affiliation{Key Laboratory of Radio Astronomy, Chinese Academy of Sciences, Nanjing 210034, China}

\author{Qiang Chang}
\affiliation{School of Physics and Optoelectronic Engineering, Shandong University of Technology, Zibo, Shandong 255000, China}

\author[0000-0002-9511-7062]{Lin Xiao}
\affiliation{College of Science, Changchun Institute of Technology, Changchun, Jilin 130012, China}

\author{Yanan Feng}
\affiliation{Xinjiang Astronomical Observatory, Chinese Academy of Sciences, 150 Science 1-Street, Urumqi, Xinjiang 830011, China}
\affiliation{ University of Chinese Academy of Sciences, Beijing 100049, China}

\author{Zhenzhen Miao}
\affiliation{Xinjiang Astronomical Observatory, Chinese Academy of Sciences, 150 Science 1-Street, Urumqi, Xinjiang 830011, China}
\affiliation{Xinjiang Key Laboratory of Radio Astrophysics, 150 Science1-Street, Urumqi, Xinjiang 830011, China}

\author{Rong Ma}
\affiliation{Xinjiang Astronomical Observatory, Chinese Academy of Sciences, 150 Science 1-Street, Urumqi, Xinjiang 830011, China}
\affiliation{ University of Chinese Academy of Sciences, Beijing 100049, China}

\author{Ryszard Szczerba}
\affiliation{Xinjiang Astronomical Observatory, Chinese Academy of Sciences, 150 Science 1-Street, Urumqi, Xinjiang 830011, China}
\affiliation{Nicolaus Copernicus Astronomical Center, Rabia\'{n}ska 8, 87-100 Toru\'{n}, Poland}

\author{Xuan Fang}
\affiliation{Xinjiang Astronomical Observatory, Chinese Academy of Sciences, 150 Science 1-Street, Urumqi, Xinjiang 830011, China}
\affiliation{National Astronomical Observatories, Chinese Academy of Sciences, A20, Datun Road, Chaoyang District, 100012 Beijing, China}
%\affiliation{Department of Physics, Faculty of Science, The University of Hong Kong, Pokfulam Road, Hong Kong, China}
\affiliation{Laboratory for Space Research, Faculty of Science, The University of Hong Kong, Hong Kong, China}

\begin{abstract}
We present an unbiased $\lambda$ 3\,mm  spectral line survey (between 84.5 and 115.8 GHz),
conducted by the Purple Mountain Observatory 13.7 meter radio telescope, together with updated modeling results,
towards the carbon-rich Asymptotic Giant Branch star, IRC\,+10216 (CW Leo).
A total of 75 spectral lines (96 transitions) are detected,
and identified to arise from 19 molecules: C$_2$H, $l$-C$_3$H, C$_4$H, CN, C$_3$N, HC$_3$N, HC$_5$N, HCN, HNC, CH$_3$CN, MgNC, CO, $c$-C$_3$H$_2$, SiC$_2$, SiO, SiS, CS, C$_2$S, C$_3$S, and their isotopologues.
Among them, one molecular emission line (H$^{13}$CCCN $J=13-12$) is discovered in IRC\,+10216 for the first time.
The excitation temperature, column density, and fractional abundance of the detected species are deduced by assuming they are in local thermodynamic equilibrium. In addition, the isotopic ratios of [$^{12}$C]/[$^{13}$C], [$^{32}$S]/[$^{34}$S], [$^{28}$Si]/[$^{29}$Si], and [$^{12}$C$^{34}$S]/[$^{13}$C$^{32}$S] are obtained and found to be consistent with previous studies.
Finally, we summarize all of the 106 species detected in IRC\,+10216 to date with their observed and modeled column densities for the convenience of future studies.
\end{abstract}

\section{Introduction}
\label{sec:intro}

One of the most important stages in the evolution of intermediate and low mass stars (0.8 $-$ 8 ${M}_{\odot}$) is the asymptotic giant branch (AGB) stage. During the AGB phase the mass-loss rate is about $10^{-8}$ $-$ $10^{-4}$ $M_{\odot}$ yr$^{-1}$ \citep{DeBeck+etal+2010}.
The ejected material forms a dense circumstellar envelope (CSE) composed of gas and dust. Based on the carbon to oxygen abundance ratio $\rm ([C]/[O])$,
AGB stars can be divided into carbon$-$rich AGB stars (C-rich, $\rm [C]/[O] > 1$), oxygen-rich AGB stars (O-rich, $\rm [C]/[O] < 1$), and S-type AGB stars $(\rm [C]/[O] \approx 1)$ \citep{Olofsson+etal+1996}.

\startlongtable
\begin{deluxetable}{cccccccc}
%\tablenum{1}
\tablecaption{A summary of the 106 detected species (excluding isotopologues) toward the CSE of the carbon-rich AGB star, IRC\,+10216, together with their total column density from both observation ($N_{\rm Obs.}$) and an astrochemical model ($N_{\rm Mod.}$) that has been updated from \cite{Li+etal+2014} and is briefly described in Section \ref{sec:updated model}.}
\label{table:1}
%Note that the ranges of $T_{\rm ex}$ and $N$, are just showing the minimum and maximum values of those results obtained from the existing observations that include results from this work  (further details are shown in Table~\ref{table:5}).}
%The molecular column densities of the model are derived from \cite{Li+etal+2014}.} \label{table:1}
\tablewidth{0pt}
\tablehead{
%\colhead{No.} &\colhead{Species} & \colhead{$T_{\rm ex,Obs}$}  & \colhead{$N_{\rm Obs.}$}         & \colhead{Ref.}   & \colhead{$N_{\rm Mod.}$}   & Note  \\   % add in something in the `Note' column.
%  xiaohu: in some cases they did not give N_obs but provided f_obs instead. this value deserves to provide in the table. We can add in them later.
%
\colhead{No.} &\colhead{Species}   & \colhead{$N_{\rm Obs.}$}         & \colhead{Ref.}   & \colhead{$N_{\rm Mod.}$}   &      \\
\colhead{ }   & \colhead{   }      & \colhead{ (cm$^{-2}$)  }         & \colhead{       }   & \colhead{ (cm$^{-2}$)     }   &
}
\startdata
 (1)   & C(CI)            &       (4.0 $-$ 7.0)     $\times$ 10$^{15}$   &     \cite{Keene+etal+1993}               &    2.19 $\times$ 10$^{16}$   &       \\
       & C (CI)           &      2.5 $\times$ 10$^{16}$           &    \cite{van+etal+1998}               &    2.19 $\times$ 10$^{16}$        &    \\
       & C (CI)           &                         &    \cite{Jeste+etal+2023}             &    2.19 $\times$ 10$^{16}$        &    \\
 (2)   & C$^+$  (CII)     &     $  (4.0 \pm 0.6)  $  $\times$ 10$^{17}$           &      \cite{Reach+etal+2022}              &    3.83 $\times$ 10$^{16}$    & \\
       &  C$^+$  (CII)     &                        &      \cite{Jeste+etal+2023}              &    3.83 $\times$ 10$^{16}$    & \\
 (3)   & C$_2$            & 7.9 $\times$ 10$^{14}$                      & \cite{Bakker+etal+1997}       & 4.79 $\times$ 10$^{15}$ &  \\
 (4)   & OH        &                                    & \cite{Ford+etal+2003}      & 2.96          $\times$ 10$^{14}$       &                \\
 (5)   & HF        &                                     & \cite{Agundez+etal+2011}        & 1.02 $\times$ 10$^{14}$  &  \\
 (6)   & AlF$^a$       & $ (0.3-1.1)  $ $\times$ 10$^{15}$      & \cite{Ziurys+etal+1994}      &      \\
       & AlF       &                                     & \cite{Cernicharo+etal+1987a}     &  \\
 (7)   & HCl       &                                     & \cite{Cernicharo+etal+2010a}    & 1.25 $\times$ 10$^{15}$ &  \\
 (8)   & KCl       &                                     & \cite{Cernicharo+etal+1987a}    & &  \\
 (9)   & NaCl      & 5 $\times$ 10$^{12}$                & \cite{Cernicharo+etal+1987a}    & &  \\
 (10)  & AlCl      &                                     & \cite{Cernicharo+etal+1987a}    & &  \\
 (11)  & CP$^a$        & {5 $\times$ 10$^{12}$}  & \cite{Halfen+etal+2008}         & {2.16 $\times$ 10$^{12}$} &  \\
       & CP        & { 8 $\times$ 10$^{13}$}  & \cite{Guelin+etal+1990}         & {2.16 $\times$ 10$^{12}$} &  \\
 (12)  & PN       & {6.3 $\times$ 10$^{12}$}  & \cite{Halfen+etal+2008}     & {3.89$\times$ 10$^{12}$}&   \\
     & PN        &                                      & \cite{Milam+etal+2008}      & {3.89 $\times$ 10$^{12}$} &  \\
     & PN         & {1.6 $\times$ 10$^{12}$} & \cite{Agundez+etal+2007}    & {3.89 $\times$ 10$^{12}$}&  \\
     & PN        &                                      & \cite{Guelin+etal+2000}     & {3.89$\times$ 10$^{12}$} &  \\
     & PN        &                                      & \cite{Cernicharo+etal+2000} & {3.89 $\times$ 10$^{12}$} &  \\
(13)  & CN        & $ (1.08\pm0.42)  $ $\times$ 10$^{15}$ &  This work                    & 3.25 $\times$ 10$^{15}$ &  \\
     & CN        & 2 $\times$ 10$^{15}$               & \cite{Agundez+etal+2010}      & 3.25 $\times$ 10$^{15}$ &  \\
     & CN        & 6.2 $\times$ 10$^{14}$             & \cite{Groesbeck+etal+1994}    & 3.25 $\times$ 10$^{15}$&   \\
     & CN        & 7.6 $\times$ 10$^{14}$             & \cite{Avery+etal+1992}        & 3.25 $\times$ 10$^{15}$&   \\
     & CN        & 1.7 $\times$ 10$^{15}$             & \cite{Wootten+etal+1982}     & 3.25 $\times$ 10$^{15}$ &  \\
     & CN        &                                    & \cite{Wilson+etal+1971}      & 3.25 $\times$ 10$^{15}$ &  \\
(14) & SiN       & {3.8 $\times$ 10$^{13}$} & \cite{Turner+etal+1992}     & {3.72 $\times$ 10$^{11}$} &  \\
(15) & CS        & $ (1.57\pm0.46)  $ $\times$ 10$^{15}$ &  This work                    & 1.02 $\times$ 10$^{16}$&   \\
     & CS        & 5.9 $\times$ 10$^{15}$             & \cite{Kawaguchi+etal+1995}    &1.02 $\times$ 10$^{16}$ &  \\
     & CS        & 3.0 $\times$ 10$^{15}$             & \cite{Groesbeck+etal+1994}    &1.02 $\times$ 10$^{16}$&  \\
     & CS$^a$    & 4 $\times$ 10$^{15}$               & \cite{Cernicharo+etal+1987}   &1.02 $\times$ 10$^{16}$ &  \\
     & CS        & 1.8 $\times$ 10$^{15}$             & \cite{Morris+etal+1975}       & 1.02 $\times$ 10$^{16}$&   \\
     & CS        &                                    & \cite{Penzias+etal+1971}      & 1.02 $\times$ 10$^{16}$ &  \\
(16) & SiC       & $ (4.9\pm2.3)  $ $\times$ 10$^{13}$    & \cite{He+etal+2008}         & 1.95 $\times$ 10$^{13}$ &  \\
     & SiC$^a$   & 6 $\times$ 10$^{13}$                & \cite{Cernicharo+etal+1989} & 1.95 $\times$ 10$^{13}$ &  \\
(17) & FeC   &                           & \cite{Koelemay+etal+2023ApJ} &                          &  \\
(18) & SiO       & $ (4.84\pm1.13)  $ $\times$ 10$^{14}$ &  This work                    & 1.74 $\times$ 10$^{15}$ &  \\
     & SiO       & 4.1 $\times$ 10$^{14}$             & \cite{Morris+etal+1975}       & 1.74 $\times$ 10$^{15}$&   \\
     & SiO       & 5.4 $\times$ 10$^{14}$             & \cite{Kawaguchi+etal+1995}    & 1.74 $\times$ 10$^{15}$ &  \\
     & SiO       & 2.1 $\times$ 10$^{14}$             & \cite{Groesbeck+etal+1994}    & 1.74 $\times$ 10$^{15}$&   \\
     & SiO       & 5.4 $\times$ 10$^{14}$             & \cite{Kawaguchi+etal+1995}    & 1.74 $\times$ 10$^{15}$&   \\
(19) & SiS       & $ (5.86\pm1.74)  $ $\times$ 10$^{15}$ &  This work                   & 2.17 $\times$ 10$^{16}$&   \\
     & SiS       & 4.7 $\times$ 10$^{15}$             & \cite{Kawaguchi+etal+1995}   & 2.17 $\times$ 10$^{16}$&   \\
     & SiS       & 3.0 $\times$ 10$^{15}$             & \cite{Groesbeck+etal+1994}   & 2.17 $\times$ 10$^{16}$ &  \\
     & SiS       & $ (6.5\pm0.5)  $ $\times$ 10$^{15}$   & \cite{Avery+etal+1992}       & 2.17 $\times$ 10$^{16}$&   \\
     & SiS       & 7 $\times$ 10$^{15}$               & \cite{Cernicharo+etal+1987}  & 2.17 $\times$ 10$^{16}$&  \\
     & SiS       & 1.6 $\times$ 10$^{15}$             & \cite{Morris+etal+1975}      & 2.17 $\times$ 10$^{16}$ &  \\
(20) & CO        & $ (4.03\pm1.51)  $ $\times$ 10$^{17}$ &  This work                    & 1.35 $\times$ 10$^{19}$&   \\
     & CO        & 2.0 $\times$ 10$^{18}$             & \cite{Groesbeck+etal+1994}   & 1.35 $\times$ 10$^{19}$&   \\
     & CO        &                                    & \cite{Solomon+etal+1971}      & 1.35 $\times$ 10$^{19}$ &  \\
(21) & H$_{2}$O  &                                     & \cite{Melnick+etal+2001}        & 3.22 $\times$ 10$^{16}$ &  \\
(22) & KCN       & $ (1\pm0.2)  $ $\times$ 10$^{12}$      & \cite{Pulliam+etal+2010}    &   \\
(23) & NaCN      & $ (8.5\pm1.7)  $ $\times$ 10$^{12}$    & \cite{Pardo+etal+2022}  &       \\
     & NaCN      & 3.8 $\times$ 10$^{14}$              & \cite{Kawaguchi+etal+1995}   &   \\
     & NaCN      & $ (2.0-3.8)  $ $\times$ 10$^{13}$      & \cite{Turner+etal+1994}    &    \\
(24) & MgCN      & $ (7.4\pm2)  $ $\times$ 10$^{11}$      & \cite{Cabezas+etal+2013}  &     \\
     & MgCN$^a$      & 2 $\times$ 10$^{12}$                & \cite{Ziurys+etal+1995}    &    \\
(25) & SiCN$^a$      & 2 $\times$ 10$^{12}$                & \cite{Guelin+etal+2000}            &     \\
(26) & FeCN      & 8.6 $\times$ 10$^{11}$              & \cite{Zack+etal+2011}      &    \\
(27) & MgNC      & $ (3.20\pm0.51)  $ $\times$ 10$^{13}$  & This work           &         \\
     & MgNC      & $ (3.9\pm1.5)  $ $\times$ 10$^{13}$    & \cite{Gong+etal+2015}    &      \\
     & MgNC      & $ (1.3\pm0.3)  $ $\times$ 10$^{13}$    & \cite{Cabezas+etal+2013}  &     \\
     & MgNC      & 7.8 $\times$ 10$^{13}$              & \cite{He+etal+2008}      &     \\
     & MgNC      & 2.5 $\times$ 10$^{13}$              & \cite{Kawaguchi+etal+1993}&    \\
     & MgNC      &                                     & \cite{Guelin+etal+1993}   &    \\
     & MgNC      &                                     & \cite{Guelin+etal+1986} &      \\
(28) & AlNC      & 9 $\times$ 10$^{11}$                & \cite{Ziurys+etal+2002}          &      \\
(29) & SiNC      & {\color{black}2 $\times$ 10$^{12}$}   & \cite{Guelin+etal+2004}         &  9.35 $\times$ 10$^{8}$  &  \\
(30) & CaNC      & 2 $\times$ 10$^{11}$                & \cite{Cernicharo+etal+2019} &   \\
(31) & HNC       & $ (1.11\pm0.08)  $ $\times$ 10$^{14}$  & This work                       & 1.17 $\times$ 10$^{14}$ &   \\
     & HNC       &                                     & \cite{Brown+etal+1976}          & 1.17 $\times$ 10$^{14}$ &    \\
(32) & HCN       & $ (7.36\pm0.92)  $ $\times$ 10$^{14}$  & This work                       & 2.51 $\times$ 10$^{17}$ &   \\
     & HCN       & 2.8 $\times$ 10$^{16}$              & \cite{Groesbeck+etal+1994}      & 2.51 $\times$ 10$^{17}$  &   \\
     & HCN       &                                     & \cite{Morris+etal+1971}         & 2.51 $\times$ 10$^{17}$     &  \\
(33)   &  MgC$_2$                   & $ (1.0\pm0.3)  $ $\times$ 10$^{12}$    & \cite{Changala+etal+2022}      &    &  \\
(34) & SiC$_2$   & $ (1.41\pm0.34)  $ $\times$ 10$^{15}$   &  This work                & 6.30 $\times$ 10$^{15}$    &  \\
     & SiC$_2$   & $ (1.87\pm1.34)  $ $\times$ 10$^{15}$   & \cite{Zhang+etal+2017}    & 6.30 $\times$ 10$^{15}$ &  \\
     & SiC$_2$   & $ (1.2\pm0.0)  $ $\times$ 10$^{15}$     & \cite{Gong+etal+2015}     & 6.30 $\times$ 10$^{15}$   &   \\
     & SiC$_2$$^b$   & $ (2.35\pm0.36)  $ $\times$ 10$^{15}$ & \cite{He+etal+2008}       & 6.30 $\times$ 10$^{15}$  &   \\
     & SiC$_2$$^c$   & $ (5.40\pm0.88)  $ $\times$ 10$^{14}$ & \cite{He+etal+2008}       & 6.30 $\times$ 10$^{15}$ &  \\
     & SiC$_2$$^d$   & 1.04 $\times$ 10$^{15}$            & \cite{He+etal+2008}       & 6.30 $\times$ 10$^{15}$   &  \\
     & SiC$_2$$^e$   & 5.3 $\times$ 10$^{15}$             & \cite{He+etal+2008}       & 6.30 $\times$ 10$^{15}$   &   \\
     & SiC$_2$   & 2.2 $\times$ 10$^{14}$             & \cite{Kawaguchi+etal+1995}    & 6.30 $\times$ 10$^{15}$ &   \\
     & SiC$_2$   & 3.7 $\times$ 10$^{14}$             & \cite{Groesbeck+etal+1994}    & 6.30 $\times$ 10$^{15}$ &  \\
     & SiC$_2$    & (cold range) 9.5 $\times$ 10$^{14}$             & \cite{Avery+etal+1992}        & 6.30 $\times$ 10$^{15}$ &  \\
     & SiC$_2$    & (warm range) 1.9 $\times$ 10$^{15}$             & \cite{Avery+etal+1992}        & 6.30 $\times$ 10$^{15}$ &  \\
     & SiC$_2$$^a$   & 2 $\times$ 10$^{15}$               & \cite{Cernicharo+etal+1987}   & 6.30 $\times$ 10$^{15}$ &  \\
     & SiC$_2$   & 1.5 $\times$ 10$^{14}$             & \cite{Thaddeus+etal+1984}     & 6.30 $\times$ 10$^{15}$&   \\
(35) & SiCSi$^f$  & $ (1.4\pm0.4)  $ $\times$ 10$^{15}$   & \cite{Cernicharo+etal+2015a}              &   \\
     & SiCSi$^g$     & $ (3.0\pm0.7)  $ $\times$ 10$^{15}$   & \cite{Cernicharo+etal+2015a}              &   \\
     & SiCSi$^h$     & $ (3.2\pm0.9)  $ $\times$ 10$^{15}$   & \cite{Cernicharo+etal+2015a}              &   \\
(36) & C$_3$       & 1 $\times$ 10$^{15}$                & \cite{Hinkle+etal+1988}       & 1.93 $\times$ 10$^{14}$ &  \\
(37) & H$_2$S$^a$      & 1 $\times$ 10$^{13}$                & \cite{Cernicharo+etal+1987}         & 5.11 $\times$ 10$^{13}$ &  \\
(38) & HCP         & $ (1.0-1.4)  $ $\times$ 10$^{14}$      & \cite{Halfen+etal+2008}       & 3.17 $\times$ 10$^{14}$ &  \\
     & HCP         & 1.3 $\times$ 10$^{15}$              & \cite{Agundez+etal+2007}      & 3.17 $\times$ 10$^{14}$ &  \\
(39) & C$_2$H      & $ (8.14\pm2.29)  $ $\times$ 10$^{15}$  & This work                    & 1.01 $\times$ 10$^{16}$  &  \\
     & C$_2$H      & $ (3.84\pm0.09)  $ $\times$ 10$^{15}$  & \cite{Beck+etal+2012}        & 1.01 $\times$ 10$^{16}$ &  \\
     & C$_2$H      & 5.0 $\times$ 10$^{15}$              & \cite{Cernicharo+etal+2000}  & 1.01 $\times$ 10$^{16}$   &  \\
     & C$_2$H      & 4.6 $\times$ 10$^{15}$              & \cite{Groesbeck+etal+1994}   & 1.01 $\times$ 10$^{16}$   &     \\
     & C$_2$H      & 5.4 $\times$ 10$^{15}$              & \cite{Avery+etal+1992}       & 1.01 $\times$ 10$^{16}$  &  \\
     & C$_2$H      &                                     & \cite{Tucker+etal+1974}      & 1.01 $\times$ 10$^{16}$  &  \\
(40) & C$_2$P      & 1.2 $\times$ 10$^{12}$              & \cite{Halfen+etal+2008}            & 8.99 $\times$ 10$^{9}$   &  \\
(41) & C$_2$S      & $ (1.59\pm0.73)  $ $\times$ 10$^{14}$  & This work                   & {1.84 $\times$ 10$^{13}$}     &  \\
     & C$_2$S     & $ (4.0\pm0.7)  $ $\times$ 10$^{13}$    & \cite{Pardo+etal+2022}      & {1.84 $\times$ 10$^{13}$}    &  \\
     & C$_2$S      & $ (2.3\pm0.8)  $ $\times$ 10$^{14}$    & \cite{Gong+etal+2015}       & {1.84 $\times$ 10$^{13}$}     &  \\
     & C$_2$S      & $ (5.0\pm0.3)  $ $\times$ 10$^{13}$    & \cite{Agundez+etal+2014}   &  1.84 $\times$ 10$^{13}$     &  \\
     & C$_2$S      & 1.5 $\times$ 10$^{14}$              & \cite{Kawaguchi+etal+1995}  & {1.84 $\times$ 10$^{13}$}    &  \\
     & C$_2$S      & $ (9\pm2)  $ $\times$ 10$^{13}$        & \cite{Bell+etal+1993}       & {1.84 $\times$ 10$^{13}$}    &  \\
     & C$_2$S      & 1.5 $\times$ 10$^{14}$              & \cite{Cernicharo+etal+1987} & {1.84 $\times$ 10$^{13}$}     &  \\
(42) & C$_2$N      & $ (1.1\pm0.6)  $ $\times$ 10$^{13}$    & \cite{Anderson+etal+2014}     & 1.46 $\times$ 10$^{13}$  &  \\
(43) & $l$-C$_3$H  & $ (1.25\pm0.22)  $ $\times$ 10$^{14}$ &  This work                    & 1.42 $\times$ 10$^{14}$   &  \\
     & $l$-C$_3$H  & $ (1.32\pm0.34)  $ $\times$ 10$^{14}$ & \cite{He+etal+2008}           & 1.42 $\times$ 10$^{14}$   &  \\
     & $l$-C$_3$H  & 7.0 $\times$ 10$^{13}$             & \cite{Cernicharo+etal+2000}   & 1.42 $\times$ 10$^{14}$   &  \\
     & $l$-C$_3$H  & 5.6 $\times$ 10$^{13}$             & \cite{Kawaguchi+etal+1995}    & 1.42 $\times$ 10$^{14}$   &  \\
     & $l$-C$_3$H  & 2.8 $\times$ 10$^{13}$             & \cite{Thaddeus+etal+1985}     & 1.42 $\times$ 10$^{14}$   &  \\
(44) & $c$-C$_3$H  &                                    & \cite{Cernicharo+etal+2000}                   &  \\
(45) & H$_2$CO     & 5 $\times$ 10$^{12}$               & \cite{Ford+etal+2004}         & 2.80 $\times$ 10$^{11}$   &  \\
(46) & NH$_{3}$  (para)     &  (2.3, 2.7)   $\times$ 10$^{14}$  & \cite{Gong+etal+2015}          & {\color{black}2.50 $\times$ 10$^{16}$}   &  \\
     & NH$_{3}$  (ortho)    &  (1.5, 2.9)   $\times$ 10$^{14}$  & \cite{Gong+etal+2015}          & {\color{black}2.50 $\times$ 10$^{16}$}   &  \\
     & NH$_{3}$          & 4.2 $\times$ 10$^{13}$        & \cite{Nguyen-Q-Rieu+etal+1984} & {\color{black}2.50 $\times$ 10$^{16}$}   &  \\
     & NH$_{3}$          & 1 $\times$ 10$^{17}$          & \cite{Betz+etal+1979}          & {\color{black}2.50 $\times$ 10$^{16}$}  &  \\
(47) & C$_3$S      & $ (3.27\pm0.16)  $ $\times$ 10$^{13}$ & This work                     & 1.55 $\times$ 10$^{13}$  &  \\
     & C$_3$S      & $ (2.58\pm3.60)  $ $\times$ 10$^{13}$ & \cite{Pardo+etal+2022}        & 1.55 $\times$ 10$^{13}$   &  \\
     & C$_3$S      & $ (2.2\pm0.4)  $ $\times$ 10$^{13}$   & \cite{Gong+etal+2015}         & 1.55 $\times$ 10$^{13}$    &  \\
     & C$_3$S      & $ (1.7\pm0.1)  $ $\times$ 10$^{13}$   & \cite{Agundez+etal+2014}      &  1.55 $\times$ 10$^{13}$  &  \\
     & C$_3$S      & $ (4.9\pm0.3)  $ $\times$ 10$^{13}$   & \cite{Kawaguchi+etal+1995}    & 1.55 $\times$ 10$^{13}$   &  \\
     & C$_3$S      & $ (5.8\pm2.7)  $ $\times$ 10$^{12}$   & \cite{Bell+etal+1993}         & 1.55 $\times$ 10$^{13}$    &  \\
     & C$_3$S      & 1.1 $\times$ 10$^{14}$             & \cite{Cernicharo+etal+1987}   & 1.55 $\times$ 10$^{13}$   &  \\
(49) & C$_3$N   & $ (4.25\pm0.81)  $ $\times$ 10$^{14}$          & This work                      & 5.85 $\times$ 10$^{14}$   &  \\
     & C$_3$N                   & $ (3.1\pm0.3)  $ $\times$ 10$^{14}$            & \cite{Gong+etal+2015}          & 5.85 $\times$ 10$^{14}$  &  \\
     & C$_3$N                   & $ (4.54\pm2.13)  $ $\times$ 10$^{14}$          & \cite{He+etal+2008}            & 5.85 $\times$ 10$^{14}$   &  \\
     & C$_3$N                  & 2.5 $\times$ 10$^{14}$                      & \cite{Cernicharo+etal+2000}    & 5.85 $\times$ 10$^{14}$  &  \\
     & C$_3$N                  & 4.1 $\times$ 10$^{14}$               & \cite{Kawaguchi+etal+1995}      & 5.85 $\times$ 10$^{14}$  &  \\
(49) & C$_2$H$_2$                & 3 $\times$ 10$^{19}$                & \cite{Ridgway+etal+1976} & 9.94 $\times$ 10$^{17}$  &  \\
(50) & H$_{2}$CS$^a$                & {\color{black}1.0 $\times$ 10$^{13}$}  & \cite{Agundez+etal+2008} & {\color{black} 5.01 $\times$ 10$^{11}$}  &  \\
(51) & SiC$_3$                  & (cold range) 4.3 $\times$ 10$^{12}$          & \cite{Apponi+etal+1999}           & 2.03 $\times$ 10$^{12}$  &  \\
     & SiC$_3$                  & (warm range) 4.3 $\times$ 10$^{12}$          & \cite{Apponi+etal+1999}           & 2.03 $\times$ 10$^{12}$  &  \\
(52) & C$_3$O                   & 1.2 $\times$ 10$^{12}$          & \cite{Tenenbaum+etal+2006}        & 1.07 $\times$ 10$^{12}$  &  \\
(53) & HC$_2$N                  & 1.4 $\times$ 10$^{13}$          & \cite{Cernicharo+etal+2004}        &  \\
     & HC$_2$N                  & 1.8 $\times$ 10$^{13}$          & \cite{Kawaguchi+etal+1995}          &  \\
     & HC$_2$N                  & 1.2 $\times$ 10$^{13}$          & \cite{Guelin+etal+1991}             &  \\
(54) & HMgNC                    & 6 $\times$ 10$^{11}$            & \cite{Cabezas+etal+2013}  &  \\
(55) & MgCCH                    & 2 $\times$ 10$^{12}$            & \cite{Agundez+etal+2014}   &  \\
     & MgCCH                    & ~2 $\times$ 10$^{12}$           & \cite{Cernicharo+etal+2019b}  &  \\
(56) & NCCP$^*$                 & 7 $\times$ 10$^{11}$            & \cite{Agundez+etal+2014}   &  \\
(57) & PH$_3$                   &                                 & \cite{Agundez+etal+2008a}  &  \\
     & PH$_3$                   &                                 & \cite{Tenenbaum+etal+2008}  &  \\
(58) & C$_5$                    & 1 $\times$ 10$^{14}$                      & \cite{Bernath+etal+1989}     & 1.61 $\times$ 10$^{14}$  &  \\
(59) & $c$-C$_3$H$_{2}$ (ortho) & $ (1.14\pm0.15)  $ $\times$ 10$^{14}$      & This work                      & 5.85 $\times$ 10$^{13}$   &  \\
     & $c$-C$_3$H$_{2}$ (para)  & $ (8.02\pm4.54)  $ $\times$ 10$^{13}$        & This work                      & 5.85 $\times$ 10$^{13}$   &  \\
     & $c$-C$_3$H$_{2}$         & $ (1.39\pm1.07)  $ $\times$ 10$^{14}$        & \cite{Zhang+etal+2017}         & 5.85 $\times$ 10$^{13}$  &  \\
     & $c$-C$_3$H$_{2}$ (ortho) & $ (1.2\pm0.3)  $ $\times$ 10$^{14}$          & \cite{Gong+etal+2015}          & 5.85 $\times$ 10$^{13}$  &  \\
     & $c$-C$_3$H$_{2}$ (para)  & $ (8.9\pm0.5)  $ $\times$ 10$^{13}$          & \cite{Gong+etal+2015}          & 5.85 $\times$ 10$^{13}$   &  \\
     & $c$-C$_3$H$_{2}$ (ortho) & $ (4.82\pm4.14)  $ $\times$ 10$^{14}$        & \cite{He+etal+2008}            & 5.85 $\times$ 10$^{13}$  &  \\
     & $c$-C$_3$H$_{2}$ (para)  & $ (1.22\pm2.03)  $ $\times$ 10$^{14}$        & \cite{He+etal+2008}            & 5.85 $\times$ 10$^{13}$  &  \\
     & $c$-C$_3$H$_{2}$         & 1.6 $\times$ 10$^{13}$                    & \cite{Kawaguchi+etal+1995}      & 5.85 $\times$ 10$^{13}$  &  \\
(60) & H$_{2}$C$_3$             & 2.6 $\times$ 10$^{12}$                          & \cite{Cernicharo+etal+1991b} & 1.20 $\times$ 10$^{13}$  &  \\
(61) & HC$_3$N                  & $ (1.18\pm0.08)  $ $\times$ 10$^{15}$ & This work               & {\color{black}4.79 $\times$ 10$^{14}$}  &  \\
     & HC$_3$N                  & $ (1.94\pm0.21)  $ $\times$ 10$^{15}$ & \cite{Zhang+etal+2017}  & {\color{black}4.79 $\times$ 10$^{14}$}  &  \\
     & HC$_3$N                  & $ (1.4\pm0.2)  $ $\times$ 10$^{15}$   & \cite{Gong+etal+2015}   & {\color{black}4.79 $\times$ 10$^{14}$}  &  \\
     & HC$_3$N                  & $ (8.0\pm1.5)  $ $\times$ 10$^{14}$   & \cite{He+etal+2008}     & {\color{black}4.79 $\times$ 10$^{14}$}  &  \\
     & HC$_3$N                  & 1.7 $\times$ 10$^{15}$             & \cite{Kawaguchi+etal+1995}& {\color{black}4.79 $\times$ 10$^{14}$}  &  \\
     & HC$_3$N                  & (cold range)  $ (7.9\pm0.2)  $ $\times$ 10$^{14}$      & \cite{Bell+etal+1993b}    & {\color{black}4.79 $\times$ 10$^{14}$}  &  \\
     & HC$_3$N                  & (warm range)  $ (2.1\pm1.4)  $ $\times$ 10$^{14}$      & \cite{Bell+etal+1993b}    & {\color{black}4.79 $\times$ 10$^{14}$}  &  \\
     & HC$_3$N                  & 5.9 $\times$ 10$^{14}$                           & \cite{Gensheimer+etal+1997} & {\color{black}4.79 $\times$ 10$^{14}$}  &  \\
     & HC$_3$N                  & 1.1 $\times$ 10$^{15}$                           & \cite{Gensheimer+etal+1997a} & {\color{black}4.79 $\times$ 10$^{14}$}  &  \\
     & HC$_3$N                  & $ (1.6\pm0.4)  $ $\times$ 10$^{15}$    & \cite{Jewell+etal+1984}    & {\color{black}4.79 $\times$ 10$^{14}$}  &  \\
     & HC$_3$N                 & 1.8 $\times$ 10$^{15}$                           & \cite{Morris+etal+1975}   & {\color{black}4.79 $\times$ 10$^{14}$}  &  \\
(62) & HNC$_3$$^a$             & 1 $\times$ 10$^{12}$                & \cite{Gensheimer+etal+1997}   & 7.32 $\times$ 10$^{11}$  &  \\
(63) & HC$_2$NC                & $2.8, 3.6$ $\times$ 10$^{12}$       & \cite{Gensheimer+etal+1997}   &  &  \\
     & HC$_2$NC$^a$            & 8.4 $\times$ 10$^{12}$              & \cite{Gensheimer+etal+1997a}  &  &    \\
(64) & CH$_{2}$CN              & $ (8.6\pm1.4)  $ $\times$ 10$^{12}$    & \cite{Agundez+etal+2008}      & 2.58 $\times$ 10$^{11}$  &  \\
(65) & CH$_{2}$NH              & {\color{black}9 $\times$ 10$^{12}$}  & \cite{Tenenbaum+etal+2010a}  & {\color{black}2.56 $\times$ 10$^{11}$}  &  \\
(66) & C$_4$H                  & $ (2.75\pm0.36)  $ $\times$ 10$^{15}$  & This work                     & 6.57 $\times$ 10$^{14}$  &  \\
     & C$_4$H                  & $ (1.84\pm0.25)  $ $\times$ 10$^{14}$  & \cite{Pardo+etal+2022}        & 6.57 $\times$ 10$^{14}$  &  \\
     & C$_4$H                  & $ (2.4\pm0.2)  $ $\times$ 10$^{15}$    & \cite{Gong+etal+2015}         & 6.57 $\times$ 10$^{14}$  &  \\
     & C$_4$H                  & $ (8.1\pm1.1)  $ $\times$ 10$^{15}$    & \cite{He+etal+2008}           & 6.57 $\times$ 10$^{14}$  &  \\
     & C$_4$H                  & 3.0 $\times$ 10$^{14}$              & \cite{Cernicharo+etal+2000}   & 6.57 $\times$ 10$^{14}$  &  \\
     & C$_4$H                  & $ (0.4-3)  $ $\times$ 10$^{15}$        & \cite{Guelin+etal+1978}       & 6.57 $\times$ 10$^{14}$  &  \\
     & C$_4$H                  & 2.4 $\times$ 10$^{15}$              & \cite{Kawaguchi+etal+1995}    & 6.57 $\times$ 10$^{14}$  &  \\
     & C$_4$H                  & 5.6 $\times$ 10$^{15}$              & \cite{Avery+etal+1992}        & 6.57 $\times$ 10$^{14}$  &  \\
(67) & CH$_4$                  & 2.5 $\times$ 10$^{17}$                          & \cite{Hall+etal+1978}      & 4.41 $\times$ 10$^{16}$  &  \\
     & CH$_4$                  & 1.8 $\times$ 10$^{16}$                          & \cite{Clegg+etal+1982}     & 4.41 $\times$ 10$^{16}$  &  \\
(68) & SiH$_4$                 & 2 $\times$ 10$^{15}$                & \cite{Goldhaber+etal+1984}    & 2.78 $\times$ 10$^{15}$   &  \\
(69) & SiC$_4$                 & $ (6.2\pm1.0)  $ $\times$ 10$^{12}$   & \cite{Pardo+etal+2022}     & 2.45 $\times$ 10$^{11}$  &  \\
     & SiC$_4$                 & $ (1.46\pm0.92)  $ $\times$ 10$^{13}$ & \cite{Zhang+etal+2017}     & 2.45 $\times$ 10$^{11}$  &  \\
     & SiC$_4$                 & $ (1.1\pm0.4)  $ $\times$ 10$^{13}$  & \cite{Gong+etal+2015}      & 2.45 $\times$ 10$^{11}$ &  \\
     & SiC$_4$                 &  7 $\times$ 10$^{12}$              & \cite{Kawaguchi+etal+1995} & 2.45 $\times$ 10$^{11}$  &  \\
     & SiC$_4$                 & $ (7\pm1)  $ $\times$ 10$^{12}$       & \cite{Ohishi+etal+1989}    & 2.45 $\times$ 10$^{11}$  &  \\
(70) & NaC$_3$N     &  $(1.2\pm0.2)$ $\times$ 10$^{11}$        & \cite{Cabezas+etal+2023}  &  \\
(71) & MgC$_3$N                & $ (5.2\pm0.6)$ $\times$ 10$^{12}$                 & \cite{Pardo+etal+2022}  &  \\
     & MgC$_3$N                & (cold range)     $ (5.7\pm1.0)  $ $\times$ 10$^{12}$        & \cite{Cernicharo+etal+2019b}  &  \\
     & MgC$_3$N                &  (warm range)     $ (3.6\pm0.6)  $ $\times$ 10$^{12}$        & \cite{Cernicharo+etal+2019b}  &  \\
(72) & MgC$_4$H                & $ (1.5\pm1.3)  $ $\times$ 10$^{13}$        & \cite{Pardo+etal+2022}   &  \\
     & MgC$_4$H                & $ (2.2\pm0.5)  $ $\times$ 10$^{13}$        & \cite{Cernicharo+etal+2019b} &  \\
(73) & HMgC$_3$N     &  $(3.0\pm0.6)$ $\times$ 10$^{12}$        & \cite{Cabezas+etal+2023}  &  \\
(74) & C$_5$H $^2\Pi_{1/2}$    & $ (2.7\pm0.5)  $ $\times$ 10$^{13}$        & \cite{Pardo+etal+2022}       & 4.40 $\times$ 10$^{13}$   &  \\
     & C$_5$H $^2\Pi_{3/2}$    & $ (2.52\pm2.48)  $ $\times$ 10$^{14}$      & \cite{Pardo+etal+2022}       & 4.40 $\times$ 10$^{13}$   &  \\
     & C$_5$H $^2\Pi_{1/2}$    & $ (2.96\pm1.21)  $ $\times$ 10$^{14}$      & \cite{Zhang+etal+2017}       & 4.40 $\times$ 10$^{13}$   &  \\
     & C$_5$H $^2\Pi_{1/2}$    & $ (2.9\pm0.6)  $ $\times$ 10$^{13}$        & \cite{Gong+etal+2015}        & 4.40 $\times$ 10$^{13}$   &  \\
     & C$_5$H $^2\Pi_{3/2}$    & $ (1.2\pm2.3)  $ $\times$ 10$^{14}$        & \cite{Gong+etal+2015}        & 4.40 $\times$ 10$^{13}$  &  \\
     & C$_5$H                  & 4.4 $\times$ 10$^{13}$                  & \cite{Cernicharo+etal+2000}  & 4.40 $\times$ 10$^{13}$   &  \\
     & C$_5$H $^2\Pi_{1/2}$    & $ (2.9\pm0.3)  $ $\times$ 10$^{14}$        & \cite{Kawaguchi+etal+1995}   & 4.40 $\times$ 10$^{13}$   &  \\
     & C$_5$H $^2\Pi_{3/2}$    & $ (2.0\pm0.1)  $ $\times$ 10$^{14}$        & \cite{Kawaguchi+etal+1995}   & 4.40 $\times$ 10$^{13}$   &  \\
     & C$_5$H                  &                                         & \cite{Cernicharo+etal+1986}  & 4.40 $\times$ 10$^{13}$    &  \\
(75) & C$_2$H$_4$              & (cold)         $ (1.7\pm0.4)  $ $\times$ 10$^{15}$        & \cite{Fonfria+etal+2017}       & 8.59 $\times$ 10$^{14}$   &  \\
     & C$_2$H$_4$              &  (warm)         $ (4.8\pm1.3)  $ $\times$ 10$^{15}$        & \cite{Fonfria+etal+2017}       & 8.59 $\times$ 10$^{14}$   &  \\
     & C$_2$H$_4$              &  4 $\times$ 10$^{15}$                    & \cite{Goldhaber+etal+1987}     & 8.59 $\times$ 10$^{14}$  &  \\
     & C$_2$H$_4$              & ~10$^{16}$                              & \cite{Betz+etal+1981}          & 8.59 $\times$ 10$^{14}$   &  \\
(76) & HC$_4$N                 & $ (3.9\pm3.7)  $ $\times$ 10$^{12}$        & \cite{Pardo+etal+2022}         & &   \\
     & HC$_4$N$^a$             & 1.5 $\times$ 10$^{12}$                  & \cite{Cernicharo+etal+2004}    &  &  \\
(77) & H$_{2}$C$_4$            & 3 $\times$ 10$^{12}$                    & \cite{Kawaguchi+etal+1995}          &    \\
     & H$_{2}$C$_4$            & 1.6 $\times$ 10$^{13}$                  & \cite{Cernicharo+etal+1991a}     &      \\
(78) & C$_5$N$^a$              & 6.3 $\times$ 10$^{12}$                  & \cite{Cernicharo+etal+2000}    & 2.91 $\times$ 10$^{13}$   &  \\
     & C$_5$N$^a$              & 3.1 $\times$ 10$^{11}$                  & \cite{Guelin+etal+1998}        & 2.91 $\times$ 10$^{13}$     &  \\
(79) & CH$_3$CN                & $ (6.81\pm3.51)  $ $\times$ 10$^{13}$      &  This work                     & 4.88 $\times$ 10$^{12}$    &  \\
     & CH$_3$CN                & $ (2.6\pm0.6)  $ $\times$ 10$^{13}$        & \cite{He+etal+2008}            & 4.88 $\times$ 10$^{12}$   &  \\
     & CH$_3$CN                & 6 $\times$ 10$^{12}$                    & \cite{Kawaguchi+etal+1995}     & 4.88 $\times$ 10$^{12}$   &  \\
     & CH$_3$CN                & 6 $\times$ 10$^{12}$                    & \cite{Guelin+etal+1991}        & 4.88 $\times$ 10$^{12}$   &  \\
     & CH$_3$CN                &                                         & \cite{Johansson+etal+1984}     & 4.88 $\times$ 10$^{12}$  &  \\
(80) & SiH$_{3}$CN             & 1 $\times$ 10$^{12}$                    & \cite{Agundez+etal+2014}  &  \\
(81) & C$_5$S                  & $ (0.2-1.4)  $ $\times$ 10$^{13}$              & \cite{Agundez+etal+2014}    &    \\
     & C$_5$S$^a$              & 3.6 $\times$ 10$^{13}$              & \cite{Bell+etal+1993}     &   \\
(82) & C$_6$H $^2\Pi_{1/2}$    & $ (1.23\pm0.18)  $ $\times$ 10$^{13}$  & \cite{Pardo+etal+2022}          & 6.01 $\times$ 10$^{14}$  &  \\
     & C$_6$H $^2\Pi_{3/2}$    & $ (1.51\pm0.14)  $ $\times$ 10$^{13}$  & \cite{Pardo+etal+2022}          & 6.01 $\times$ 10$^{14}$   &  \\
     & C$_6$H $^2\Pi_{1/2}$    & $ (1.85\pm0.90)  $ $\times$ 10$^{13}$  & \cite{Zhang+etal+2017}          & 6.01 $\times$ 10$^{14}$  &  \\
     & C$_6$H $^2\Pi_{1/2}$    & $ (1.0\pm0.2)  $ $\times$ 10$^{14}$    & \cite{Gong+etal+2015}           & 6.01 $\times$ 10$^{14}$  &  \\
     & C$_6$H $^2\Pi_{3/2}$    & $ (1.0\pm0.1)  $ $\times$ 10$^{14}$    & \cite{Gong+etal+2015}           & 6.01 $\times$ 10$^{14}$  &  \\
     & C$_6$H $^2\Pi_{3/2}$    & $ (6.6\pm2.0)  $ $\times$ 10$^{13}$    & \cite{Cernicharo+etal+2007}     & 6.01 $\times$ 10$^{14}$  &  \\
     & C$_6$H                  & 5.5 $\times$ 10$^{13}$              & \cite{Cernicharo+etal+2000}     & 6.01 $\times$ 10$^{14}$  &  \\
     & C$_6$H $^2\Pi_{1/2}$            & $ (1.13\pm0.06)  $ $\times$ 10$^{14}$  & \cite{Kawaguchi+etal+1995}      & 6.01 $\times$ 10$^{14}$  &  \\
     & C$_6$H $^2\Pi_{3/2}$            & $ (1.67\pm0.07)  $ $\times$ 10$^{14}$  & \cite{Kawaguchi+etal+1995}      & 6.01 $\times$ 10$^{14}$ &  \\
     & C$_6$H $^2\Pi_{1/2}$             & $ (3\pm1)  $ $\times$ 10$^{13}$        & \cite{Saito+etal+1987}          & 6.01 $\times$ 10$^{14}$  &  \\
     & C$_6$H $^2\Pi_{3/2}$            & $ (1^{+1.5}_{-0.5})  $ $\times$ 10$^{13}$  & \cite{Saito+etal+1987}      & 6.01 $\times$ 10$^{14}$  &  \\
     & C$_6$H$^a$                         & 3 $\times$ 10$^{14}$                & \cite{Cernicharo+etal+1987b}    & 6.01 $\times$ 10$^{14}$    &  \\
(83) & HC$_5$N                      & $ (5.41\pm0.56)  $ $\times$ 10$^{14}$  & This work                       & 1.39 $\times$ 10$^{14}$   &  \\
     & HC$_5$N                     & $ (4.2\pm0.4)  $ $\times$ 10$^{14}$    & \cite{Pardo+etal+2022}          & 1.39 $\times$ 10$^{14}$  &  \\
     & HC$_5$N                      & $ (5.47\pm0.77)  $ $\times$ 10$^{14}$  & \cite{Zhang+etal+2017}          & 1.39 $\times$ 10$^{14}$ &  \\
     & HC$_5$N                      & $ (4.6\pm0.2)  $ $\times$ 10$^{14}$    & \cite{Gong+etal+2015}           & 1.39 $\times$ 10$^{14}$ &  \\
     & HC$_5$N                         & $ (2.3-5.47)  $ $\times$ 10$^{14}$     & \cite{Kawaguchi+etal+1995}       & 1.39 $\times$ 10$^{14}$  &  \\
     & HC$_5$N                    &  (warm range)    $ (2.3-5.47)  $ $\times$ 10$^{14}$   & \cite{Bell+etal+1993b}       & 1.39 $\times$ 10$^{14}$  &  \\
     & HC$_5$N                    & (cold range)    3.2 $\times$ 10$^{14}$             & \cite{Bell+etal+1992}       & 1.39 $\times$ 10$^{14}$   &  \\
     & HC$_5$N                    & (warm range)    3.7 $\times$ 10$^{14}$             & \cite{Bell+etal+1992}       & 1.39 $\times$ 10$^{14}$  &  \\
     & HC$_5$N                    & 4 $\times$ 10$^{14}$                & \cite{Winnewisser+etal+1978}    & 1.39 $\times$ 10$^{14}$  &  \\
     & HC$_5$N                    &                                     & \cite{Churchwell+etal+1978}     & 1.39 $\times$ 10$^{14}$   &  \\
(84) & SiC$_{6}^*$             & 1.3 $\times$ 10$^{12}$              & \cite{Pardo+etal+2022}    &  \\
(85) & CH$_{2}$CHCN             & 5.5 $\times$ 10$^{12}$              & \cite{Agundez+etal+2008} & {\color{black}4.71 $\times$ 10$^{11}$}   &  \\
(86) & CH$_{3}$CCH$^a$                & 1.8 $\times$ 10$^{13}$              & \cite{Agundez+etal+2008} & {\color{black}1.44 $\times$ 10$^{12}$}   &  \\
(87) & MgC$_5$N                    & 4.7 $\times$ 10$^{12}$              & \cite{Pardo+etal+2022,Pardo+etal+2021}     &     \\
(88) & MgC$_6$H                     & 2.0 $\times$ 10$^{13}$              & \cite{Pardo+etal+2022,Pardo+etal+2021}       &   \\
(89) & CH$_{3}$SiH$_{3}$              &                                     & \cite{Cernicharo+etal+2017}                          &    \\
(90) & C$_{7}$H $^2\Pi_{1/2}$    & $ (5.1\pm1.0)  $ $\times$ 10$^{12}$ & \cite{Pardo+etal+2022}    & {\color{black}7.11 $\times$ 10$^{13}$}   &  \\
     & C$_{7}$H $^2\Pi_{3/2}$       & $ (3.2\pm2.0)  $ $\times$ 10$^{12}$    & \cite{Pardo+etal+2022}   & {\color{black}7.11 $\times$ 10$^{13}$}   &  \\
     & C$_{7}$H                             & 2.2 $\times$ 10$^{12}$              & \cite{Cernicharo+etal+2000} & {\color{black}7.11 $\times$ 10$^{13}$}  &  \\
     & C$_{7}$H                             & 2 $\times$ 10$^{12}$                & \cite{Guelin+etal+1997}     & {\color{black}7.11 $\times$ 10$^{13}$}    &  \\
(91) & C$_{6}$H$_{2}$                        &                                     & \cite{Guelin+etal+2000,Guelin+etal+1997}  & 6.96 $\times$ 10$^{14}$    &  \\
(92) & HC$_7$N                 & $ (1.95\pm0.11)  $ $\times$ 10$^{14}$  &\cite{Pardo+etal+2022}         & 4.61 $\times$ 10$^{13}$   &   \\
     & HC$_7$N                       & $ (1.52-6.21)  $ $\times$ 10$^{14}$    & \cite{Zhang+etal+2017}        & 4.61 $\times$ 10$^{13}$   &   \\
     & HC$_7$N                     & $ (3.7\pm0.4)  $ $\times$ 10$^{14}$    & \cite{Gong+etal+2015}         & 4.61 $\times$ 10$^{13}$   &  \\
     & HC$_7$N                         & $ (1.29\pm0.06)  $ $\times$ 10$^{14}$  & \cite{Kawaguchi+etal+1995}    & 4.61 $\times$ 10$^{13}$   &  \\
     & HC$_7$N$^a$                           & 2 $\times$ 10$^{14}$                & \cite{Winnewisser+etal+1978}  & 4.61 $\times$ 10$^{13}$    &  \\
(93) & C$_8$H $^1\Pi_{3/2}$           & $ (1.15\pm1.38)  $ $\times$ 10$^{13}$  & \cite{Pardo+etal+2022}        & 1.38 $\times$ 10$^{14}$  &  \\
     & C$_8$H $^2\Pi_{3/2}$       & $ (8.9\pm0.9)  $ $\times$ 10$^{12}$    & \cite{Pardo+etal+2022}        & 1.38 $\times$ 10$^{14}$  &  \\
     & C$_8$H $^2\Pi_{3/2}$           & $ (5.00\pm1.008)  $ $\times$ 10$^{13}$ & \cite{Zhang+etal+2017}        & 1.38 $\times$ 10$^{14}$   &  \\
     & C$_8$H $^2\Pi_{3/2}$           & $ (8.4\pm1.4)  $ $\times$ 10$^{13}$    & \cite{Gong+etal+2015}         & 1.38 $\times$ 10$^{14}$  &  \\
     & C$_8$H                          & $ (8\pm3)  $ $\times$ 10$^{12}$        & \cite{Remijan+etal+2007}      & 1.38 $\times$ 10$^{14}$   &  \\
     & C$_8$H                               & 1.0 $\times$ 10$^{13}$              & \cite{Cernicharo+etal+2000}   & 1.38 $\times$ 10$^{14}$   &  \\
     & C$_8$H                               & 5.5 $\times$ 10$^{12}$              & \cite{Cernicharo+etal+1996}   & 1.38 $\times$ 10$^{14}$   &  \\
(94) & HC$_{9}$N                   & $ (4.5\pm0.6)  $ $\times$ 10$^{13}$    & \cite{Pardo+etal+2022}        & 1.31 $\times$ 10$^{13}$   &  \\
     & HC$_{9}$N                  & $ (5.88\pm1.28)  $ $\times$ 10$^{13}$  & \cite{Zhang+etal+2017}        & 1.31 $\times$ 10$^{13}$   &  \\
     & HC$_{9}$N                    & $ (2.5\pm1.4)  $ $\times$ 10$^{13}$    & \cite{Gong+etal+2015}         & 1.31 $\times$ 10$^{13}$   &  \\
     & HC$_{9}$N                      & $ (2.7\pm0.9)  $ $\times$ 10$^{13}$    & \cite{Kawaguchi+etal+1995}    & 1.31 $\times$ 10$^{13}$   &  \\
     & HC$_{9}$N                            & 4 $\times$ 10$^{13}$                & \cite{Bell+etal+1992a}        & 1.31 $\times$ 10$^{13}$    &  \\
(95) & CN$^{-}$                            & 5 $\times$ 10$^{12}$                & \cite{Agundez+etal+2010}    & 5.56 $\times$ 10$^{10}$ &  \\
(96) & HCO$^{+}$                       &                                     & \cite{Pulliam+etal+2011}    & 1.98 $\times$ 10$^{12}$   &  \\
     & HCO$^{+}$                         &                                     & \cite{Tenenbaum+etal+2010}  & 1.98 $\times$ 10$^{12}$  &  \\
(97) & C$_3$N$^{-}$                         & 1.6 $\times$ 10$^{12}$              & \cite{Thaddeus+etal+2008}   & 3.07 $\times$ 10$^{11}$   &  \\
(98) & C$_4$H$^{-}$                       & $ (7.1\pm2.0)  $ $\times$ 10$^{11}$   & \cite{Cernicharo+etal+2007} & 1.28 $\times$ 10$^{13}$   &  \\
(99)& MgC$_3$N$^{+}$ & $ (1.2\pm0.2)  $ $\times$ 10$^{11}$    & \cite{Cernicharo+etal+2023a}      &   &  \\
(100) &  MgC$_4$H$^{+}$ & $ (4.8\pm0.3)  $ $\times$ 10$^{11}$    & \cite{Cernicharo+etal+2023a}      &   &  \\
(101) & C$_5$N$^{-}$                 & $ (4.9\pm1.6)  $ $\times$ 10$^{12}$    & \cite{Pardo+etal+2022}      & 3.58 $\times$ 10$^{12}$   &  \\
     & C$_5$N$^{-}$                  & $ (3.4-4.9)  $ $\times$ 10$^{12}$      & \cite{Cernicharo+etal+2008} & 3.58 $\times$ 10$^{12}$   &  \\
&  C$_5$N$^{-}$        & $(5.7\pm0.2)$ $\times$  10$^{12}$     & \cite{Cernicharo+etal+2023} & 3.58 $\times$ 10$^{12}$   &  \\
(102) & C$_6$H$^{-}$                 & $ (4.3\pm0.6)  $ $\times$ 10$^{12}$    & \cite{Pardo+etal+2022}      & 8.71 $\times$ 10$^{13}$   &  \\
     & C$_6$H$^{-}$                  & $ (6.11\pm2.99)  $ $\times$ 10$^{12}$  & \cite{Zhang+etal+2017}      & 8.71 $\times$ 10$^{13}$  &  \\
     & C$_6$H$^{-}$                  & $ (5.8\pm0.5)  $ $\times$ 10$^{12}$    & \cite{Gong+etal+2015}       & 8.71 $\times$ 10$^{13}$   &  \\
     & C$_6$H$^{-}$                  &  $(6.1-8.0)$   $\times$ 10$^{12}$     & \cite{Kasai+etal+2007}      & 8.71 $\times$ 10$^{13}$  &  \\
     & C$_6$H$^{-}$                  & $ (4.1\pm1.5)  $ $\times$ 10$^{12}$    & \cite{Cernicharo+etal+2007} & 8.71 $\times$ 10$^{13}$  &  \\
     & C$_6$H$^{-}$                  & 4 $\times$ 10$^{12}$                & \cite{McCarthy+etal+2006}   & 8.71 $\times$ 10$^{13}$   &  \\
(103) & MgC$_5$N$^{+}$ & $ (1.1\pm0.2)  $ $\times$ 10$^{11}$    & \cite{Cernicharo+etal+2023a}      &   &  \\
(104)  & MgC$_6$H$^{+}$ & $ (2.5\pm0.3)  $ $\times$ 10$^{11}$    & \cite{Cernicharo+etal+2023a}      &   &  \\
(105)  &  C$_7$N$^{-}$                     & $ (2.4\pm0.2)  $ $\times$ 10$^{12}$    & \cite{Cernicharo+etal+2023}      &   &  \\
(106) & C$_8$H$^{-}$                     & $ (1.1\pm0.4)  $ $\times$ 10$^{12}$    & \cite{Pardo+etal+2022}      & 3.30 $\times$ 10$^{12}$  &  \\
     & C$_8$H$^{-}$                   & $ (8\pm3)  $ $\times$ 10$^{12}$        & \cite{Remijan+etal+2007}    & 3.30 $\times$ 10$^{12}$  &  \\
     & C$_8$H$^{-}$               & $ (1.5\pm0.6)  $ $\times$ 10$^{12}$    & \cite{Kawaguchi+etal+2007}  & 3.30 $\times$ 10$^{12}$  &  \\
\enddata
\tablecomments{
 $^{\it (a)}$ Assumed rotational temperature.
 $^{\it (*)}$ The detection of SiC$_{6}$ is marginal and the identification of NCCP is tentative.
 $^{\it (b)}$ The range of data for SiC$_2$ fitting is $J_{(up)}=6, 7$; $K_a=0, 2, 4$.
 $^{\it (c)}$ The range of data for SiC$_2$ fitting is $J_{(up)}=9, 10, 11, 12$; $K_a=0, 2, 4$.
 $^{\it (d)}$ The range of data for SiC$_2$ fitting is $J_{(up)}=10, 11$; $K_a=6$.
 $^{\it (e)}$ The range of data for SiC$_2$ fitting is $J_{(up)}=10, 11$; $K_a=8$.
 $^{\it (f)}$ The range of upper level energy for SiCSi fitting is $E\rm_u < 60 $\,K.
 $^{\it (g)}$ The range of upper level energy for SiCSi fitting is $60\,\rm K < E\rm_u < 200 $\,K.
 $^{\it (h)}$ The range of upper level energy for SiCSi fitting is $E\rm_u > 200 $\,K.}
\end{deluxetable}

IRC +10216 (CW Leo), a C-rich AGB star, is the brightest astronomical source at 5 $\mu$m outside the Solar System \citep{Becklin+etal+1969}. Its mass-loss rate is about 2 $\times$ 10$^{-5}$ $M_{\odot}$ yr$^{-1}$ at a distance of  130\,pc to Earth \citep{Crosas+Menten+1997, Menten+etal+2012}. As summarized in Table~\ref{table:1},  106 molecular species (note that these are 35\% of the overall-known 300 molecules in space$^{a}$\footnote{\tt $^{(a)}$ https://cdms.astro.uni-koeln.de/classic/molecules/}), have been detected in the CSE of IRC\,+10216 \citep{Agundez+etal+2014, Cernicharo+etal+2017, Fonfria+etal+2018, McGuire+etal+2018, Cernicharo+etal+2019, Cernicharo+etal+2019b,  McGuire+etal+2022, Pardo+etal+2022}.
 Molecules that were observed in the CSE of IRC\,+10216 and the column densities of these molecules calculated by our CSE chemical model (see Sect.~\ref{sec:updated model}) are listed in Table~\ref{table:1}.
The detected molecules include a large number of linear carbon-chain molecules such as C$_{n}$H ($n=2-8$) \citep{Tucker+etal+1974, Guelin+etal+1978, Thaddeus+etal+1985, Cernicharo+etal+1986, Saito+etal+1987, Cernicharo+etal+1996, Guelin+etal+1997}, HC$_{n}$N ($n = 1, 2, 3, 4, 5, 7, 9$) \citep{Morris+etal+1971, Morris+etal+1975, Churchwell+etal+1978, Winnewisser+etal+1978, Guelin+etal+1991, Bell+etal+1992, Cernicharo+etal+2004}, C$_{n}$S  ($n= 1, 2, 3, 5$)  \citep{Penzias+etal+1971, Cernicharo+etal+1987, Bell+etal+1992}, negatively and positively charged molecules such as C$_{n}$H$^{-}$  ($n = 4, 6, 8$) \citep{McCarthy+etal+2006, Cernicharo+etal+2007, Remijan+etal+2007}, C$_{n}$N$^{-}$  ($n = 1, 3, 5$)  \citep{Cernicharo+etal+2008, Thaddeus+etal+2008, Agundez+etal+2010} and HCO$^{+}$ \citep{Tenenbaum+etal+2010},
metal cyanides/isocyanides such as KCN, CaNC, NaCN, MgNC, MgCN, AlNC, FeCN, SiCN, SiNC, HMgNC and SiH$_{3}$CN \citep{Kawaguchi+etal+1993, Turner+etal+1994, Ziurys+etal+1995, Ziurys+etal+2002, Guelin+etal+2000,  Guelin+etal+2004, Pulliam+etal+2010, Zack+etal+2011, Cabezas+etal+2013, Cernicharo+etal+2017, Cernicharo+etal+2019},
and even metal halides such as NaCl, AlCl, KCl and AlF \citep{Cernicharo+etal+1987a}. Some of these molecules, such as the metal halides and the metal-bearing molecules, were identified during line surveys, covering a significant part of the radio and (sub-)millimeter spectrum \citep{Cernicharo+etal+2011}. Therefore, unbiased spectral line surveys are of fundamental importance to the systematic evaluation of the molecular composition and in particular to those species whose chemistries are poorly studied. For example, the first identified Ca-bearing species in space, CaNC, was discovered by a full-spectral line survey despite not being expected from chemical models \citep{Cernicharo+etal+2019}.

\begin{deluxetable*}{ccc}
%\tablenum{2}
\tablecaption{Published radio line surveys toward IRC\,+10216.$^{*}$
Note that a collection of line surveys of evolved stars at $\lambda$ 3\,mm  wavelength band can be found in Table \ref{table:3mmsurvery}.}
\label{table:2}
\tablewidth{0pt}
\tablehead{
\colhead{Covered Frequencies} & \colhead{Telescope} & \colhead{Reference}  \\
\colhead{ (GHz)  }               & \colhead{ }         & \colhead{        }
}
%\decimalcolnumbers
\startdata
4 -- 6                       &    Arecibo--305\,m       & \cite{Araya+etal+2003}          \\
13.3 -- 18.5                 & 	TMRT--65\,m           & \cite{Zhang+etal+2017}          \\
17.8 -- 26.3                 &	Effelsberg--100\,m    & \cite{Gong+etal+2015}           \\
18 -- 50                     &    VLA                   & \cite{Keller+etal+2015}         \\
20 -- 25	                 &	Nobeyama--45\,m	      & \cite{Zhang+etal+2020}          \\
28 -- 50                     &	Nobeyama--45\,m       & \cite{Kawaguchi+etal+1995}      \\
31.3 -- 50.6                 &	Yebes--40\,m	      & \cite{Tercero+etal+2021}        \\
31.3 -- 50.3                 &	Yebes--40\,m	      & \cite{Pardo+etal+2022}        \\
72.2 -- 91.1                 &	Onsala--20\,m         & \cite{Johansson+etal+1984, Johansson+etal+1985}    \\
84.5 -- 115.8                &	PMO--13.7\,m          &  This work                       \\
129.0 -- 172.5               &	IRAM--30\,m           & \cite{Cernicharo+etal+2000}     \\
131.2 -- 160.3               &	ARO--12\,m            & \cite{He+etal+2008}             \\
214.5 -- 285.5               &	SMT--10\,m            & \cite{Tenenbaum+etal+2010}      \\
219.5 -- 245.5 and 251.5 -- 267.5        &	SMT--10\,m            & \cite{He+etal+2008}             \\
222.4 -- 267.92 $^{a}$       & 	JCMT--15\,m	          & \cite{Avery+etal+1992}         \\
253 -- 261                   &	CARMA                 & \cite{Fonfria+etal+2014}        \\
255.3 -- 274.8               &	ALMA                  & \cite{Cernicharo+etal+2013}     \\
255.3 -- 274.8               &	ALMA                  & \cite{Velilla+Prieto+etal+2015}     \\
293.9 -- 354.8               &	SMA                   & \cite{Patel+etal+2011}          \\
330.2 -- 358.1               &	CSO--10.4\,m          & \cite{Groesbeck+etal+1994}      \\
339.6 -- 364.6               & 	JCMT--15\,m	          & \cite{Avery+etal+1992}     \\
554.5 -- 636.5               &	Herschel/HIFI         & \cite{Cernicharo+etal+2010}     \\
\enddata
\tablecomments{
$^{(*)}$ Unpublished work is not included  in this table; for instance, the 3\,mm IRAM--30\,m and ALMA surveys toward IRC\,+10216 mentioned in the papers of \cite{ Agundez+etal+2014, Agundez+etal+2017}.
$^{ (a)  }$ The line survey was discontinuously conducted in the listed frequency range and contains significant gaps.}
\end{deluxetable*}

Multiple transition lines of each molecular species observed in different spectral line surveys have greatly improved estimates of molecular abundances and excitation temperatures, thus better estimating the physical conditions in which each  molecular species occurs.
By fitting several spectral lines of C$_3$S, HC$_3$N, and HC$_5$N in the CSE of IRC\,+10216 in different wavebands, it was found that these molecules all have both cold and warm excitation temperature components \citep{Bell+etal+1992, Bell+etal+1993b, Bell+etal+1993}. Later, \cite{Zhang+etal+2017} fitted the spectral line survey data of HC$_5$N, HC$_7$N, and HC$_{9}$N in three different frequency ranges and  found that the high rotational transitions of these molecules trace the warmer molecular regions and that the thermal structure is stratified, which is consistent with the results for HC$_5$N from the CSE of the C-rich AGB star CIT6 \citep{Chau+etal+2012}.

Molecular line surveys are a powerful tool to analyzing the physical and chemical parameters of astronomical sources. IRC\,+10216 is the definitive source for studying the circumstellar chemistry in the late stages of stellar evolution \citep{Zhang+etal+2020}, and there have been many reports on its detailed and high sensitivity surveys at centimeter, millimeter, and submillimeter wavelengths. Based on the works of \cite{Gong+etal+2015}, and \cite{Zhang+etal+2017}, we collected the published line surveys toward IRC\,+10216 and list them in Table~\ref{table:2}. We find that \cite{Johansson+etal+1984} carried out the first survey toward IRC\,+10216 with the OSO--20\,m telescope at $\sim$3.8\,mm (72.2 -- 91.1\,GHz), and they reviewed the excitation conditions and relative abundances of molecules in the CSE of IRC\,+10216. \cite{Agundez+etal+2014} mainly reconfirmed three transitions of C$_5$S and preliminarily identified four transitions of MgCCH, three transitions of NCCP, and six transitions of SiH$_{3}$CN within the $\lambda$ $\sim$ 3\,mm region of IRC\,+10216, while \cite{Agundez+etal+2017} mainly explored the growth mechanism of carbon chain molecules in IRC\,+10216 based on data in this band.

Note, however, that a systematic spectral line survey in the frequency range from 91\,GHz to 116\,GHz is still missing, with the original survey of \cite{Johansson+etal+1984, Johansson+etal+1985} only covering frequencies up to 91\,GHz (see Table~\ref{table:2}). Therefore, we present a spectral line survey of the C-rich AGB star IRC\,+10216 in the $\lambda$ 3\,mm  (84.5 -- 115.8 GHz)  region using the Purple Mountain Observatory 13.7 meter radio telescope (PMO--13.7\,m).

We describe the observations and data reduction process in Section~\ref{sec:Observations and data reduction}.
and present the identification of the emission lines in Section~\ref{sec:Results}.
Section~\ref{sec:DISCUSSION} contains the observationally-deduced and modeled excitation temperatures, column densities, and the fractional abundances of detected species, together with their isotopic ratios.
The conclusions are given in Section~\ref{sec:Conclusions}. Further information is included in the six Appendices. The complete spectra of the survey of IRC\,+10216 range from 84.5 to 115.8 GHz and are shown in Appendix~\ref{Zoom-in plots of observed spectra}. The spectral line parameters of all the identified species from this survey are presented in Appendix~\ref{Line parameters of transitions}. All the observed spectral profiles are included in Appendix~\ref{Line spectrum}. A detailed description of the detected molecules from this survey, as well as from previous observations, is included in Appendix~\ref{Detected species}. Appendix~\ref{Rotational diagrams} presents the rotational diagrams for the identified species, while the stratified thermal structures of a few molecules can be found in Appendix~\ref{Appendix Comparison of rotational diagrams}. Detailed comparisons of emission lines obtained with different angular resolutions are included in Appendix~\ref{comparisons of line shapes}.

\section{Observations and data reduction}
\label{sec:Observations and data reduction}
The spectral line survey of IRC\,+10216 between 84.5 and 115.8\,GHz was carried out in August 2019 using the PMO 13.7\,m millimeter-wave radio telescope of the Purple Mountain Observatory at Delingha in China. The typical system temperature was about 150 -- 300\,K \citep{Li+etal+2013}. The frontend of the receiver uses a nine-beam sideband separation Superconduction Spectroscopic Array Receiver system, and the backend consists of 18 high resolution digital spectrum analyzer Fast Fourier Transform Spectrometers (FFTSs). Each FFTS provides 16384 channels with a total bandwidth of 1\,GHz \citep{Shan+etal+2012}. The beam separation of the nine beams of the PMO-13.7\,m telescope is 3\,$^\prime$. The observations adopted the position switching mode using the standard chopper wheel calibration method. The position switching mode uses only one beam, while the other beams point to other parts of the sky. On- and off-source integration times were one minute each per scan. The temperature scale is antenna temperature ($T_{\rm A}^{*}$)  after correction of atmospheric absorption and ohmic loss. The $T_{\rm A}^{*}$ scale is related to the main beam brightness temperature ($T_{\rm mb}$) via the main beam efficiency (${\eta}$), i.e. $T_{\rm mb}$ = $T_{\rm A}^{*}/{\eta}$; the value of ${\eta}$ is $\sim$ 0.6 $^{b}$\footnote{\tt $^{(b)}$ http://www.radioast.nsdc.cn/zhuangtaibaogao.php}. The half-power bandwidths (HPBWs) are ${51}^{\prime\prime}$ and ${49}^{\prime\prime}$ at 110\,GHz and 115\,GHz, respectively. The velocity resolution is $\sim$ 0.16 -- 0.17\,km s$^{-1}$. The total observing time of the spectral line survey is about 32.5\,hours, which includes overhead due to telescope movement. The pointing accuracy was better than ${5}^{\prime\prime}$.

The data were reduced using the GILDAS $^{c}$\footnote{\tt $^{(c)}$ GILDAS is developed and distributed by the Observatoire de Grenoble and IRAM.} software package, including CLASS and GREG. Linear baseline subtractions were used for all the spectra. Since the edges of the spectral range may have unknown defects, channels closer than 150\,MHz to each edge of a spectral window were discarded. Velocity is given relative to the local standard of rest. Except for CO, CS, and HCN, molecular lines were smoothed  (combined channels) to improve  signal-to-noise (S/N) ratios for individual channels. The rms noise of the smoothed spectra is about 0.014  (or $\sim$ 0.135) K in 1.759  (or $\sim$ 0.402) km s$^{-1}$ wide  channels for the detected species  (or CO, CS, and HCN).

\section{Results}
\label{sec:Results}

\subsection{The overall survey}

\begin{figure*}
\gridline{\fig{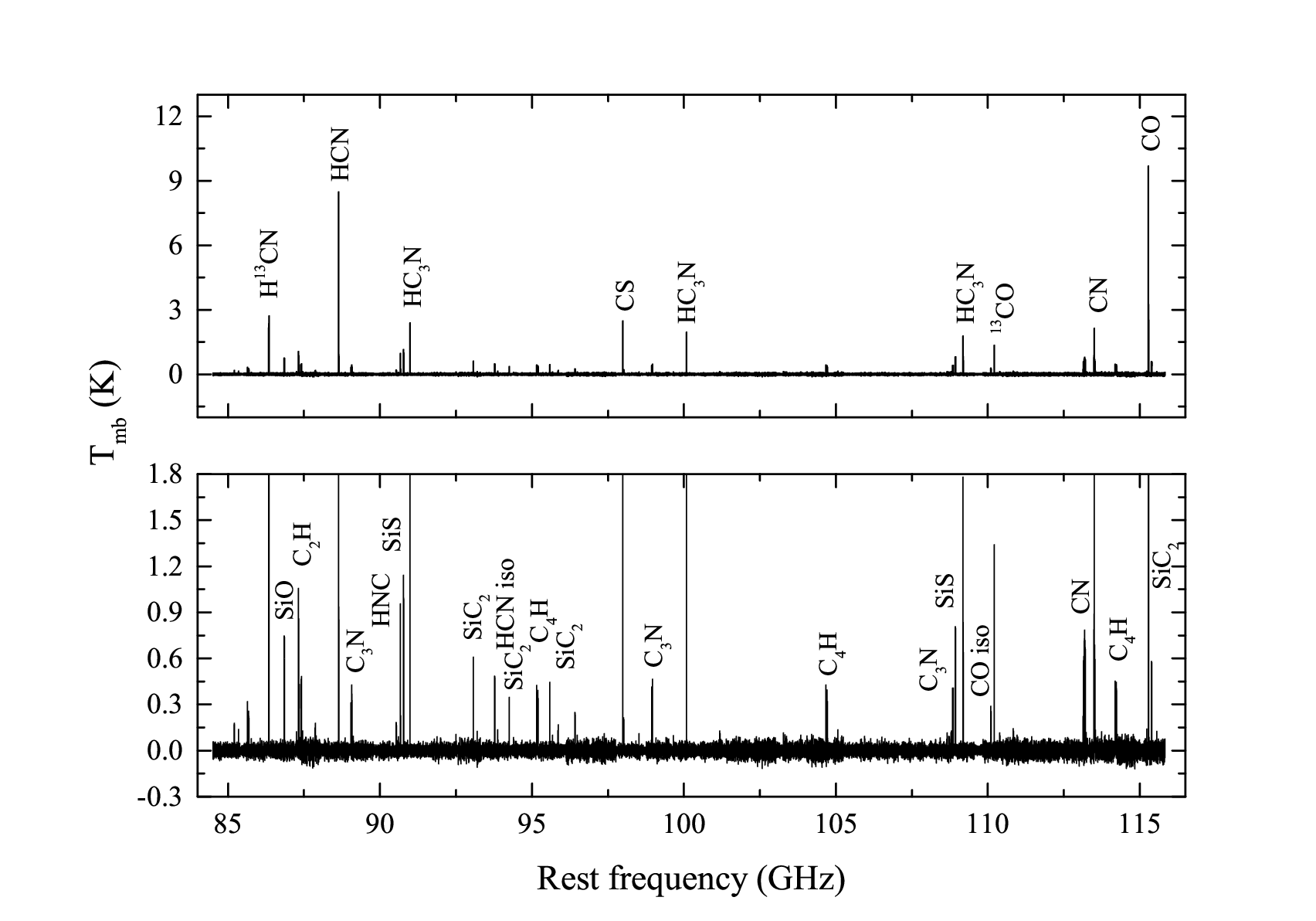}{1.\textwidth}{ }
          }
   \caption{An overall spectral line survey towards IRC\,+10216 in the  3\,mm band by the PMO--13.7\,m radio telescope. The rest frequency scale was obtained by adopting a radial velocity of $\rm -26.5\,km\,s^{-1}$. The X-axis denotes the rest frequency (GHz), and the Y-axis depicts $T_{\rm mb}$ (K). On the upper panel, spectral lines with $T_{\rm mb}$ greater than 1.3\,K are marked, while those with $T_{\rm mb}$ greater than 350\,mK are marked on the lower panel.}
   \label{fig:survey}
\end{figure*}

%{\color{green} HCN iso and CO iso represent mirror images of HCN and CO, respectively.}

Figure~\ref{fig:survey} presents an overview of the spectral line survey toward IRC\,+10216 in $\lambda$ 3\,mm band,
which is displayed in consecutive 475 MHz frequency segments with 0.244 MHz spectral resolution in Appendix~\ref{Zoom-in plots of observed spectra}.
We find 96 transitions, and the identified lines are assigned to 19 different molecular species and  their isotopologues. The line identification was performed by using the Jet Propulsion Laboratory $^{d}$\footnote{\tt $^{(d)}$ http://spec.jpl.nasa.gov}  \cite[JPL,][]{Pickett+etal+1998}, Cologne Database for Molecular Spectroscopy catalogs $^{e}$\footnote{\tt $^{(e)}$ http://www.astro.uni-koeln.de/cdms/catalog} \cite[CDMS,][]{Muller+etal+2005}, and splatalogue databases $^{f}$\footnote{\tt $^{(f)}$ https://splatalogue.online//advanced.php} as well as the online Lovas line list $^{g}$\footnote{\tt $^{(g)}$ http://www.nist.gov/pml/data/micro/index.cfm} \citep{Lovas+2004} for astronomical spectroscopy. The local standard of rest radial velocity  ($V_{\rm lsr}$)   of -26.5\, km s$^{-1}$ is adopted to derive the rest frequency of the observed lines \citep{Zhang+etal+2017}. This paper discusses the lines detected with S/N of at least three \citep{Fonfria+etal+2014}.

We used the SHELL fitting routine in CLASS to fit the spectral line profiles and obtain the relevant parameters \citep{Gong+etal+2015}. The fitted spectral line parameters include peak intensity, integrated intensity, and expansion velocity, defined as the halfwidth at zero power \citep{Gong+etal+2015, Zhang+etal+2017}. For the lines with hyperfine structure (hfs, like HCN, H$^{13}$CN, etc.), only their primary components are fitted with the routine to obtain their expansion velocities, and the integrated intensity is a sum over all hyperfine components. All detected transitions and the line parameters are listed in Table~\ref{table:3}. The rms noise on a main beam brightness temperature ($T_{\rm mb}$) scale is obtained from the statistics in regions free of noticeable spectral lines, and the measurement uncertainty of the integrated intensity given in this work is calculated as the product of the rms noise and the line width \citep{Zhang+etal+2017}.

\startlongtable
\begin{deluxetable*}{cccccccc}
%\tablenum{4}
\tablecaption{An overview over previous studies on molecules identified within our measured $\lambda$  3\,mm band towards IRC\,+10216. Note that in this work we have provided all of the information on all of the species listed here.
 \label{table:4}}
\tablewidth{0pt}
\tablehead{
\colhead{No.}  &\colhead{Molecule} & \colhead{Rest Freq.}  & \colhead{Detected?} & \colhead{Spectrum?}  & \colhead{$T_{\rm ex}$} obtained?   & \colhead{$N$ or $f^{a}$ obtained?}            & \colhead{Ref.}\\
\colhead{ }  &\colhead{ } & \colhead{ (MHz)  }  & \colhead{ } & \colhead{ }  & \colhead{ }     & \colhead{ }            & \colhead{ }
}
\startdata
(1)  &  HC$_5$N	        & 85201.346  & $\checkmark$   & $\checkmark$    & $\checkmark$  & $\checkmark$  & 1, 2, 3, 4, 5 \\
(2)  & $c$-C$_3$H$_{2}$ & 85338.906  & $\checkmark$   & $\checkmark$    &               & $\checkmark$  & 2, 6, 7\\
(3)  &  C$_4$H	        & 85634.000	 & $\checkmark$	  & $\checkmark$	& $\checkmark$  & $\checkmark$  & 2, 5, 7, 8\\
(4)  &  C$_4$H	        & 85672.570	 & $\checkmark$	  & $\checkmark$	& $\checkmark$  & $\checkmark$  & 2, 5, 8\\
(5)  &  $^{29}$SiO      & 85759.188	 & $\checkmark$	  & $\checkmark$	& $\checkmark$  & $\checkmark$  & 2, 9, 10, 11, 12\\
(6)  &  H$^{13}$CN      & 86338.737  & $\checkmark$	  & $\checkmark$	&               &               & 4, 6, 9, 13 \\
(7)  &  H$^{13}$CN      & 86340.176  & $\checkmark$	  & $\checkmark$	&               & $\checkmark$  & 4, 6, 9, 13, 14, 15, 16 \\ 		
(8)  &  H$^{13}$CN      & 86342.255  & $\checkmark$	  & $\checkmark$	&               &               & 4, 6, 9, 13 \\
(9)  &  SiO	            & 86846.995	 & $\checkmark$	  & $\checkmark$	& $\checkmark$  & $\checkmark$  & 2, 6, 9, 11, 12, 17 \\
(10) &  C$_2$H	        & 87284.156	 & $\checkmark$	  & $\checkmark$    & $\checkmark$  & $\checkmark$  & 2, 4, 10, 18, 19 \\
(11) &  C$_2$H	        & 87316.925	 & $\checkmark$	  & $\checkmark$    & $\checkmark$  & $\checkmark$  & 2, 4, 10, 16, 18, 19, 20, 21 \\
(12) &  C$_2$H	        & 87328.624	 & $\checkmark$	  & $\checkmark$    & $\checkmark$  & $\checkmark$  & 2, 4, 5, 10, 16, 18, 19, 20 \\
(13) &  C$_2$H	        & 87402.004	 & $\checkmark$	  & $\checkmark$    & $\checkmark$  & $\checkmark$  & 2, 4, 10, 16, 18, 19, 20 \\
(14) &  C$_2$H	        & 87407.165	 & $\checkmark$	  & $\checkmark$    & $\checkmark$  & $\checkmark$  & 2, 4, 10, 18, 19, 20  \\
(15) &  C$_2$H	        & 87446.512	 & $\checkmark$	  & $\checkmark$    & $\checkmark$  & $\checkmark$  & 2, 4, 10, 18, 19 \\
(16) &  HC$_5$N	        & 87863.630	 & $\checkmark$	  & $\checkmark$    & $\checkmark$  & $\checkmark$  & 1, 2, 3, 5, 9\\
(17) &  H$^{13}$CCCN    & 88166.808  & $\checkmark$	  & $\checkmark$	&               &               & 2, 3, 4, 9 \\
(18) &  HCN	            & 88630.416	 & $\checkmark$	  & $\checkmark$	&               &               & 2, 4, 6, 9, 13  \\
(19) &  HCN             & 88631.847	 & $\checkmark$	  & $\checkmark$	& $\checkmark$  & $\checkmark$  & 2, 4, 6, 9, 13, 14, 16, 15 \\
(20) &  HCN             & 88633.936	 & $\checkmark$	  & $\checkmark$	&               &               & 2, 4, 6, 9, 13, 22  \\
(21) &  C$_3$N	        & 89045.590	 & $\checkmark$	  & $\checkmark$    & $\checkmark$  & $\checkmark$  & 2, 5, 23, 24 \\
(22) &  C$_3$N	        & 89064.360	 & $\checkmark$	  & $\checkmark$    & $\checkmark$  & $\checkmark$  & 4, 5, 23, 24 \\
(23) &  $^{29}$SiS      & 89103.720	 & $\checkmark$	  & $\checkmark$	& $\checkmark$  & $\checkmark$  & 4, 10, 11, 12 \\
(24) &  HC$_5$N	        & 90525.890	 & $\checkmark$	  & $\checkmark$	& $\checkmark$  & $\checkmark$  & 2, 3, 5, 25, 26 \\
(25) &  HC$^{13}$CCN    & 90593.059	 & $\checkmark$	  & $\checkmark$	&               & $\checkmark$  & 2, 4, 6 \\
(26) &  HCC$^{13}$CN    & 90601.791	 & $\checkmark$	  & $\checkmark$    &               & $\checkmark$  & 2, 4, 6 \\
(27) &  HNC	            & 90663.450	 & $\checkmark$	  & $\checkmark$	&               &               & 2, 4, 6, 9, 27, 28\\ 		
(28) &  HNC	            & 90663.574	 & $\checkmark$	  & $\checkmark$	& $\checkmark$  & $\checkmark$  & 2, 4, 6, 9, 16, 21, 27, 28 \\
(29) &  HNC	            & 90663.656	 & $\checkmark$	  & $\checkmark$	&               &               & 2, 4, 6, 9, 27, 28\\	
(30) &  SiS	            & 90771.561	 & $\checkmark$	  & $\checkmark$	& $\checkmark$  & $\checkmark$  & 2, 6, 9, 11, 12, 16, 17, 22, 29, 30 \\
(31) &  HC$_3$N	        & 90978.989  & $\checkmark$	  & $\checkmark$    & $\checkmark$  & $\checkmark$  & 2, 4, 5, 6, 9 16, 17, 21 \\
(32) &  $^{13}$CS       & 92494.270	 & $\checkmark$	  & $\checkmark$    & $\checkmark$  & $\checkmark$  & 11, 12, 15, 31\\	
(33) &  SiC$_2$	        & 93063.639	 & $\checkmark$	  & $\checkmark$	& $\checkmark$  & $\checkmark$  & 25, 32, 33, 34, 35 \\
(34) &  HC$_5$N	        & 93188.123	 & $\checkmark$	  & $\checkmark$	& $\checkmark$  & $\checkmark$  & 3, 5, 36 \\
(35) &  C$_4$H          & 93863.300  & $\checkmark$	  & $\checkmark$	&               &               & 37 \\
(36) &  C$_2$S	        & 93870.098	 & $\checkmark$	  & $\checkmark$	& $\checkmark$  & $\checkmark$  & 31, 38 \\
(37) &  SiC$_2$	        & 94245.393	 & $\checkmark$	  & $\checkmark$	& $\checkmark$  & $\checkmark$  & 7, 25, 30, 32, 33, 35\\
(38) &  C$_4$H	        & 95150.320	 & $\checkmark$	  & $\checkmark$	& $\checkmark$  & $\checkmark$  & 5, 6, 8, 34 \\
(39) &  C$_4$H	        & 95188.940	 & $\checkmark$	  & $\checkmark$	& $\checkmark$  & $\checkmark$  & 5, 8, 34 \\
(40) &  SiC$_2$	        & 95579.381	 & $\checkmark$	  & $\checkmark$	& $\checkmark$  & $\checkmark$  & 25, 32, 33, 35, 39 \\
(41) &  HC$_5$N	        & 95850.335	 & $\checkmark$	  & $\checkmark$	& $\checkmark$  & $\checkmark$  & 3, 5\\
(42) &  C$^{34}$S	    & 96412.950	 & $\checkmark$	  & $\checkmark$	& $\checkmark$  & $\checkmark$  & 4, 10, 11, 12\\
(43) &  CS	            & 97980.953  & $\checkmark$	  & $\checkmark$	&               & $\checkmark$  & 4, 10, 11, 12, 15, 16, 22, 30, 40, 41\\
(44) &  $l$-C$_3$H      & 97995.212	 & $\checkmark$	  & $\checkmark$    & $\checkmark$  & $\checkmark$  & 2, 4, 40 \\
(45) &  $l$-C$_3$H      & 97995.951	 & $\checkmark$	  & $\checkmark$    & $\checkmark$  & $\checkmark$  & 2, 4, 6, 40 \\
(46) &  $l$-C$_3$H      & 98011.649	 & $\checkmark$	  & $\checkmark$	& $\checkmark$  & $\checkmark$  & 2, 4, 34, 40 \\
(47) &  $l$-C$_3$H      & 98012.576	 & $\checkmark$	  & $\checkmark$    & $\checkmark$  & $\checkmark$  & 2, 4, 34, 40 \\
(48) &  HC$_5$N	        & 98512.524	 & $\checkmark$	  & $\checkmark$	& $\checkmark$  & $\checkmark$  & 1, 3, 5  \\
(49) &  C$_3$N	        & 98940.020	 & $\checkmark$	  & $\checkmark$    & $\checkmark$  & $\checkmark$  & 5, 23, 24, 42 \\
(50) &  C$_3$N	        & 98958.780	 & $\checkmark$	  & $\checkmark$    & $\checkmark$  & $\checkmark$  & 5, 23, 24, 42 \\
(51) &  HC$^{13}$CCN    & 99651.863	 & $\checkmark$	  & $\checkmark$    &               &               & 7 \\
(52) &  HCC$^{13}$CN    & 99661.471	 & $\checkmark$	  & $\checkmark$    &               &               & 7 \\
(53) &  HC$_3$N	        & 100076.386 & $\checkmark$	  & $\checkmark$	& $\checkmark$  &  $\checkmark$ & 5, 16 \\
(54) &  HC$_5$N	        & 101174.677 & $\checkmark$	  &                 &               &               & 5 \\
(55) &  C$_4$H          & 103266.081 & $\checkmark$	  &              	&               &               & 37 \\
(56) &  $l$-C$_3$H      & 103319.278 & $\checkmark$	  & $\checkmark$	&               &               & 37 \\
(57) &  $l$-C$_3$H      & 103319.818 & $\checkmark$	  & $\checkmark$	&               &               & 37 \\
(58) &  $l$-C$_3$H      & 103372.506 & $\checkmark$	  & $\checkmark$	&               &               & 37 \\
(59) &  $l$-C$_3$H      & 103373.129 & $\checkmark$	  & $\checkmark$	&               &               & 37 \\
(60) &  HC$_5$N	        & 103836.817 & $\checkmark$	  &                 &               &               & 5 \\
(61) &  C$_3$S	        & 104048.451 & $\checkmark$	  & $\checkmark$    & $\checkmark$  & $\checkmark$  & 31, 42 \\
(62) &  C$_4$H	        & 104666.560 & $\checkmark$	  & $\checkmark$	& $\checkmark$  & $\checkmark$  & 5, 8 \\
(63) &  C$_4$H	        & 104705.100 & $\checkmark$	  & $\checkmark$	& $\checkmark$  & $\checkmark$  & 5, 8 \\
(64) &  C$_4$H	        & 105838.000 & $\checkmark$	  &               	&               &               & 37 \\
(65) &  Si$^{34}$S      & 105941.503 & $\checkmark$	  &             	&               &               & 11, 12 \\
(66) &  C$_4$H          & 106132.800 & $\checkmark$	  &                 &               &               & 37 \\
(67) &  C$_2$S	        & 106347.740 & $\checkmark$	  &          	    & $\checkmark$  & $\checkmark$  & 31, \\
(68) &  HC$_5$N	        & 106498.910 & $\checkmark$	  & $\checkmark$	& $\checkmark$  & $\checkmark$  & 3, 5    \\
(69) &  MgNC	        & 107399.420 & $\checkmark$	  & $\checkmark$	& $\checkmark$  & $\checkmark$  & 43 \\
(70) &  HC$^{13}$CCN    & 108710.523 & $\checkmark$   &                 &               & $\checkmark$  & 6 \\
(71) &  HCC$^{13}$CN    & 108721.008 & $\checkmark$   &                 &               & $\checkmark$  & 6 \\
(72) &  C$_3$N	        & 108834.270 & $\checkmark$	  & $\checkmark$    & $\checkmark$  & $\checkmark$  & 2, 5, 6, 22, 23, 24 \\
(73) &  C$_3$N          & 108853.020 & $\checkmark$	  & $\checkmark$    & $\checkmark$  & $\checkmark$  & 2, 5, 22, 23, 24 \\
(74) &  SiS	            & 108924.297 & $\checkmark$	  & $\checkmark$	&               & $\checkmark$  & 2, 6, 11, 12, 17, 22, 29, 44 \\
(75) &  HC$_3$N	        & 109173.638 & $\checkmark$	  & $\checkmark$	& $\checkmark$  & $\checkmark$  & 2, 5, 6, 9, 15, 16\\
(76) &  $^{13}$CO       & 110201.354 & $\checkmark$	  & $\checkmark$	&               &               & 6, 9, 15, 16, 41, 45, 46\\
(77) &  CH$_3$CN	    & 110375.049 & $\checkmark$	  & $\checkmark$    & $\checkmark$  & $\checkmark$  & 7 \\
(78) &  CH$_3$CN	    & 110381.400 & $\checkmark$	  & $\checkmark$    & $\checkmark$  & $\checkmark$  & 7 \\
(79) &  CH$_3$CN	    & 110383.518 & $\checkmark$   & $\checkmark$    & $\checkmark$  & $\checkmark$  & 7 \\
(80) &  C$_4$H          & 112922.500 & $\checkmark$	  &                 &               &               & 37 \\
(81) &  CN	            & 113123.337 & $\checkmark$	  & $\checkmark$    &               &               & 16, 18 \\
(82) &  CN	            & 113144.192 & $\checkmark$	  & $\checkmark$    &               & $\checkmark$  & 16, 18 \\
(83) &  CN	            & 113170.528 & $\checkmark$	  & $\checkmark$    &               & $\checkmark$  & 16, 18 \\
(84) &  CN	            & 113191.317 & $\checkmark$	  & $\checkmark$    &               & $\checkmark$  & 16, 18 \\
(85) &  C$_4$H          & 113265.900 & $\checkmark$	  & $\checkmark$ 	&               &               & 37, 47 \\
(86) &  CN	            & 113488.140 & $\checkmark$	  & $\checkmark$    &               &               & 7, 14, 16, 18 \\	
(87) &  CN	            & 113490.982 & $\checkmark$	  & $\checkmark$    &               & $\checkmark$  & 6, 14, 16, 18, 41 \\	
(88) &  CN	            & 113499.639 & $\checkmark$	  & $\checkmark$    &               &               & 5, 14, 16, 18 \\
(89) &  CN	            & 113508.944 & $\checkmark$	  & $\checkmark$    &               &               & 14, 16, 18 \\
(90) &  CN	            & 113520.414 & $\checkmark$	  & $\checkmark$    &               &               & 14, 16, 18 \\
(91) &  C$_4$H	        & 114182.510 & $\checkmark$	  & $\checkmark$	& $\checkmark$  & $\checkmark$  & 5, 8 \\
(92) &  C$_4$H	        & 114221.040 & $\checkmark$	  & $\checkmark$	& $\checkmark$  & $\checkmark$  & 5, 8 \\
(93) &  H$^{13}$CCCN    & 114615.021 &                &                 &               &               &      \\
(94) &  C$_4$H	        & 115216.800 & $\checkmark$	  &                 &               &               & 37 \\
(95) &  CO   	        & 115271.202 & $\checkmark$	  & $\checkmark$	& $\checkmark$  & $\checkmark$  & 6, 9, 15, 16, 41, 48, 49, 50, 45\\
(96) &  SiC$_2$	        & 115382.375 & $\checkmark$	  & $\checkmark$	& $\checkmark$  & $\checkmark$  & 32, 33, 35, 51 \\
\enddata
\tablecomments{
$^{(a)}$$N$ and $f$ depict the total column density and fractional abundance of the molecular species.
Since there exist numerous observations toward IRC\,+10216, here we only provide a few key references containing molecular spectral line profiles, excitation temperature, and column density, which are:
(1) \cite{Bujarrabal+etal+1981}; (2) \cite{Johansson+etal+1984}; (3) \cite{Cernicharo+etal+1986a}; (4) \cite{Kahane+etal+1988}; (5) \cite{Agundez+etal+2017}; (6) \cite{Nyman+etal+1993}; (7) \cite{Agundez+etal+2008}; (8) \cite{Guelin+etal+1978}; (9) \cite{Olofsson+etal+1982}; (10) \cite{Groesbeck+etal+1994}; (11) \cite{Agundez+etal+2012};
(12) \cite{Velilla-Prieto+etal+2019}; (13) \cite{Morris+etal+1971}; (14) \cite{Dayal+etal+1995}; (15) \cite{Wannier+etal+1978}; (16) \cite{Park+etal+2008}; (17) \cite{Morris+etal+1975}; (18) \cite{Truong-Bach+etal+1987}; (19) \cite{Beck+etal+2012}; (20) \cite{Tucker+etal+1974}; (21) \cite{Bieging+etal+1988}; (22) \cite{Henkel+etal+1985}; (23) \cite{Guelin+etal+1977}; (24) \cite{Thaddeus+etal+2008}; (25) \cite{Cernicharo+etal+1986c}; (26) \cite{Agundez+etal+2014}; (27) \cite{Brown+etal+1976}; (28) \cite{Daniel+etal+2012}; (29) \cite{Sahai+etal+1984}; (30) \cite{Lucas+etal+1995}; (31) \cite{Cernicharo+etal+1987}; (32) \cite{Thaddeus+etal+1984}; (33) \cite{Snyder+etal+1985}; (34) \cite{Cernicharo+etal+1986}; (35) \cite{Cernicharo+etal+2018}; (36) \cite{Turner+etal+1994}; (37) \cite{Yamamoto+etal+1987}; (38) \cite{Cernicharo+etal+2019b}; (39) \cite{Guelin+etal+1990}; (40) \cite{Thaddeus+etal+1985}; (41) \cite{Wilson+etal+1971}; (42) \cite{Cernicharo+etal+1987a}; (43) \cite{Guelin+etal+1986}; (44) \cite{Bieging+etal+1989}; (45) \cite{Kuiper+etal+1976}; (46) \cite{knapp+etal+1985}; (47) \cite{Cernicharo+etal+2019}; (48) \cite{Solomon+etal+1971}; (49) \cite{Fong+etal+2003}; (50) \cite{Fong+etal+2006}; (51) \cite{Tenenbaum+etal+2006}.}
\end{deluxetable*}
%\end{longrotatetable}

All detected transitions are shown in Figs:~\ref{fig:S1-HC5N+SiO+12CO+13CO}, ~\ref{fig:C2H+C3H}, ~\ref{fig:C4H}, ~\ref{fig:C2H+C4H+C2S}, ~\ref{fig:CN}, ~\ref{fig:C3N+c-C3H2}, ~\ref{fig:HC3N}, ~\ref{fig:HC5N}, ~\ref{fig:HCN+HNC+29SiO}, ~\ref{fig:SiS+SiC2}, and ~\ref{fig:CS+MgNC+CH3CN}.  To clearly present the spectral line profile and $V_{lsr}$ of each molecule, the X-axis range of the spectral line figure we show varies. For most spectral lines, the velocity range is set from -100 to 50\,km\,s$^{-1}$, while the velocity range from -50 to 0\,km\,s$^{-1}$ is assumed for transitions shown in Figs.~\ref{fig:C2H+C3H},  ~\ref{fig:CN}  (except for the top right panel where it extends from -200 to 50\,km\,s$^{-1}$), and ~\ref{fig:C3N+c-C3H2}  (except for the bottom right panel,  where the velocity range is the standard one).

The detected lines include six transitions each for C$_2$H and C$_3$N, eight transitions for $l$-C$_3$H as well as C$_4$H\, (${\nu  = 0}$), seven transitions of C$_4$H\, (${\nu = 1}$), nine transitions for CN as well as HC$_5$N, two transitions each for H$^{13}$CCCN, SiS, and C$_2$S, three transitions each for HC$^{13}$CCN, HCC$^{13}$CN, HNC, HCN, H$^{13}$CN, CH$_3$CN, and HC$_3$N, one transition for $c$-C$_3$H$_2$, CO, $^{13}$CO, SiO, $^{29}$SiO, {Si}$^{34}$S, $^{29}$SiS, C$_3$S, CS, $^{13}$CS, C$^{34}$S, and MgNC, and four transitions of SiC$_2$.

Table~\ref{table:4} summarizes previous 3\,mm observations towards IRC\,+10216.
We checked if the previous work had detected all of the species we identified, or if they had provided all of the necessary information about these species, such as spectral line profile,  excitation temperature ($T_{\rm ex}$), and the molecular column density  ($N$)  or the fractional abundance ($f$). Clearly, none of them have provided all of these parameters. Therefore, a detailed complete study of the 3\,mm molecules in this famous source is highly desirable. Here, we find that the $J=13-12$ transition of H$^{13}$CCCN is detected in IRC\,+10216 for the first time.

\subsection{Typical spectral line profiles}

% Detailed description of the molecular line profiles detected in our survey and their comparison with previous observations is presented in this section.

\begin{figure*}
\gridline{\fig{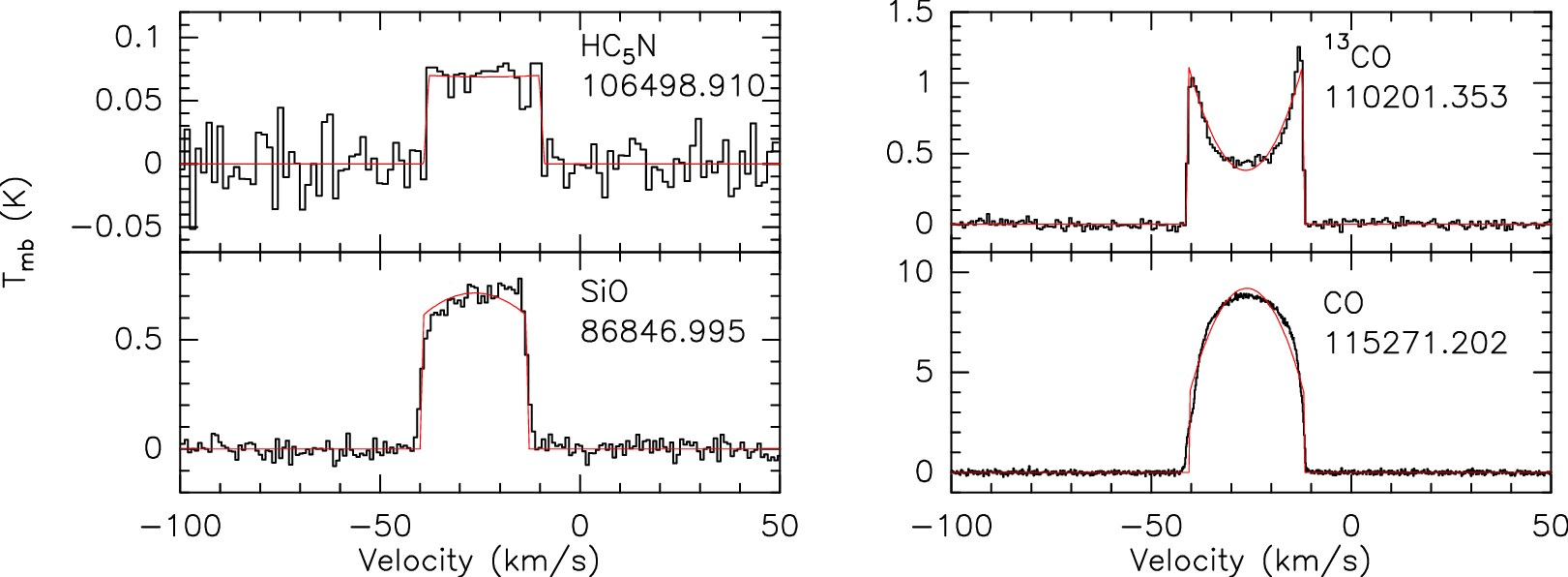}{0.86\textwidth}{ }
          }
\caption{Illustrations of the four typical shapes of the spectral line profiles observed in IRC\,+10216: flat-topped (e.g. HC$_5$H, optically thin, spatially unresolved), double-peaked (e.g. $^{13}$CO, optically thin, spatially resolved), truncated parabolic (e.g. SiO, optically thick, spatially resolved), and parabolic (e.g. CO, optically thick, spatially unresolved). The solid black line shows the observed spectrum, and the solid red line represents the spectral line profile fitted by the shell function. The rest frequency of separable spectral lines are in MHz, and the molecule's name is displayed in the upper right corner of each panel.
All of the detected spectra from this survey are included in Appendix \ref{Line spectrum}.
}
\label{fig:S1-HC5N+SiO+12CO+13CO}
\end{figure*}

Typically, in the CSEs of AGB stars, there are four types of molecular spectral line profiles:  double-peaked, parabolic, flat topped, and truncated parabolic line shapes (see Figure~\ref{fig:S1-HC5N+SiO+12CO+13CO}). The different shapes of the observed spectral lines are mainly determined by the optical depth of the transition and the ratio of the angular area of the emitting region to the beam size of the telescope \citep{Zuckerman+etal+1987}. Optically thin unresolved and resolved emission lines show flat-topped and double-peaked profiles, while optically thick unresolved and resolved emission lines have parabola and truncated parabola profiles. \citep{Morris+etal+1975, Zuckerman+etal+1987, Habing+etal+2004}.

The profiles of the C$_2$H \citep{Johansson+etal+1984}, $l$-C$_3$H, C$_4$H\, (${\nu=0}$) \citep{Olofsson+etal+1982, Johansson+etal+1984}, C$_4$H\,(${\nu=1}$), C$_3$N \citep{Johansson+etal+1984}, $c$-C$_3$H$_2$ \citep{Zhang+etal+2017}, $^{13}$CO, H$^{13}$CCCN \citep{Zhang+etal+2017}, HC$_5$N \citep{Gong+etal+2015, Zhang+etal+2017}, MgNC, CH$_3$CN, HNC, SiC$_2$, $^{13}$CS, C$^{34}$S \citep{Groesbeck+etal+1994}, C$_2$S and C$_3$S emission lines appear to be double-peaked or/and flat-topped, indicating that they are optically thin. The profiles of the HCN, H$^{13}$CN, and CO emission lines appear to be parabolic, while the CS line is a truncated parabola profile, which implies that they are all optically thick.

Interestingly, the CN\,$N  = 1-0$, $J  = 1/2-1/2$, $F= 1/2-1/2$ profile is double-peaked, while other $N =1-0$, $J=1/2-1/2$ hyperfine transitions of CN ($F=1/2-3/2$, $F=3/2-1/2$, $F=3/2-3/2$) show truncated parabolic profiles (see Figure~\ref{fig:CN}). \cite{Groesbeck+etal+1994} find that CN ($N=3-2$) lines have prominent double-peaked profiles indicative of low optical depth.

\cite{Zhang+etal+2017} found that HC$_3$N produced a double-peaked profile and was rated to be optically thin. However, we find that the spectral lines of HC$_3$N are truncated parabola (see Figure~\ref{fig:HC3N}), which indicates that HC$_3$N is at least partially saturated \citep{Olofsson+etal+1982}.

Our fitting results for SiO  (see Figure~\ref{fig:S1-HC5N+SiO+12CO+13CO}) showed a truncated  parabolic shape, indicating that the line is slightly optically thick. $^{29}$SiO shows a double-peaked structure (see Figure~\ref{fig:HCN+HNC+29SiO}), which indicates that this molecule is optically thin. Judging from the average profile of the $^{29}$SiO and $^{30}$SiO lines, \cite{Kahane+etal+1988} pointed out that the $J=2-1$ rotational transition of SiO seems to be optically thick. However, \cite{Olofsson+etal+1982} noted that the SiO $J=2-1$ transition has a flat-topped appearance, albeit that the red-shifted emission is slightly stronger than the blue-shifted emission and resembles the typical rectangular line profile of an  unresolved, optically thin molecular shell with a constant velocity gas flow. On the other hand, the  spectral lines of the SiO $J=2-1$ to $J=8-7$ transitions observed by \cite{Agundez+etal+2012} have a slight double-peaked structure, but the spectral line profiles simulated by them using radiative transfer are parabolic. This may indicate that the envelope of SiO is slightly unresolved \citep{Agundez+etal+2012}, or more likely that an enhanced density shell should be considered in the IRC\,+10216 model  (see \cite{Beck+etal+2012} and \cite{Daniel+etal+2012}), or that more infrared pumping should be added in the model \citep{Agundez+etal+2012}.

The SiS\,$J  = 6-5$ spectrum (see Figure.~\ref{fig:SiS+SiC2})   shows an optically thick emission with a parabolic shape. This result is consistent with observations by \cite{Johansson+etal+1984} and \cite{Fonfria+etal+2018}. \cite{Johansson+etal+1984} and \cite{Olofsson+etal+1982} speculated that the SiS\,$J=5-4$ double-peaked profile might be due to maser enhanced line wings. \cite{Fonfria+etal+2018} find through monitoring of the SiS\,$J=5-4$ and SiS\,$J=6-5$ lines in IRC\,+10216 that the SiS\,$J=5-4$ spectral line shows narrow peaks at high expansion velocities, which vary strongly with the pulsation of the star, while the remaining line profile is approximately constant  over the stellar pulsation period. On the contrary, SiS\,$J=6-5$ shows neither narrow peaks nor a constant profile over the stellar period at high expansion velocities.

Apparently the line shape of a certain transition of a molecule could be different when being observed by different telescopes. We compared and summarized all of the line profiles related to this work in Appendix~\ref{comparisons of line shapes}.

\section{Discussion}
\label{sec:DISCUSSION}
\subsection{Rotation-diagram Analysis}
\subsubsection{Molecular excitation temperature and column density}
\label{sec:Excitation temperature and column density}
Assuming local thermal equilibrium  (LTE), the column densities  ($N_{\rm{tot}}$)   and excitation temperatures of detected lines were derived by using the following equation:
\begin{equation}
  \ln\frac{N_{\rm u}}{g_{\rm u}}=\ln\frac{3kW}{8\pi^3{\nu}S\mu^2}  = \ln\frac{N_{\rm{tot}}}{Q (T_{\rm{ex}})}-\frac{E_{\rm u}}{kT_{\rm{ex}}}
\label{eq:1},
\end{equation}
where, $N_{\rm u}$ and $g_{\rm u}$ are the population and statistical weights of the upper level, $k$ is the Boltzmann constant, $W=\int T_{\rm R}dv$ is the integral of the source brightness temperature ($T_{\rm R}$) over the line's velocity range, $\nu$ is the rest frequency of the line, $S$ is the transition's intrinsic strength, $\mu$ is the permanent dipole moment, and $Q$ is the partition function, which is  related to $T_{\rm ex}$, and $E_{\rm u}$, the excitation energy of the upper level. The values of $Q$, $E_{\rm u}$/${k}$ and $S\mu^{2}$ are taken from the ``Splatalogue" spectral line catalogs.

In actual astronomical measurements, the observed molecules often do not fill the telescope beam, so it is necessary to consider the correction of the main beam brightness temperature and source brightness temperature by using the beam dilution correction of the molecular source (${\eta}_{\rm BD}$).

The relationship between the main beam brightness temperature and the source brightness temperature is as follows:
\begin{equation}
  T_{\rm R}  = \frac{T_{\rm mb}}{{\eta}_{\rm BD}}
\label{eq:2},
\end{equation}
where
\begin{equation}
  {\eta}_{\rm BD}  = \frac{\theta_{\rm s}^{2}}{\theta_{s}^{2}+\theta_{\rm beam}^{2}}
\label{eq:3}.
\end{equation}
Here, ${\eta}_{\rm BD}$ is the beam dilution factor or beam filling factor, $\theta_{\rm beam}$ is the half power beam width of the antenna (HPBW), and $\theta_{\rm s}$ is the source size.
In our calculations, ${\eta}_{\rm BD}$ of all the identified unblended molecular lines and the supplemented data are calculated via equation (\ref{eq:3}), while data from the blended lines are excluded.
When a molecule's excitation temperature is obtained using rotational diagrams, it is required that at least two unblended transition lines of the same molecule, with significant upper energy difference have been observed. Unfortunately, this does not hold for all our detections. We only
observed one line in case of CS, C$^{34}$S, $^{13}$CS, C$_3$S, CO, $^{13}$CO, SiO, $^{29}$SiS, Si$^{34}$S, MgNC, and $c$-C$_3$H$_{2}$. For molecules like HNC, HCN, H$^{13}$CN, and CH$_3$CN, we only observed the hyperfine structure of one rotational transition. Because the frequency interval of these fine structure lines is too small, their spectral lines are blended. For CN,  SiC$_2$, $l$-C$_3$H, C$_3$N, H$^{13}$CCCN, and C$_2$H, at least two transitions have been observed, but the energy level spacing of these lines is minimal. It has been reported that the excitation temperatures of HC$_3$N and HC$_5$N are related to their energy level transitions \citep{Bell+etal+1992}. Therefore, to calculate these molecules' excitation temperature and column density, we supplemented our data with spectral line surveys of other research groups. The additional frequency ranges with observational data includes 13.3 to 18.5\,GHz \citep{Zhang+etal+2017}, 17.8 to 26.3\,GHz \citep{Gong+etal+2015}, 28 to 50\,GHz \citep{Kawaguchi+etal+1995}, 129.0 to 172.5\,GHz \citep{Cernicharo+etal+2000}, 131.2 to 160.3\,GHz \citep{He+etal+2008}, 219.5 to 267.5\,GHz \citep{He+etal+2008}, and 330 to 358\,GHz \citep{Groesbeck+etal+1994}. In addition, supplementary data for HNC come from \cite{Daniel+etal+2012}. For CS, $^{13}$CS, C$^{34}$S, SiO, $^{29}$SiO, SiS, $^{29}$SiS, and Si$^{34}$S, we have also supplemented the data using observations of \cite{Agundez+etal+2012}. Data for HCN from \cite{Nyman+etal+1993} are also included. We processed the data of \cite{Groesbeck+etal+1994}, \cite{Kawaguchi+etal+1995}, \cite{He+etal+2008}, \cite{Gong+etal+2015}, and \cite{Zhang+etal+2017}, calculating the uncertainty of the integrated intensity as the product of the rms noise and the line width. For other supplementary data, we adopted the integrated intensity and uncertainty values as reported in the original literature.
%{\color{green}We have considered the influence of main beam efficiency and beam dilution correction for all the supplementary data}.}
%% xiaohu:
%% Moreover, different transitions of the same molecule have different source sizes \citep{He+etal+2008}. {\color{green} We neglected this factor in the calculations of the molecules' excitation temperature and column density}  given the lack of available high-resolution data and detailed radiative transfer modelling, including radiative pumping, which could allow one to determine the radial extent of the observed transitions.
%%
To derive the excitation temperature and column density of a molecule, one needs to know the spatial  distribution of the molecular source or the source size. \cite{Agundez+etal+2017} find that the observational results of HC$_5$N ($J=32-31$ to $J=43-42$) with ALMA are not showing a radial shift in the emission peak between low-J and high-J lines as might reasonably be expected, and all emission distributions of HC$_5$N appear quite similar. The same is found for HC$_3$N, C$_4$H, and C$_3$N in the $\lambda$ 3\,mm  band \citep{Agundez+etal+2017}. The source size employed in this work is taken from previous high resolution maps in the $\lambda$ 3\,mm  band of the relevant molecule. If such a study does not exist, the source size is obtained from chemically related species \citep{He+etal+2008, Gong+etal+2015}. The employed sizes of the  identified molecules, $\theta_s$, and the references are listed in Table~\ref{table:5}.

%{\color{blue} Therefore, the source sizes of HC$_3$N, HC$_5$N, $l$-C$_3$H, C$_4$H, CH$_3$CN, and MgNC are 30$^{\prime\prime}$ \citep{Keller+etal+2015, Agundez+etal+2017, Agundez+etal+2015, Guelin+etal+1993}, source size of C$_2$H is 32$^{\prime\prime}$ \citep{Guelin+etal+1999}, source size of C$_3$N is 36$^{\prime\prime}$ \citep{Bieging+etal+1993}, source size of SiC$_2$ is 27$^{\prime\prime}$ \citep{Lucas+etal+1995}, source size of $^{12}$CO is 200$^{\prime\prime}$ \citep{Fong+etal+2006}, source sizes of HCN, H$^{13}$CN and CN are 64$^{\prime\prime}$, 56$^{\prime\prime}$ and 100$^{\prime\prime}$, respectively \citep{Dayal+etal+1995}. Source size of HNC is 40$^{\prime\prime}$ \citep{Guelin+etal+1997}. Source sizes of SiS, $^{29}$SiS, Si$^{34}$S, SiO, and $^{29}$SiO are 22$^{\prime\prime}$, and of CS, $^{13}$CS, and C$^{34}$S are 40$^{\prime\prime}$ \citep{Velilla-Prieto+etal+2019}. For the H$^{13}$CCCN molecule, we assumed the same size as its main isotopomer, which is 30$^{\prime\prime}$ \citep{He+etal+2008}. The source size of $c$-C$_3$H$_2$ is taken to be the same as that of C$_4$H, which is 30$^{\prime\prime}$ \citep{Gong+etal+2015}. We assume that $^{13}$CO and $^{12}$CO have the same size. The source size of C$_2$S and C$_3$S are taken to be the same as that of CS, $^{13}$CS, and C$^{34}$S, which are 40$^{\prime\prime}$ \citep{Velilla-Prieto+etal+2019}.}

We used the least squares method to perform the linear fitting for the rotational diagrams of these molecules and their isotopes. The fitting results are shown in Figs:~\ref{fig:T1}, ~\ref{fig:T2}, ~\ref{fig:T3}, ~\ref{fig:T4}, and ~\ref{fig:T5}. According to the rotational diagrams, it can be seen that most spectra have large dispersion, especially CS, $^{13}$CS, and C$^{34}$S. The derived excitation temperatures, column densities, and the molecular fractional abundances relative to H$_{2}$  (see Section~\ref{sec:Molecular fractional abundances})   together with observational results from the literature (including line surveys in the 18, 13, 8, 2, and 1.3\,mm bands), are summarized in Table~\ref{table:5}. The excitation temperatures of the molecules range from 5.3 to $73.4\,$K and their column densities range from 3.27 $\times$ 10$^{13}$ to 4.15 $\times$ 10$^{17}$ cm$^{-2}$.

Compared to observational results in the literature, the molecules with large differences in excitation temperatures include $l$-C$_3$H, C$_4$H  ($\nu=0$), C$_4$H  ($\nu=1$), HC$_5$N, $c$-C$_3$H$_2$, C$_2$S, and CH$_3$CN. We also find that column densities, except for C$_4$H, HCN, and CH$_3$CN, agree with previously published observational results.

The excitation temperatures of $l$-C$_3$H in the $^{2}\Pi_{1/2}$ and $^{2}\Pi_{3/2}$ states are 8.5 K \citep{Thaddeus+etal+1985, Kawaguchi+etal+1995} and 52 K \citep{Thaddeus+etal+1985}, respectively. \cite{Thaddeus+etal+1985} pointed out that the low excitation temperature (8.5 K) of the $^{2}\Pi_{1/2}$ state of $l$-C$_3$H is due to rapid radiative decay, while the high excitation temperature  (52 K)   of the $ ^{2}\Pi_{3/2}$ state is a consequence of the fact that interconnecting far-IR radiative transitions are only weakly permitted. When we adopt an $l$-C$_3$H diameter of $ 30^{\prime\prime}$, the excitation temperatures of the $^{2}\Pi_{1/2}$ and $ ^{2}\Pi_{3/2}$ states of $l$-C$_3$H are 27.4 $\pm$ 9.3 and  14.2 $\pm$ 1.7\,K, respectively. %It is possible to use the temperature between the two spin states  ($ ^{2}\Pi_{1/2}$ and $^{2}\Pi_{3/2}$ states) of $l$-C$_3$H to measure the kinetic temperature of the region in which this radical exists since the cross-ladder transitions between $^{2}\Pi_{1/2}$ and $^{2}\Pi_{3/2}$ in $l$-C$_3$H have a tiny transition dipole moment \citep{Kawaguchi+etal+1995, Kalluru+etal+2022, Ditchburn+etal+1976}.

The excitation temperature of C$_4$H ($\nu=0$) obtained in our work is 45.9 $\pm$ 3.1\,K, and the excitation temperature of C$_4$H deduced by \cite{Avery+etal+1992} is 48 $\pm$ 4\,K. Comparing the excitation temperatures and column densities in Table~\ref{table:5}, we find that the excitation temperatures derived by others are lower than those derived by us except for that of \cite{Avery+etal+1992} and \cite{He+etal+2008}. For \cite{Kawaguchi+etal+1995}, and \cite{Gong+etal+2015}, the difference even amounts to a factor of two. \cite{Pardo+etal+2022} derive both the lowest excitation temperature, 4.9 $\pm$ 0.6\,K, and column density, (1.84 $\pm$ 0.25) $\times$ $10^{14}$\,cm$^{-2}$, the latter about an order of magnitude lower than the other values. The excitation temperatures and column densities of the vibrationally excited C$_4$H ($\nu=1$) derived by us and \cite{He+etal+2008} are obviously much larger than those derived by \cite{Pardo+etal+2022}. In contrast, \cite{Pardo+etal+2022} observed low-J lines of C$_4$H  ($\nu=1$),  which trace the cold component of the gas (see Appendix~\ref{Appendix Comparison of rotational diagrams} for a detailed discussion), resulting in a lower excitation temperature. Although they also used a rotational diagram to derive the excitation temperature and column density of C$_4$H ($\nu=1$), they did not consider the effect of the beam dilution factor. The large excitation temperatures and column densities derived for the $\nu=0$ and $\nu=1$ states of C$_4$H indicate that infrared pumping could be very efficient \citep{Avery+etal+1992, He+etal+2008}.

\startlongtable
\begin{deluxetable*}{cccccccccccccc}
%\tablenum{5}
\tablecaption{A summary of column density ($N$), excitation temperature ($T_{\rm ex}$), source size, ($\theta_s$) and fractional abundance relative to H$_{2}$ ($f = N/N_{\rm H_{2}} $), of the detected species in this work and the literature.
The derivation of the fractional abundance is discussed in Sec~\ref{sec:Molecular fractional abundances}. \label{table:5}}
\tablewidth{0pt}
\tablehead{
  &  \multicolumn{2}{c}{This work} &  &  \multicolumn{10}{c}{Lit. observational}   \\
\cline{2-5} \cline{7-8}
\colhead{Molecule} & \colhead{$T_{\rm ex}$} & \colhead{$N$} & \colhead{$f$} & \colhead{$\theta_{s}$} & \colhead{Ref. } & \colhead{$T_{\rm ex}$}  & \colhead{$N$} & \colhead{Ref.}\\
        & \colhead{(K)}  & \colhead{(cm$^{-2}$)} &      & \colhead{($^{\prime\prime}$)} & \colhead{(for ${\theta_{s}}$)} & \colhead{(K)}  &\colhead{ (cm$^{-2}$)} &   \colhead{(for ${T_{\rm ex},  N}$)}}
\startdata
C$_2$H	    & 17.0 (2.0) & 8.14 (2.29) $\times$ $10^{15}$    & 3.37 $\times$ $10^{-6}$      & 32 & 1 & 16         & 4.6 $\times$ $10^{15}$         & 11  \\
$l$-C$_3$H  & 14.3 (0.9) & 1.25 (0.22) $\times$ $10^{14}$    & 4.84 $\times$ $10^{-8}$      & 30 & 2 & 8.5        & 5.6 $\times$ $10^{13}$         & 11  \\
 	        &            & 	                                 &                              &    &   & 25 (5)     & 1.32 (0.35) $\times$ $10^{14}$ & 12  \\
C$_4$H ($\nu=0$)& 45.9 (3.1)& 2.75 (0.36) $\times$ $10^{15}$ & 1.07 $\times$ $10^{-6}$      & 30 & 2 & 18.5 (7.6) & 2.4 (0.2)   $\times$ $10^{15}$ & 13 \\
 	        &            & 	                                 &                              &    &   & 15         & 2.4 $\times$ $10^{15}$         & 11   \\
 	        &            & 	                                 &                              &    &   & 4.9 (0.6)  & 1.84 (0.25)   $\times$ $10^{14}$& 14  \\
 	        &            & 	                                 &                              &    &   & 53 (3)     & 8.1 (1.1)   $\times$ $10^{15}$  & 12   \\
C$_4$H ($\nu=1$)& 64.3 (2.8)& 3.66 (0.74) $\times$ $10^{16}$ & 1.42 $\times$ $10^{-5}$      & 30 & 3 & 53 (4) & 5.9 (2.1)   $\times$ $10^{15}$  & 12   \\
             &           &                                    &                             &    &   & 10 (5) & 6.6 (2.0)   $\times$ $10^{13}$  & 14  \\
CN	         & 9.8 (2.0) & $>$ 1.08 (0.42) $\times$ $10^{15}$ & $>$ 1.4 $\times$ $10^{-6}$  & 100 & 4 & 8.7   & 6.2 $\times$ $10^{14}$               & 15   \\
C$_3$N	     & 27.6 (1.9)& 4.25 (0.81) $\times$ $10^{14}$   & 1.98 $\times$ $10^{-7}$       & 36 & 5 & 20.2 (1.1) & 3.1 (0.3)   $\times$ $10^{14}$   & 13   \\
 	         &           & 	                                 &                              &    &   & 15         & 4.1 $\times$ $10^{14}$           & 11   \\
HC$_3$N      & 21.2 (0.5) & $>$ 1.18 (0.08) $\times$ $10^{15}$ & $>$ 4.59 $\times$ $10^{-7}$ & 30 & 6 & 27.2 (19.4)& 1.94 (0.21)   $\times$ $10^{15}$ & 16 \\	
 	         &            & 	                             &                              &    &   & 24.7 (18.5)& 1.4 (0.2)   $\times$ $10^{15}$   & 13   \\
 	         &            & 	                             &                              &    &   & 26         & 1.7 $\times$ $10^{15}$           & 11   \\
 	         &            & 	                             &                              &    &   & 28 (1)     & 8.0 (1.5)   $\times$ $10^{14}$   & 12   \\
H$^{13}$CCCN & 15.9 (0.9) & 4.38 (0.88) $\times$ $10^{13}$   & 1.70 $\times$ $10^{-8}$      & 30 & 3 & 18.7 (3.3) & 3.25 (1.46)   $\times$ $10^{13}$ & 16 \\
 	         &            & 	                             &                              &    &   & 27         & 2 $\times$ $10^{13}$             & 12    \\
HC$_5$N      & 23.7 (0.5) & 5.41 (0.56) $\times$ $10^{14}$   & 2.10 $\times$ $10^{-7}$     & 30 & 6 & 13.6 (2.1) & 5.47 (0.77)   $\times$ $10^{14}$ & 16 \\
 	         &            & 	                             &                             &    &   & 18.8 (1.3) & 4.6 (0.2)   $\times$ $10^{14}$   & 13   \\
 	         &            & 	                             &                             &    &   & 27 (5)     & 2.7 (0.2)   $\times$ $10^{14}$   & 11  \\
 	         &            & 	                             &                             &    &   & 10.1 (0.6) & 4.2 (0.4)   $\times$ $10^{14}$   & 14   \\
HNC          & 12.1 (0.1) & 1.11 (0.08)   $\times$ $10^{14}$   & 5.76 $\times$ $10^{-8}$     & 40 & 7 &          &                                       \\
HCN          & 16.0 (1.4) & $>$ 7.36 (0.92) $\times$ $10^{14}$ & $>$ 6.10 $\times$ $10^{-7}$ & 64 & 4 &          & 2.8 $\times$ $10^{16}$        &15     \\	
H$^{13}$CN   & 20.8 (5.5) & $>$ 3.92 (1.17) $\times$ $10^{14}$ & $>$ 2.84 $\times$ $10^{-7}$ & 56 & 4 &          & 2.8 $\times$ $10^{14}$        &15      \\		
CO	         & 31.9 (10.9) & $>$ 4.03 (1.51) $\times$ $10^{17}$ & $>$ 1.04 $\times$ $10^{-3}$ & 200 & 8 &         & 2.0 $\times$ $10^{18}$        &15      \\
$^{13}$CO	 & 30.2 (5.4)  & 4.12 (0.86) $\times$ $10^{16}$    & 1.07 $\times$ $10^{-4}$     & 200 & 3 & 17       & 5.0 $\times$ $10^{16}$        &15    \\
c-C$_3$H$_2$ (ortho) & 5.3 (0.2) & 1.14 (0.15) $\times$ $10^{14}$ & 4.44 $\times$ $10^{-8}$  & 30  & 3 & 5.7 (1.4) & 4.82 (4.14) $\times$ $10^{14}$ &12    \\
c-C$_3$H$_2$ (para)      & 5.7 (0.9) & 8.02 (4.54) $\times$ $10^{13}$ & 3.11 $\times$ $10^{-8}$     & 30  & 3 & 8 (6)     & 1.22 (2.03) $\times$ $10^{14}$ &12    \\
 	        &            & 	                                  &                              &     &   & 6.1 (2.8) & 1.39 (1.07) $\times$ $10^{14}$ & 16 \\
 	        &            & 	                                  &                              &     &   & 5.5 (0.6) & 2.1 (0.4) $\times$ $10^{14}$   &13    \\
  	        &            & 	                                  &                            &    &   & 20         & 1.6 $\times$ $10^{13}$        &11    \\
SiO         & 29.2 (3.9) & $>$ 4.84 (1.13) $\times$ $10^{14}$ & $>$ 1.38 $\times$ $10^{-7}$ & 22 & 9 & 29         & 5.4 $\times$ $10^{14}$        &11    \\
            &            & 	                                  &                              &    &   & 29         & 2.1 $\times$ $10^{14}$        &15    \\
$^{29}$SiO  & 27.7 (3.5) & 3.95 (0.91) $\times$ $10^{13}$     & 1.13 $\times$ $10^{-8}$      & 22 & 9 & 29         & 2.5 $\times$ $10^{13}$        &11    \\
 	        &            & 	                                  &                              &    &   & 29         & 1.9 $\times$ $10^{13}$        &15    \\
$^{29}$SiS	& 64.8 (4.6) & 4.39 (0.75) $\times$ $10^{14}$     & 1.25 $\times$ $10^{-7}$     & 22 & 9 & 85         & 2.1 $\times$ $10^{14}$        &15    \\
Si$^{34}$S	& 72.5 (5.2) & 3.35 (0.59) $\times$ $10^{14}$     & 9.54 $\times$ $10^{-8}$     & 22 & 9 & 79         & 1.8 $\times$ $10^{14}$        &15    \\
SiS         & 73.4 (11.3)& $>$ 5.86 (1.74) $\times$ $10^{15}$ & $>$ 1.70 $\times$ $10^{-6}$ & 22 & 9 & 70         & 4.7 $\times$ $10^{15}$        &11    \\
            &            & 	                                  &                              &    &   & 70           & 3 $\times$ $10^{15}$          &15    \\
CS	        & 22.9 (4.3) & $>$ 1.57 (0.46) $\times$ $10^{15}$ & $>$ 8.12 $\times$ $10^{-7}$ & 40 & 9 & 28          & 5.9 $\times$ $10^{15}$        &11       \\	
            &            &                                   &                              &    &    &             & 3.0 $\times$ $10^{15}$        &15      \\
$^{13}$CS	& 51.4 (15.8)& 8.09 (2.80) $\times$ $10^{13}$   & 4.19 $\times$ $10^{-8}$       & 40 & 9  &             &                                      \\	
C$^{34}$S	& 26.1 (3.3) & 1.47 (0.26) $\times$ $10^{14}$   & 7.63 $\times$ $10^{-8}$       & 40 & 9  & 28          & 1.5 $\times$ $10^{14}$         &15    \\	
C$_2$S	    & 11.2 (1.2) & 1.59 (0.73) $\times$ $10^{14}$   & 8.22 $\times$ $10^{-8}$       & 40 & 3  & 39.2 (135.3)& 2.3 (0.8)   $\times$ $10^{14}$ &13   \\
 	        &            & 	                                &                               &    &    & 14          & 1.5 $\times$ $10^{14}$         & 11   \\
 	        &            & 	                                &                               &    &    & 7.9 (1.1)   & 4.0 (0.7)   $\times$ $10^{13}$   &14    \\
C$_3$S      & 34.0 (1.2) & 3.27 (0.16) $\times$ $10^{13}$   & 1.70 $\times$ $10^{-8}$       & 40 & 3  & 20.4 (13.7) & 2.2 (0.4)   $\times$ $10^{13}$   &13    \\
 	        &            & 	                                &                               &    &    & 33          & 4.9 $\times$ $10^{13}$        &11   \\
 	        &            & 	                                &                               &    &    & 23.6 (96.85)& 2.58 (3.60)   $\times$ $10^{13}$ &14    \\
MgNC	    & 18.9 (1.2) & 3.20 (0.51) $\times$ $10^{13}$   & 1.24 $\times$ $10^{-8}$       & 30 & 2  & 25.8 (54.9) & 3.9 (1.5)   $\times$ $10^{13}$    &13    \\	
 	        &            & 	                                &                               &    &    & 15          & 2.5 $\times$ $10^{13}$         &11    \\
 	        &            & 	                                &                               &    &    & 8.6         & 7.8 $\times$ $10^{13}$         &12    \\
CH$_3$CN	& 37.4 (6.4) & 6.81 (3.51) $\times$ $10^{13}$   & 2.64 $\times$ $10^{-8}$       & 30 & 3  & 16          & 6 $\times$ $10^{12}$           &11, 17     \\
 	        &            & 	                                &                               &    &    & 35 (4)      & 2.6  (0.6)   $\times$ $10^{13}$   &12    \\
SiC$_2$	    & 34.9 (5.5) & 1.41 (0.34) $\times$ $10^{15}$   & 4.92 $\times$ $10^{-7}$       & 27 & 10 & 40.1 (36.2) & 1.87 (1.34)   $\times$ $10^{15}$  & 16   \\
 	        &            & 	                                &                               &    &    & 31.8 (0.9)  & 1.2 (0.0)   $\times$ $10^{15}$    &13    \\
\enddata
\tablecomments{
The values in parentheses in this table are the 1 $\sigma$ errors associated with the fits to our rotational diagrams. The corresponding references are: (1) \cite{Guelin+etal+1999};  (2)   \cite{Guelin+etal+1993};  (3) The source size assumed to be the same as its main isotopomer or obtained from chemically related species  \citep{He+etal+2008, Gong+etal+2015};  (4) \cite{Dayal+etal+1995};  (5) \cite{Bieging+etal+1993};  (6) \cite{Keller+etal+2015};  (7) \cite{Guelin+etal+1997};  (8)   \cite{Fong+etal+2006};  (9) \cite{Velilla-Prieto+etal+2019};  (10) \cite{Lucas+etal+1995};  (11)   \cite{Kawaguchi+etal+1995};  (12) \cite{He+etal+2008};  (13) \cite{Gong+etal+2015};  (14) \cite{Pardo+etal+2022}; (15) \cite{Groesbeck+etal+1994};  (16) \cite{Zhang+etal+2017}; (17) \cite{Guelin+etal+1991}.}
\end{deluxetable*}

The excitation temperature of HC$_5$N obtained by our rotational diagram analysis is close to that of \cite{Kawaguchi+etal+1995}, but more than twice that of \cite{Pardo+etal+2022}. In fact, the excitation temperatures of HC$_5$N are stratified (see Appendix~\ref{Appendix Comparison of rotational diagrams} for a detailed discussion). The excitation temperature of HC$_5$N obtained by us is the result of fitting multiple rotational transitions, including low and high rotational transitions, which may result in our excitation temperature being close to the one from \cite{Kawaguchi+etal+1995}. \cite{Kawaguchi+etal+1995} derived a higher excitation temperature than \cite{Pardo+etal+2022}, likely because the latter did not consider the beam dilution factor. \cite{Bell+etal+1992} pointed out that the moderately high excitation temperature of HC$_5$N, 25\,K, is probably due to the influence of infrared excitation. Warm HC$_3$N ($T_{\rm ex}=48$ K) and warm HC$_5$N ($T_{\rm ex}=35$ K) in the CSE of IRC\,+10216 may indicate that the excitation is almost entirely due to infrared radiation \citep{Bell+etal+1993b}.

For $^{13}$CO, the excitation temperature and column density obtained by us are consistent with the results of \cite{Groesbeck+etal+1994}. The column density of CO with opacity corrections obtained by \cite{Groesbeck+etal+1994} is about five times larger than our value.  Our value of $N$ (CO) in Table~\ref{table:5} is derived from a rotation diagram and is a lower limit as CO is optically thick.

%They derived their value by first using the rotation diagram method to obtain $T_{\rm ex}$, and} $N$ ($^{13}$CO) and using a [$^{12}$C]/[$^{13}$C] ratio of 44 $\pm$ 3} to derive $N (\rm CO)=1.8 \times$ 10$^{18}$ or 2.2 $\times$ 10$^{18}$ cm$^{-2}$. Here, the value of 44 $\pm$ 3 is obtained from observations of optically thin, isotopic lines of SiC$_2$ \citep{Cernicharo+etal+1991} and CS \citep{Kahane+etal+1988}. If we use $^{13}$CO and a [$^{12}$C]/[$^{13}$C] ratio of 44 $\pm$ 3, we found the column density of CO is 1.82 $\times$ 10$^{18}$ cm$^{-2}$, consistent with the results of \cite{Groesbeck+etal+1994}.

%{\color{blue}At face value, these are huge values for $N$ (12CO) and $N$ (13CO)  -the 12CO fractional abundance is above 10$^{-3}$. However, it would be good to compare this to Fronfrias new value for $X$ (CO) =   (6.7 $\pm$ 1.4)   $\times$ 10$^{-4}$ using their new H$_2$ observations. So you are close to their values.  QUESTION  how does your beam-averaged of N (H2)   used to determine X (12CO) compare with Fronfrias value for N (H$_2$)  ?  Since you are using a large 200 source size, and Fronfria is detecting absorption of H2, the values should, in principle be close.}

Our HCN column density is much different from that calculated by \cite{Groesbeck+etal+1994}. Our column density of H$^{13}$CN is slightly larger than their value, but our value for $N$ (HCN) is only 2.6\% of theirs. They derived their value by first using the rotation diagram method  for HC$^{15}$N to obtain $T_{\rm ex}$ and $N$ (HC$^{15}$N). Using a [$^{12}$C]/[$^{13}$C] ratio of 44 \citep{Kahane+etal+1988, Cernicharo+etal+1991} and a [$^{14}$N]/[$^{15}$N] = 4000 \citep{Kahane+etal+1988}, they derived $N$ (HCN) and $N$ (H$^{13}$CN). We detected HC$^{15}$N with a S/N of less than three, so our observations of HCN and H$^{13}$CN, which are both optically thick, give lower limits to their column densities in Table~\ref{table:5}. Similarly, using the rotation diagram method, \cite{Qiu+etal+2022} derived the column density of HCN in C-rich post-AGB star CRL 2688, and they find that this molecule exhibits a much lower observed abundance, which could be attributed to the effect of optical depth.

For CS, \cite{Groesbeck+etal+1994} used the same isotopic rotational diagram method to derive its column density using observations of C$^{34}$S and a [$^{32}$S]/[$^{34}$S] ratio of 20.2, derived by \cite{Kahane+etal+1988} from Si$^{32}$S and Si$^{34}$S. We used the same method and our determination of $N$ (C$^{34}$S)  to derive $N (\rm CS)=3.0 \times$ 10$^{15}$ cm$^{-2}$, agreeing with the value derived by \cite{Groesbeck+etal+1994}. The excitation temperature of C$_2$S derived by us is in good agreement with that obtained by \cite{Kawaguchi+etal+1995}, but the excitation temperature derived by \cite{Gong+etal+2015} is three times that of ours. In the work of \cite{Gong+etal+2015}, the uncertainty in the C$_2$S excitation temperature is large because the C$_2$S transitions only cover a narrow energy range. We find that the excitation temperature of C$_3$S is three times that of C$_2$S, \cite{Agundez+etal+2014} find that the rotational temperature of C$_3$S is more than twice that of C$_2$S, which points to differences in the excitation and may indicate that radiative pumping to vibrationally excited states plays an important role in the case of C$_3$S \citep{Agundez+etal+2008}.

In thermal equilibrium, ortho-to-para ratios (OPRs) for species containing two protons with parallel and anti-parallel nuclear spins range from their statistical weight limit of 3:1 at high temperatures (T\,${\geq}$ 30\,K) to infinity or zero as the temperature decreases, depending on the symmetry of the rotational and electronic wave functions \citep{Decin+etal+2010, Khouri+etal+2014, LeGal+etal+2017, Chapovsky+etal+2021, Tsuge+etal+2021}. The statistical OPR of $c$-C$_3$H$_2$ derived by \cite{He+etal+2008} is approximately 4 (3.95 $\pm$ 3.16) while we derived a ratio of 1.42 $\pm$ 0.58, consistent with the result estimated by \cite{Gong+etal+2015}. However, compared to \cite{He+etal+2008} and \cite{Gong+etal+2015}, we have collected and employed more data, and the excitation temperatures (less than 10\,K) of para and ortho-$c$-C$_3$H$_2$ are fitted independently. The amounts of ortho-$c$-C$_3$H$_2$ and para-$c$-C$_3$H$_2$ formed are influenced by the OPR of H$_2$ and by ion-neutral reactions of C$_3$H$_2$ with H$^+$ and H$_3^+$ that can exchange protons with it. Both \cite{Park+etal+2006} and \cite{Morisawa+etal+2006} have observed and modeled the OPR of $c$-C$_3$H$_2$ in the dark cloud TMC-1. The latter authors show that this OPR is dependent on the age of the cloud as abundances of species change with time. The situation in IRC\,+10216 is in part simpler since the OPR in H$_2$ is likely to be 3 since it forms at high temperature and its flow time through the envelope is too short to change it to any extent. On the other hand, the changing physical conditions of temperature and density, ion abundances, and collision times in the outflow add complexity to the state-selective chemistry that ultimately sets the OPR of molecules such as $c$-C$_3$H$_2$ that form in the outer envelope. A detailed study of the ortho-para chemistry of $c$-C$_3$H$_2$ in the CSE is needed to understand the observed ratio.

%{\color{blue} The dissociative recombination of C$_3$H$^+_3$ can form $c$-C$_3$H$_2$, and the C$_3$H$^+_3$ is the product of the radiative association of C$_3$H$^+$ and H$_2$ \citep{Herbst+etal+1984, Millar+etal+1997, Nejad+etal+1987}. The formed ortho-$c$-C$_3$H$_2$ and para-$c$-C$_3$H$_2$ can be influenced by the ortho-H$_2$ and para-H$_2$.The OPR of $c$-C$_3$H$_2$ in TMC-1 and other dark clouds were also less than 3 \citep{Takakuwa+etal+2001}. \cite{Takakuwa+etal+2001} pointed out that this is due to the OPR of H$_2$ is less than its statistical ratio, resulting in a small OPR of $c$-C$_3$H$_2$, which can be attributed to the conservation of nuclear spin angular momentum in chemical reactions. \cite{Park+etal+2006} and \cite{Morisawa+etal+2006} studied the OPR of $c$-C$_3$H$_2$ in TMC-1 by using the gas phase chemical model and statistical equilibrium analysis respectively. They found that the OPR of $c$-C$_3$H$_2$ is affected by the concentration of H$_2$, any protonated ion [XH$^+$], the reaction rate of formed ortho-$c$-C$_3$H$_2$ and para-$c$-C$_3$H$_2$, as well as the concentrations of electrons and neutral atoms \citep{Park+etal+2006, Morisawa+etal+2006}. This is related to the conservation of nuclear spin angular momentum in chemical reactions. A detailed study of the ortho-para chemistry of $c$-C$_3$H$_2$ in IRC\,+10216 is needed to understand the observed ratio.}

For SiC$_2$, we have not added more data. \cite{Avery+etal+1992} carefully studied transitions with upper level energies of 10 -- 300\,K and later \cite{He+etal+2008} those with values in the range 20 -- 200 \,K. The excitation temperature and column density of SiC$_2$ obtained by us agrees with the observed results collected in Table~\ref{table:5}.

For the column density of CH$_3$CN, the result we obtained is in good agreement with that of \cite{Agundez+etal+2008}, \cite{He+etal+2008}, and \cite{Agundez+etal+2015}, but it is one order of magnitude higher than that estimated by \cite{Kawaguchi+etal+1995}. Unfortunately, we are unable to provide any insight on why we have a larger column density than \cite{Kawaguchi+etal+1995}. His values are actually taken from \cite{Guelin+etal+1991}. However, these authors simply give the values in the text of the article and give no information on how these were derived.

We have discussed all spectral lines with S/N of at least three. Spectral lines whose S/N are between three and five will affect the calculated results of molecular excitation temperature and column density \citep{Gong+etal+2015}. Moreover, different transitions of the same molecule have different source sizes \citep{He+etal+2008}. We neglected this factor in the calculations of the molecules' excitation temperature and column density given the lack of available high-resolution data and detailed radiative transfer modelling, including radiative pumping, which could allow one to determine the radial extent of the observed transitions. In addition, significant uncertainties will arise when one deals with the optically thick lines by employing the rotational diagram method. These factors will cause uncertainties in the observationally deduced excitation temperature and column density of the molecules.

\subsubsection{Impact of the source size on molecular excitation temperature and column density}

%Different transitions of the same molecule may lead to different molecular sizes \citep{He+etal+2008}. Here, we discuss the influence of different source sizes  (10$^{\prime\prime}$, 15$^{\prime\prime}$, 18$^{\prime\prime}$, 30$^{\prime\prime}$, 50$^{\prime \prime}$, 80$^{\prime\prime}$, 120$^{\prime\prime}$, and 200$^{\prime\prime}$) on the excitation temperatures and column densities of all the molecules obtained using the rotational diagram method. {\color{blue} Here, we only varied this one parameter, which gradually increased from small to large.} It is {\color{blue} find} that with an increase of the source size, the excitation temperatures of a molecule are gradually increasing, while the column densities are gradually decreasing, and the column densities of the molecule vary greatly between source sizes of 10$^{\prime\prime}$ and 50$^{\prime\prime}$. The excitation temperatures of molecules vary by a factor of two and column densities by up to three orders of magnitude. The CN, HC$_5$N, and SiO results are shown in Figure~\ref{fig:Size}.
%{\color{blue}Say something here about coupling line surveys with high-resolution imaging?}

\begin{figure*}
\gridline{\fig{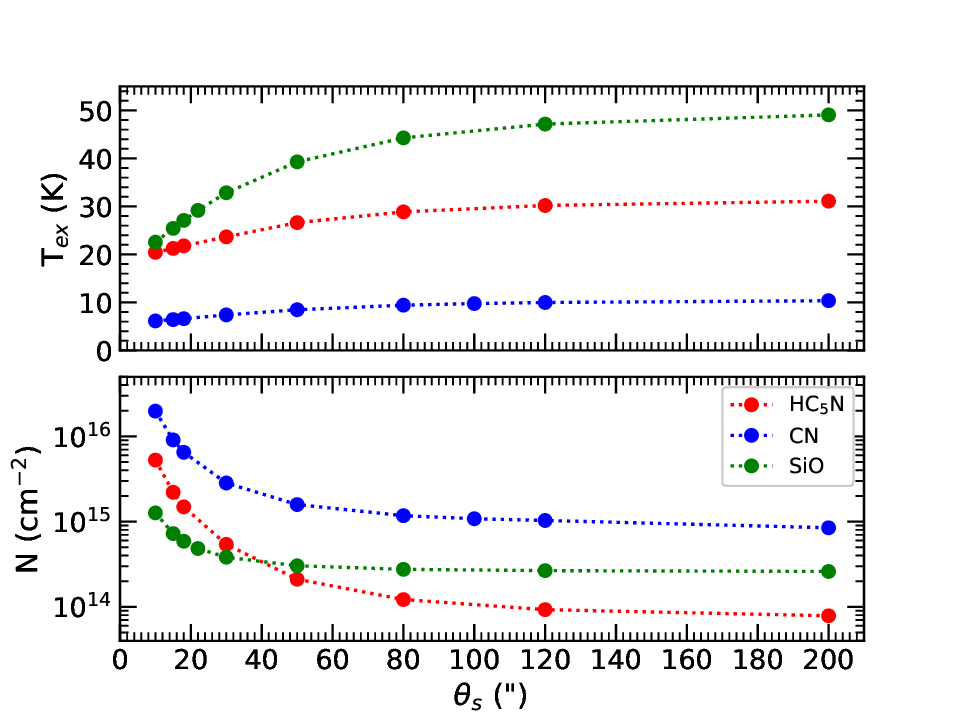}{0.7\textwidth}{ }
          }
   \caption{Illustrations for the impact of the employed source size ($\theta_{s}$) on the calculated excitation temperature ($T_{\rm ex}$) and column density ($N$) of HC$_5$N (in red), CN (in blue), and SiO (in green).
}
   \label{fig:Size}
\end{figure*}

In order to get some handle on the effect of the influence of source size on $T\rm_{ex}$ and column density, we use different source sizes (10$^{\prime\prime}$, 15$^{\prime\prime}$, 18$^{\prime\prime}$, 30$^{\prime\prime}$, 50$^{\prime \prime}$, 80$^{\prime\prime}$, 120$^{\prime\prime}$, and 200$^{\prime\prime}$) for all molecules for which we determine these quantities through the rotational diagram method. It is found that as source size increases so do the excitation temperatures while column densities decrease. Figure~\ref{fig:Size} shows the results for CN, HC$_5$N, and SiO. Molecular column densities decrease rapidly between souce sizes of 10$^{\prime\prime}$ and 50$^{\prime\prime}$. The excitation temperatures of molecules vary by a factor of two, while column densities vary by up to three orders of magnitude.

\subsection{Observationally deduced fractional abundances}
\label{sec:Molecular fractional abundances}
Once the column density of H$_2$ is known, the molecular fractional abundances relative to H$_2$ can be obtained. The beam-averaged H$_2$ column density  ($N (\rm H_2)  $)   can be calculated from the mass loss rate via the formula \citep{Gong+etal+2015}:
\begin{equation}
  N_{{\rm H}_{2}} = \frac{\dot{M}R/V_{\rm exp}}{\pi R^{2}m_{{\rm H}_{2}}} = \frac{\dot{M}}{\pi RV_{\rm exp}\mu m_{\rm H}}
\label{eq:LebsequeI},
\end{equation}
where ${\dot{M}}$, R, $V_{\rm exp}$, and $m_{\rm H}$ are the mass-loss rate, the radius of the molecular emission, expansion velocity, and the mass of a hydrogen atom, respectively. For IRC\,+10216, ${\dot{M}}$ is 2 $\times$ 10$^{-5}$ ${M}_{\odot}$ yr$^{-1}$ \citep{Crosas+Menten+1997, Menten+etal+2012}; $V_{\rm exp}$ is 14.5 km s$^{-1}$ \citep{Cernicharo+etal+2000}; $\mu$ is 2.3 amu \citep{Massalkhi+etal+2018}. According to Equation~(\ref{eq:LebsequeI}), the beam-averaged H$_2$ column density varies with radius (the location of the molecular emission region in the CSE). Due to the large differences in the observed sizes of molecules (e.g., the emission size of CO is over 100$^{\prime\prime}$ \citep{Fong+etal+2006} while that of HC$_3$N is about 15$^{\prime\prime}$ \citep{Keller+etal+2015}), $N(\rm H_2)$ obtained at different radii can be quite different (up to an order of magnitude \citep{Gong+etal+2015}). We thus calculated the beam-averaged H$_2$ column density at different radii to obtain the molecular fractional abundances relative to H$_{2}$.
\begin{figure*}
\gridline{\fig{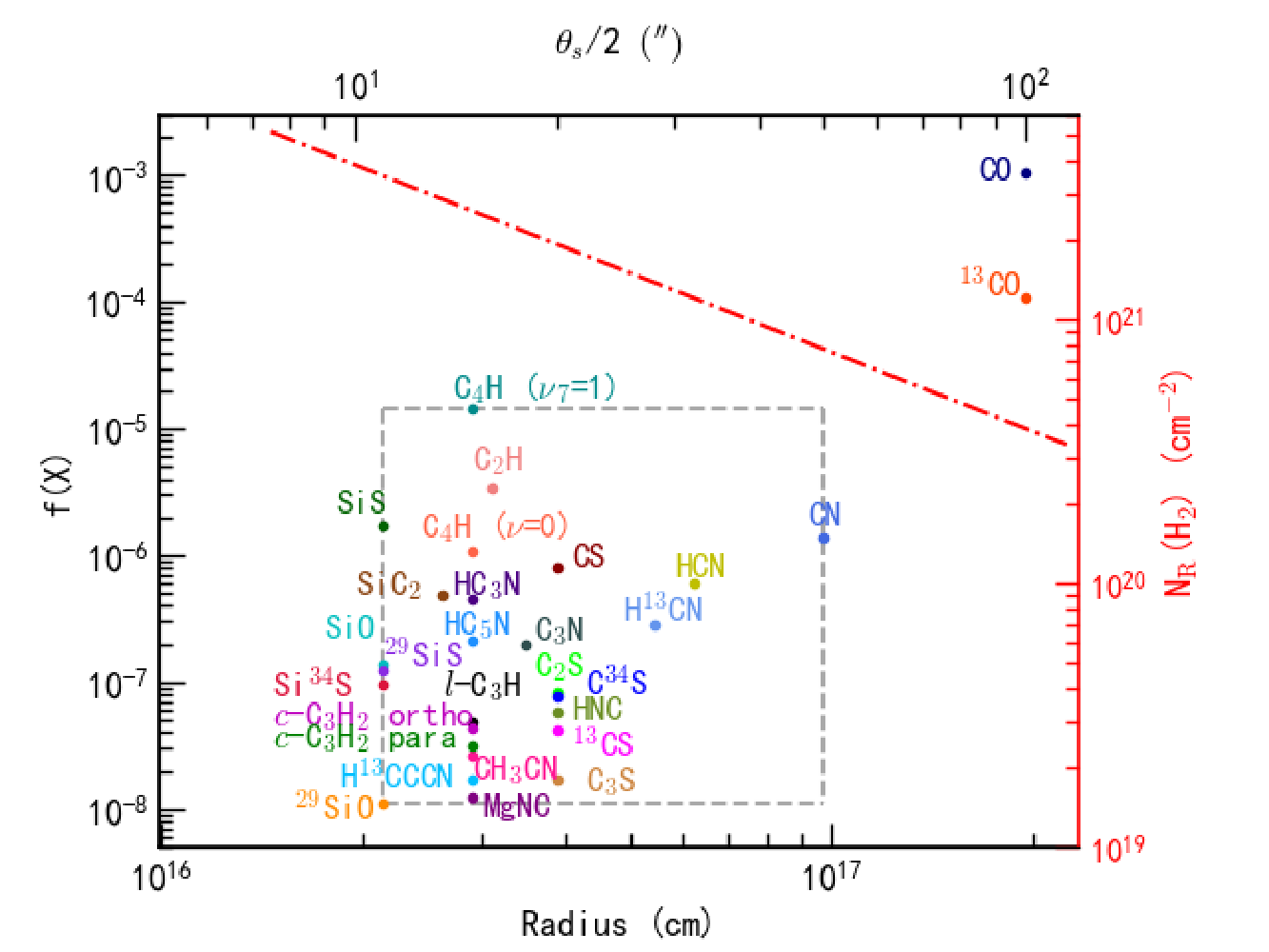}{0.85\textwidth}{ }
          }
   \caption{Observationally deduced fractional abundances ($f$), relative to H$_2$, of all the $\lambda$ 3\,mm band detected species (X) toward IRC\,+10216 by the PMO--13.7\,m, as a function of radius ($R$). In the plot, $f$(X) = $N$(X) / $N_{R}$(H$_{2}$), where $N$(X) depict the total column density of the species (X), and  $N_{R}$(H$_{2}$) (the red dash-dotted line) represents the beam-averaged H$_{2}$ column density at radius R. Except for CO and $^{13}$CO, the detected molecules are located within the area marked by the gray-dashed lines.}
   \label{fig:fN+NNH2}
\end{figure*}

The observationally-deduced molecular abundance relative to H$_{2}$, along with the beam-averaged H$_{2}$ column density that varies with radius, are summarised in Table~\ref{table:5} and plotted in Figure~\ref{fig:fN+NNH2}.
Here, $f$(X)=$N$(X) / $N_{R}$(H$_{2}$), where $N$(X) depicts the total column density of the species (X), and $N_{R}$(H$_{2}$) represents the beam-averaged H$_{2}$ column density at radius R.
For those parent species that peak on the central star, the corresponding $N_{R}$(H$_{2}$) values are calculated at the maximum emission radii of these species. In the case of daughter species that are distributed in a shell-like structure around the central star, $N_{R}$(H$_{2}$) is calculated at the peak emission radii of the molecules.

It is found that in IRC\,+10216, the fractional abundances of the majority of detected species are between 10$^{-8}$ and 10$^{-6}$ in the $\lambda$ 3\,mm wavelength band.
Some parent species, however, are more abundant.
\cite{Agundez+etal+2012} used the Large Velocity Gradient (LVG) radiative transfer model to deduce
the fractional abundances of CS, SiO, SiS, and CO, which are 7 $\times$ 10$^{-7}$, 1.8 $\times$ 10$^{-7}$, 1.3 $\times$ 10$^{-6}$, and 6 $\times$ 10$^{-4}$, respectively. These results are very close to our observational results.

%%%%%%%%%%%%%%%%%%%%%%%%%%%%%%
\subsection{Updated CSE model} \label{sec:updated model}
%%%%%%%%%%%%%%%%%%%%%%%%%%%%%%
\begin{table*}
  \caption{ Initial abundances of parent species relative to H$_2$, updated from the CSE model for IRC\,+10216 \citep{Li+etal+2014}. }
\label{tab:InitialABofParentSpecies}
%\scalebox{1}
\centering
\resizebox{0.5\linewidth}{!}{ %%  %% to make talbe fit the scale that we are happy with.
\begin{tabular}{llll}
\hline\hline
Species & Abundance   & Species & Abundance  \\
\hline
 N$_2$      & $4.0\times10^{-5}$         & SiS   &          $1.7\times10^{-6}$$^{ (a) }$      \\ % 1
 NH$_3$     & $6.0\times10^{-8}$         & CH$_4$     &     $3.5\times10^{-6}$       \\ %2
 HCN        & $2.0\times10^{-5}$         & H$_2$O   &       $2.6\times10^{-6}$       \\ %3
% He        & $1.0\times10^{-1}$         & Mg         & $1.0\times10^{-5}$  \\ % 4
 He         & $1.7\times10^{-1}$         & PN         & 		 $3.0\times10^{-10}$    \\ % 4
 HF         & $8.0\times10^{-9}$         & C$_2$H$_4$ & 	 $2.0\times10^{-8}$      \\ %5
 C$_2$H$_2$ & $8.0\times10^{-5}$         & SiH$_4$    &		 $2.2\times10^{-7}$     \\ %6
 CO         & $1.0\times10^{-3}$$^{ (a) }$ & HCl        & 		 $1.0\times10^{-7}$     \\ % 7
 H$_2$S     & $4.0\times10^{-9}$         & HCP        & 		 $2.5\times10^{-8}$       \\ %8
 CS         & $8.1\times10^{-7}$$^{ (a) }$ & SiC$_2$   & 		 $4.9\times10^{-7}$$^{ (a) }$     \\ %9
 SiO        & $1.4\times10^{-7}$$^{ (a) }$ &            &            \\ %10
\hline
\end{tabular}
} %%
%\tablefoot{
%\tablefoottext{a}{Notation a(b) indicates $a\times 10^{b}$}. }
\tablecomments{
$^{ (a) }${The observationally-deduced abundance from this work.}
}
\end{table*}
%%%%%%%%%%%%%%%%%%%%%%%%%

Good astrochemical modeling results are valuable in explaining, comparing with, and creating observations. As one of the best-observed ``molecular factories'' in space, IRC\,+10216, was frequently modeled for various purposes, e.g.:
\cite{Millar+etal+2000},
\cite{Agundez+2006},
\cite{Cordiner+etal+2009},
\cite{Li+etal+2014},
\cite{Agundez+etal+2017},
\cite{VandeSande+etal+2019},
\cite{Millar+2020},
\cite{VandeSande+etal+2021},
\cite{VandeSande+etal+2022},
\cite{Reach+etal+2022}.
We employ the CSE model described in \cite{Li+etal+2014}, with an update of the initial fractional abundances of several parent species listed in Table \ref{tab:InitialABofParentSpecies}.
Unless stated elsewhere, the assumptions and parameters are the same as those described in the model of \cite{Li+etal+2014}
that accurately calculates the photodissociation of CO and N$_{2}$ ( the two most abundant molecules after H$_{2}$) in the full 3D spherical geometry with isotropic incident radiation calculated by the \cite{Draine78} radiation field.
Initially, parent species are injected at a radius of $r_i$= 1 $\times$ 10$^{15}$ cm, where the molecular hydrogen number density is $n_{H_{2}}$ = 1.3 $\times$ 10${^7}$cm$^{-3}$,  the visual extinction is $A_V$ = 13.8 mag, and the kinetic temperature of the gas is $T$ = 575\,K.
The final radius of the CSE is set at $r_f$ = 2 $\times$ 10$^{17}$ cm, where the gas density decreases to $n_{H_{2}}$ = 26 cm$^{-3}$,  with $T$ = 10\,K, and $A_V$ = 0.02 mag. In the calculations, the mass-loss rate of the star is 1.5 $\times$ 10$^{-5}$ M$_\odot$ yr$^{-1}$ and the outflow expansion velocity is 14.5 km s$^{-1}$. The gas-phase reaction network of the UMIST database for astrochemistry 2012 \citep{McElroy+etal+2013} is employed.
Reasonable good agreements are found between the simulated and observed column densities (within an order of magnitude) for most of the detected species ($>$ 70\%), as summarised in Table \ref{table:1}.

\subsection{Isotopic ratios} \label{subsec-isotopic-ratio}
%%%%%%%%%%%%%%%%%%%
\begin{deluxetable*}{cccccc}
%\tablenum{6}
\tablecaption{Isotopic ratios obtained in IRC\,+10216, as described in Section~\ref{subsec-isotopic-ratio}.   }  \label{table:6}
\tablewidth{0pt}
\tablehead{
\colhead{Isotope ratio}      & \colhead{Method}                   & \colhead{Value} &  \colhead{Local ISM$^{(i)}$}    & \colhead{Solar$^{(j)}$}
}
\startdata
[$^{28}$Si]/[$^{29}$Si] & SiS             	    &  17.3 $\pm$ 5.3               &         & 19.7    \\
                        & SiO           	    & $>$ 10.9 $\pm$ 3.7            &         &        \\
                        & SiS and  SiC$_{2}$    & 18.7$_{-1.0}^{+1.3}$ $^{(b)}$ &         &         \\
                        & SiS and  SiC$_{2}$    & $>$ 15.4 $\pm$ 1.1 $^{(c)}$   &            &         \\
                        & SiC$_{2}$             & 17.2 $\pm$ 1.1  $^{(d)}$      &            &         \\
                        & SiS and SiO           & 18 $\pm$ 2  $^{(e)}$          &            &         \\
                        & SiS                   & $>$ 15.1 $\pm$ 0.7 $^{(f)}$   &            &         \\
                        & SiS                   & 17$_{-4}^{+5}$ $^{(g)}$       &            &         \\\
[$^{32}$S]/[$^{34}$S]   & SiS                   & $>$ 11.4 $\pm$ 3.5            & 24 $\pm$ 4 & 22.1    \\
                        & CS                    & $>$ 10.6 $\pm$ 1.3            &            &         \\
                        & SiS                   & 20.2$_{-2.1}^{+2.6}$ $^{(b)}$ &            &         \\
                        & CS and SiS            & 21.8 $\pm$ 2.6 $^{(c)}$       &            &         \\
                        & $^{29}$Si$^{32}$S/$^{28}$Si$^{34}$S, $^{30}$Si$^{32}$S/$^{28}$Si$^{34}$S     & 18.9 $\pm$ 1.3 $^{(d)}$   & &   \\
                        & SiS                     & 22 $\pm$ 2.5   $^{(e)}$       &             &         \\
                        & SiS                     & 19.6 $\pm$ 1.3 $^{(f)}$       &             &         \\
                        & SiS                     & $>$ 14$_{-4}^{+6}$ $^{(g)}$   &             &         \\
                        & CS                      & 22 $\pm$ 4 $^{(h)}$           &             &         \\\
[$^{12}$C]/[$^{13}$C]   & $^{12}$C$^{34}$S/$^{13}$C$^{32}$S$^{(a)}$    & 46.4 $\pm$ 0.1                & 54 $\pm$ 10 & 89.4    \\
                        & HC$_3$N                        & $>$ 24.1 $\pm$ 6.9   &             &     \\
                        & CO                             & $>$ 9.6 $\pm$ 0.3    &             &     \\
                        & HCN                            & $>$ 2.7 $\pm$ 0.2    &             &     \\
                        & CS                             & $>$ 23.1 $\pm$ 6.3  &             &     \\
                        & $^{12}$C$^{34}$S/$^{13}$C$^{32}$S            & 47$_{-5}^{+6}$ $^{(b)}$       &             &         \\
                        & CS                                           & 45 $\pm$ 3     $^{(c)}$       &             &         \\
                        & SiC$_{2}$                                    & 34.7 $\pm$ 4.3 $^{(d)}$       &             &         \\
                        & CS                                           & 35 $\pm$ 3.5   $^{(e)}$       &             &         \\
                        & CS                                           & 40$_{-10}^{+18}$ $^{(f)}$     &             &         \\
\enddata
\tablecomments{
$^{(a)}$ Assuming that the $^{34}$S/$^{32}$S ratio is solar.
$^{(b)}$ \cite{Kahane+etal+1988}.
$^{(c)}$ \cite{Cernicharo+etal+2000}.
$^{(d)}$ \cite{He+etal+2008}.
$^{(e)}$ \cite{Agundez+etal+2012}.
$^{(f)}$ \cite{Patel+etal+2009}.
$^{(g)}$ \cite{Fonfria+etal+2015}.
$^{(h)}$ \cite{Wannier+etal+1978}.
$^{(i)}$ The local ISM values refer to 7.5\,kpc $\leq$ $R\rm_{GC}$ $\leq$ 8.5\,kpc, and the data come from \cite{Yan+etal+2023}.
$^{(j)}$ \cite{Asplund+etal+2009}.
}
\end{deluxetable*}
%%%%%%%%%%%%%%%%%%%%%
Isotopic ratios are closely related to the physical conditions in the star's core  where the isotopes are formed. Therefore, isotopic ratios provide information on understanding the nucleosynthesis and the Galactic chemical evolution as well as the abundances of the interstellar medium \citep{Kahane+etal+1988, Wilson+etal+1994, Herwig+etal+2005, Peng+etal+2013}. AGB stars play an important role in the chemical evolution of galaxies. Thermal pulsing on the AGB creates a convective envelope, which transfers the products of He-shell burning (light elements, such as C, N and F) and the elements produced by the slow neutron-capture process (s-process) (half of the elements heavier than Fe) to the stellar surface through the third dredge-up (TDU). This alters the composition (isotopic ratios) of the envelope \citep{Busso+etal+1999, Herwig+etal+2005, Cristallo+etal+2009, Zhang+etal+2013, Wasserburg+etal+2017, Battino+etal+2022}. \cite{Peng+etal+2013} pointed out that the mass and metallicity of a star will affect the composition of its ejected material.
This will influence its surrounding interstellar medium and the isotope ratios
of its gas.
In fact, IRC\,+10216 is a much younger star (IRC\,+10216 was born $\sim$ 1 -- 5 $\times$ 10$^8$ years ago, assuming masses of 3 -- 5 $M_{\odot}$) than the Sun (see \cite{Peng+etal+2013} and \cite{Portinari+etal+1998}). Although the carbon-rich AGB star IRC\,+10216 is located in the solar neighbourhood, extensive observational results show that its isotopic composition is remarkably non-solar \citep{Kahane+etal+1988, Kahane+etal+1992, Cernicharo+etal+2000}. This is likely due to the fact that the shell of IRC\,+10216 is highly enriched in processed material after the ultimate dredge-ups \citep{Cernicharo+etal+2000}.

Here, we also observed some rotational transitions of rare isotopes. We calculated the observed isotopic ratios of these molecules using the integrated intensities ratio of spectral lines and considering the correction factor ${\nu}^{-2}$ \citep{Kahane+etal+1988, Cernicharo+etal+2000} since column density is proportional to ${\nu}^{-2}$, where ${\nu}$ represents the rest frequency \citep{Linke+etal+1977}. Our results and some results from the literature are listed in Table~\ref{table:6}. We have provided only a few key references, which include research results from the millimeter, submillimeter, and mid-infrared wavelength ranges. For optically thin molecules, the results obtained by this method are close to the isotopic ratios.
Optically thick molecules may lead to a lower limit. In this work, the [$^{28}$Si]/[$^{29}$Si] ratio is derived to be 17.3 $\pm$ 5.3 from the $^{28}$SiS/$^{29}$SiS ($J=5-4$) transitions, but is only 10.9 $\pm$ 3.7 from $^{28}$SiO/$^{29}$SiO ($J=2-1$). Thus the ratio from the SiS ($J=5-4$) transitions is close to the solar system value, 19.7 \citep{Asplund+etal+2009}, while the ratio from SiO ($J=2-1$) is considerably smaller. In the following we therefore emphasize ratios (see Table 6), where the lines used to determine isotope ratios are both (almost) optically thin. For the [$^{32}$S]/[$^{34}$S] ratio with a solar system ratio of 22.1 \citep{Asplund+etal+2009}, this holds when analysing the sulfur isotopic ratios which yield calculated by optically thin spectral lines and are thus also close to the ratios obtained in the local ISM \citep{Cernicharo+etal+2000, He+etal+2008, Patel+etal+2009, Agundez+etal+2012}. \cite{Fonfria+etal+2015} used the Texas Echelon-cross-Echelle Spectrograph at the NASA Infrared Telescope Facility to obtain high resolution mid-infrared spectra of SiS in the innermost envelope of IRC\,+10216. They find that the [$^{28}$Si]/[$^{29}$Si] and [$^{32}$S]/[$^{34}$S] ratios in the vicinity of the star are compatible with the previously measured values for the outer shells of the envelope.

We have also derived [$^{12}$C]/[$^{13}$C] ratios from four carbon-bearing molecules: HC$_3$N, CO, HCN, and CS, and obtain lower limits in that these species are optically thick. The [$^{12}$C]/[$^{13}$C] ratios deduced from HC$_3$N/H$^{13}$CCCN ($J=10-9$), $^{12}$CO/$^{13}$CO ($J=1-0$), H$^{12}$CN/H$^{13}$CN ($J=1-0$), and $^{12}$CS/$^{13}$CS are 24.1 $\pm$ 6.9, 9.6 $\pm$ 0.3, 2.7 $\pm$ 0.2, and 23.1 $\pm$ 6.3, respectively. Unlike the optically thick species, we also observed C$^{34}$S and $^{13}$CS, both of which are optically thin. Therefore we are able to deduce a reliable isotopic ratio of [$^{12}$C$^{34}$S]/[$^{13}$C$^{32}$S] = 2.1 $\pm$ 0.7, rather than a lower limit.
Because the [$^{32}$S]/[$^{34}$S] ratio in IRC\,+10216 is confirmed to be consistent with the solar value, 22.1, \citep{Asplund+etal+2009},
we estimate a [$^{12}$C]/[$^{13}$C] ratio of 46.4 $\pm$ 0.1.
This value is in good agreement with existing results \citep{Kahane+etal+1988, Kahane+etal+1992, Cernicharo+etal+2000, Agundez+etal+2012}. Therefore, we confirm that the [$^{12}$C]/[$^{13}$C] ratio in IRC\,+10216 is much smaller than the solar value  \citep[89.4, see][]{Asplund+etal+2009} and local ISM value \citep[$54 \pm 10$, see][]{Yan+etal+2023}. This indicates that there is a nonstandard mixing mechanism or cool bottom processing \citep{Charbonnel+etal+1995, Zhang+etal+2013} in the CSE of IRC\,+10216. A more detailed discussion on the effect of [$^{12}$C]/[$^{13}$C] for AGB nucleosynthesis can be find in \cite{Milam+etal+2009}. All of the isotopic ratios discussed here are included in Table~\ref{table:6}.

In IRC\,+10216, the sulfur and silicon isotope ratio calculated by the optically thin spectral lines are close to solar while the carbon isotope ratio is non-solar \citep{Cernicharo+etal+2000, He+etal+2008}. \cite{Fonfria+etal+2015} pointed out that in the past 1000\,yr, the nucleosynthesis of the most abundant silicon and sulfur isotopes are not expected to change significantly. In this work, we detected in several cases the same transitions and obtained close results to those described by \cite{Kahane+etal+1988}, \cite{Kahane+etal+1992}, and \cite{Park+etal+2008}. Note that we ignored the impact of the beam dilution and did not consider the time-variation of the intensities of the spectra. These effects might be important in some cases \citep{He+etal+2017, He+etal+2019, Pardo+etal+2018}, and might lead to discrepancies in the observationally-deduced isotopic ratios.

\subsection{Line surveys toward other evolved stars}
%%%%%%%%%%%%%%%%%%%%%%%%%%%%%%%%%%

\begin{deluxetable*}{cccccc}
%\tablenum{7}
\tablecaption{Published radio line surveys of evolved stars around the $\lambda$ 3\,mm band.
Note that the published line surveys toward IRC\,+10216 are summarised in Table \ref{table:2}. }
\label{table:3mmsurvery}
\tablewidth{0pt}
\tablehead{
\colhead{Sources Name} & Evolutiona stage & Chemical type &\colhead{Covered Frequencies} & \colhead{Telescope} & \colhead{Reference}  \\
\colhead{  }           &      &           &\colhead{ (GHz)  }               & \colhead{ }         & \colhead{        }
}
%\decimalcolnumbers
\startdata
IK\,Tau           & AGB        & O-rich   & 79 -- 116                    & IRAM--30\,m    &  \cite{Prieto+etal+2017}    \\
IRC\,+10216       & AGB        & C-rich   & 84.5 -- 115.8                & PMO--13.7\,m   &  This work                \\
IRC\,+10216       & AGB        & C-rich   & 72.2 -- 91.1                 & Onsala--20\,m  &  \cite{Johansson+etal+1984, Johansson+etal+1985}    \\
II\,Lup           & AGB        & C-rich   & 80.5 -- 115.5$^c$            & Mopra--22\,m   &  \cite{Smith+etal+2015}    \\
RAFGL\,4211       & AGB        & C-rich   & 80.5 -- 115.5$^c$            & Mopra--22\,m   &  \cite{Smith+etal+2015}    \\
AI\,Vol           & AGB        & C-rich   & 80.5 -- 115.5$^c$            & Mopra--22\,m   &  \cite{Smith+etal+2015}    \\
RAFGL\,4211       & AGB        & C-rich   & $\sim$ 85.10 -- 115.75$^d$   & SEST--15\,m    &  \cite{Woods+etal+2003}    \\
RAFGL\,4078       & AGB        & C-rich   & $\sim$ 85.10 -- 115.75$^d$   & SEST--15\,m    &  \cite{Woods+etal+2003}    \\
CIT\,6            & AGB        & C-rich   & $\sim$ 85.10 -- 115.75$^d$   & SEST--15\,m    &  \cite{Woods+etal+2003}    \\
CIT\,6            & AGB        & C-rich   & 90 -- 116                    & ARO--12\,m     &  \cite{Yang+etal+2023}    \\
AFGL\,3068        & AGB        & C-rich   & $\sim$ 85.10 -- 115.75$^d$   & SEST--15\,m    &  \cite{Woods+etal+2003}    \\
IRC\,+40540       & AGB        & C-rich   & $\sim$ 85.10 -- 115.75$^d$   & SEST--15\,m    &  \cite{Woods+etal+2003}    \\
II\,Lup           & AGB        & C-rich   & $\sim$ 85.10 -- 115.75$^d$   & SEST--15\,m    &  \cite{Nyman+etal+1993}    \\
Red\,Rectangle    & post-AGB   & O-rich   & 84.50 -- 92.25               & IRAM--30\,m    &  \cite{Gallardo+etal+2022} \\
HD\,52961         & post-AGB   & O-rich   & 84.40 -- 92.15               & IRAM--30\,m    &  \cite{Gallardo+etal+2022} \\
AI\,Canis Minoris & post-AGB   & O-rich   & 84.40 -- 92.15               & IRAM--30\,m    &  \cite{Gallardo+etal+2022} \\
IRAS\,20056\,+1834 & post-AGB   & O-rich   & 84.40 -- 92.20               & IRAM--30\,m    &  \cite{Gallardo+etal+2022} \\
R\,Scuti          & post-AGB   & O-rich   & 83.40 -- 92.30               & IRAM--30\,m    &  \cite{Gallardo+etal+2022} \\
89\,Herculis      & post-AGB   & C-rich   & 83.40 -- 92.50               & IRAM--30\,m    &  \cite{Gallardo+etal+2022} \\
CRL\,2688   &  post-AGB  & C-rich  & 75.7--83.5 and 91.4--99.2 & IRAM--30\,m    & \cite{Qiu+etal+2022} \\
CRL\,2688   &  post-AGB  & C-rich  & 71--111 & ARO--12\,m    & \cite{Zhang+etal+2013} \\
CRL\,2688         & post-AGB   & C-rich   & 85--116                    & TRAO--14\,m    &  \cite{Park+etal+2008}      \\
CRL\,618          & post-AGB   & C-rich   & 80.25--115.75              & IRAM--30\,m    &  \cite{Pardo+etal+2007}     \\
AC\,Herculis      & post-AGB   & O-rich   & 84.40--92.15               & IRAM--30\,m    &  \cite{Gallardo+etal+2022} \\
IRAS\,19157\,-0257  & post-AGB   &  & 84.40--92.20               & IRAM--30\,m    &  \cite{Gallardo+etal+2022} \\
IRAS\,18123\,+0511  & post-AGB   &   & 84.40--92.15               & IRAM--30\,m    &  \cite{Gallardo+etal+2022} \\
IRAS\,19125\,+0343  & post-AGB   &   & 84.40--92.15               & IRAM--30\,m    &  \cite{Gallardo+etal+2022} \\
 NGC\,7027 &  PN  & C-rich  & 71--111 & ARO--12\,m    & \cite{Zhang+etal+2008}\\
IC\,4406          & PN         & O-rich   & 80.5--115.5$^c$            & Mopra--22\,m   &  \cite{Smith+etal+2015}    \\
NGC\,6537         & PN         & O-rich   & 80.5--115.5$^c$            & Mopra--22\,m   &  \cite{Smith+etal+2015}    \\
NML\,Cyg           & RSG$^a$    & O-rich   & 68--116                    & OSO--20\,m     &  \cite{Andrews+etal+2022} \\
IRC\,+10420         & YHG$^b$    & N-rich   & $\sim$ 83.2--117.1         & IRAM--30\,m    &  \cite{Quintana-Lacaci+etal+2016} \\
\enddata
\tablecomments{
$^{a}$ Red supergiant  (RSG)  star.
$^{b}$ Yellow hypergiant  (YHG) star.
$^{c}$ Note that 98.5 -- 107.5\ GHz is uncovered in the observation.
$^{d}$ Frequency range is directly taken from the reference.}
\end{deluxetable*}

In addition to the widely studied AGB star IRC\,+10216, detailed studies of other carbon stars have also been reported. For instance, IRAS 15194-5115 (II\,Lup), the brightest carbon star in the Southern hemisphere at 12 $\mu$m, and also the third brightest in both hemispheres, with only IRC\,+10216 and CIT 6 being brighter \citep{Epchtein+etal+1987}, was also well investigated. \cite{Nyman+etal+1993} used the 15-meter Swedish-ESO Submillimeter Telescope (SEST) to make a molecular line survey in the 3 and 1. 3\,mm bands toward II\,Lup, and find that the gas composition of this object is very similar to that of IRC\,+10216. Subsequently, \cite{Woods+etal+2003} conducted a detailed study of the molecules in seven carbon stars with high mass-loss rate (IRAS\,15082\,-4808, IRAS\,07454\,-7112, CIT\,6, AFGL\,3068, IRC\,+40540, IRC\,+10216, and II\,Lup) in the northern and southern skies. They found that the chemical compositions of these carbon stars were similar, too. However, there was a significant difference in the [$^{12}$C]/[$^{13}$C] ratio, which is related to nucleosynthesis and reflects the evolutionary status of these stars.
Moreover, \cite{Zhang+etal+2009a, Zhang+etal+2009, Zhang+etal+2020} conducted spectral line surveys toward CIT6 and CRL 3068, and find that only the most abundant molecules could be detected and that the gas compositions of these two sources were very similar to IRC\,+10216, too.

In short, it is found that the gas compositions of the evolved stars are highly determined by their chemical types (namely their [C]/[O] ratios), and could be simply classified as C-rich, O-rich, and S-type cases. In principle, the gas compositions of all three types of AGB stars will be well simulated by the corresponding chemical models based on the parent species and physical conditions. This conclusion is further confirmed by the results of other line surveys toward other types of evolved stars in the $\lambda$ 3\,mm band, as summarized in Table~\ref{table:3mmsurvery}.

%%%%%%%%%%%%%%%%
\section{Conclusions}
\label{sec:Conclusions}
%%%%%%%%%%%%%%%%

%We present an unbiased} $\lambda$ 3mm spectral line survey (between 84.5 and 115.8 GHz),
%conducted by the Purple Mountain Observatory 13.7 meter radio telescope, together with updated modeling results,
%towards the carbon-rich Asymptotic Giant Branch star, IRC\,+10216 (CW Leo).

 We present the first unbiased $\lambda$ 3\,mm (between 84.5 and 115.8 GHz) line survey toward the carbon-rich envelope of IRC\,+10216 using the PMO--13.7\,m telescope, and provide a comprehensive analysis of the molecular spectra by making use of the results from previous studies. A total of 75 spectral lines were detected and assigned to 19 different molecules. The detected molecules are C$_2$H, $l$-C$_3$H, C$_4$H, CN, C$_3$N, HC$_3$N, HC$_5$N, HCN, HNC, CH$_3$CN, MgNC, CO, $c$-C$_3$H$_{2}$, SiC$_2$, SiO, SiS, CS, C$_2$S, C$_3$S. The $J  = 13-12$ transition of H$^{13}$CCCN is detected in IRC\,+10216 for the first time.  The excitation temperatures and column densities of these molecules are derived by assuming LTE with excitation temperatures ranging from 5.3 to 73.4\, K, and molecular column densities ranging from $3.27 {\times} 10^{13}$ to $4.03 {\times} 10^{17}$\,cm$^{-2}$. Except for HCN, the observationally deduced excitation temperatures and column densities roughly agree with previous studies. We find that the fractional abundances of the detected species range between $1.14 {\times} 10^{-8}$ and $1.04 {\times} 10^{-3}$. We also find that HC$_5$N, HC$_7$N and  HC$_{9}$N, together with higher J transitions of HC$_3$N, C$_3$S, CH$_3$CN, SiS, C$_4$H\, (${\nu=0}$) and C$_4$H\, (${\nu=1}$), trace the warmer molecular layers in IRC\,+10216.
 %{\color{green} Thanks to previous high-resolution maps observation results, we found that radial shift in the emission peak between low-J and high-J lines. With the gradual increase of rotational energy level, the molecular excitation moves inward, and the tracer region becomes warmer and warmer.} {\color{blue} For CH$_3$CN molecule, it is necessary to make high resolution maps observation of its low rotationally transitions. It is possible that this molecule is similar to the spatial distribution of C$_4$H\, (${\nu  = 0}$), which can be used to better constrain the chemical model and reveal the formation mechanism of CH$_3$CN based on the map observation results.}
%Moreover, we detected several isotopologues and derived their isotopic ratios. The ratios for [$^{12}$C]/[$^{13}$C], [$^{32}$S]/[$^{34}$S] and [$^{28}$Si]/[$^{29}$Si] are $46.4\pm0.1$, $11.0\pm0.6$ and $14.1\pm4.5$, respectively, which are roughly consistent with earlier determinations and are all smaller than their solar system values. This is most likely due to the optical depth effects.
Moreover, we detected several isotopologues and derived their isotopic ratios. The [$^{28}$Si]/[$^{29}$Si] ratio, for example, is found to be $17.3\pm5.3$, which is roughly consistent with earlier determinations and close to solar system value.
Last but not least, we have summarised all of the 106 species detected in IRC\,+10216 to date with their observed and modeled column densities for the convenience of future studies.

%%%%%%%%%%%%%%%%%%%
\begin{acknowledgments}

We are extremely grateful to the anonymous reviewer for the fruitful and insightful comments, along with valuable inputs to this manuscript, which greatly helped improving the quality of this paper. We thank Prof. Y. Xu (PMO) and Prof. C. Walsh (Univ. of Leeds) for their constructive suggestions on this work. We appreciate the assistance of the PMO--13.7\,m telescope operators during the observations. X. Li acknowledges support from the Xinjiang Tianchi Talent project  (2019). This work was also funded by National Key R\&D Program of China under grant No.2022YFA1603103, the NSFC under grants 11973076, 12173023, 11973075, the Natural Science Foundation of Xinjiang Uygur Autonomous Region 2022D01B221. Y. Gao thanks the Project of Xinjiang Uygur Autonomous Region of China for Flexibly Fetching in Upscale Talents. TJM thanks the Leverhulme Trust for the award of an Emeritus Fellowship. Astrophysics at QUB is supported by the STFC through grant ST/T000198/1. Y. Zhang and X. Fang thank the Xinjiang Tianchi Talent Program (2023).

\end{acknowledgments}
%%%%%%%%%%%%%%%%%%%%

%%%%%%%%%%%%%%%%%%%%%%%%%%%%%%%%%%%%%%%%%%%%%%%
%              Appendix
%%%%%%%%%%%%%%%%%%%%%%%%%%%%%%%%%%%%%%%%%%%%%%%
\appendix

%%%%%%%%%%%%%%%%%%%%%%%%%%%%%%%%%%%%%%%%%%%%%%%
%              Appendix  A
%%%%%%%%%%%%%%%%%%%%%%%%%%%%%%%%%%%%%%%%%%%%%%%
\section{Zoom-in plots of observed spectra} \label{Zoom-in plots of observed spectra}

This appendix includes the spectra of the survey of IRC\,+10216 ranging from 84.5 to 115.8\,GHz, displayed in consecutive 475\,MHz frequency segments with 0.244 MHz frequency resolution. The observed molecular spectral lines whose S/N is greater than 3 are marked in blue. Molecules marked with a ? indicate that the S/N of the spectral line is between 2 and 3.
%%%%%%%%%%%%%%%%%%%%%%%%%
\renewcommand\thefigure{\Alph{section}\arabic{figure}}
\setcounter{figure}{0}

\begin{figure*}
\gridline{\fig{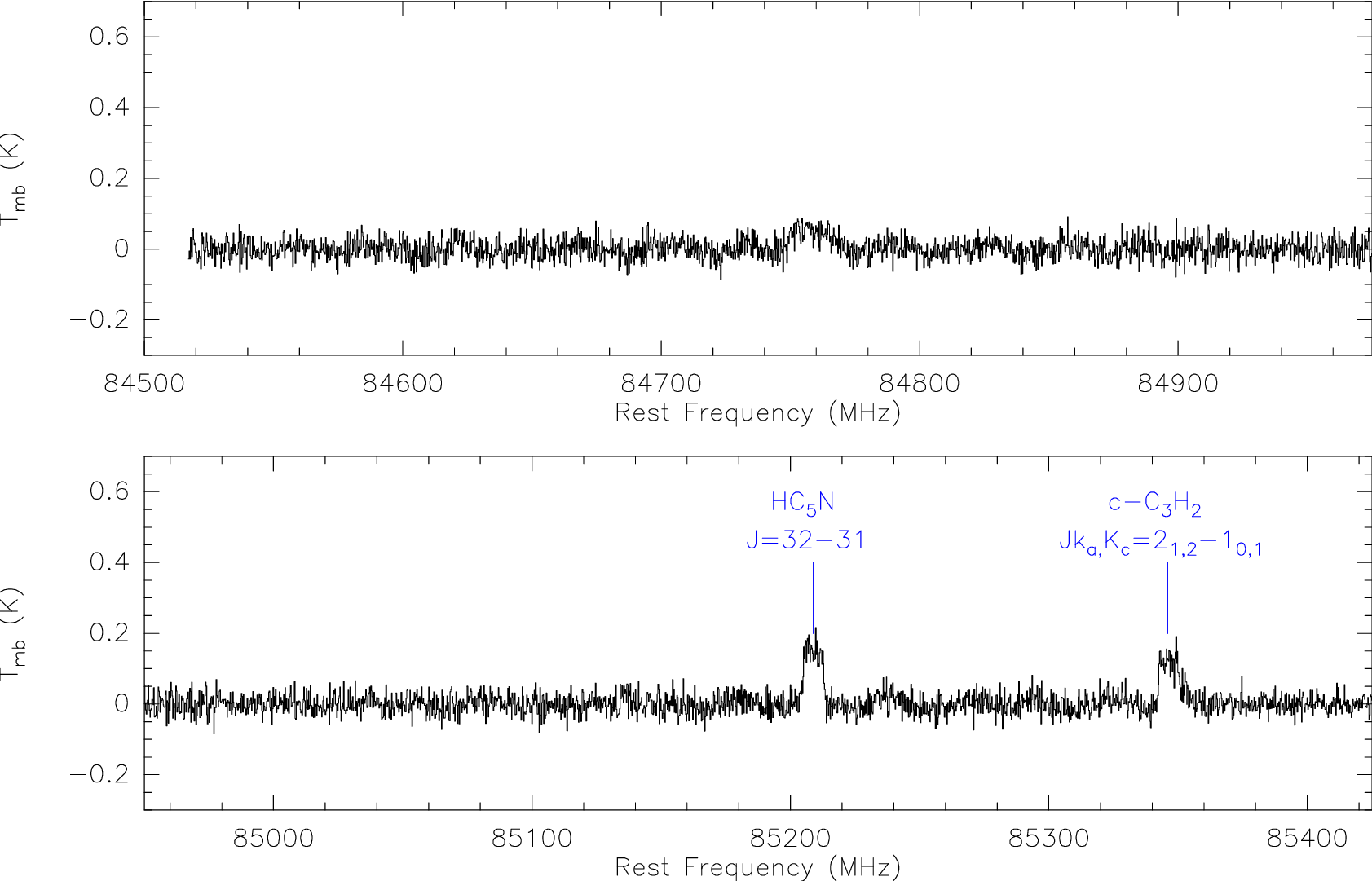}{0.85\textwidth}{ }
          }
\gridline{\fig{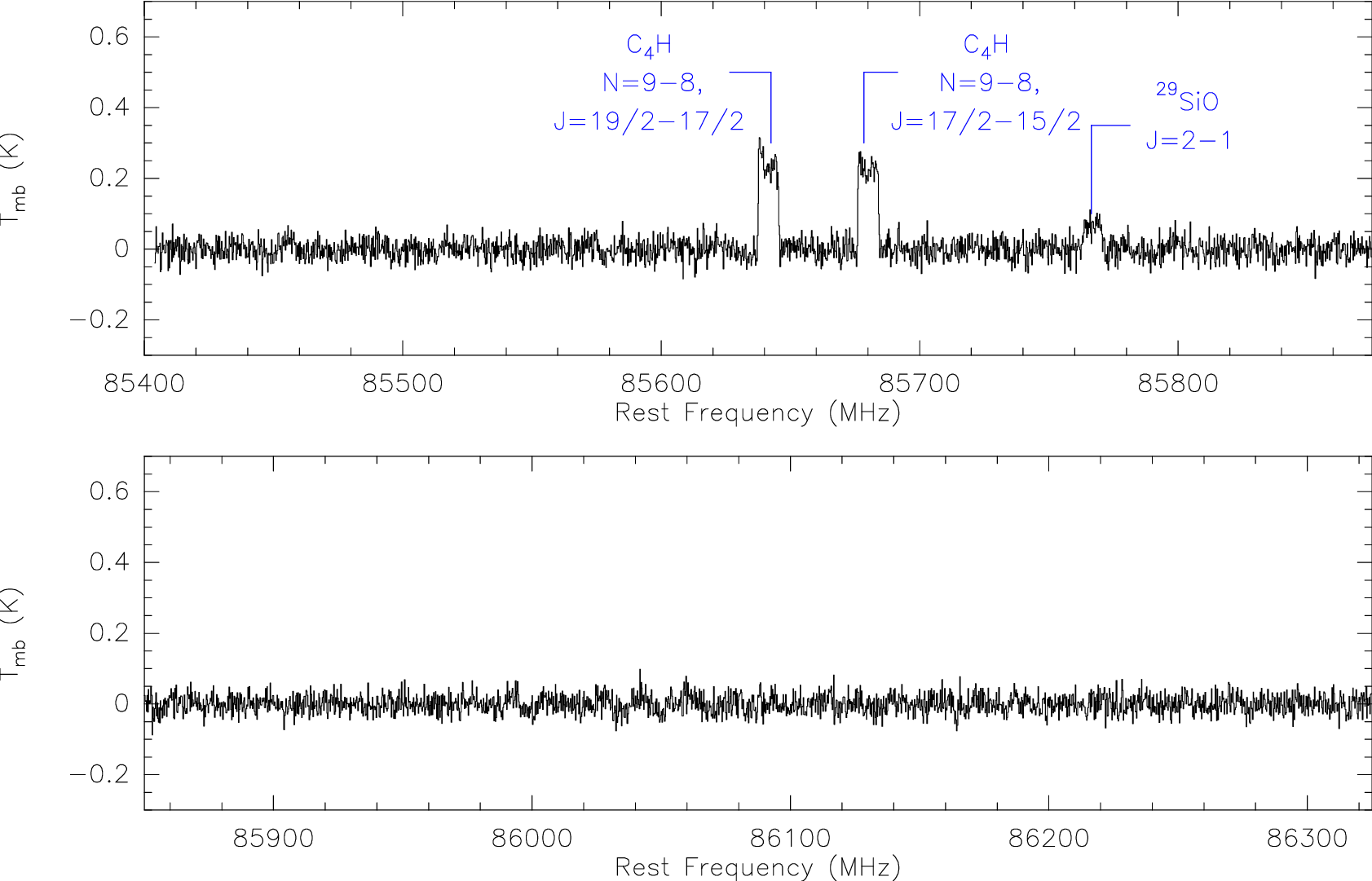}{0.85\textwidth}{ }
          }
\caption{84.5 -- 115.8 GHz spectrum of IRC +10216 displayed in consecutive 475 MHz frequency segments with 0.244 MHz spectral resolution.
\label{fig:F1}}
\end{figure*}

\begin{figure*}
\gridline{\fig{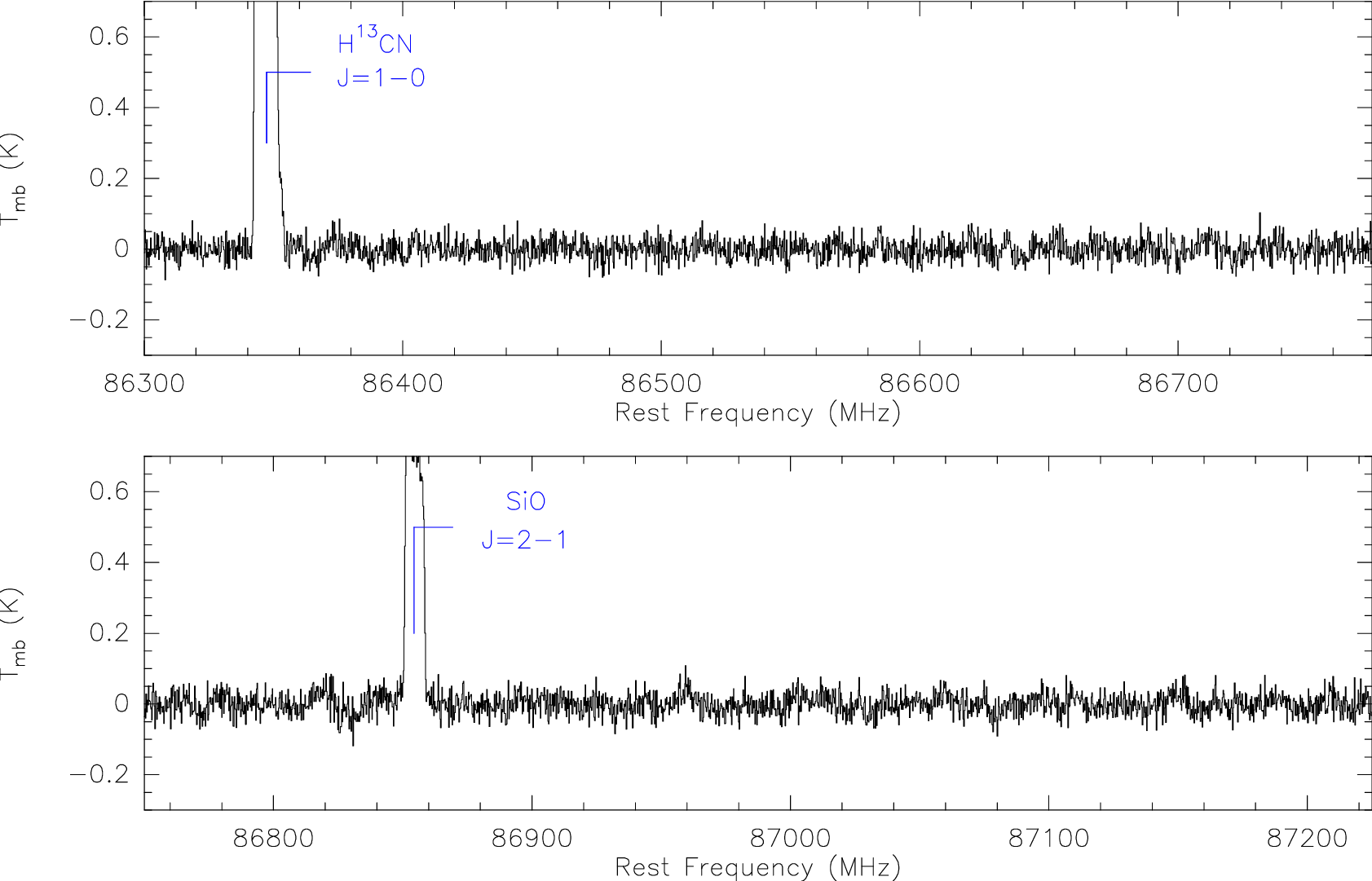}{0.85\textwidth}{ }
          }
\gridline{\fig{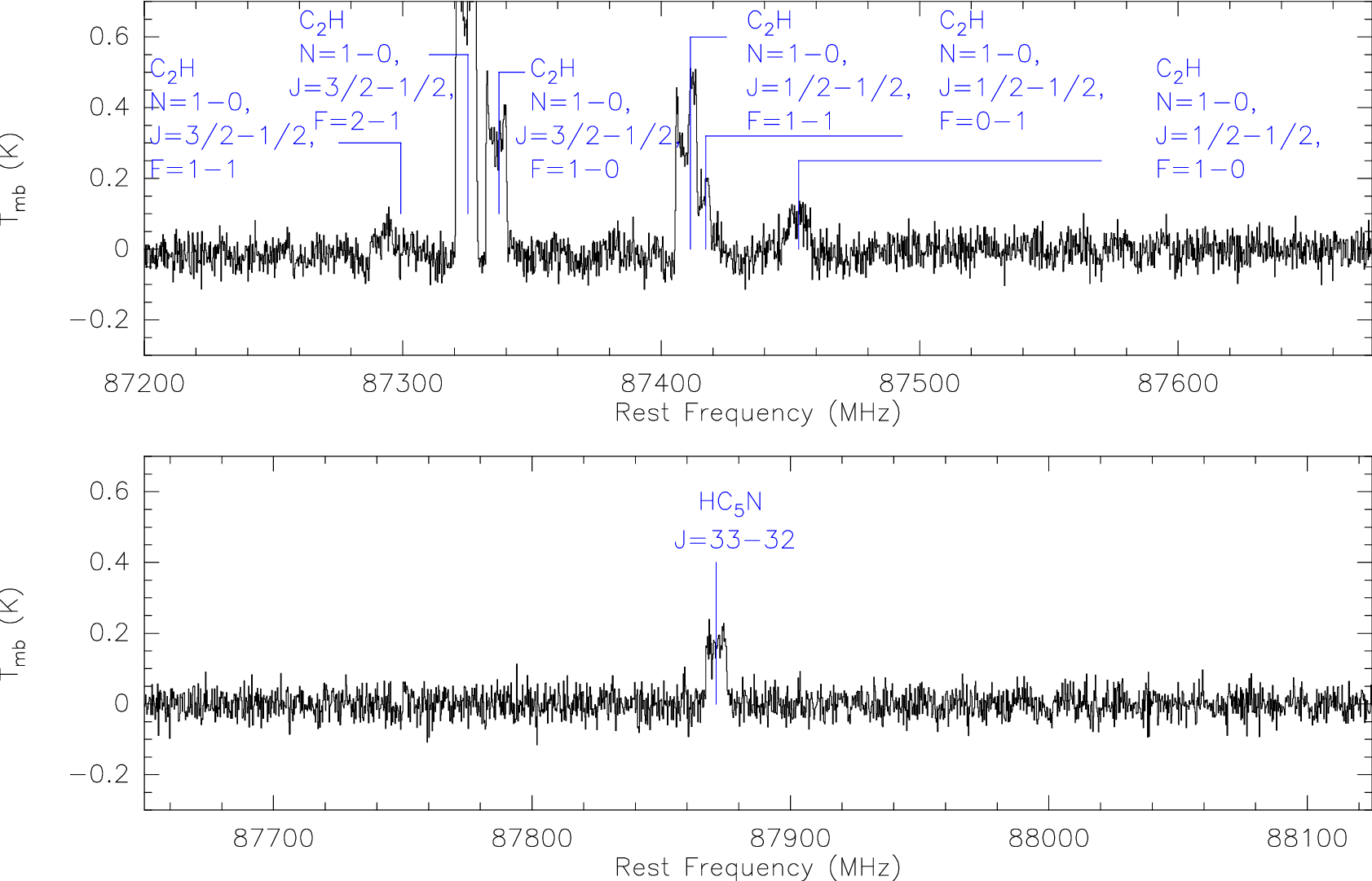}{0.85\textwidth}{ }
          }
\caption{Same as Figure~\ref{fig:F1}.
\label{fig:F2}}
\end{figure*}

\begin{figure*}
\gridline{\fig{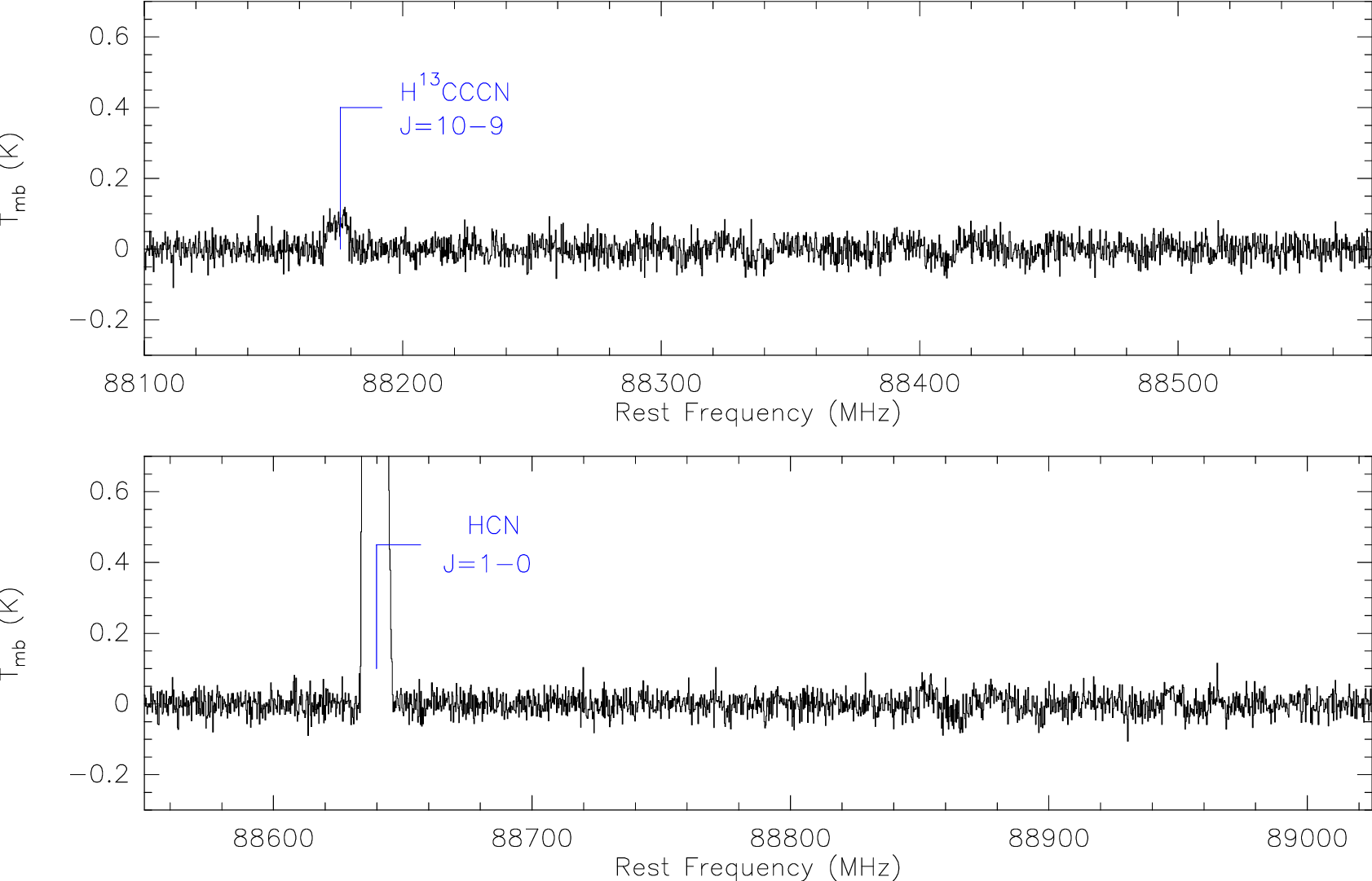}{0.85\textwidth}{ }
          }
\gridline{\fig{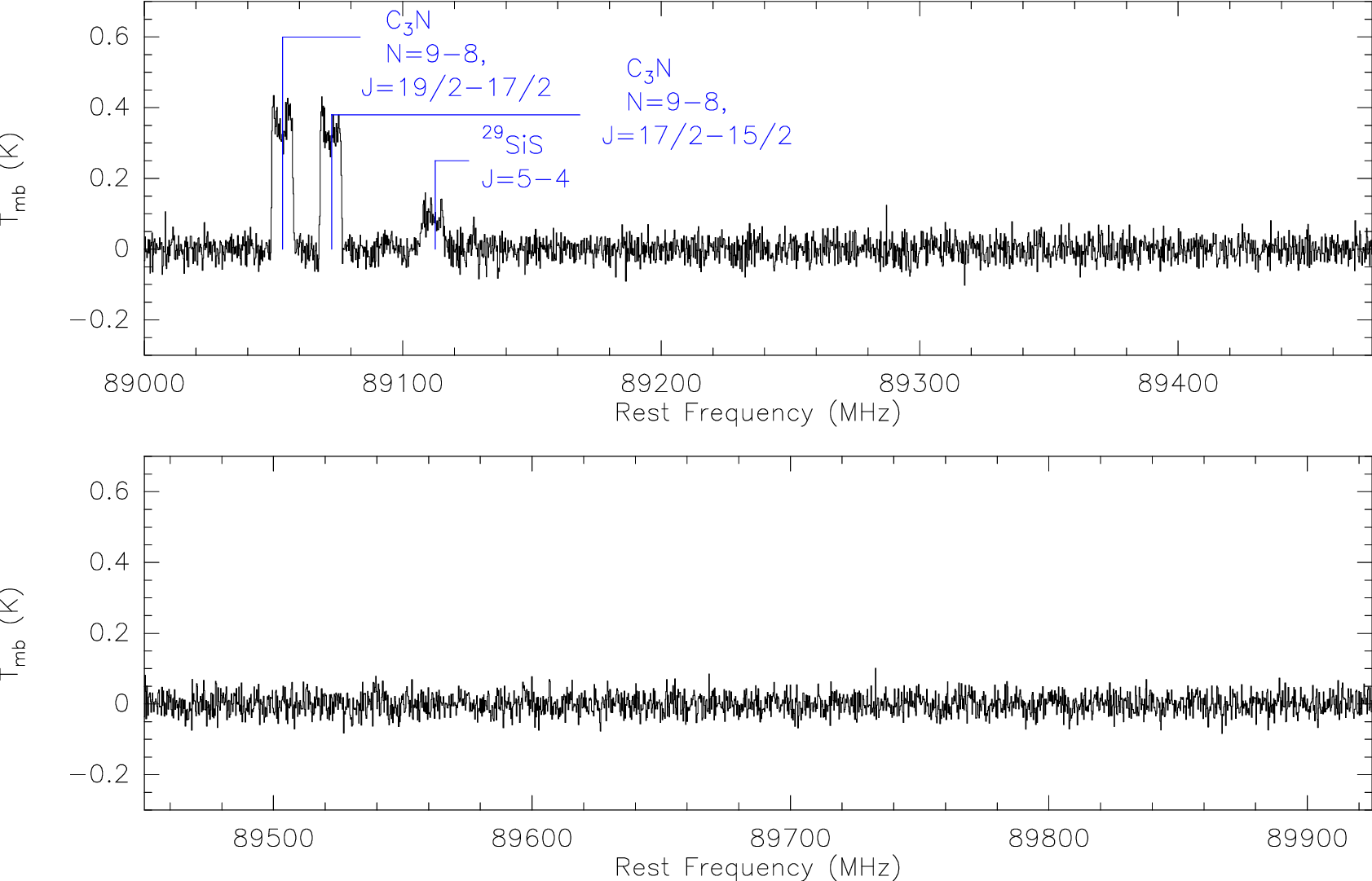}{0.85\textwidth}{ }
          }
\caption{Same as Figure~\ref{fig:F1}.
\label{fig:F3}}
\end{figure*}

\begin{figure*}
\gridline{\fig{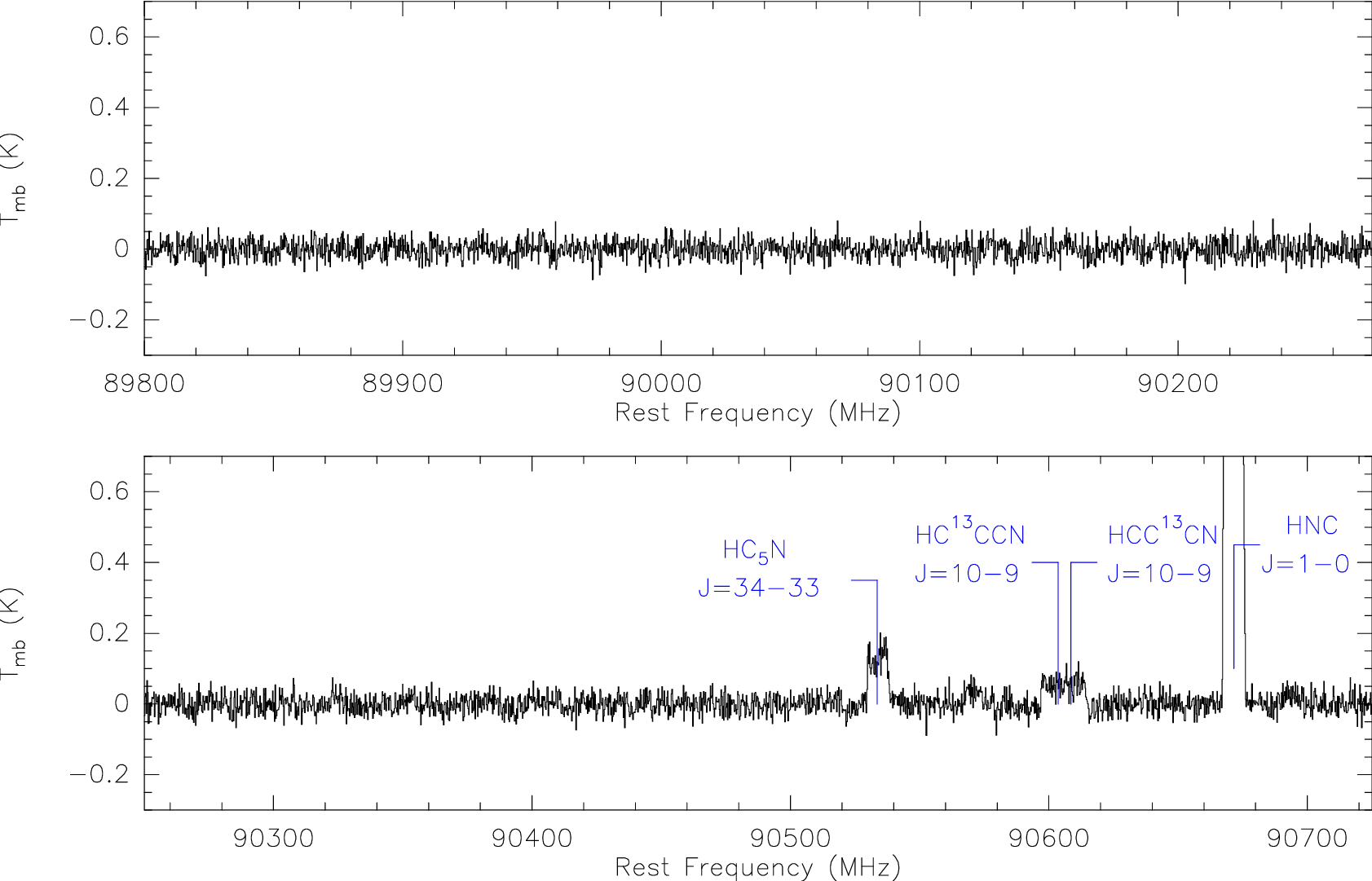}{0.85\textwidth}{ }
          }
\gridline{\fig{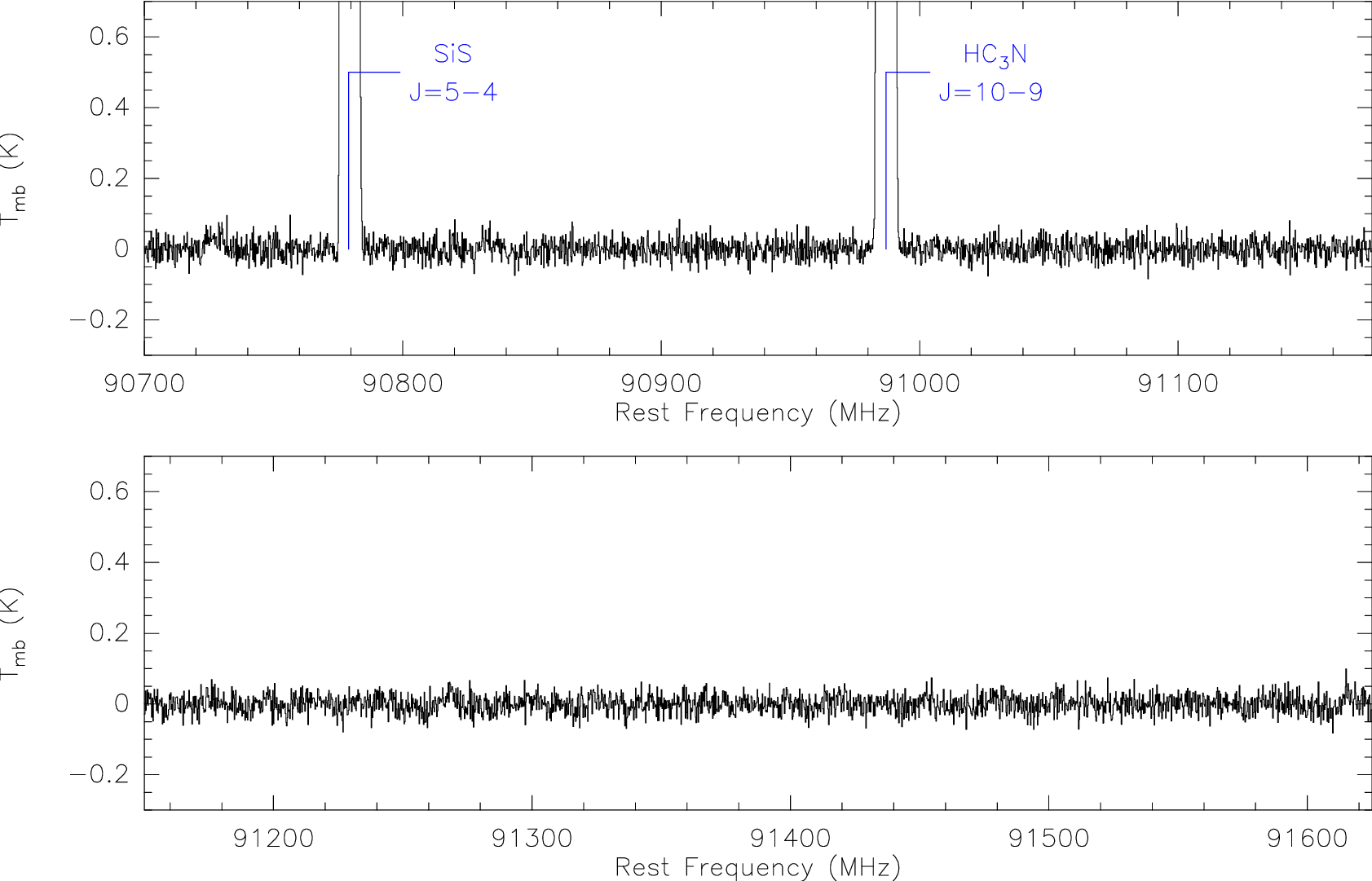}{0.85\textwidth}{ }
          }
\caption{Same as Figure~\ref{fig:F1}.
\label{fig:F4}}
\end{figure*}

\begin{figure*}
\gridline{\fig{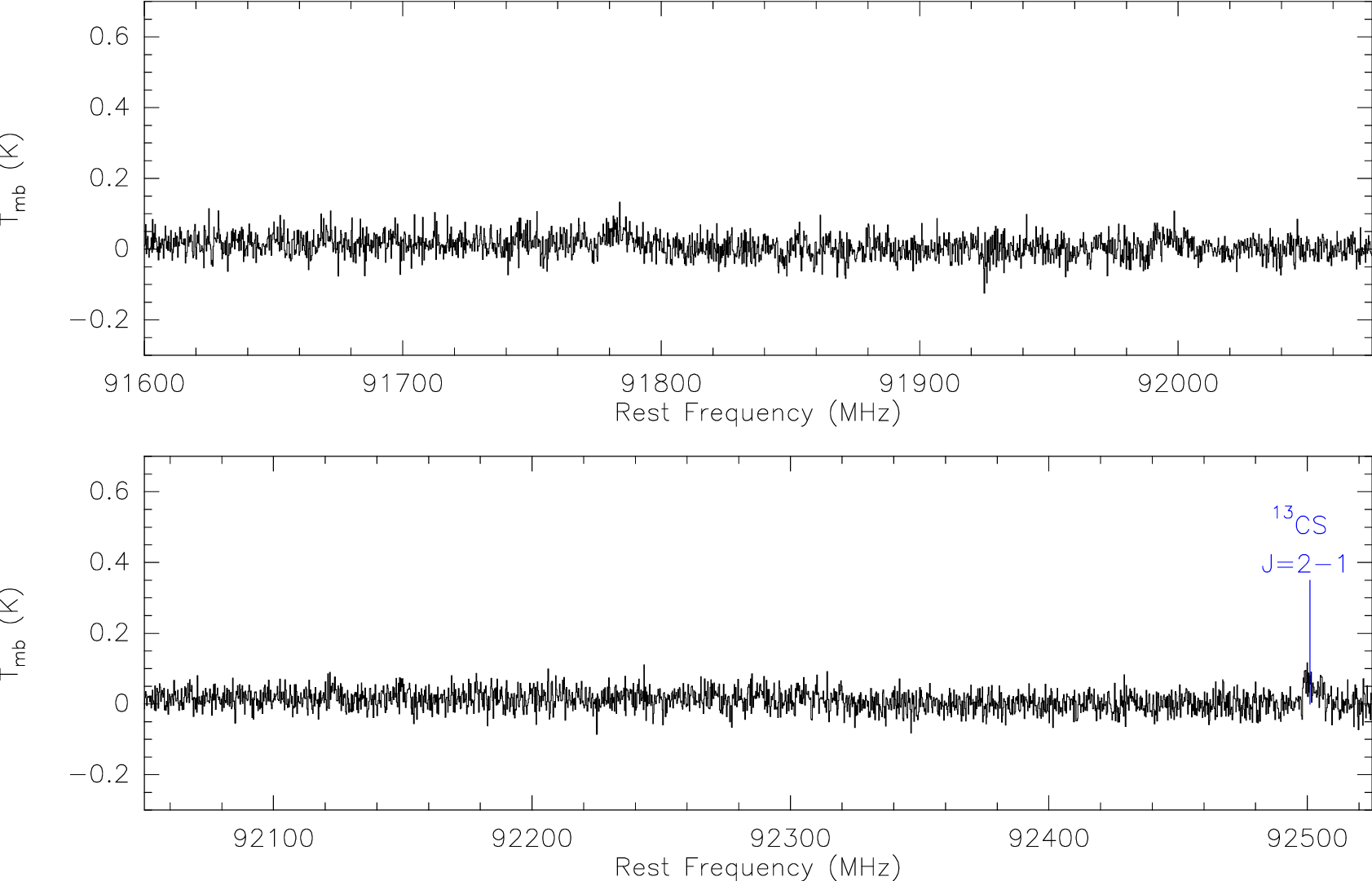}{0.85\textwidth}{ }
          }
\gridline{\fig{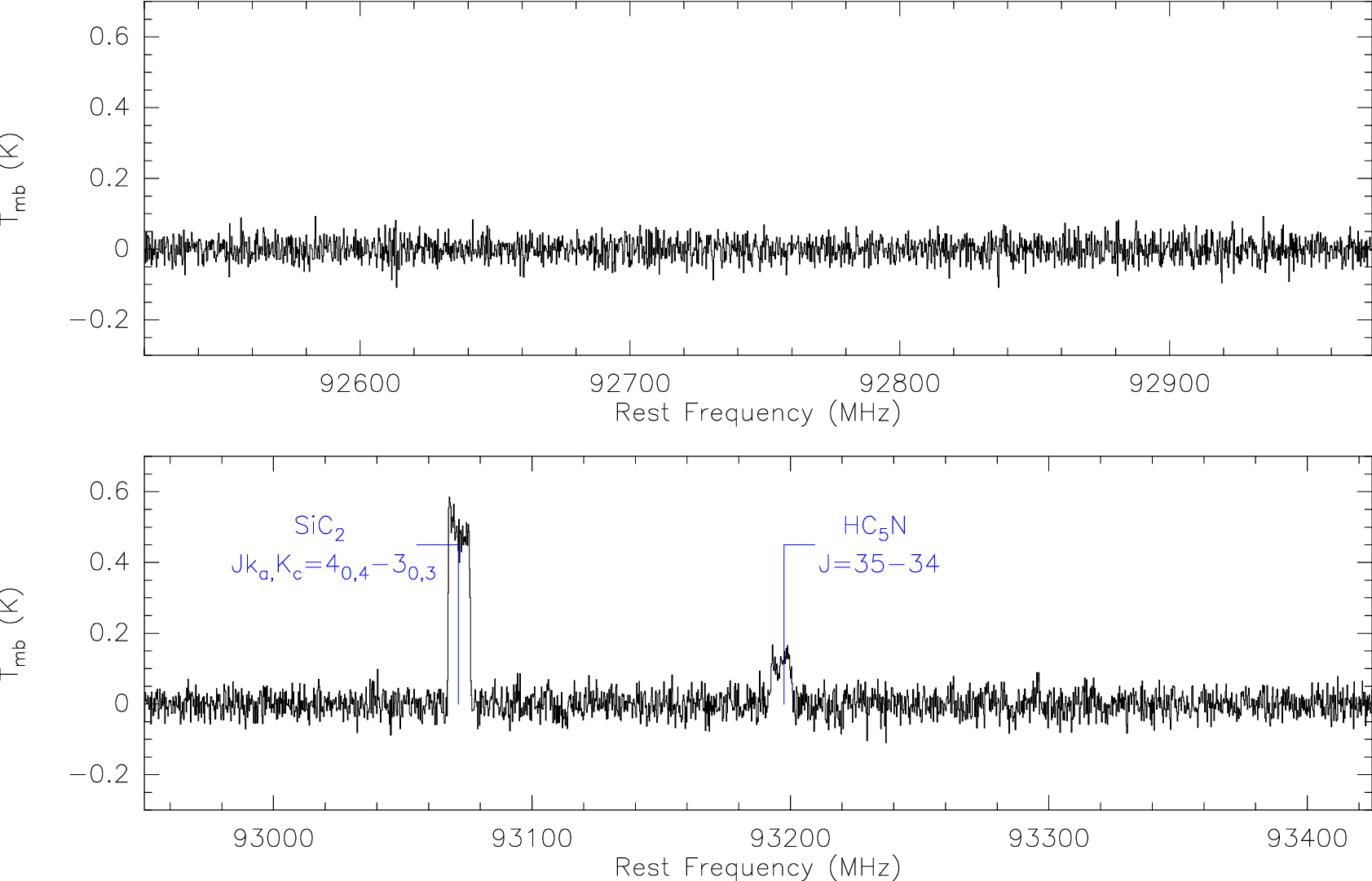}{0.85\textwidth}{ }
          }
\caption{Same as Figure~\ref{fig:F1}.
\label{fig:F5}}
\end{figure*}

\begin{figure*}
\gridline{\fig{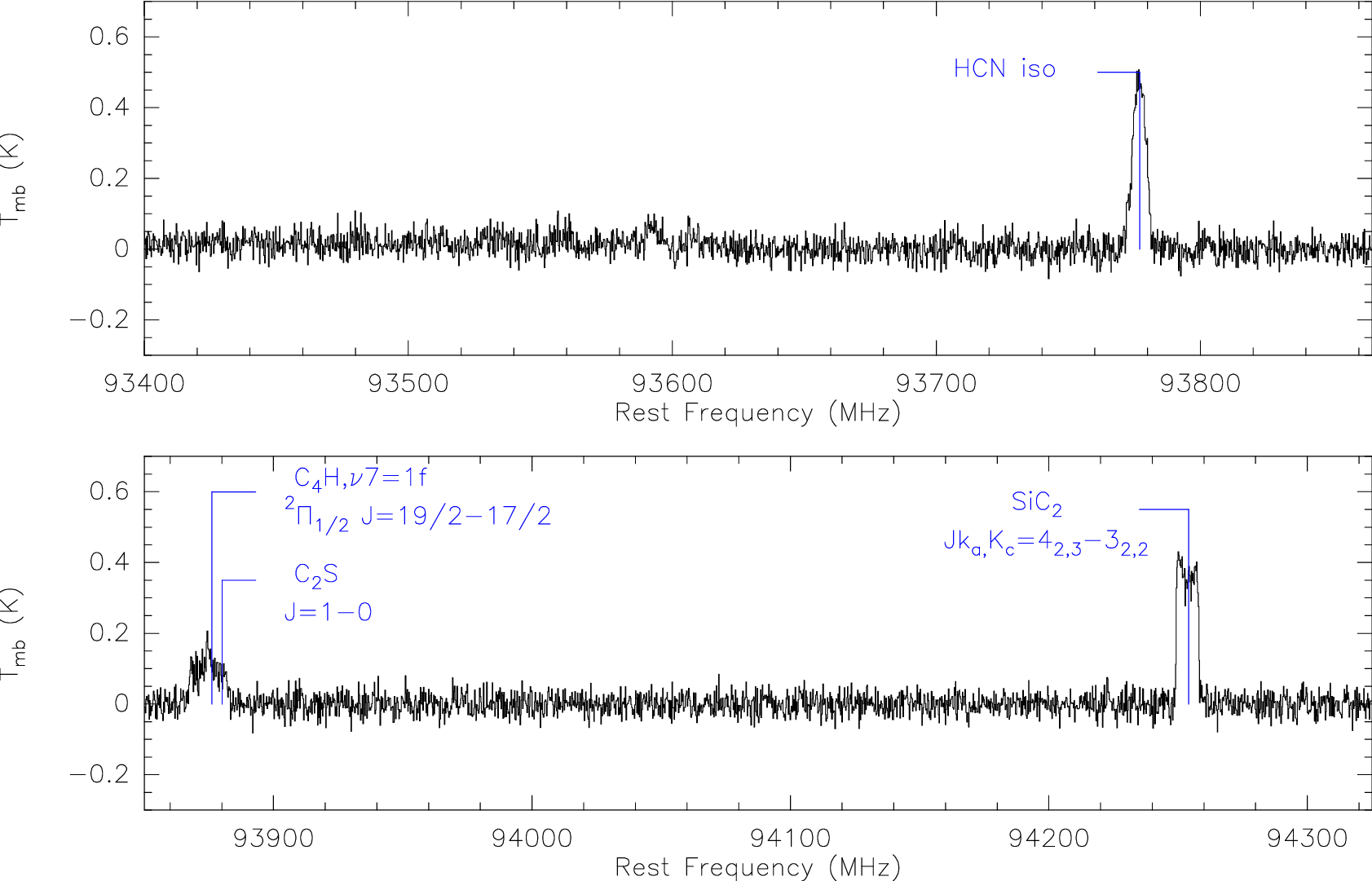}{0.85\textwidth}{ }
          }
\gridline{\fig{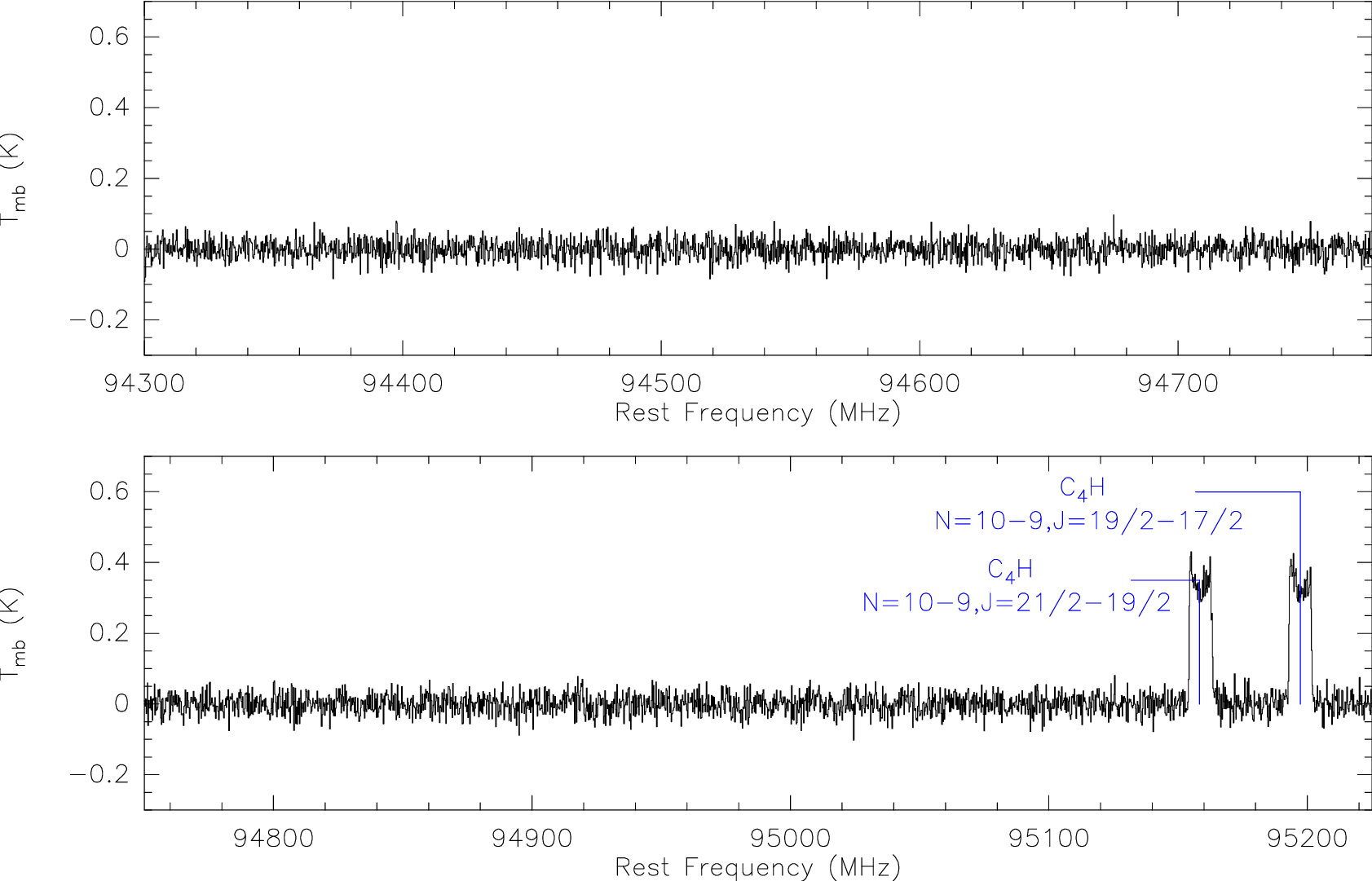}{0.85\textwidth}{ }
          }
\caption{Same as Figure~\ref{fig:F1}.
\label{fig:F6}}
\end{figure*}

\begin{figure*}
\gridline{\fig{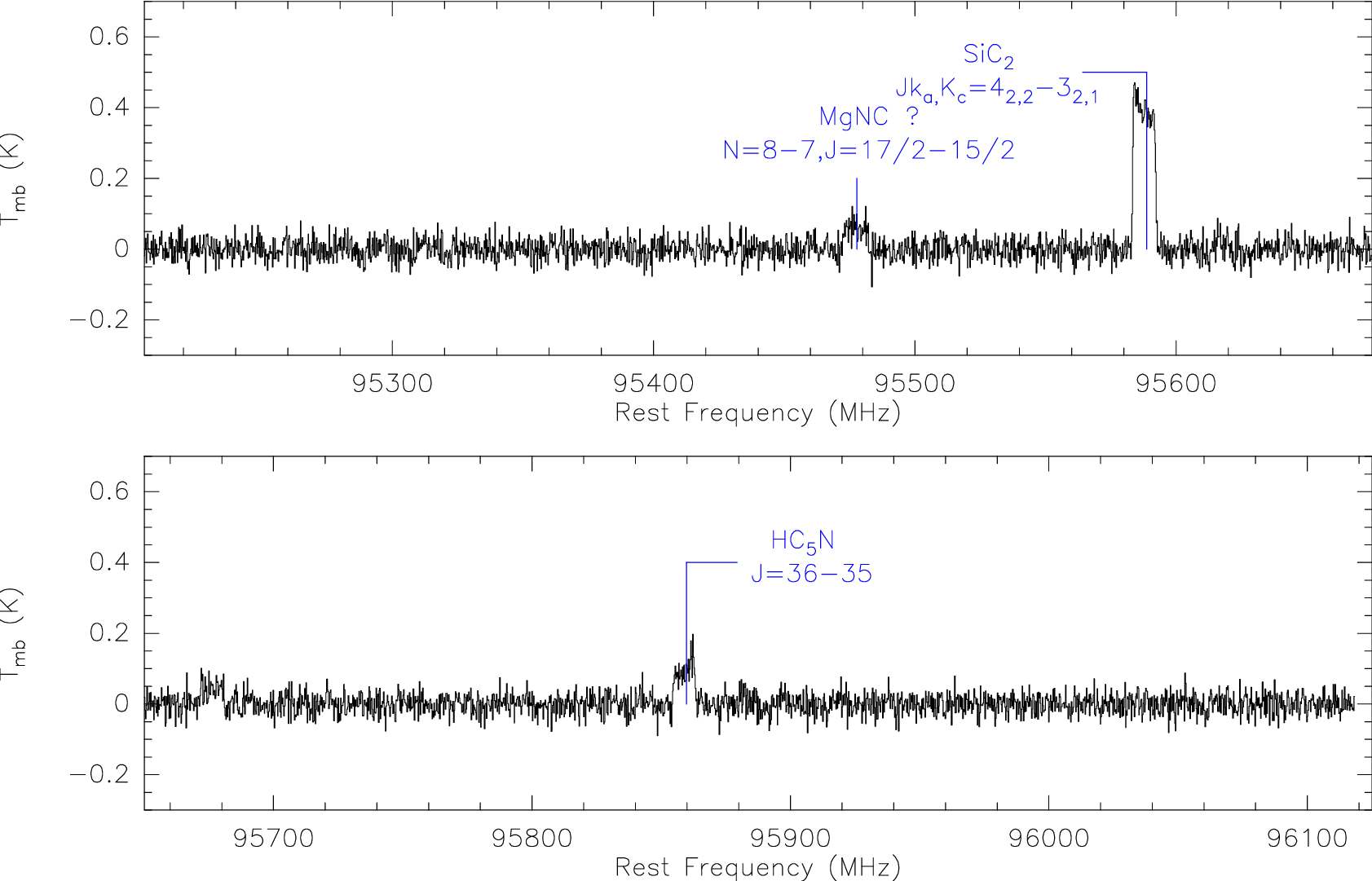}{0.85\textwidth}{ }
          }
\gridline{\fig{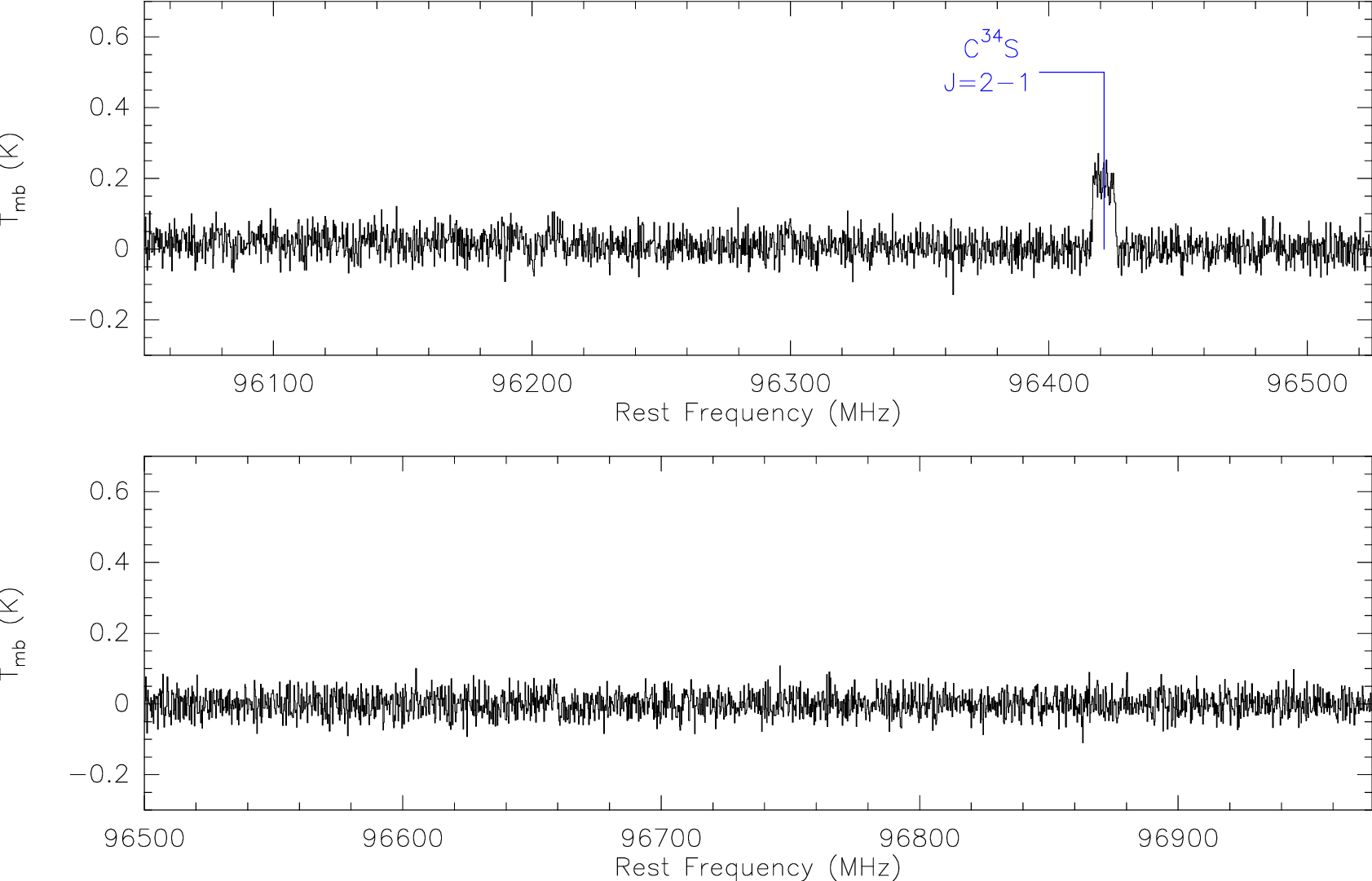}{0.85\textwidth}{ }
          }
\caption{Same as Figure~\ref{fig:F1}.
\label{fig:F7}}
\end{figure*}

\begin{figure*}
\gridline{\fig{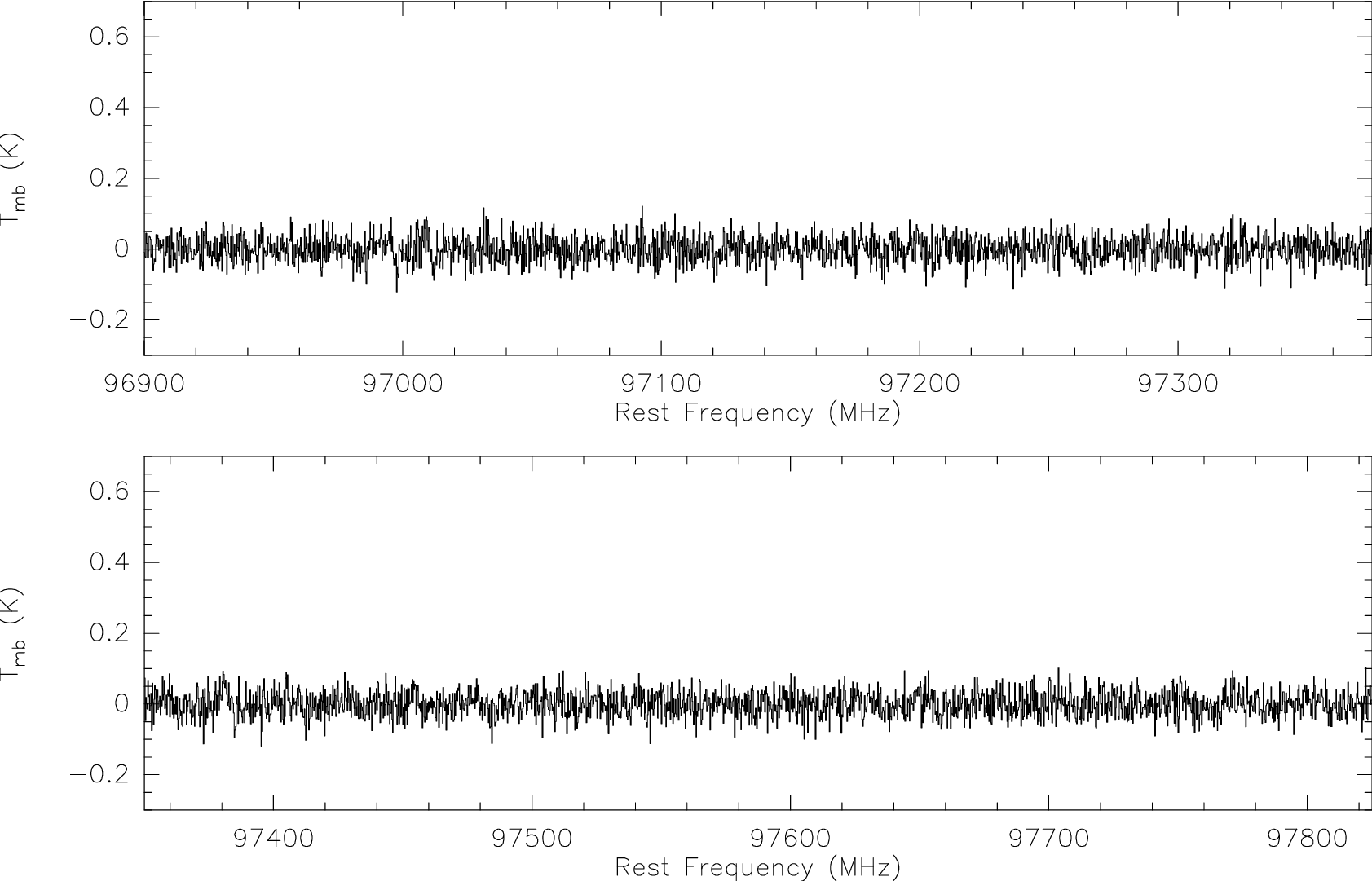}{0.85\textwidth}{ }
          }
\gridline{\fig{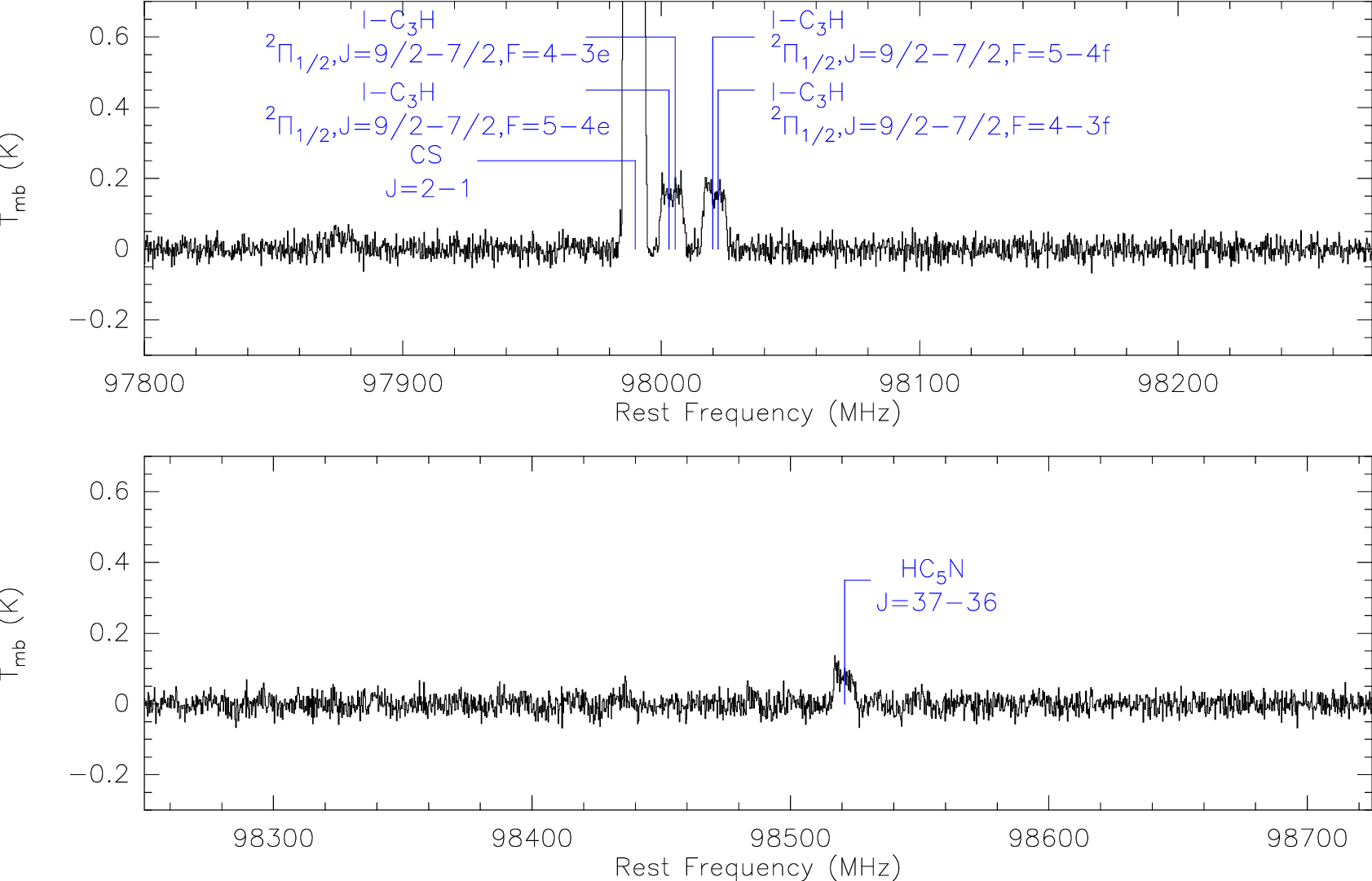}{0.85\textwidth}{ }
          }
\caption{Same as Figure~\ref{fig:F1}.
\label{fig:F8}}
\end{figure*}

\begin{figure*}
\gridline{\fig{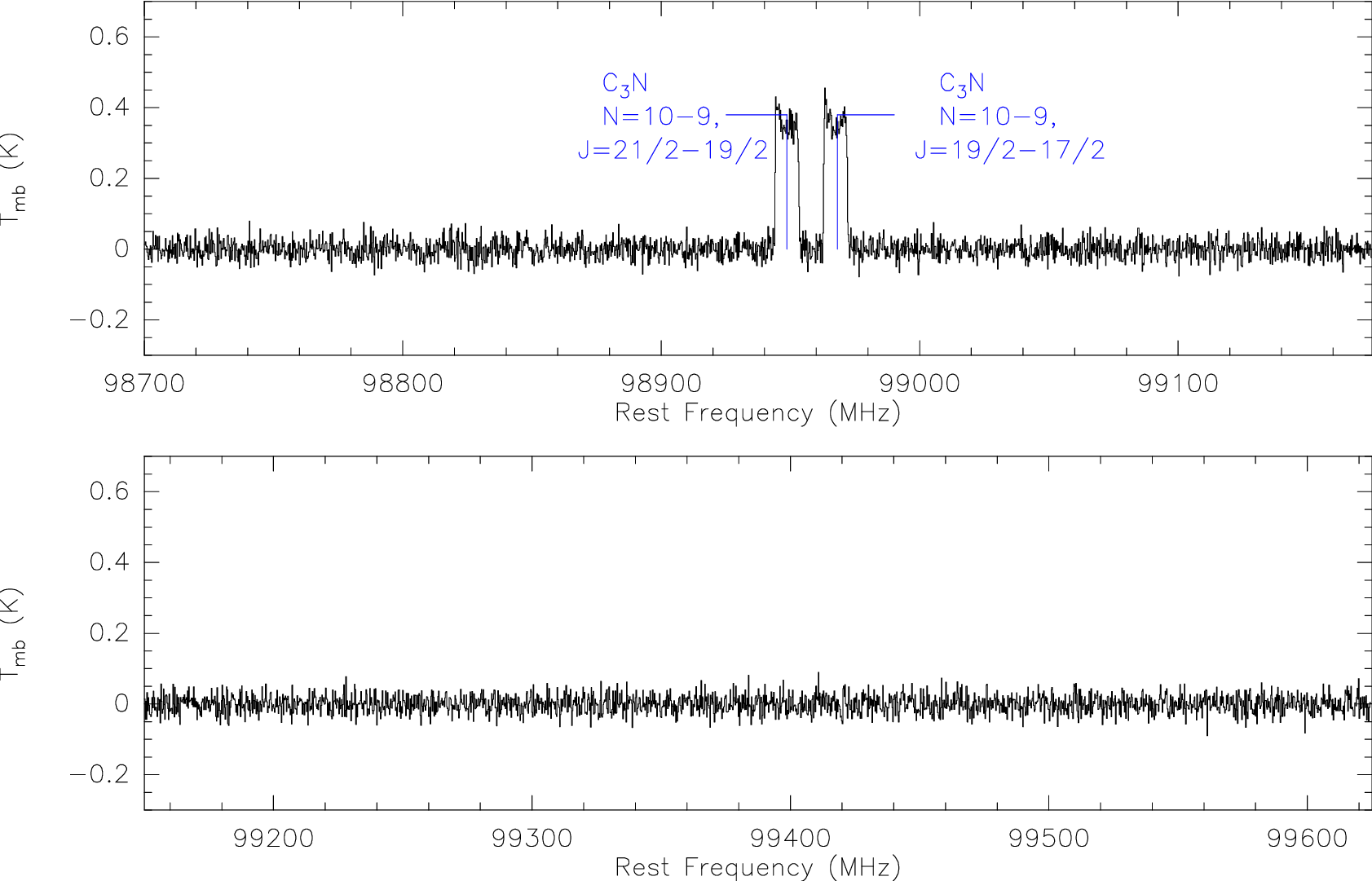}{0.85\textwidth}{ }
          }
\gridline{\fig{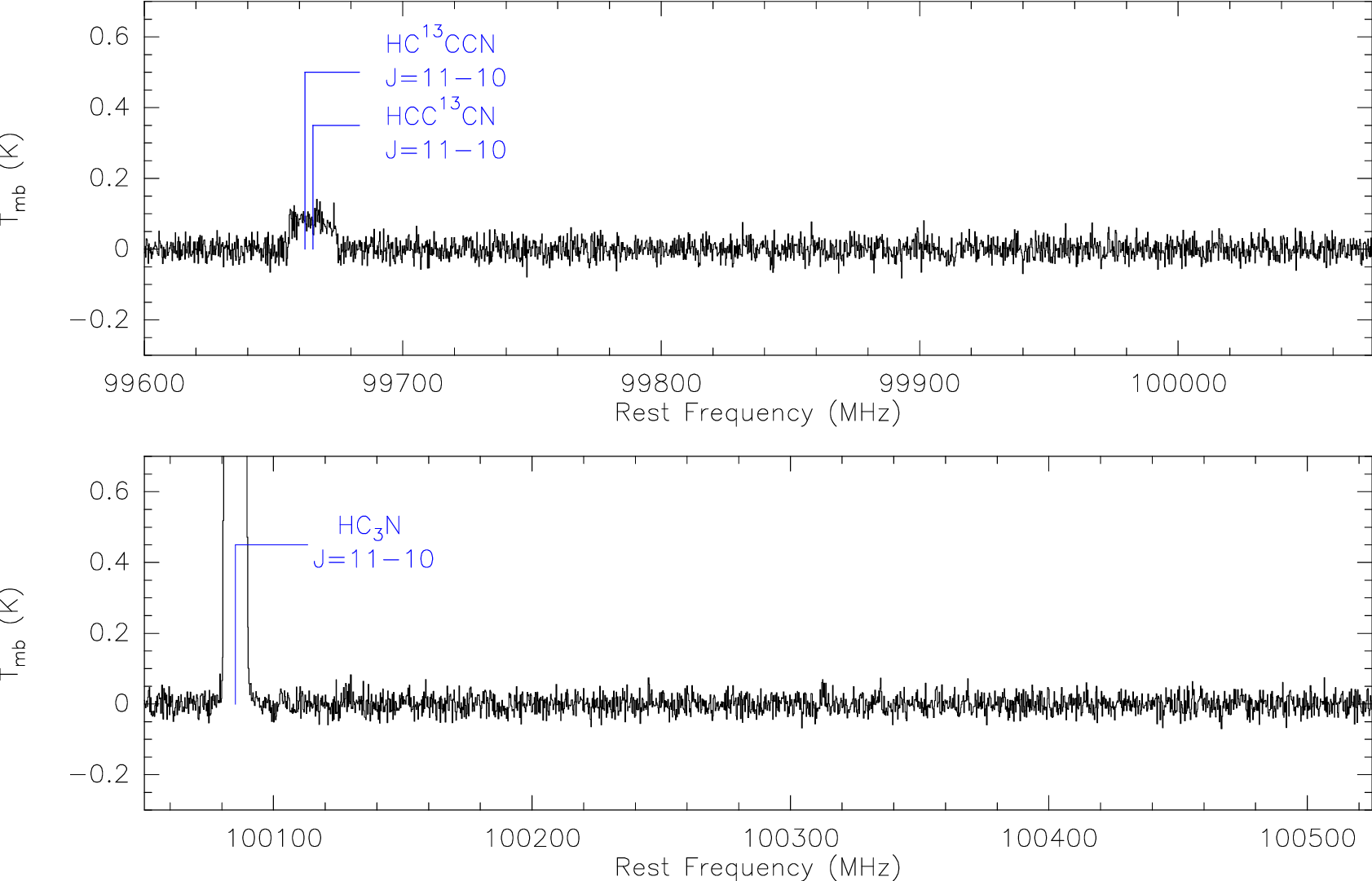}{0.85\textwidth}{ }
          }
\caption{Same as Figure~\ref{fig:F1}.
\label{fig:F9}}
\end{figure*}

\begin{figure*}
\gridline{\fig{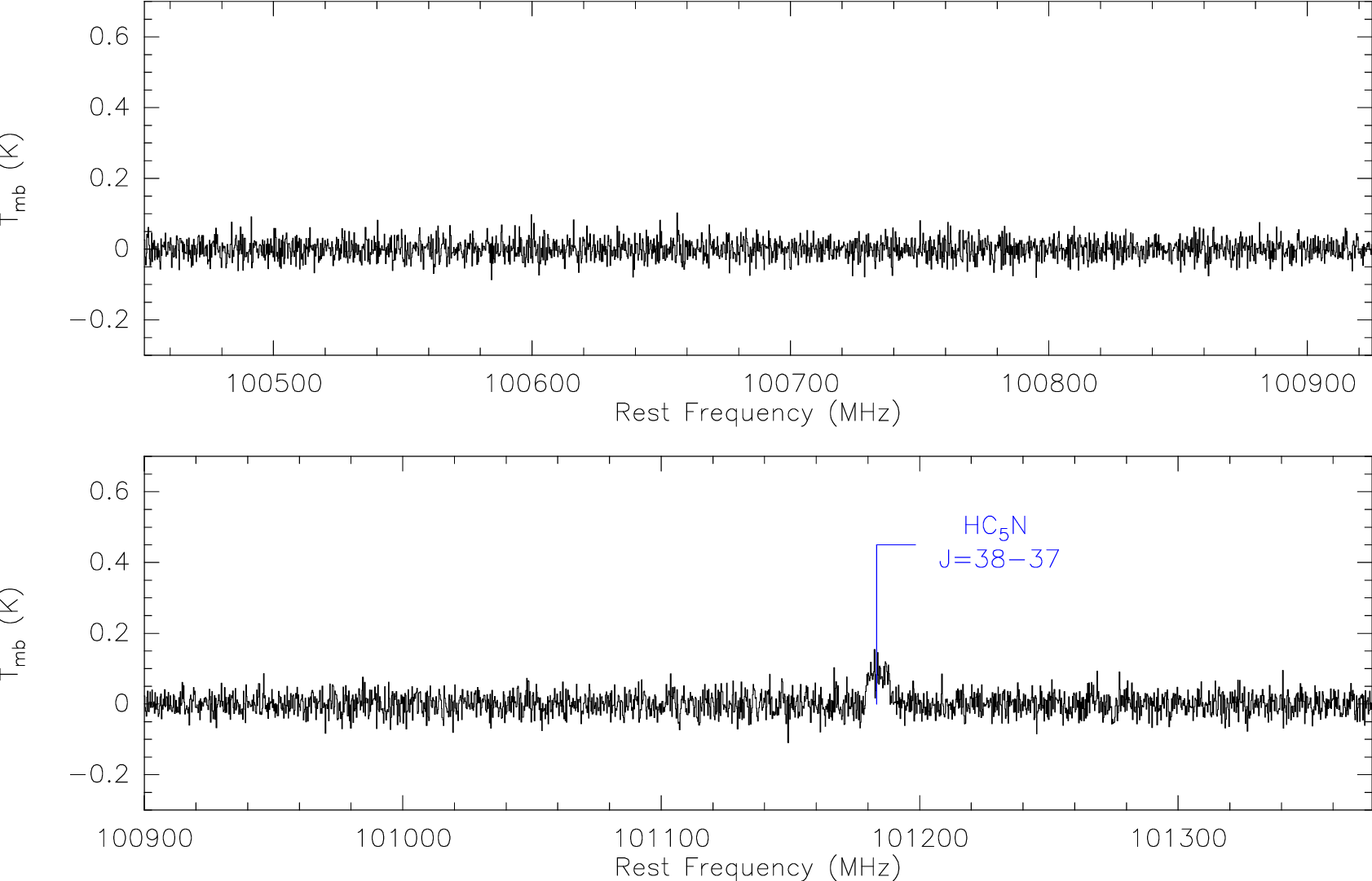}{0.85\textwidth}{ }
          }
\gridline{\fig{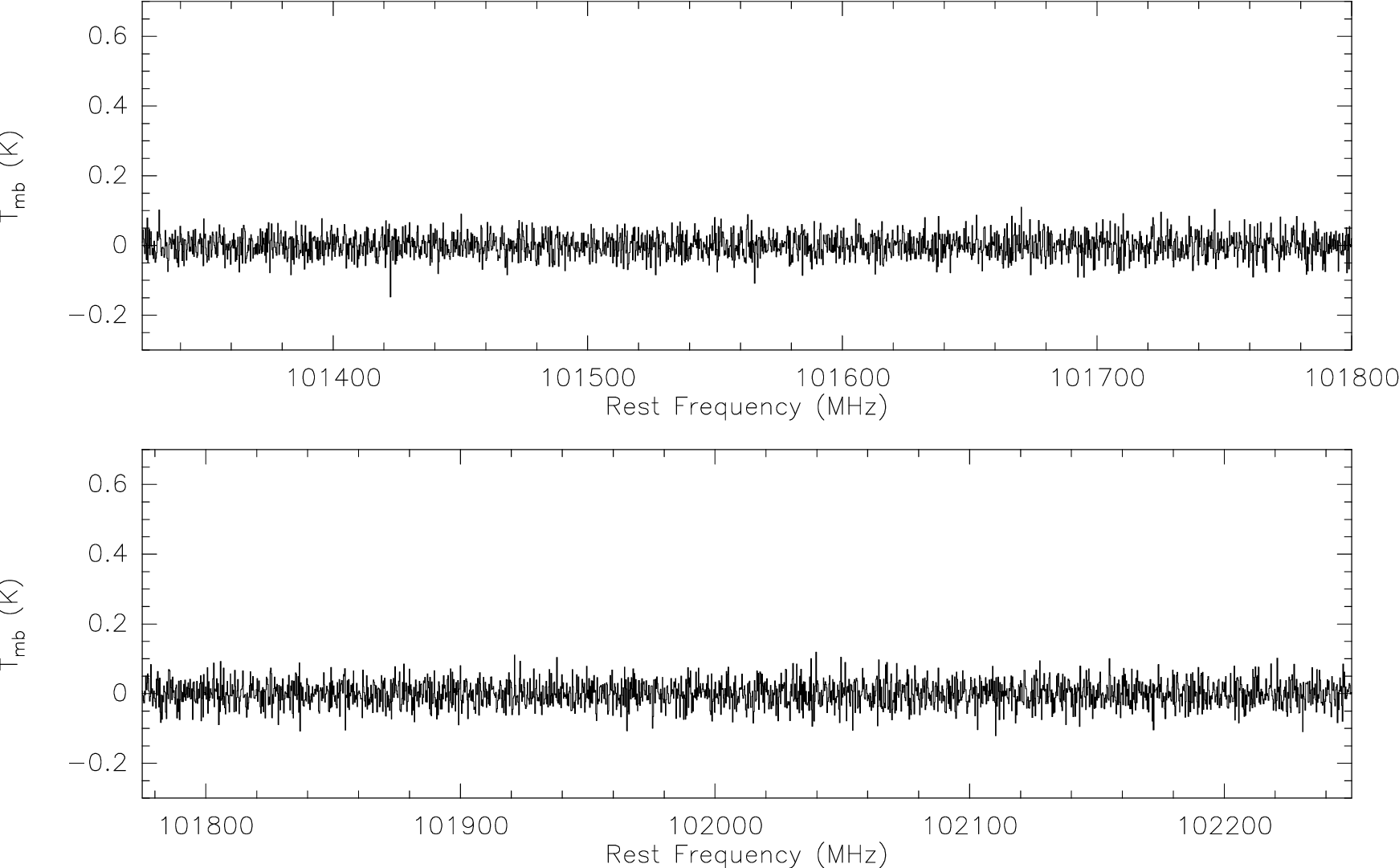}{0.85\textwidth}{ }
          }
\caption{Same as Figure~\ref{fig:F1}.
\label{fig:F10}}
\end{figure*}

\begin{figure*}
\gridline{\fig{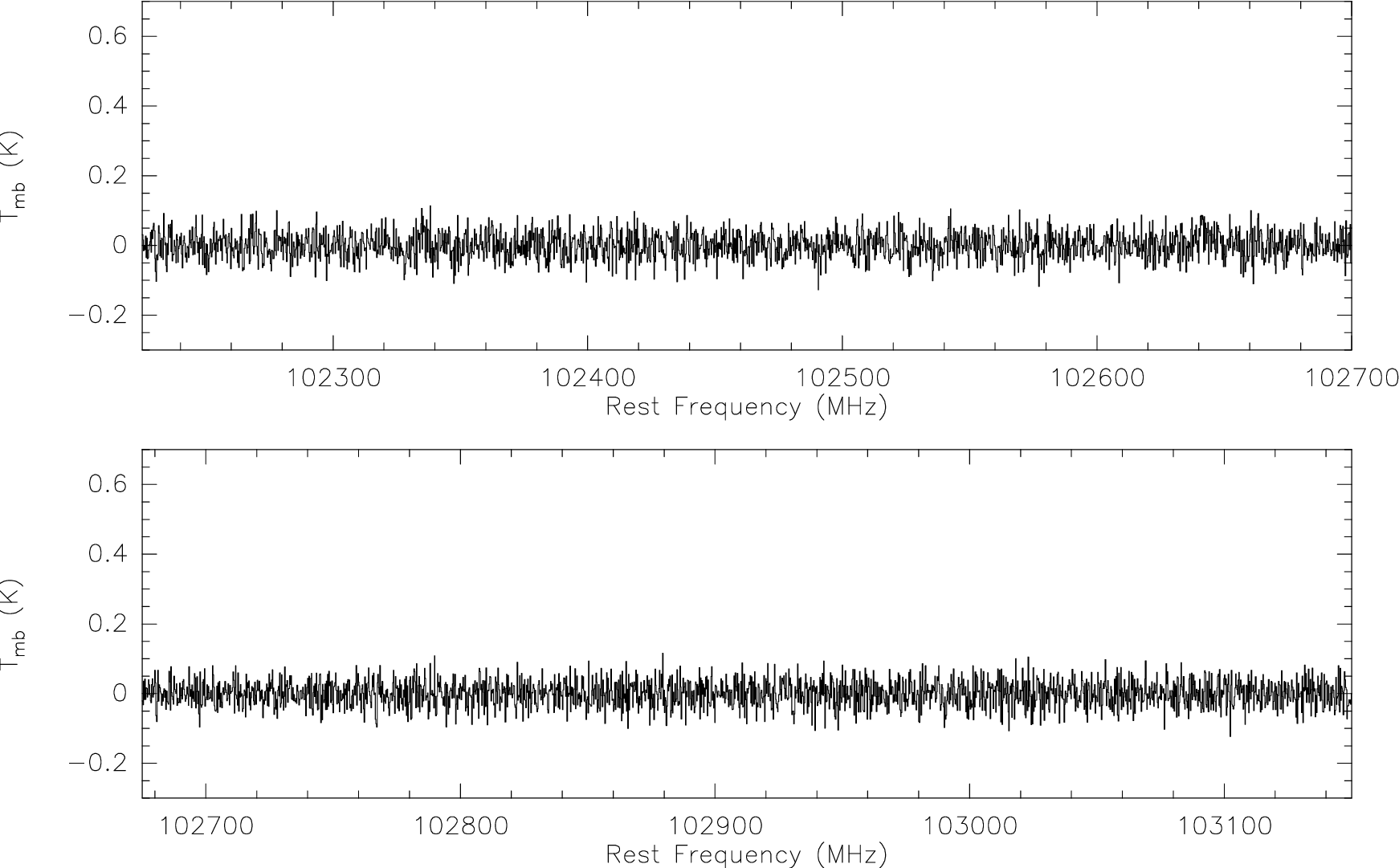}{0.85\textwidth}{ }
          }
\gridline{\fig{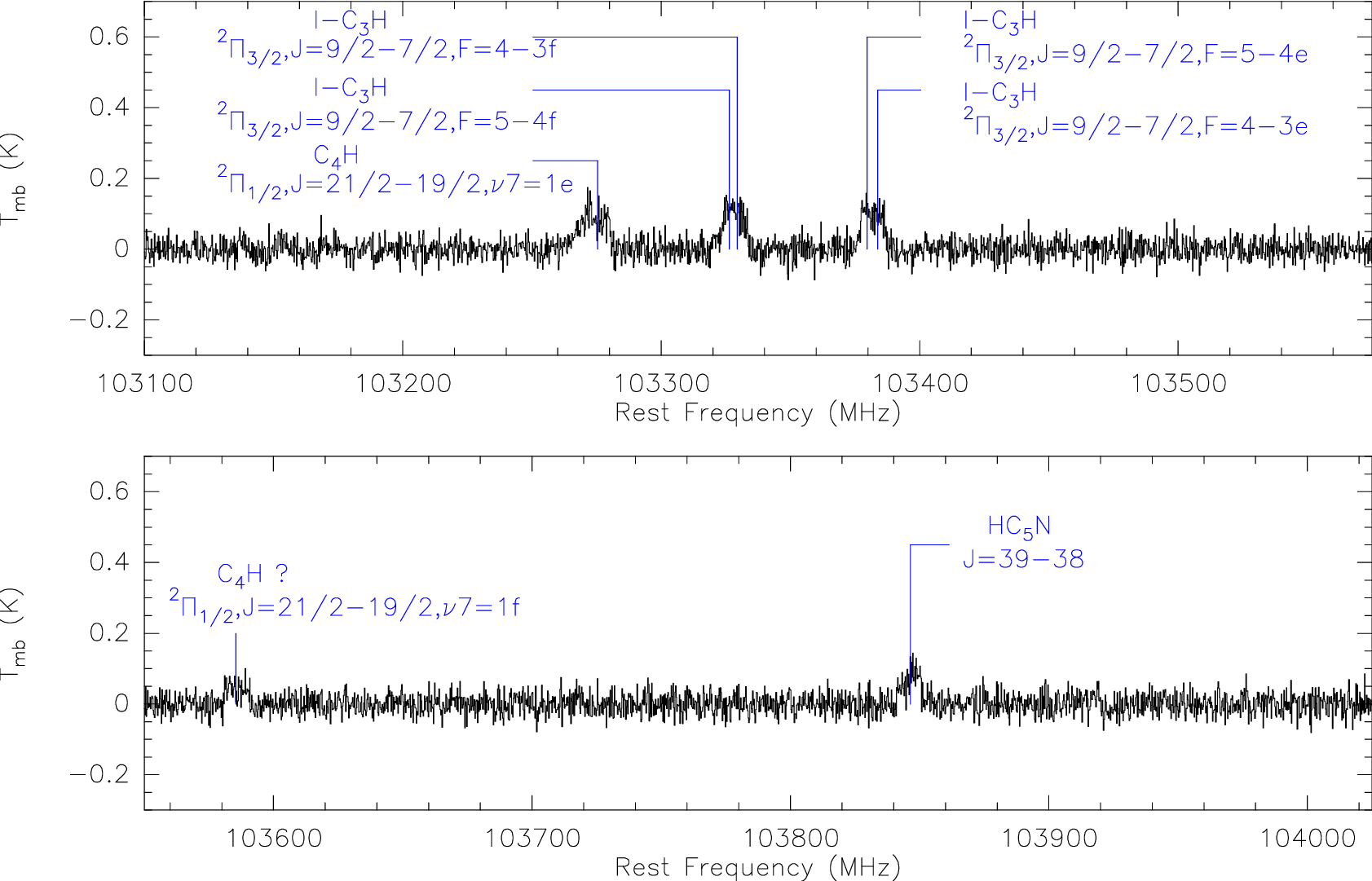}{0.85\textwidth}{ }
          }
\caption{Same as Figure~\ref{fig:F1}.
\label{fig:F11}}
\end{figure*}

\begin{figure*}
\gridline{\fig{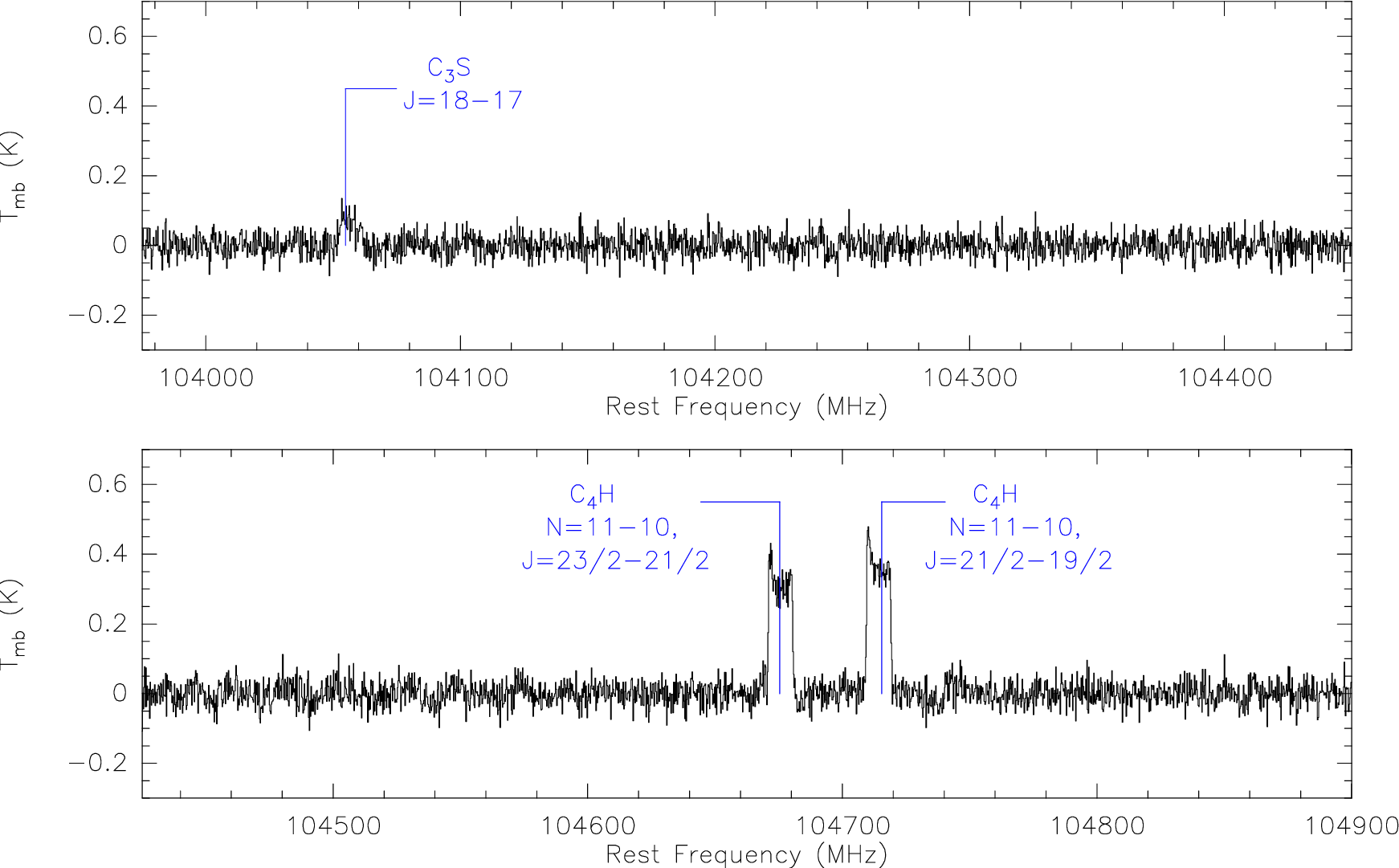}{0.85\textwidth}{ }
          }
\gridline{\fig{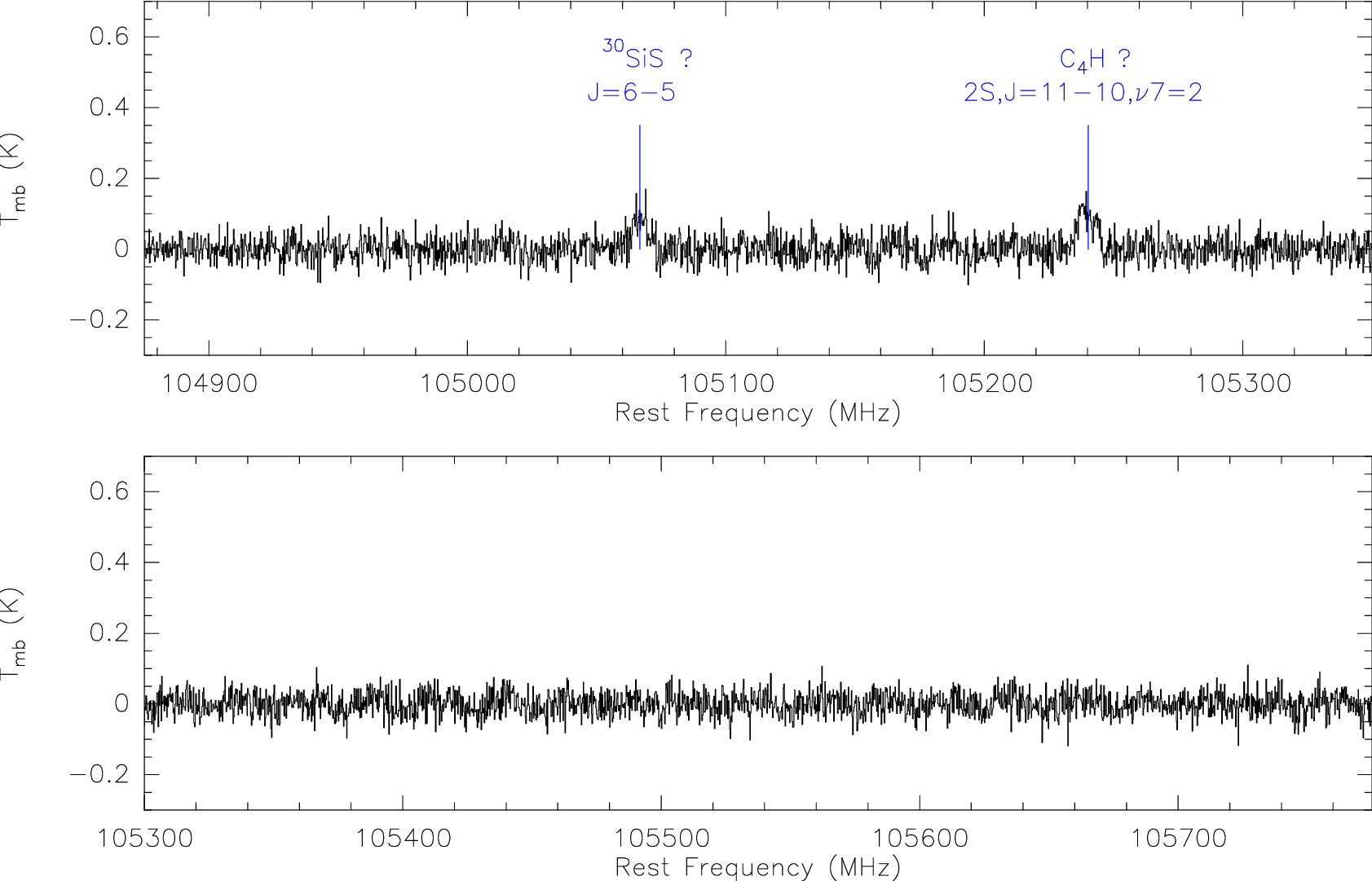}{0.85\textwidth}{ }
          }
\caption{Same as Figure~\ref{fig:F1}.
\label{fig:F12}}
\end{figure*}

\begin{figure*}
\gridline{\fig{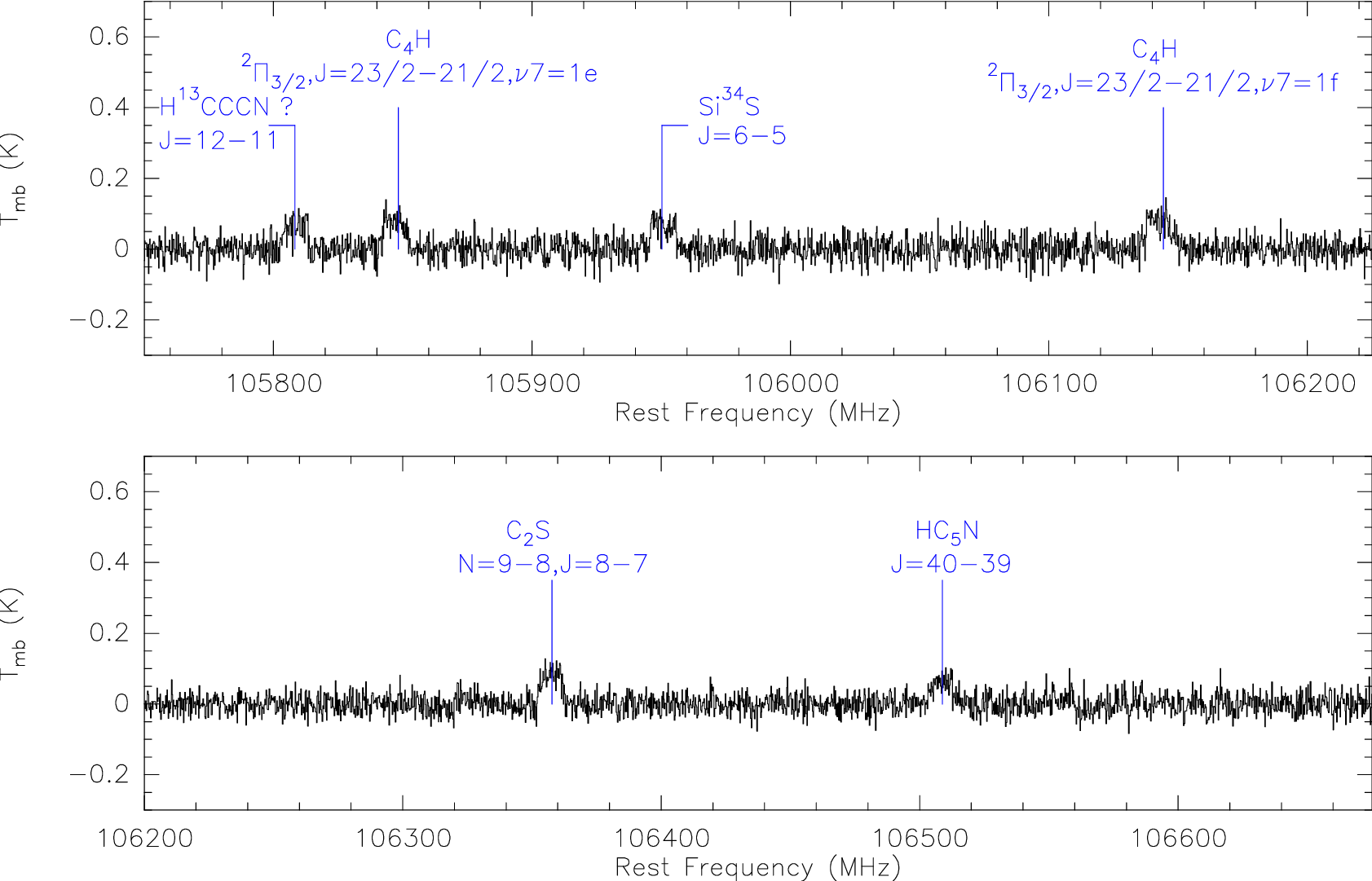}{0.85\textwidth}{ }
          }
\gridline{\fig{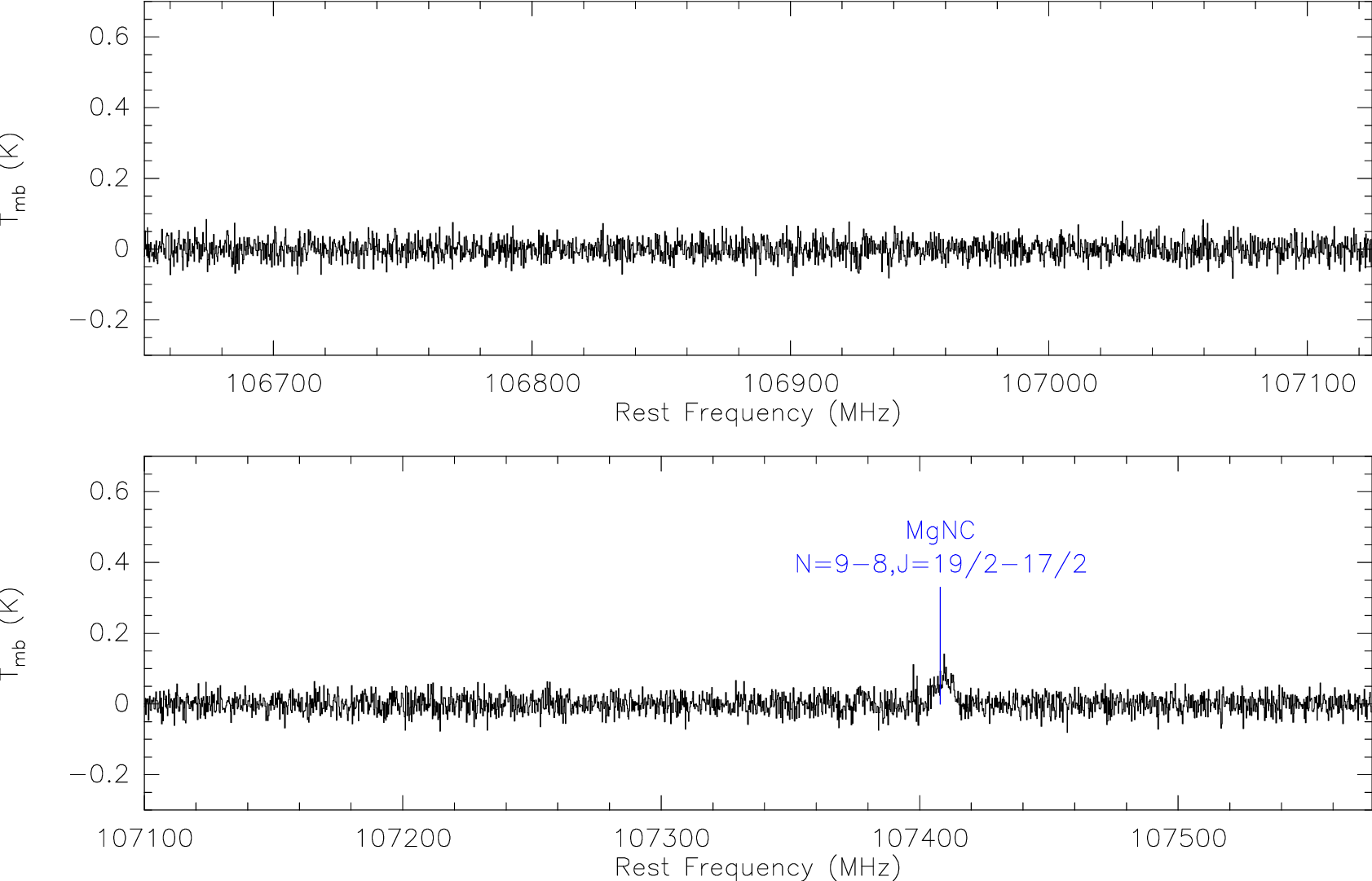}{0.85\textwidth}{ }
          }
\caption{Same as Figure~\ref{fig:F1}.
\label{fig:F13}}
\end{figure*}

\begin{figure*}
\gridline{\fig{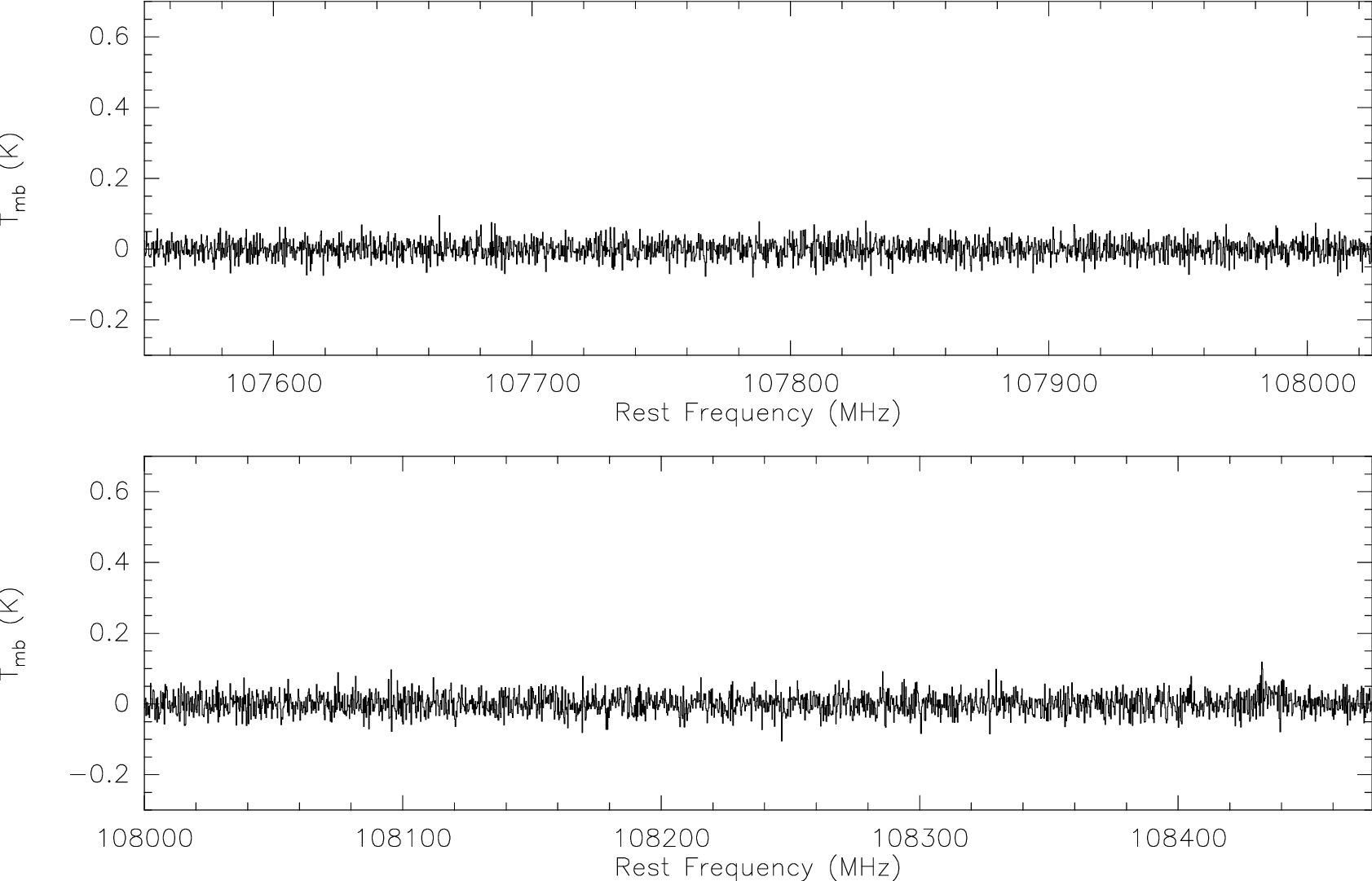}{0.85\textwidth}{ }
          }
\gridline{\fig{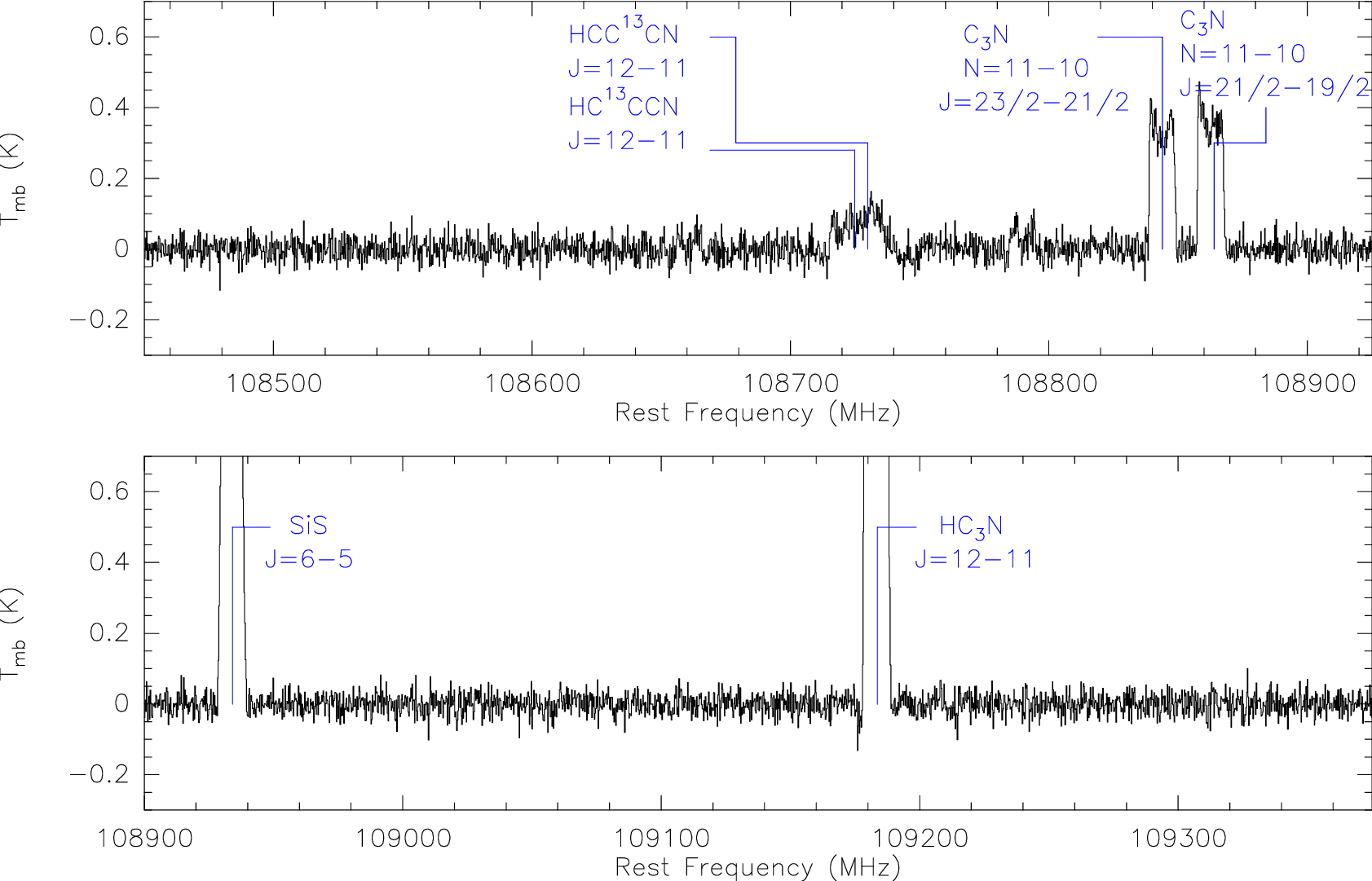}{0.85\textwidth}{ }
          }
\caption{Same as Figure~\ref{fig:F1}.
\label{fig:F14}}
\end{figure*}

\begin{figure*}
\gridline{\fig{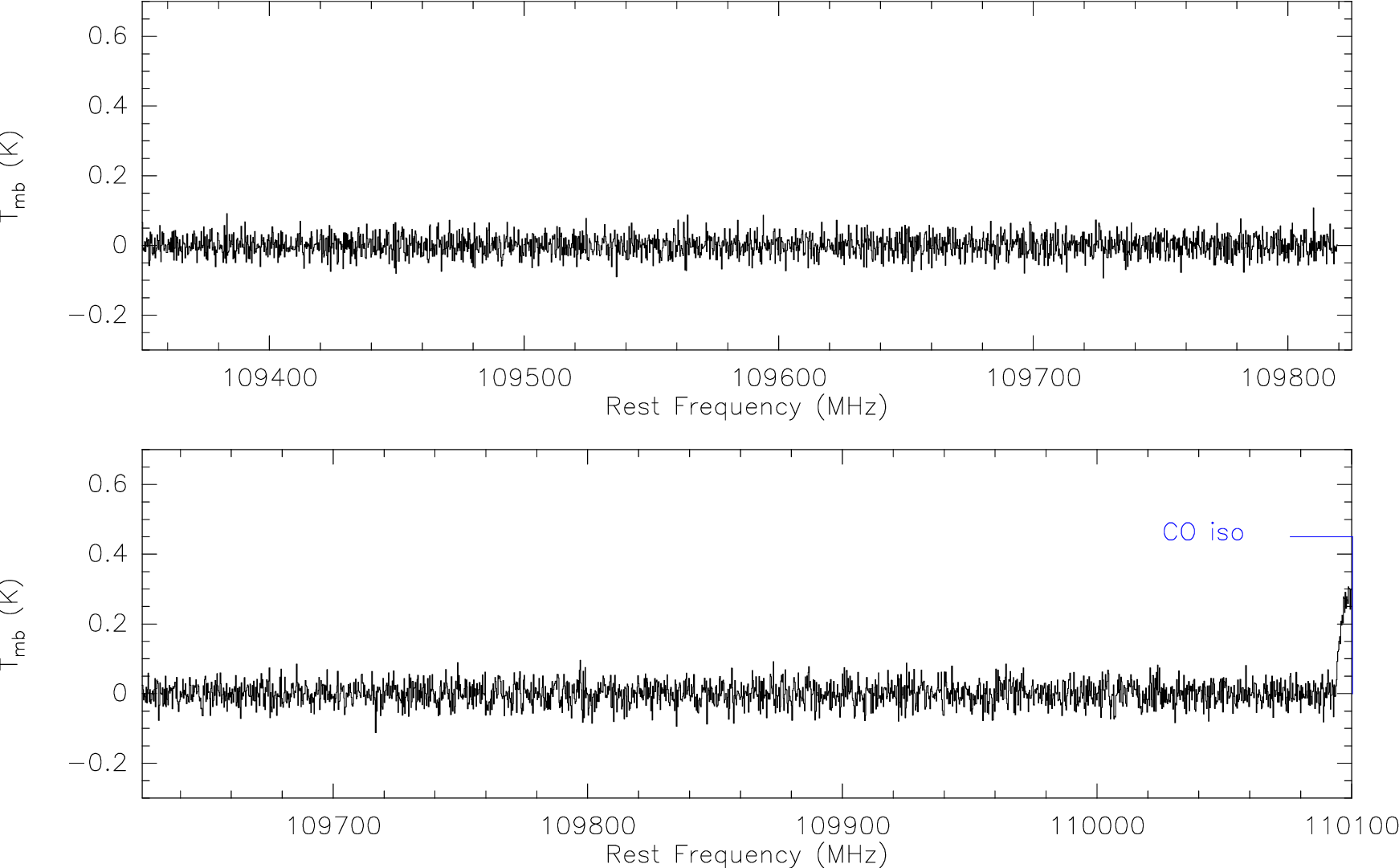}{0.85\textwidth}{ }
          }
\gridline{\fig{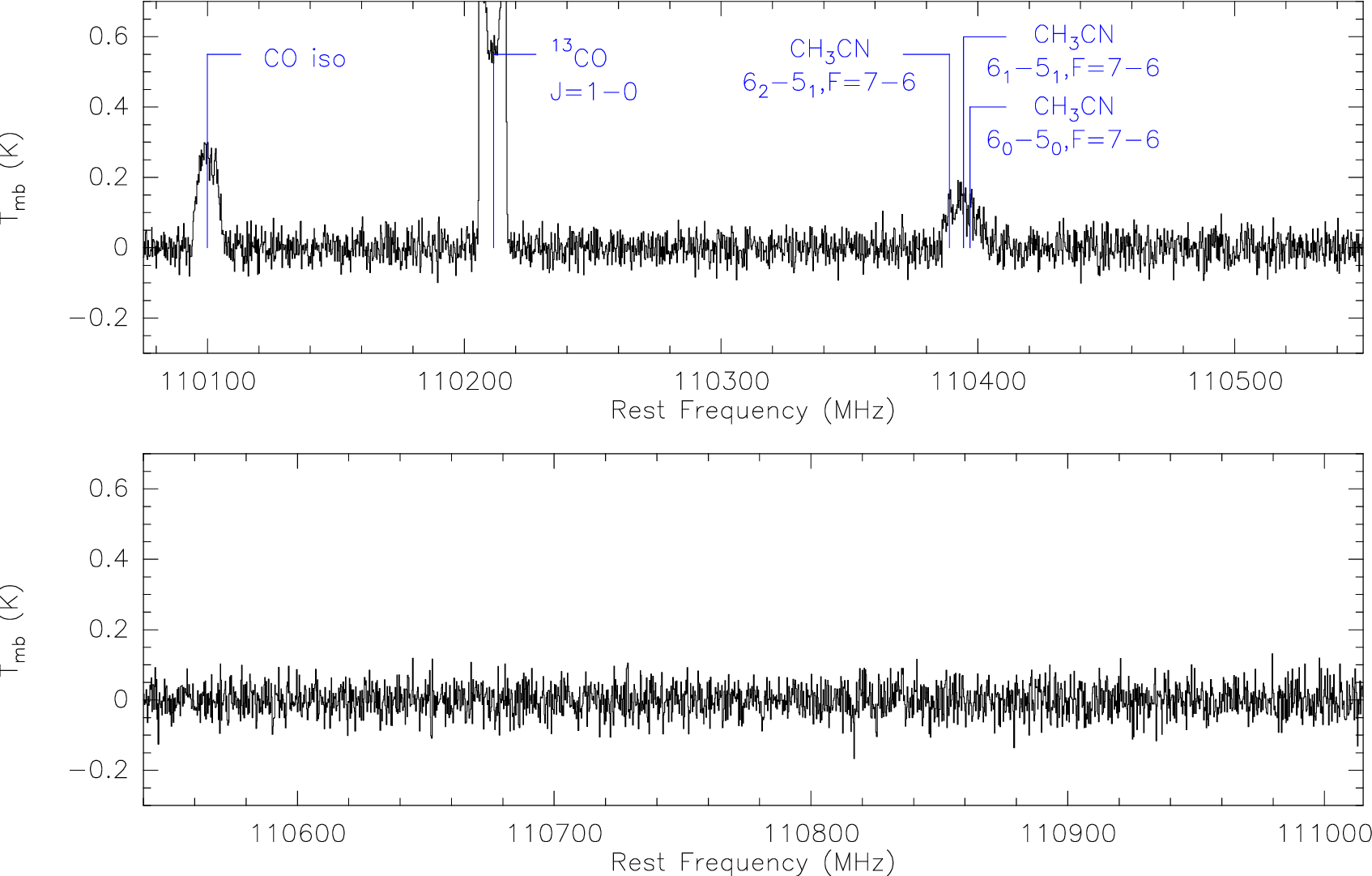}{0.85\textwidth}{ }
          }
\caption{Same as Figure~\ref{fig:F1}.
\label{fig:F15}}
\end{figure*}

\begin{figure*}
\gridline{\fig{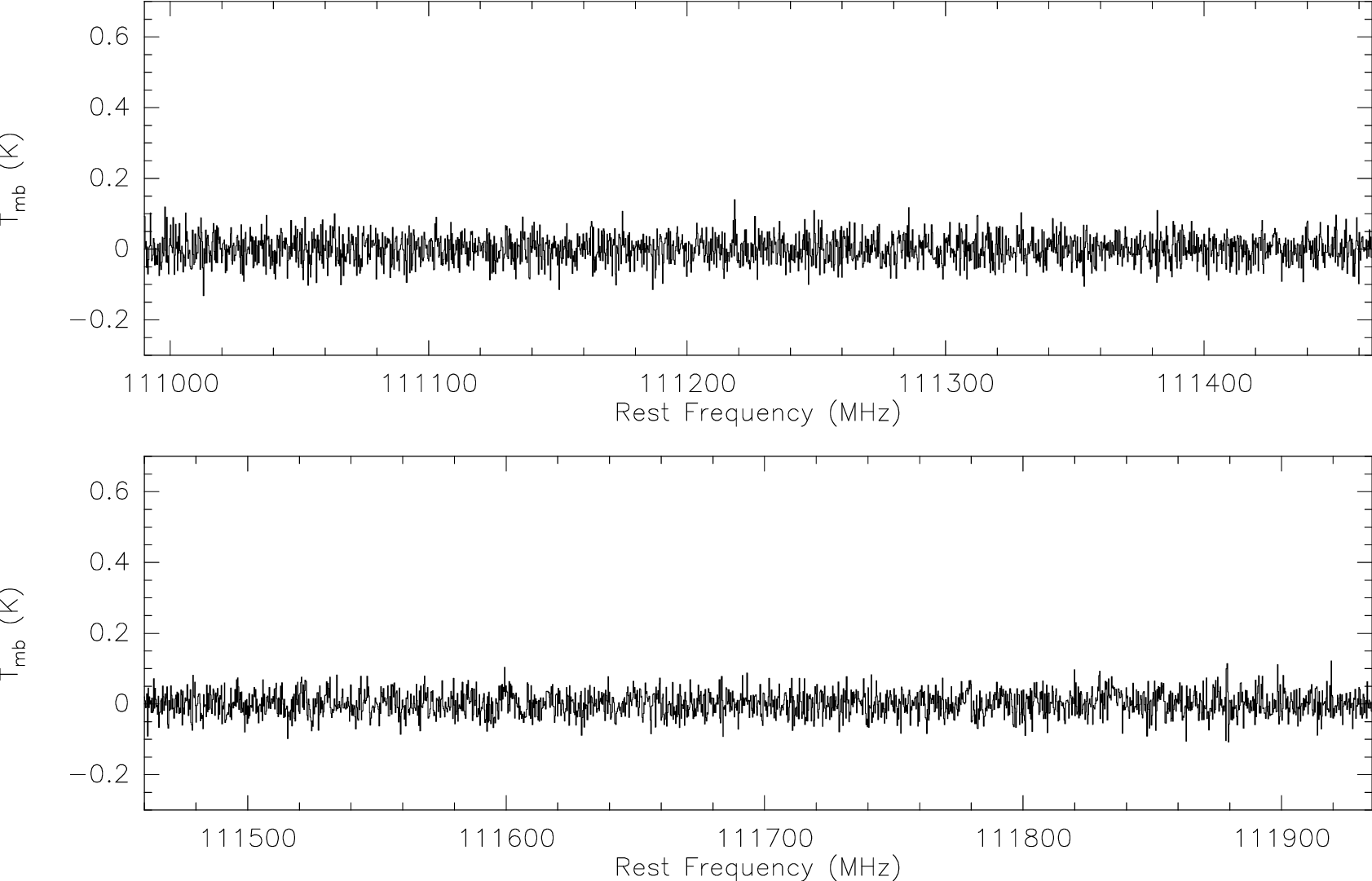}{0.85\textwidth}{ }
          }
\gridline{\fig{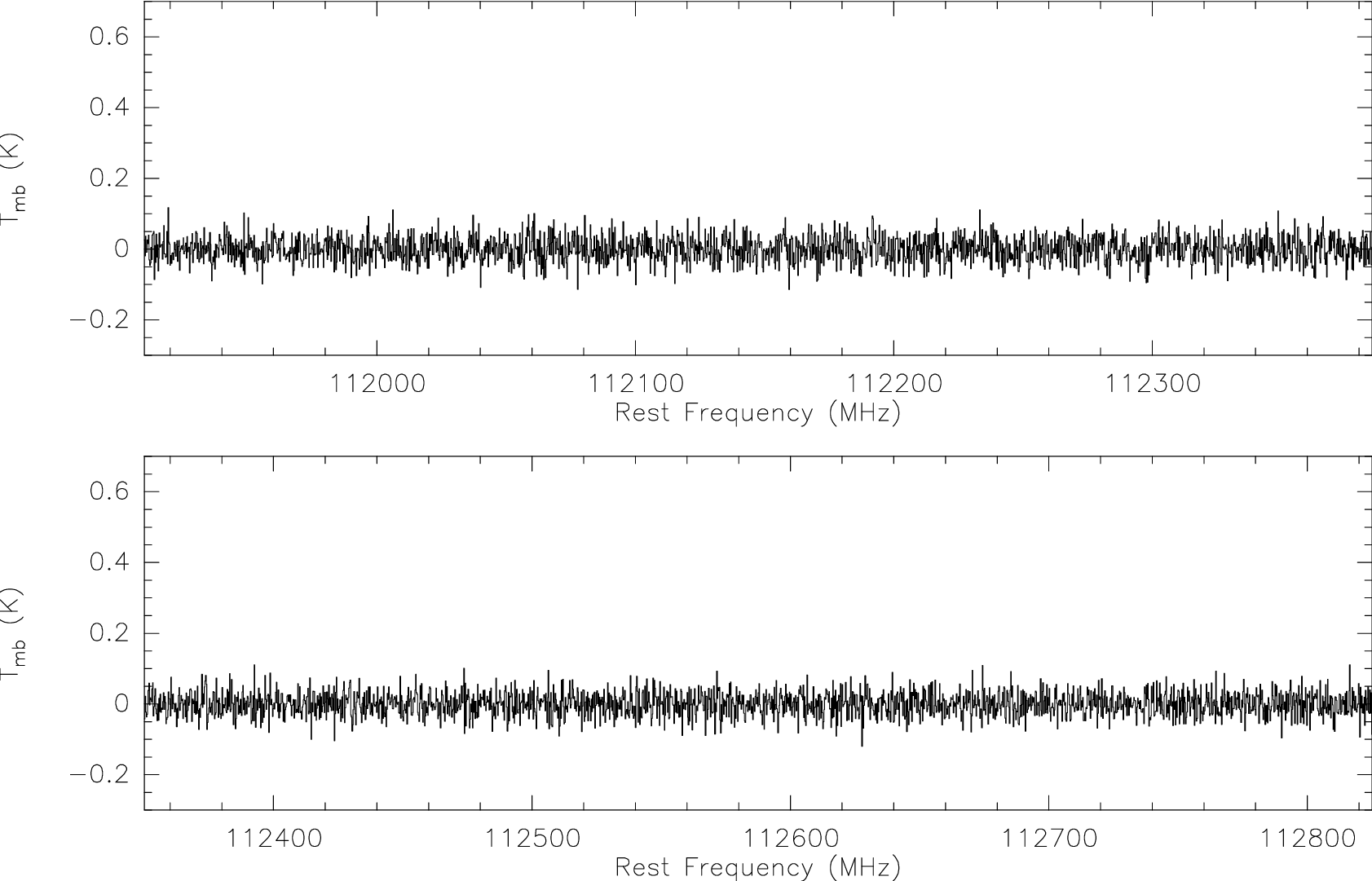}{0.85\textwidth}{ }
          }
\caption{Same as Figure~\ref{fig:F1}.
\label{fig:F16}}
\end{figure*}

\begin{figure*}
\gridline{\fig{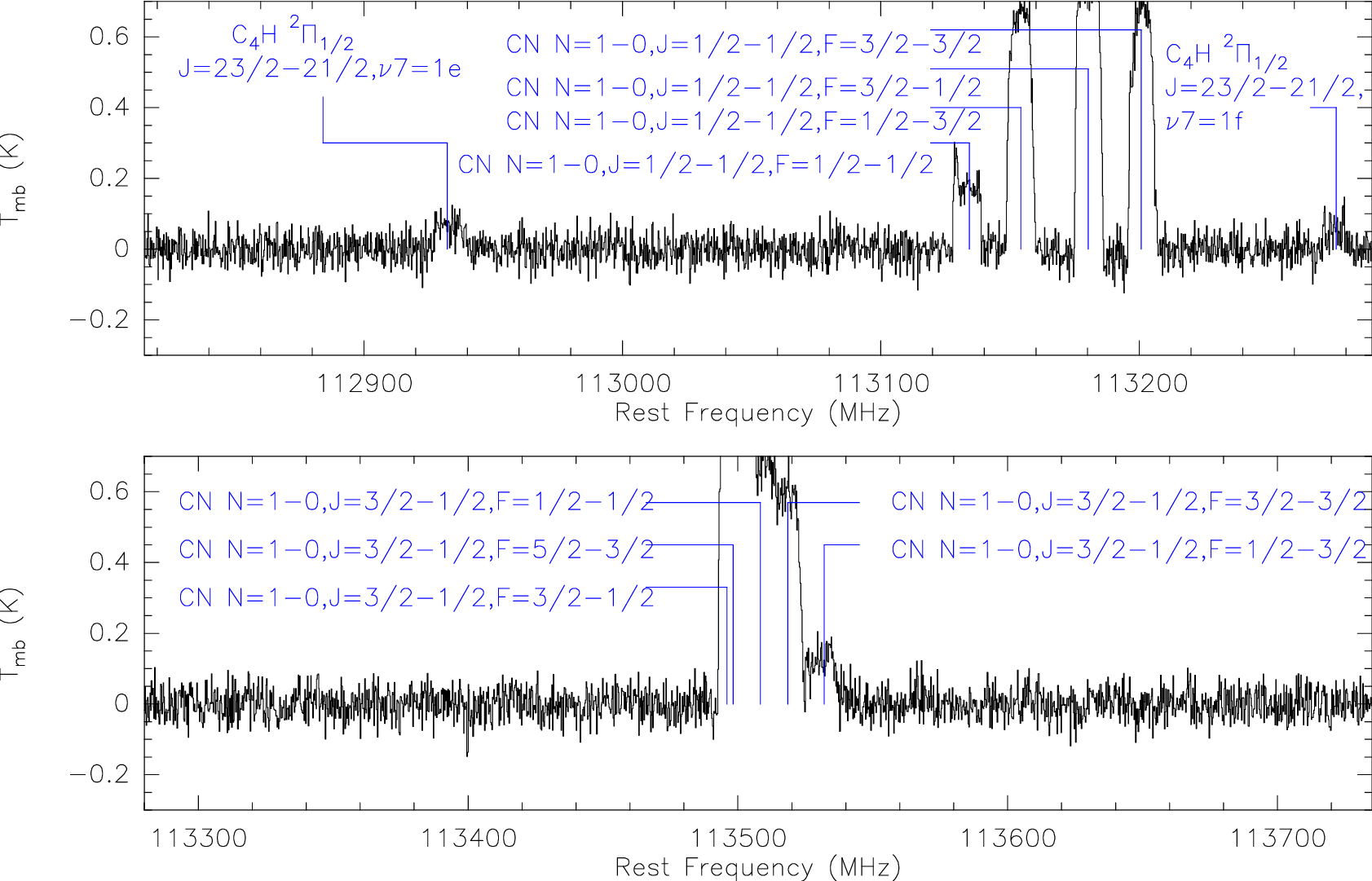}{0.85\textwidth}{ }
          }
\gridline{\fig{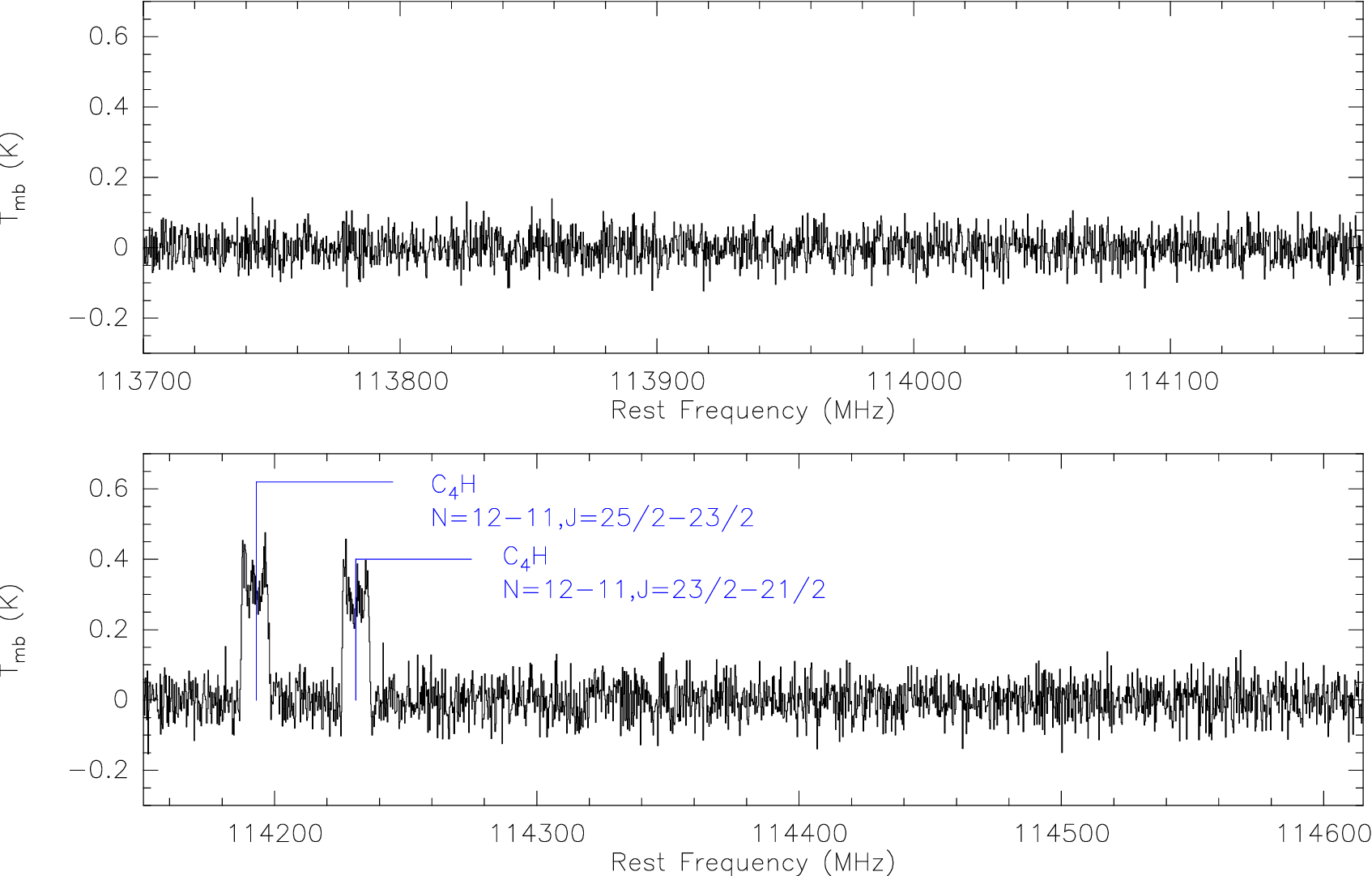}{0.85\textwidth}{ }
          }
\caption{Same as Figure~\ref{fig:F1}.
\label{fig:F17}}
\end{figure*}

\begin{figure*}
\gridline{\fig{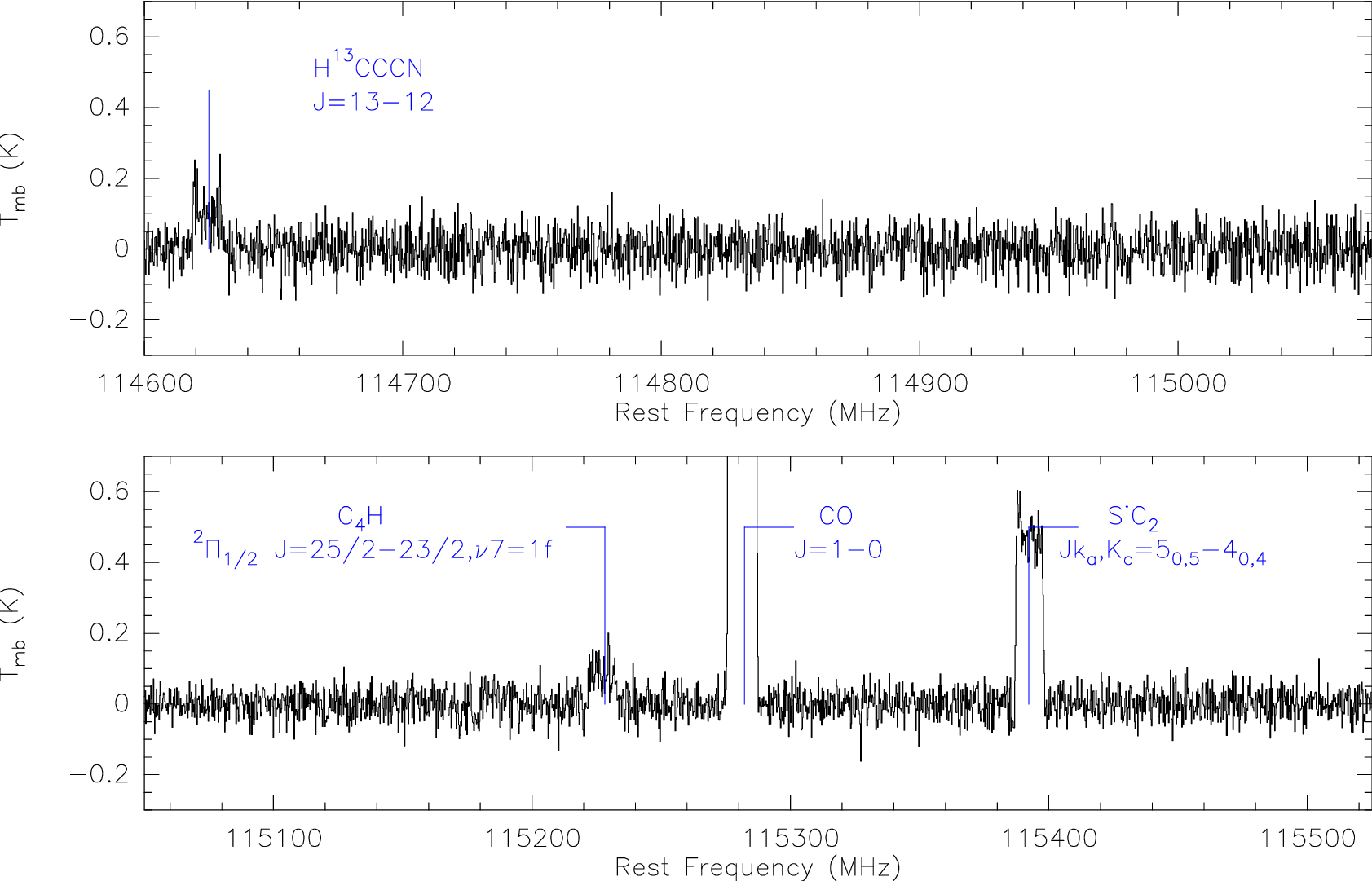}{0.85\textwidth}{ }
          }
\gridline{\fig{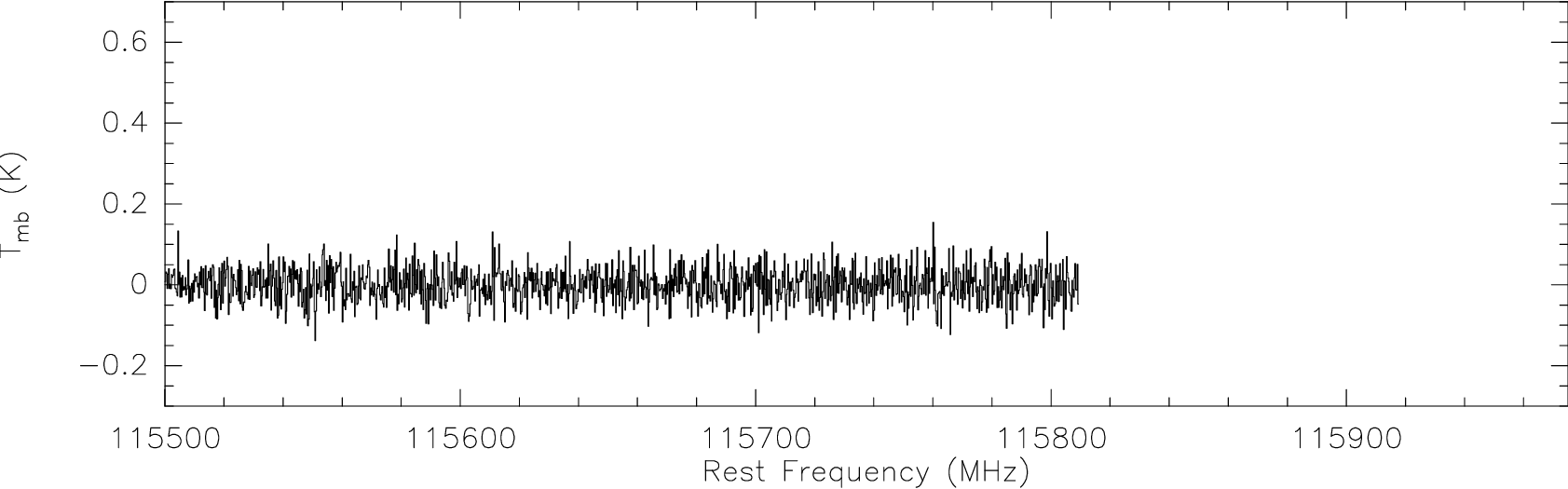}{0.85\textwidth}{ }
          }
\caption{Same as Figure~\ref{fig:F1}.
\label{fig:F18}}
\end{figure*}

%%%%%%%%%%%%%%%%%%%%%%%%%%%%%%%%%%%%%%%%%%%%%%%
%              Appendix  B
%%%%%%%%%%%%%%%%%%%%%%%%%%%%%%%%%%%%%%%%%%%%%%%
\section{Spectral line parameters of all the identified species} \label{Line parameters of transitions}

%Here provides the spectral line parameters of all the identified species from this survey towards IRC\,+10216, see Table \ref{table:3}.

The spectral line parameters of all the identified species from this survey towards IRC\,+10216 are listed in Table \ref{table:3}.

\renewcommand\thetable{\Alph{section}\arabic{table}}
\setcounter{table}{0}
\begin{longrotatetable}
%\tablenum{3}
\begin{deluxetable*}{lcclccccc}
\tablecaption{Spectral line parameters of all the identified species from this survey towards IRC\,+10216. \label{table:3}}
\tablewidth{700pt}
\tabletypesize{\scriptsize}
\tablehead{
\colhead{No.} &\colhead{Molecule} & \colhead{Transitions} & \colhead{Rest Freq.} & \colhead{$\int T_{\rm mb}dv$} & \colhead{$V_{\rm lsr}$} & \colhead{$V_{\rm exp}$} & \colhead{T$_{\rm mb}$} & \colhead{Notes} \\
\colhead{  }      &\colhead{  }      & \colhead{  }           & \colhead{(MHz)}      & \colhead{(K km s$^{-1}$)}    & \colhead{(km s$^{-1}$)} & \colhead{(km s$^{-1}$)} & \colhead{(K)} &\colhead{  }
}
\startdata
(1)  &  C$_2$H	    &$N=1-0$, $J=3/2-1/2$, $F=1-1$	            &87284.156	     & 2.15(0.606)	  &  -25.96(0.839)   & 14.988(0.157) & 0.066(0.021)      \\
(2)  &  C$_2$H	    &$N=1-0$, $J=3/2-1/2$, $F=2-1$	            &87316.925	     & 22.00(0.904)   &  -25.75(0.419)   & 14.035(0.094) & 0.693(0.031)      \\
(3)  &  C$_2$H	    &$N=1-0$, $J=3/2-1/2$, $F=1-0$	            &87328.624	     & 9.38(1.070)	  &  -25.44(0.419)   & 13.830(0.028) & 0.281(0.037)      \\
(4)  &  C$_2$H	    &$N=1-0$, $J=1/2-1/2$, $F=1-1$	            &87402.004	     & 13.40(0.837)   &	 *               &    *          & 0.341(0.037)     & 1 \\
(5)  &  C$_2$H	    &$N=1-0$, $J=1/2-1/2$, $F=0-1$	            &87407.165	     &	$\cdots$      &	  $\cdots$       &   $\cdots$    &    $\cdots$      & 1 \\	
(6)  &  C$_2$H	    &$N=1-0$, $J=1/2-1/2$, $F=1-0$	            &87446.512	     & 1.86(0.599)	  &  -23.94(0.837)   & 14.461(0.008) & 0.070(0.021)      \\
(7)  &  l-C$_3$H	&$^{2}\Pi_{1/2}$, $J=9/2-7/2$, $F=5-4e$	    &97995.212	     & 4.01(0.599)	  &  -26.87(0.747)   & 14.137(0.094) & 0.142(0.021)  & 1 \\
(8)  &  l-C$_3$H	&$^{2}\Pi_{1/2}$, $J=9/2-7/2$, $F=4-3e$	    &97995.951	     &	$\cdots$      &  $\cdots$        &   $\cdots$    &   $\cdots$         \\
(9)  &  l-C$_3$H	&$^{2}\Pi_{1/2}$, $J=9/2-7/2$, $F=5-4f$	    &98011.649	     & 4.60(0.667)	  &  -26.79(0.747)	 & 14.552(0.080) & 0.167(0.023)  & 1 \\
(10) &  l-C$_3$H	&$^{2}\Pi_{1/2}$, $J=9/2-7/2$, $F=4-3f$	    &98012.576	     &	$\cdots$	  &  $\cdots$        &   $\cdots$    &   $\cdots$           \\
(11) &  l-C$_3$H	&$^{2}\Pi_{3/2}$, $J=9/2-7/2$, $F=5-4f$	    &103319.278	     & 2.31(0.667)	  &  -26.67(0.708)	 & 13.754(0.181) & 0.079(0.023)  &1  \\
(12) &  l-C$_3$H	&$^{2}\Pi_{3/2}$, $J=9/2-7/2$, $F=4-3f$	    &103319.818      &	$\cdots$	  &    $\cdots$      & $\cdots$      &   $\cdots$          \\
(13) &  l-C$_3$H	&$^{2}\Pi_{3/2}$, $J=9/2-7/2$, $F=5-4e$	    &103372.506	     & 2.24(0.725)	   &  -24.96(0.708)  & 13.452(0.145) & 0.075(0.025)  &1  \\
(14) &  l-C$_3$H	&$^{2}\Pi_{3/2}$, $J=9/2-7/2$, $F=4-3e$	    &103373.129      &	$\cdots$	   &   $\cdots$      & $\cdots$      &   $\cdots$      \\
(15) &  C$_4$H	    &$N=9-8$, $J=19/2-17/2$	                    &85634.000	     & 6.73(0.552)	   &  -25.73(1.709)	 & 13.692(0.068)    & 0.282(0.019)      \\
(16) &  C$_4$H	    &$N=9-8$, $J=17/2-15/2$	                    &85672.570	     & 6.26(0.651)	   &  -26.63(1.709)	 & 14.227(0.102)    & 0.217(0.022)      \\
(17) &  C$_4$H	    &$N=10-9$, $J=21/2-19/2$	                &95150.320	     & 9.69(0.623)	   &  -26.45(1.528)	 & 14.332(0.068)    & 0.326(0.021)      \\
(18) &  C$_4$H	    &$N=10-9$, $J=19/2-17/2$	                &95188.940	     & 8.63(0.605)	   &  -25.94(1.528)	 & 13.840(0.071)    & 0.289(0.021)      \\
(19) &  C$_4$H	    &$N=11-10$, $J=23/2-21/2$	                &104666.560	     & 8.17(0.893)	   &  -26.31(0.699)  & 13.894(0.046)    & 0.262(0.031)      \\
(20) &  C$_4$H	    &$N=11-10$, $J=21/2-19/2$	                &104705.100	     & 10.3(0.824)	   &  -25.62(0.699)  & 13.924(0.038)    & 0.351(0.028)      \\
(21) &  C$_4$H	    &$N=12-11$, $J=25/2-23/2$	                &114182.510      & 8.99(0.984)	   &  -26.11(1.282)  & 13.666(0.089)    & 0.291(0.034)      \\
(22) &  C$_4$H	    &$N=12-11$, $J=23/2-21/2$	                &114221.040	     & 9.75(1.046)	   &  -26.19(1.282)	 & 13.859(0.079)    & 0.327(0.036)      \\
(23) &  C$_4$H      &$^{2}\Pi_{1/2}$ $J=19/2-17/2$, $\nu7=1f$   &93863.300        & 3.36(0.551)	   &   $\times$	     &  $\times$        & 0.080(0.019)  & B1\\
(24) &  C$_4$H      &$^{2}\Pi_{1/2}$ $J=21/2-19/2$, $\nu7=1e$   &103266.081       & 2.66(0.609)	   &  -27.09(1.418)	 & 14.154(0.289)    & 0.090(0.021)      \\
(25) &  C$_4$H      &$^{2}\Pi_{1/2}$ $J=23/2-21/2$, $\nu7=1e$   &112922.500       & 1.59(0.580)	   &  -26.95(1.296)	 & 13.801(0.339)    & 0.061(0.020)      \\
(26) &  C$_4$H      &$^{2}\Pi_{1/2}$ $J=23/2-21/2$, $\nu7=1f$   &113265.900       & 2.48(0.667)	   &  -26.49(1.292)	 & 13.919(0.233)    & 0.084(0.023)      \\
(27) &  C$_4$H	    &$^{2}\Pi_{3/2}$ $J=23/2-21/2$, $\nu7=1e$   &105838.000       & 1.96(0.475)	   &  -25.52(1.383)	 & 13.801(0.193)    & 0.090(0.016)      \\
(28) &  C$_4$H      &$^{2}\Pi_{3/2}$ $J=23/2-21/2$, $\nu7=1f$   &106132.800       & 1.48(0.551)	   &  -26.54(1.379)	 & 13.048(0.286)    & 0.057(0.019)      \\
(29) &  C$_4$H	    &$^{2}\Pi_{3/2}$ $J=25/2-23/2$, $\nu7=1f$   &115216.800       & 2.09(0.638)	   &  -25.68(1.271)	 & 14.674(0.331)    & 0.068(0.022)      \\
(30) &  CN	            &$N=1-0$, $J=3/2-1/2$, $F=3/2-1/2$	    &113488.140	     & 86.4(0.841)	   &  $\cdots$       & $\cdots$         & 1.916(0.029)  & 1 \\	
(31) &  CN	            &$N=1-0$, $J=3/2-1/2$, $F=5/2-3/2$	    &113490.982	     & $\cdots$        &  $\cdots$		 & $\cdots$         & $\cdots$          \\	
(32) &  CN	            &$N=1-0$, $J=3/2-1/2$, $F=1/2-1/2$	    &113499.639	     & $\cdots$        &  $\cdots$		 & $\cdots$         & $\cdots$          \\	
(33) &  CN	            &$N=1-0$, $J=3/2-1/2$, $F=3/2-3/2$	    &113508.944	     & $\cdots$        &  $\cdots$	     & $\cdots$         & $\cdots$          \\	
(34) &  CN	            &$N=1-0$, $J=3/2-1/2$, $F=1/2-3/2$	    &113520.414  	 & $\cdots$        &  $\cdots$	     & $\cdots$         & $\cdots$          \\	
(35) &  CN	            &$N=1-0$, $J=1/2-1/2$, $F=1/2-1/2$	    &113123.337	     & 6.24(0.696)    &  -25.91(0.647)	 & 14.363(0.051)    & 0.235(0.024)    \\
(36) &  CN	            &$N=1-0$, $J=1/2-1/2$, $F=1/2-3/2$	    &113144.192	     & 20.00(0.957)   &  -26.01(0.647)	 & 13.350(0.057)    & 0.789(0.033)   \\
(37) &  CN	            &$N=1-0$, $J=1/2-1/2$, $F=3/2-1/2$	    &113170.528	     & 15.80(0.783)   &  -25.96(0.647)	 & 13.594(0.061)    & 0.679(0.027)    \\
(38) &  CN	            &$N=1-0$, $J=1/2-1/2$, $F=3/2-3/2$	    &113191.317	     & 18.60(1.218)   &  -25.81(0.647)	 & 13.264(0.080)    & 0.804(0.042)     \\
(39) &  C$_3$N	        &$N=9-8$, $J=19/2-17/2$	                &89045.590	     & 9.71(0.879)	  &  -26.28(0.822)	 & 13.807(0.049)    & 0.320(0.030)    \\
(40) &  C$_3$N	        &$N=9-8$, $J=17/2-15/2$	                &89064.360	     & 9.41(0.963)	  &  -25.80(0.822)	 & 13.891(0.053)    & 0.315(0.033)    \\
(41) &  C$_3$N        	&$N=10-9$, $J=21/2-19/2$	            &98940.020	     & 8.84(0.883)	  &  -25.67(0.740)	 & 13.734(0.049)    & 0.297(0.031)    \\
(42) &  C$_3$N	        &$N=10-9$, $J=19/2-17/2$	            &98958.780	     & 10.40(0.835)   &  -25.88(0.740)	 & 13.781(0.043)    & 0.358(0.029)    \\
(43) &  C$_3$N	        &$N=11-10$, $J=23/2-21/2$	            &108834.270	     & 5.63(0.546)	  &  -25.81(0.673)	 & 14.027(0.049)    & 0.193(0.019)    \\
(44) &  C$_3$N          &$N=11-10$, $J=21/2-19/2$	            &108853.020	     & 5.51(0.491)	  &  -26.09(0.672)	 & 14.365(0.041)    & 0.185(0.017)    \\
(45) &  HC$_3$N	        &$J=10-9$	                            &90978.989       & 56.30(1.018)   &  -25.88(0.402)  & 13.854(0.014)    & 2.448(0.035)     \\
(46) &  HC$_3$N	        &$J=11-10$	                            &100076.386      & 45.20(0.744)   &  -25.22(1.463)  & 13.399(0.047)    & 1.938(0.026)    \\
(47) &  HC$_3$N	        &$J=12-11$	                            &109173.638      & 41.10(0.745)   &  -25.96(1.341)  & 13.435(0.047)    & 1.764(0.026)     \\
(48) &  H$^{13}$CCCN    &$J=10-9$                               &88166.808       & 2.19(0.626)	  &  -27.68(1.660)  & 14.177(0.311)    & 0.076(0.022)    \\
(49) &  H$^{13}$CCCN    &$J=13-12$                              &114615.021      & 3.17(0.918)	  &  -27.34(1.277)  & 14.173(0.271)    & 0.094(0.031)   & NS\\
(50) &  HC$^{13}$CCN    &$J=10-9$	                            &90593.059	     &  $\times$      &    $\times$     &    $\times$      &  $\times$      & B2 \\
(51) &  HCC$^{13}$CN    &$J=10-9$	                            &90601.791	     & 4.56(0.551)    &    $\times$     &    $\times$      & 0.077(0.019)   & B2 \\
(52) &  HC$^{13}$CCN    &$J=11-10$	                            &99651.863	     & 2.55(0.435)    &    $\times$     &    $\times$      & 0.045(0.015)   & B3 \\
(53) &  HCC$^{13}$CN    &$J=11-10$	                            &99661.471	     &	$\times$      &    $\times$     &    $\times$      &  $\times$      & B3 \\
(54) &  HC$^{13}$CCN    &$J=12-11$	                            &108710.523	     & 2.63(0.406)    &    $\times$     &    $\times$      & 0.048(0.014)   & B4 \\
(55) &  HCC$^{13}$CN    &$J=12-11$	                            &108721.008	     &	$\times$      &    $\times$     &    $\times$      &  $\times$      & B4 \\
(56) &  HC$_5$N	        &$J=32-31$	                            &85201.346	     & 4.43(0.434)	  &  -26.10(1.718)  &  14.345(0.130)   & 0.148(0.015)    \\
(57) &  HC$_5$N	        &$J=33-32$	                            &87863.630	     & 4.75(0.645)	  &  -26.30(1.666)  &  14.042(0.160)   & 0.166(0.022)    \\
(58) &  HC$_5$N	        &$J=34-33$	                            &90525.890	     & 3.54(0.460)	  &  -26.81(1.617)  &  13.911(0.156)   & 0.124(0.016)    \\
(59) &  HC$_5$N	        &$J=35-34$	                            &93188.123	     & 3.22(0.750)	  &  -26.45(1.571)  &  13.336(0.230)   & 0.118(0.026)     \\
(60) &  HC$_5$N	        &$J=36-35$	                            &95850.335	     & 3.30(0.641)	 &  -28.05(1.527)  &  13.111(0.181)   & 0.112(0.022)  \\
(61) &  HC$_5$N	        &$J=37-36$	                            &98512.524	     & 1.96(0.414)	 &  -25.21(1.486)  &  13.377(0.172)   & 0.074(0.014)  \\
(62) &  HC$_5$N	        &$J=38-37$	                            &101174.677	     & 2.66(0.577)	 &  -27.46(1.447)  &  13.756(0.232)   & 0.096(0.020)  \\
(63) &  HC$_5$N	        &$J=39-38$	                            &103836.817	     & 2.25(0.513)	 &  -26.59(1.410)  &  13.266(0.181)   & 0.089(0.018)  \\
(64) &  HC$_5$N	        &$J=40-39$	                            &106498.910	     & 2.42(0.530)	 &  -25.02(1.374)  &  14.428(0.288)   & 0.067(0.018)  \\
(65) &  $c$-C$_3$H$_{2}$ & $J_{\rm K_a,K_c}=2_{1,2}-1_{0,1}$    &85338.906       & 2.97(0.509)   &  -25.77(1.715)  &  13.869(0.202)   & 0.099(0.018)  \\
(66) &  HNC	            &$J=1-0$, $F=0-1$	                    &90663.450	     &	$\cdots$     &  $\cdots$       &  $\cdots$        & $\cdots$  \\		
(67) &  HNC	            &$J=1-0$, $F=2-1$	                    &90663.574	     & 22.30(0.702)  &  -25.93(0.807)  & 13.740(0.019)    & 0.199(0.024) &1  \\		
(68) &  HNC	            &$J=1-0$, $F=1-1$	                    &90663.656	     &	$\cdots$     &  $\cdots$       &  $\cdots$        & $\cdots$   \\			
(69) &  HCN	            &$J=1-0$, $F=1-1$	                    &88630.416	     &	$\cdots$     &  $\cdots$       &  $\cdots$        & $\cdots$ \\		
(70) &  HCN             &$J=1-0$, $F=2-1$	                    &88631.847	     & 174(9.918)    &  -22.87(0.159)  & 16.211(0.052)    & 8.135(0.342) &1  \\		
(71) &  HCN             &$J=1-0$, $F=0-1$	                    &88633.936	     &	$\cdots$     &   $\cdots$      &  $\cdots$        & $\cdots$          \\	
(72) &  H$^{13}$CN  	&$J=1-0$, $F=1-1$	                    &86338.737       &	$\cdots$     &    $\cdots$     &  $\cdots$        & $\cdots$   \\
(73) &  H$^{13}$CN	    &$J=1-0$, $F=2-1$                       &86340.176       & 60.70(3.915)  &  -24.16(0.424)  & 16.925(0.055)    & 2.645(0.135) &1\\	
(74) &  H$^{13}$CN  	&$J=1-0$, $F=0-1$	                    &86342.255       &	$\cdots$     &    $\cdots$     &  $\cdots$        & $\cdots$     \\	
(75) &  CO	            &$J=1-0$                                &115271.202      & 228.00(2.697) &  -26.02(0.159)  & 14.220(0.006)    & 9.673(0.093) \\
(76) &  $^{13}$CO	    &$J=1-0$                                &110201.354      & 21.80(0.732)  &  -26.07(0.664)  & 14.547(0.009)    & 0.455(0.025)  \\
(77) &  SiO	            &$J=2-1$, $\nu=0$	                    &86846.995	     & 18.20(0.921)  &  -25.81(0.843)  & 13.518(0.040)    & 0.722(0.032)  \\
(78) &  $^{29}$SiO	    &$J=2-1$, $\nu=0$	                    &85759.188	     & 1.63(0.551)	 &  -27.88(1.707)  & 13.460(0.047)    & 0.058(0.019)  \\
(79) &  $^{29}$SiS	    &$J=5-4$, $\nu=0$	                    &89103.720	     & 1.70(0.551)	 &  -25.84(1.643)  & 14.044(0.266)    & 0.065(0.019)  \\
(80) &  Si$^{34}$S      &$J=6-5$, $\nu=0$	                    &105941.503	     & 1.97(0.609)	 &  -24.42(1.382)  & 14.679(0.231)    & 0.070(0.021)  \\
(81) &  SiS	            &$J=5-4$, $\nu=0$	                    &90771.561	     & 30.40(0.725)  &  -25.93(0.843)  & 13.742(0.018)    & 1.102(0.025)  \\
(82) &  SiS	            &$J=6-5$, $\nu=0$	                    &108924.297	     & 23.80(0.551)  &  -26.22(1.344)  & 13.466(0.043)    & 1.009(0.019)  \\	
(83) &  SiC$_2$	        & $J_{\rm K_a,K_c} = 4_{0,4}-3_{0,3}$	    &93063.639	     & 14.30(0.986)  &  -26.95(0.787)  &  13.845(0.037)   & 0.501(0.034)  \\
(84) &  SiC$_2$	        & $J_{\rm K_a,K_c} = 4_{2,3}-3_{2,2}$	    &94245.393	     & 9.44(0.838)	 &  -25.40(1.553)  &  13.987(0.069)   & 0.332(0.029)  \\
(85) &  SiC$_2$	        & $J_{\rm K_a,K_c} = 4_{2,2}-3_{2,1}$	    &95579.381	     & 10.30(0.559)  &  -25.98(0.153)  &  13.815(0.059)   & 0.335(0.019)  \\
(86) &  SiC$_2$	        & $J_{\rm K_a,K_c} = 5_{0,5}-4_{0,4}$	    &115382.375	     & 13.40(0.870)  &  -25.92(1.269)  &  13.855(0.029)   & 0.468(0.030)  \\
(87) &  CS	            &$J=2-1$	                            &97980.953	     & 61.20(4.421)  &  -25.85(0.187)  &  13.275(0.015)   &   2.522(0.049)  \\
(88) &  $^{13}$CS	    &$J=2-1$	                            &92494.270	     & 2.36(0.696)	 &  -23.24(1.583)  &  13.908(0.310)   &   0.079(0.024)  \\
(89) &  C$^{34}$S	    &$J=2-1$	                            &96412.950	     & 5.46(0.638)	 &  -25.59(1.518)  &  14.292(0.134)   &   0.197(0.022)  \\
(90) &  C$_2$S	        &$N=8-7$, $J=7-6$                       &93870.098	 	 &	$\times$     &   $\times$      &    $\times$      &    $\times$     & B1 \\	
(91) &  C$_2$S	        &$N=9-8$, $J=8-7$                       &106347.740	     & 1.94(0.405)	 &  -27.16(1.376)  &  13.352(0.253)   &   0.070(0.014)  \\
(92) &  C$_3$S	        &$J=18-17$	                            &104048.451	     & 1.71(0.535)	 &  -23.46(1.759)  &  13.553(0.394)   &   0.057(0.018)  \\
(93) &  MgNC	        &$N=9-8$, $J=19/2-17/2$	                &107399.420	     & 1.67(0.609)   &  -25.97(0.682)  &  14.001(0.230)   &   0.064(0.021)  \\
(94) &  CH$_3$CN	    & $J_{\rm K}=6_{2}-5_{2}, F=7-6$	                &110375.049	     & 2.01(0.783)	 &	$\cdots$       &     $\cdots$     &   0.081(0.026)  &1 \\
(95) &  CH$_3$CN	    & $J_{\rm K}=6_{1}-5_{1}, F=7-6$	                &110381.400	     &	$\cdots$     &  $\cdots$       &     $\cdots$     &   $\cdots$   \\
(96) &  CH$_3$CN	    & $J_{\rm K}=6_{0}-5_{0}, F=7-6$	                &110383.518	     &	$\cdots$     &  $\cdots$       &     $\cdots$     &   $\cdots$    \\
\enddata
\end{deluxetable*}
\tablecomments{The numbers in parentheses represent the 1\,$\sigma$ error. The rms noise on a main beam brightness temperature scale is obtained by statistics in areas without clear spectral lines, and the measurement uncertainty of the integrated intensities is the product of the rms noise and the width of the line, about 29\,km\,s$^{-1}$. The error in the expansion velocity is obtained by shell function fitting, and the error in the LSR velocity is the velocity width of a channel. The information for each column in the table is as follows, Col. (1) : Number; Col. (2) : Molecule name; Col. (3) : Quantum numbers; Col. (4) : Rest frequency (Rest frequency are adopted from JPL \citep{Pickett+etal+1998}, CDMS \citep{Muller+etal+2005}, and splatalogue databases as well as the online Lovas line list \citep{Lovas+2004} for astronomical spectroscopy.); Col. (5) : Integrated intensity (in main beam brightness temperature); Col. (6) : LSR velocity; Col. (7) : Expansion velocity; Col. (8) : Main beam brightness temperature. $\cdots$ means the values cannot be obtained because of unresolved hyperfine structure, $\times$ implies that the values cannot be obtained due to two blended spin-rotation components. $*$ addresses values that cannot be derived from our detection. ({\bf 1}) The hyperfine structure is not resolved. (B1) $\rm^{2}\Pi_{1/2}$\,C$_4$H\,$J=19/2-17/2$, ${\nu7=1f}$ blended with C$_2$S\,$N=8-7$, $J=7-6$. (B2) HC$^{13}$CCN\,$J=10-9$ is blend with HCC$^{13}$CN\,$J=10-9$. (B3) HC$^{13}$CCN\,$J=11-10$ is blended with HCC$^{13}$CN\,$J=11-10$. (B4) HC$^{13}$CCN\,$J=12-11$ is blended with HCC$^{13}$CN\,$J=12-11$. (NS) Transitions that are detected for the first time toward the source are marked with ``NS".}
\end{longrotatetable}

%%%%%%%%%%%%%%%%%%%%%%%%%%%%%%%%%%%%%%%%%%%%%%%
%              Appendix C
%%%%%%%%%%%%%%%%%%%%%%%%%%%%%%%%%%%%%%%%%%%%%%%
\section{Spectral line profiles of the detected species} \label{Line spectrum}

In this appendix, we give all the observed spectral profiles of the identified species in  Figs.~\ref{fig:C2H+C3H} $-$ \ref{fig:CS+MgNC+CH3CN}. %Note that these line profiles are classified into four kinds of typical spectra that illustrated in  Figs.~\ref{fig:S1-HC5N+SiO+12CO+13CO}.
Note that we have fitted each spectrum by one of the four line profiles shown in Figure~\ref{fig:S1-HC5N+SiO+12CO+13CO}.

\renewcommand\thefigure{\Alph{section}\arabic{figure}}
\setcounter{figure}{0}
\begin{figure*}
\gridline{\fig{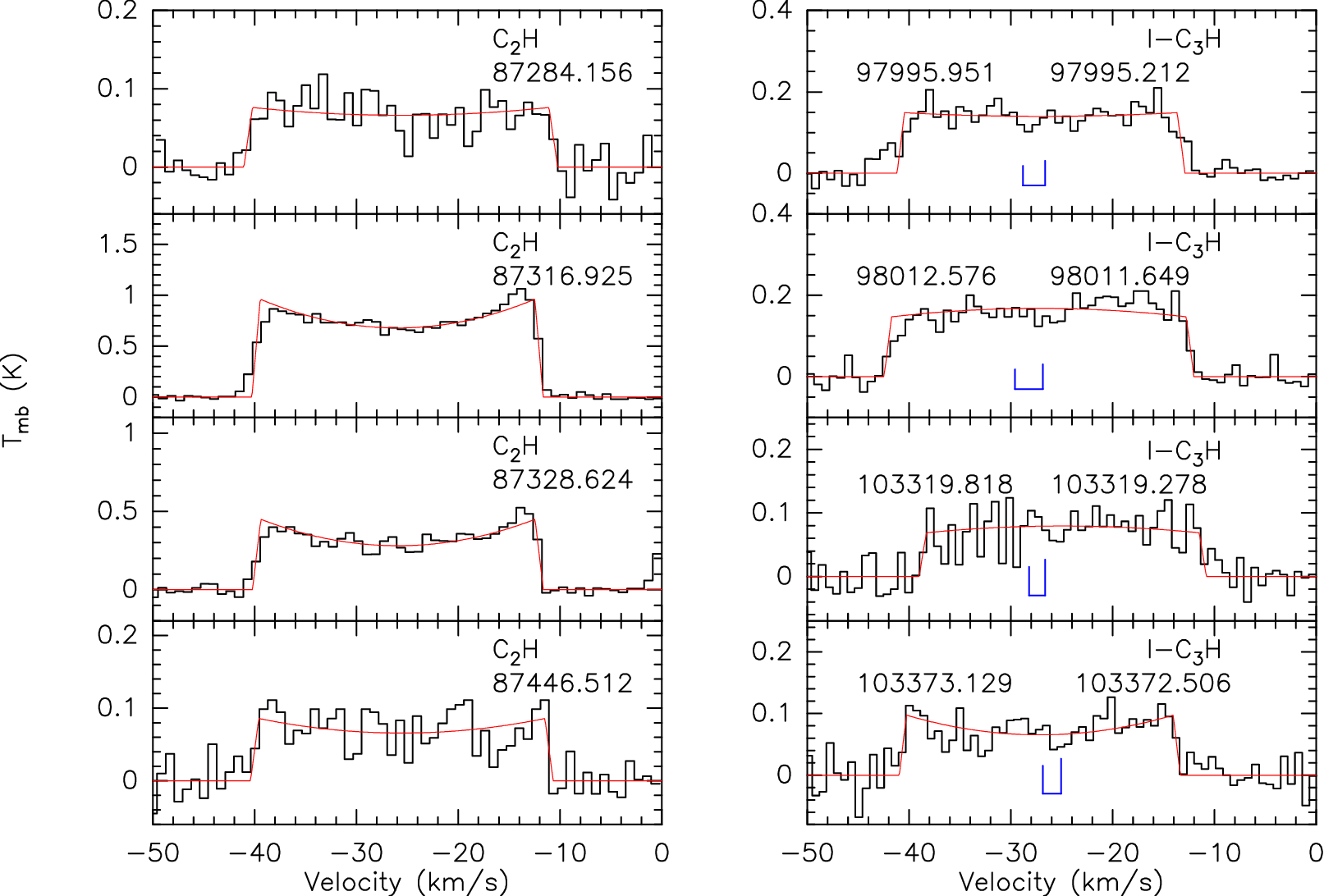}{0.72\textwidth}{ }
          }
\caption{Same as Figure~\ref{fig:S1-HC5N+SiO+12CO+13CO} but for C$_2$H and $I$-C$_3$H. The solid blue line indicates the hyperfine structure components of the molecule. The rest frequencies in MHz are shown in each panel's upper right corner for single, and in the upper left and right corners for lines with two fine structure components.}
\label{fig:C2H+C3H}
\end{figure*}

\begin{figure*}
\gridline{\fig{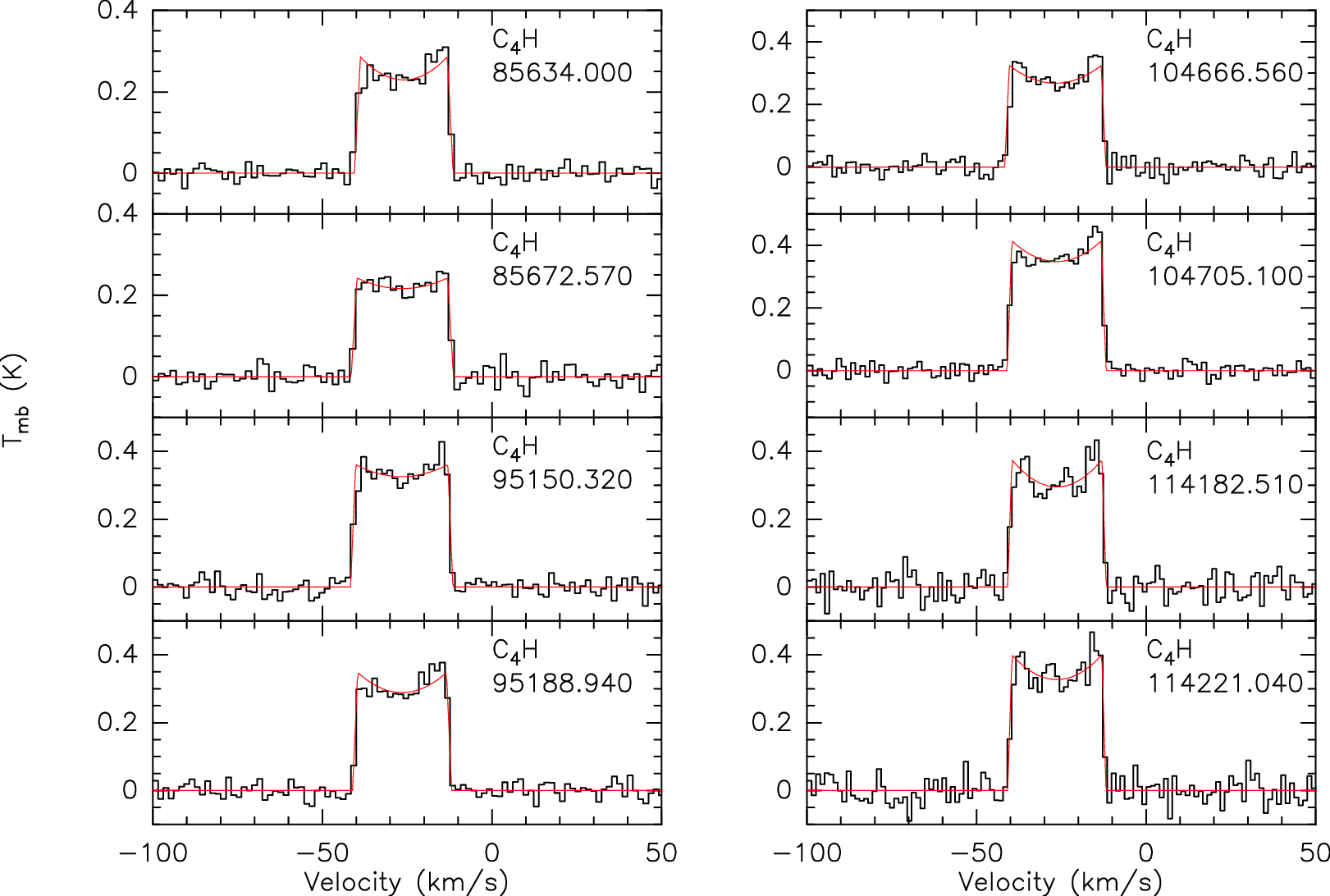}{0.72\textwidth}{ }
          }
\caption{Same as Figure~\ref{fig:S1-HC5N+SiO+12CO+13CO} but for C$_4$H.
\label{fig:C4H}}
\end{figure*}

\begin{figure*}
\gridline{\fig{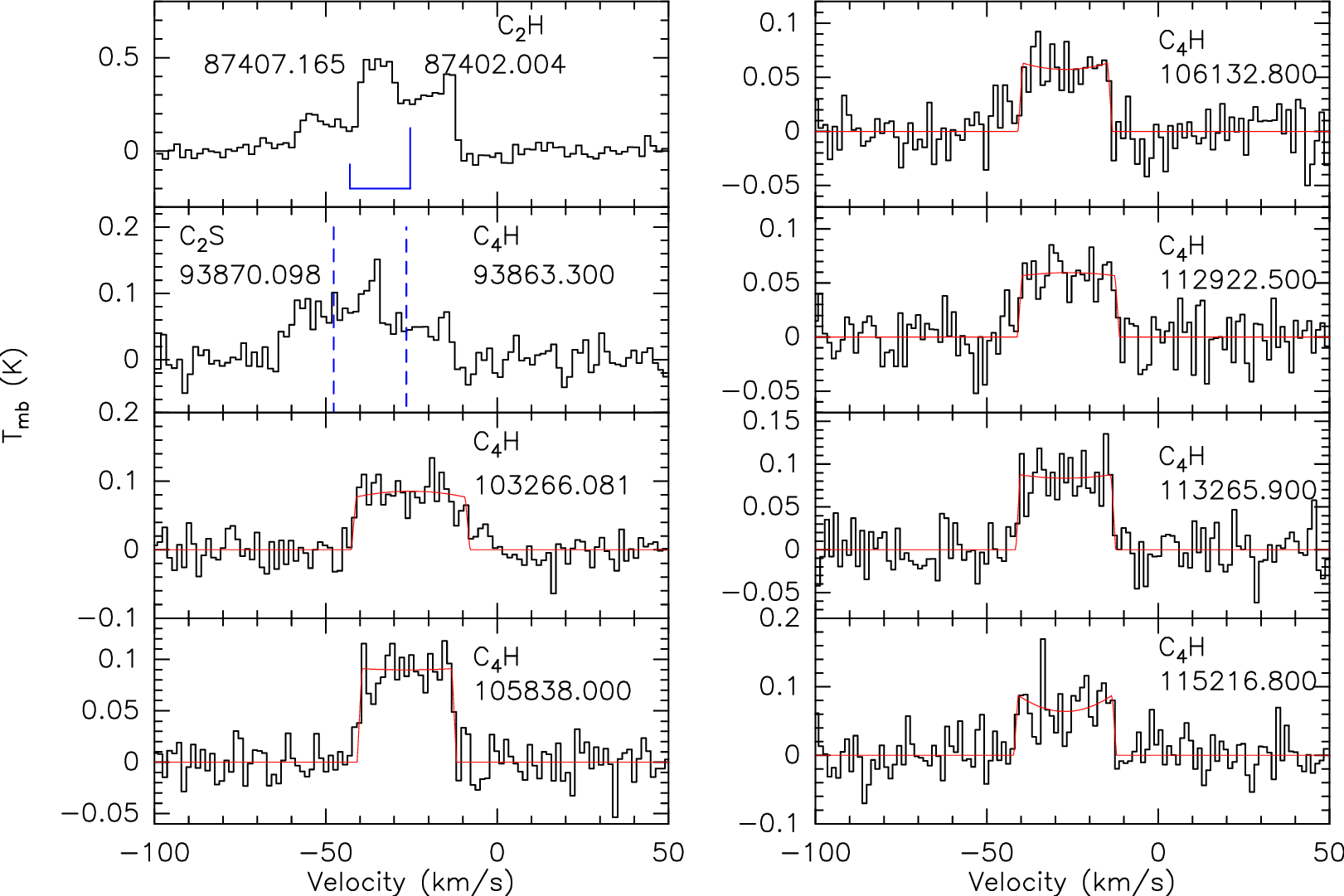}{0.72\textwidth}{ }}
\caption{Same as Figure~\ref{fig:S1-HC5N+SiO+12CO+13CO} and Figure~\ref{fig:C2H+C3H} but for C$_2$H, C$_4$H\,(${\nu=1}$), C$_2$S, and CN. The blue dashed lines are the blended spectral lines.
\label{fig:C2H+C4H+C2S}}
\end{figure*}

\begin{figure*}
\gridline{\fig{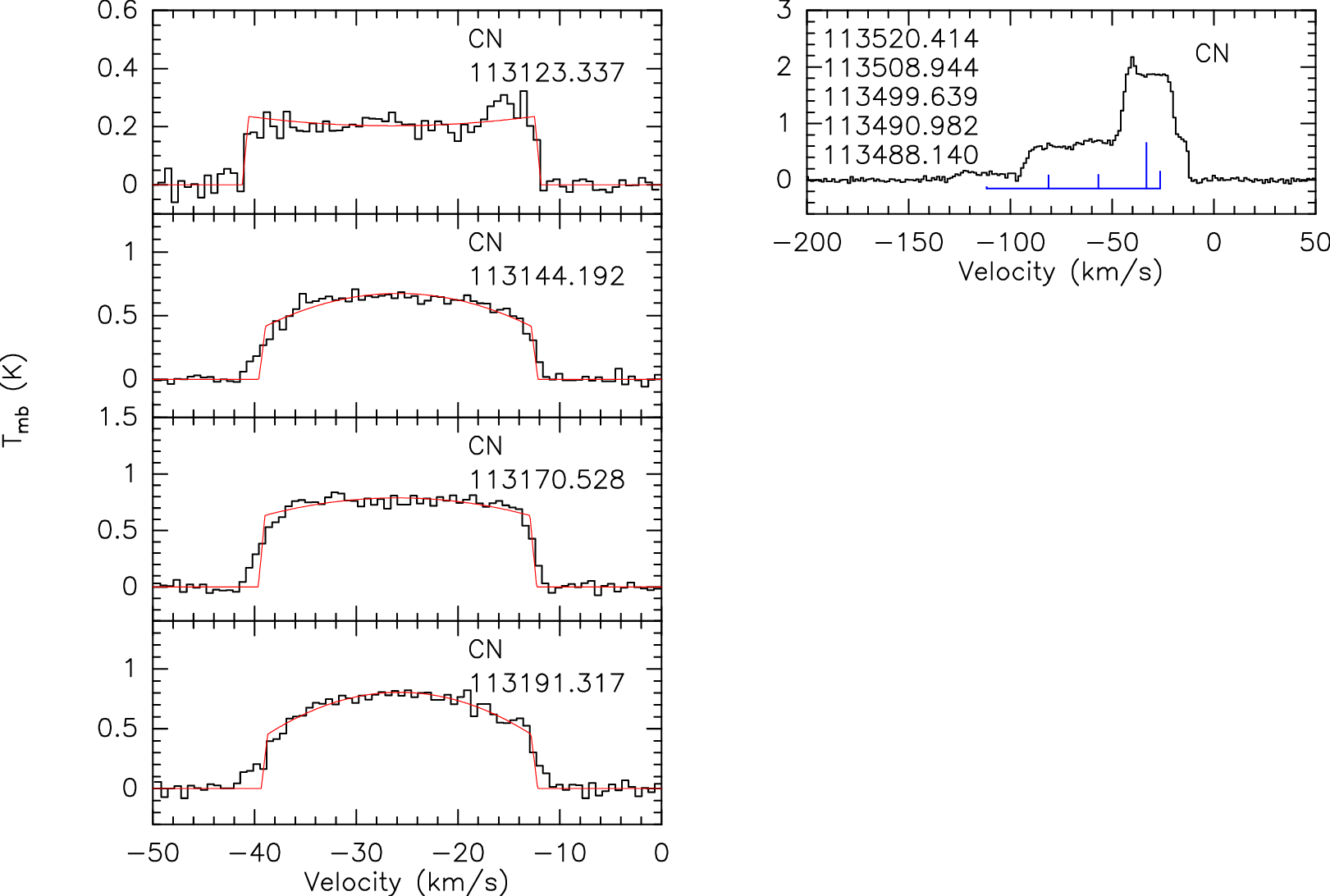}{0.72\textwidth}{ }
          }
\caption{Same as Figure~\ref{fig:S1-HC5N+SiO+12CO+13CO} and Figure~\ref{fig:C2H+C3H}  but for CN. The rest frequencies in MHz of spectral lines of more than two fine structure components are displayed in the upper left corner of each panel.
\label{fig:CN}}
\end{figure*}

\begin{figure*}
\gridline{\fig{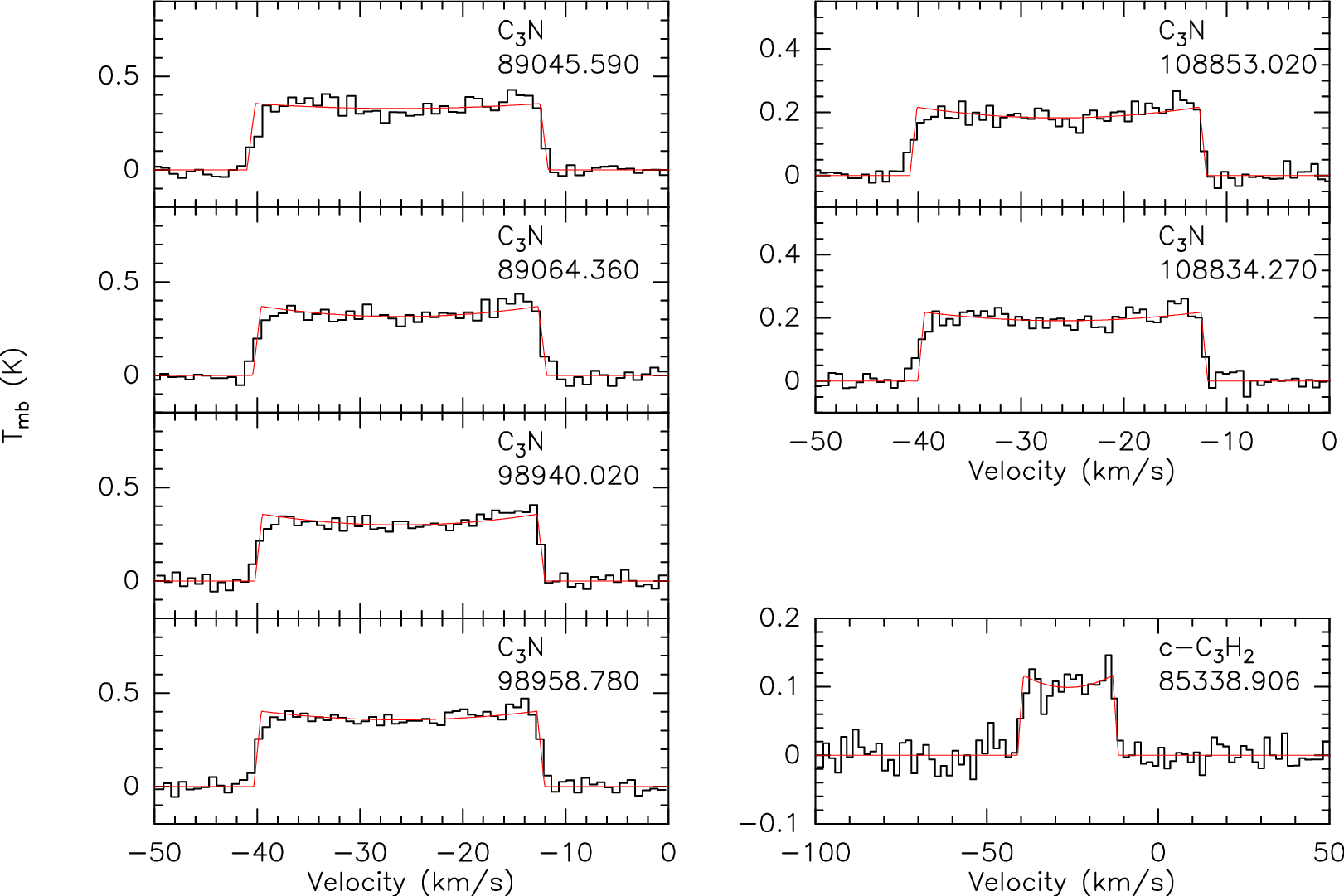}{0.72\textwidth}{ }
          }
\caption{Same as Figure~\ref{fig:S1-HC5N+SiO+12CO+13CO} but for C$_3$N and $c$-C$_3$H$_2$.
\label{fig:C3N+c-C3H2}}
\end{figure*}

\begin{figure*}
\gridline{\fig{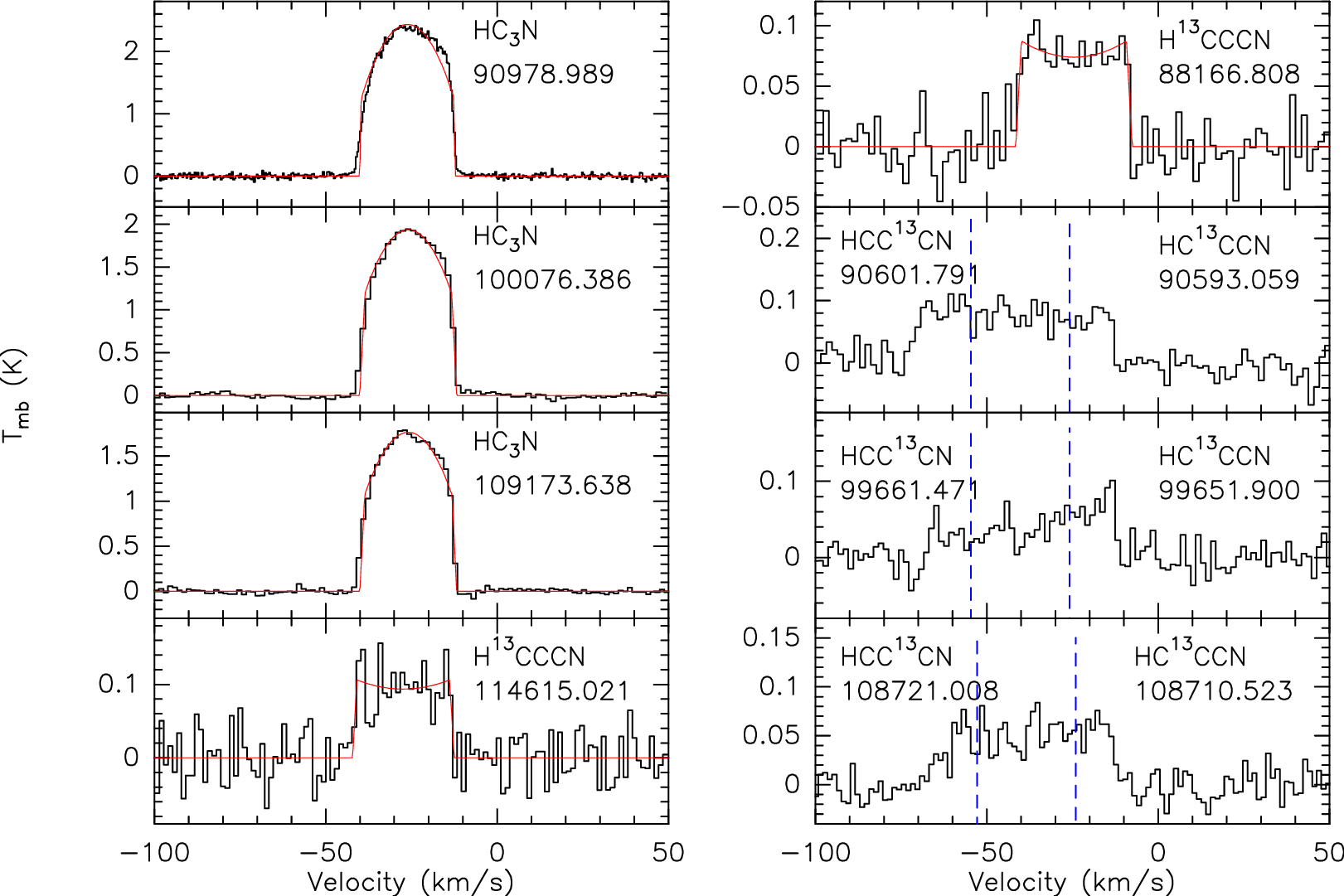}{0.72\textwidth}{ }
          }
\caption{Same as Figure~\ref{fig:S1-HC5N+SiO+12CO+13CO} and Figure~\ref{fig:C2H+C4H+C2S} but for HC$_3$N, and its isotopologues.
\label{fig:HC3N}}
\end{figure*}

\begin{figure*}
\gridline{\fig{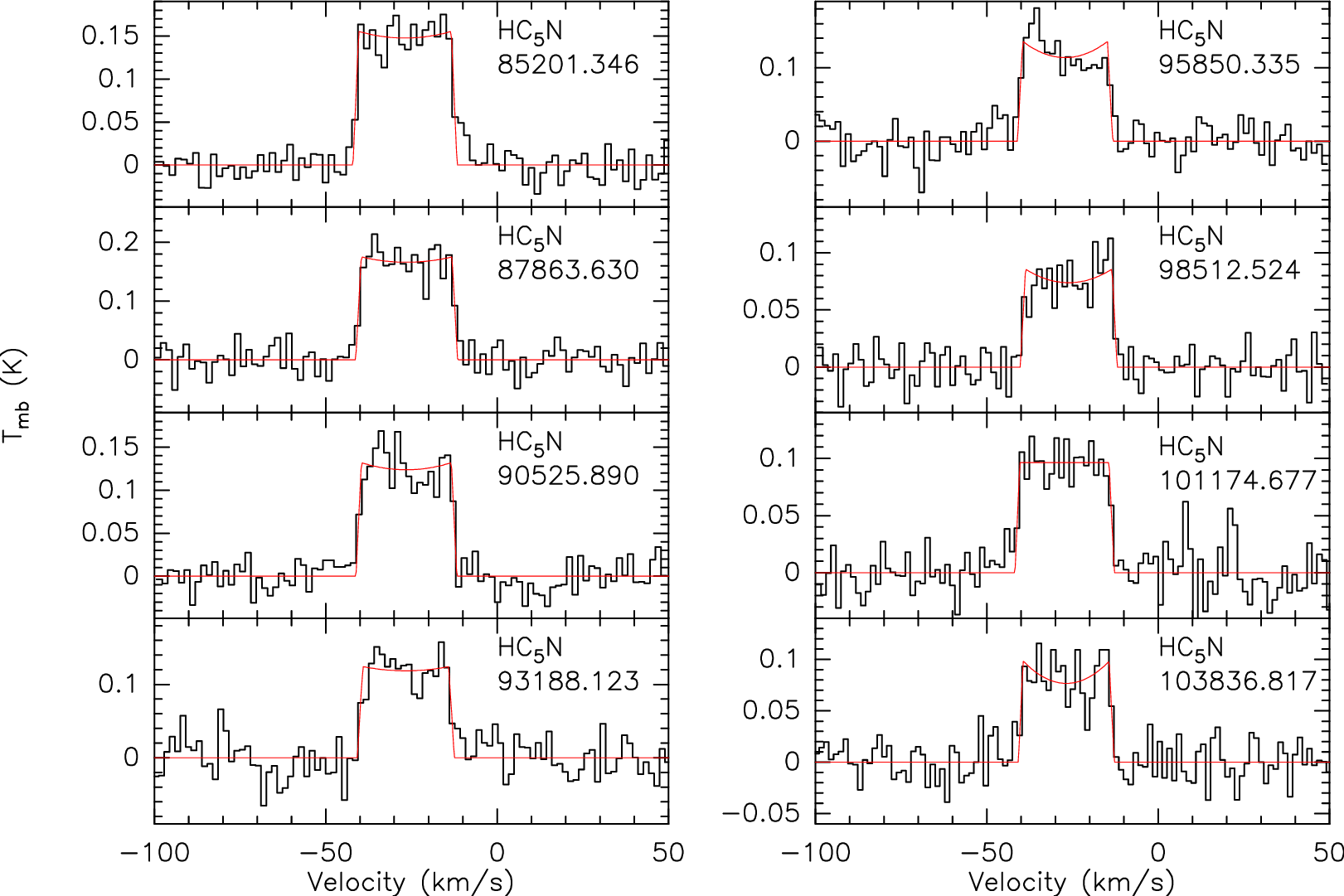}{0.72\textwidth}{ }
          }
\caption{Same as Figure~\ref{fig:S1-HC5N+SiO+12CO+13CO} but for HC$_5$N.
\label{fig:HC5N}}
\end{figure*}

\begin{figure*}
\gridline{\fig{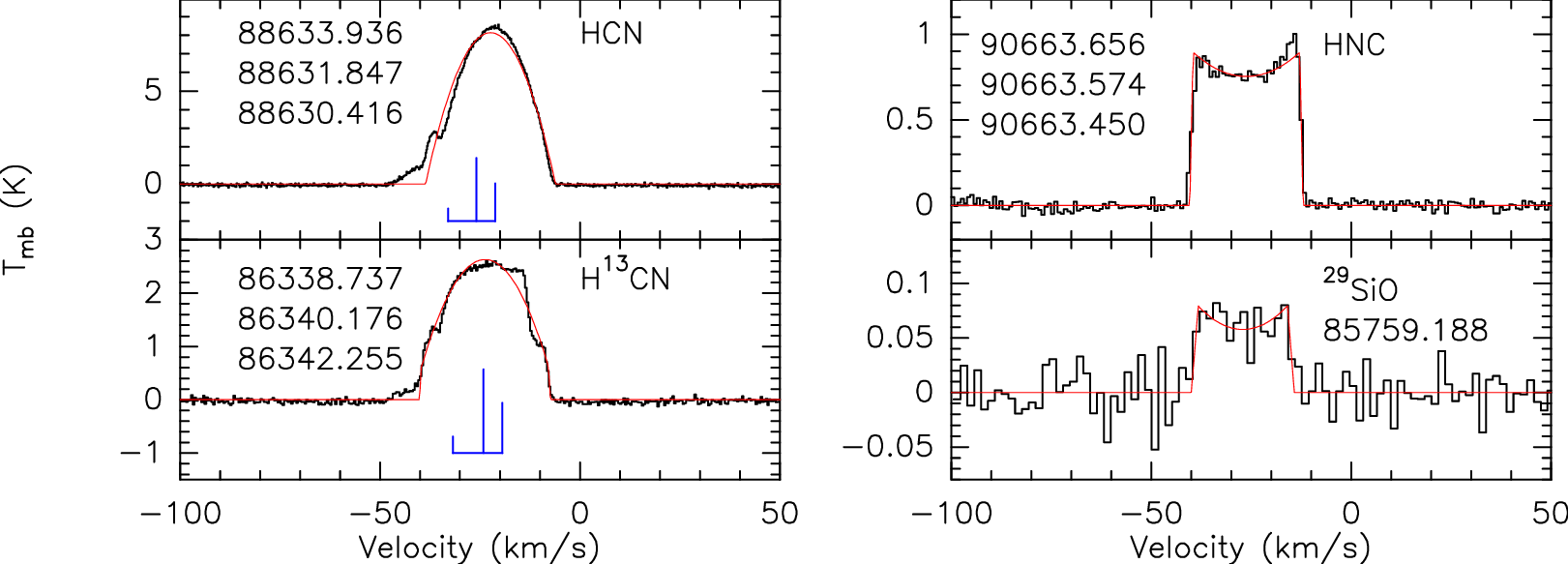}{0.72\textwidth}{ }
          }
\caption{Same as Figure~\ref{fig:S1-HC5N+SiO+12CO+13CO} and Figure~\ref{fig:C2H+C3H} but for HCN, H$^{13}$CN, HNC, and $^{29}$SiO.
\label{fig:HCN+HNC+29SiO}}
\end{figure*}

\begin{figure*}
\gridline{\fig{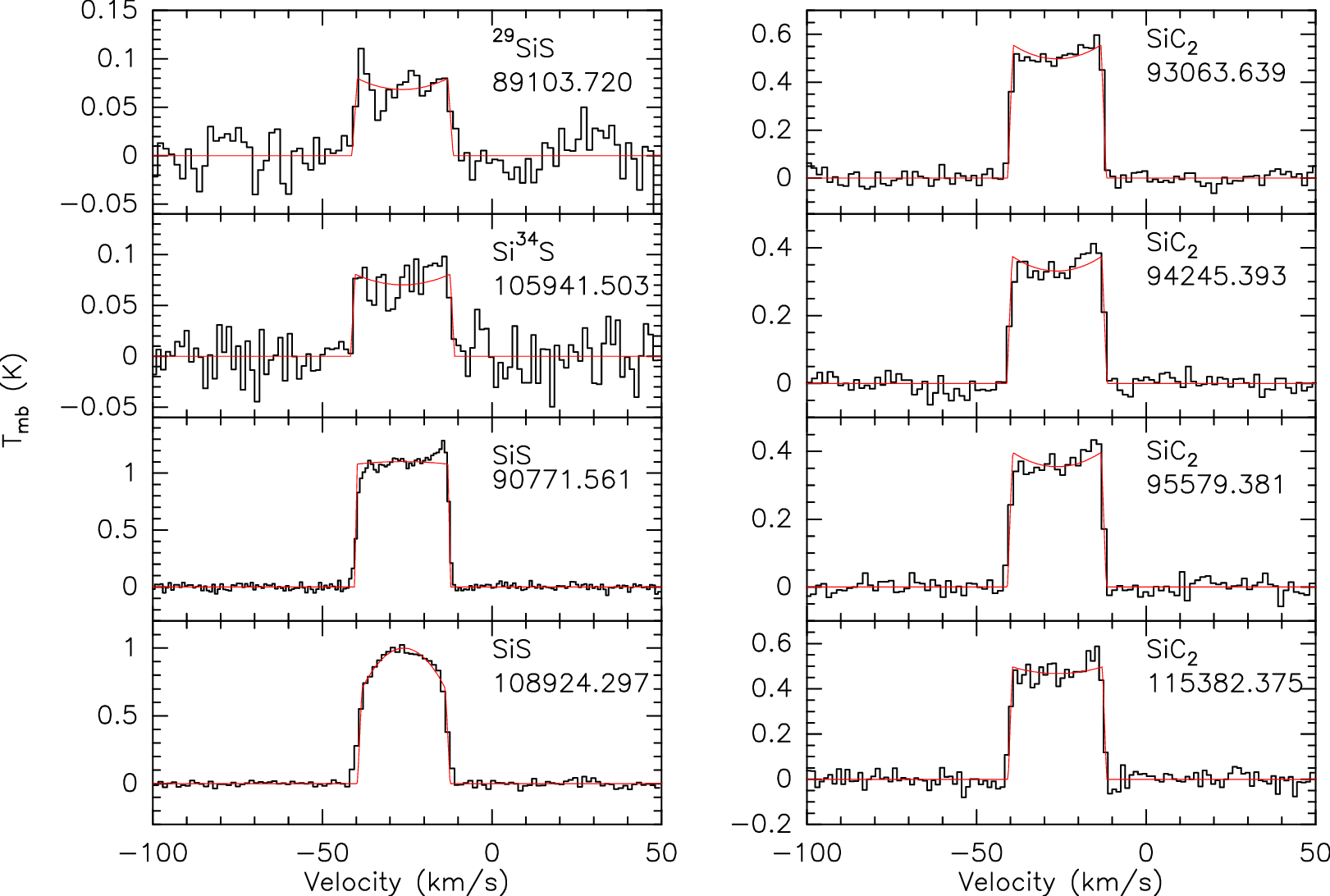}{0.72\textwidth}{ }
          }
\caption{Same as Figure~\ref{fig:S1-HC5N+SiO+12CO+13CO} but for SiS, $^{29}$SiS, Si$^{34}$S, and SiC$_2$.
\label{fig:SiS+SiC2}}
\end{figure*}

\begin{figure*}
\gridline{\fig{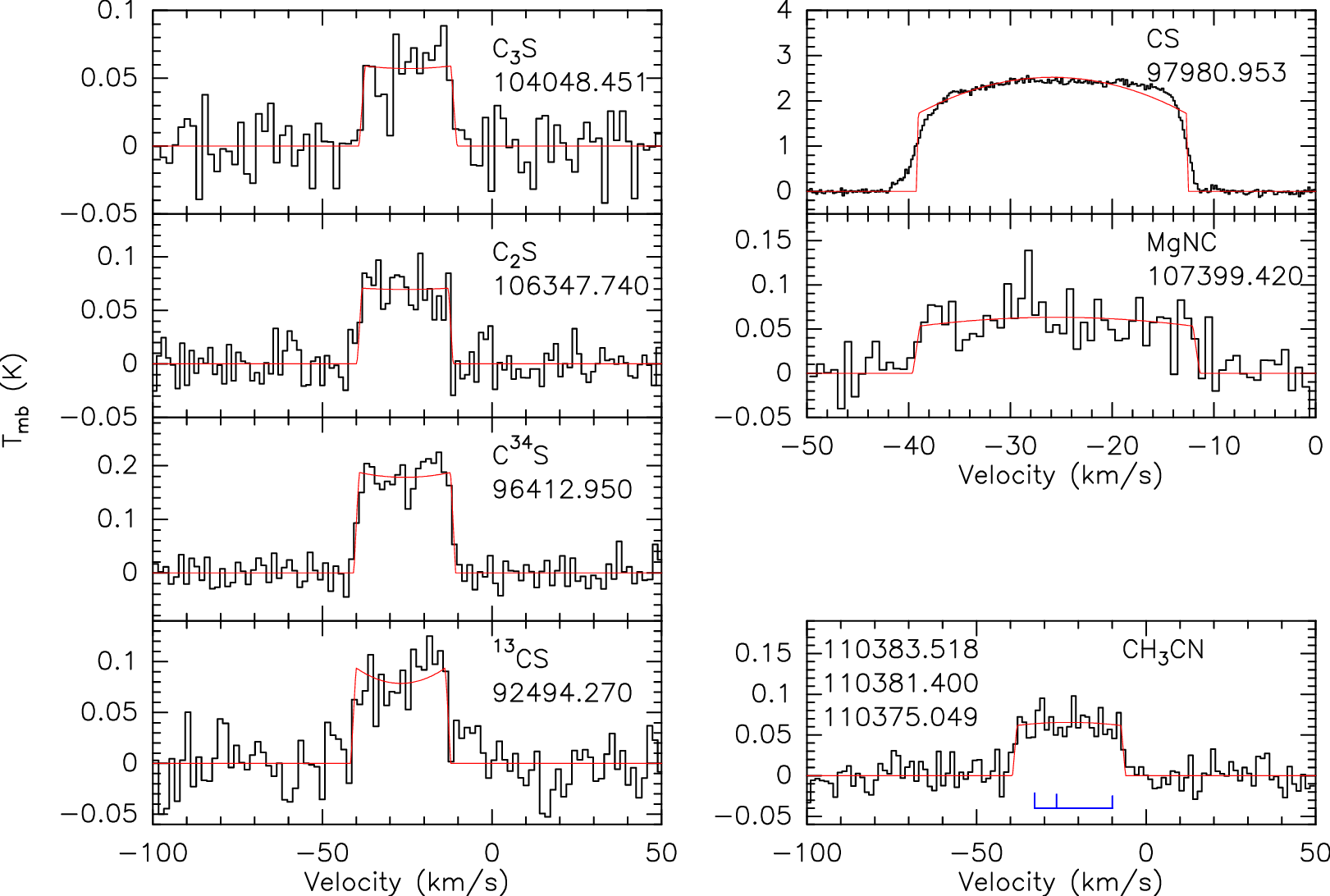}{0.72\textwidth}{ }
}

\caption{Same as Figure~\ref{fig:S1-HC5N+SiO+12CO+13CO} and Figure~\ref{fig:C2H+C3H} but for CS, $^{13}$CS, C$^{34}$S, C$_2$S, C$_3$S, MgNC, and CH$_3$CN.
\label{fig:CS+MgNC+CH3CN}}
\end{figure*}

%%%%%%%%%%%%%%%%%%%%%%%%%%%%%%%%%%%%%%%%%%%%%%%
%              Appendix  D
%%%%%%%%%%%%%%%%%%%%%%%%%%%%%%%%%%%%%%%%%%%%%%%
\section{Detected species}  \label{Detected species}
A detailed description of the detected molecules, together with their previous observations, is provided in this appendix.

{\bf C$_2$H.}
The ethynyl radical C$_2$H was first detected and identified by \cite{Tucker+etal+1974} in several galactic sources, including IRC\,+10216. In this source, the $N=1-0$ to $N=4-3$ and $N=6-5$ to $N=9-8$ transitions of C$_2$H have been reported by \cite{Beck+etal+2012}. We detected the two spin-rotation fine structure components of C$_2$H\,$N=1-0$  ($J=3/2-1/2$ and $J=1/2-1/2$) in which the nuclear spin of hydrogen further causes the spin-rotation to split into six hyperfine components: ($J=3/2-1/2$, $F=1-1$), ($J=3/2-1/2$, $F=2-1$),  ($J=3/2-1/2$, $F=1-0$), ($J=1/2-1/2$, $F=1-1$ and $F=0-1$) and ($J=1/2-1/2$, $F=1-0$). Among these components, C$_2$H  ($J=1/2-1/2$, $F=1-1$) is blended with C$_2$H ($J=1/2-1/2$, $F=0-1$), producing two blended U-shaped spectral lines with the overlapping part showing stronger intensity than both sides. The transitions are presented in the left panel of Figure~\ref{fig:C2H+C3H} and the top left panel of Figure~\ref{fig:C2H+C4H+C2S}. These spectral lines have also been detected by \cite{Johansson+etal+1984}, \cite{Truong-Bach+etal+1987}, and \cite{Kahane+etal+1988}.

{\bf $l$-C$_3$H.}
The linear propynylidyne radical $l$-C$_3$H was initially identified by \cite{Thaddeus+etal+1985} in IRC\,+10216 and TMC-1. Interestingly, \cite{Johansson+etal+1984} had previously detected $l$-C$_3$H at frequencies of 76202, 97995, and 98011 MHz, but had classified these spectral lines as U-lines and presumptively attributed them to transitions of $l$-C$_3$H. The $J=3/2-1/2$, $J=13/2-11/2$ and $J=21/2-19/2$ transitions of this molecule have also been detected in other line surveys in this source \citep{Kawaguchi+etal+1995, He+etal+2008}. The right panel of Figure~\ref{fig:C2H+C3H} shows eight hyperfine components that we have identified from the $l$-C$_3$H rotational lines $\rm^{2}{\Pi}_{1/2}$\,$J=9/2-7/2$ ($F=5-4e$, $F=4-3e$, $F=5-4f$, $F=4-3f$) and $\rm^{2}{\Pi}_{3/2}$\,$J=9/2-7/2$ ($F=5-4f$, $F=4-3f$, $F=5-4e$, $F=4-3e$). These transitions have been previously reported by \cite{Thaddeus+etal+1985}, \cite{Yamamoto+etal+1987}, and \cite{Nyman+etal+1993}.

{\bf C$_4$H.}
The butadiynyl radical C$_4$H\,(${\nu=0}$) was first identified by \cite{Guelin+etal+1978} in the CSE of IRC\,+10216. \cite{Yamamoto+etal+1987} subsequently confirmed the transitions of C$_4$H in the CSE of IRC\,+10216 caused by the $\rm^{2}{\Pi}_{1/2}$, $\rm^{2}{\Pi}_{3/2}$, and ${\Delta}$ states through the use of experimental microwave spectroscopy. Further, other transitions of C$_4$H have also been detected in this source such as $N =2-1$, $N=3-2$ through $N=5-4$, $N=8-7$, $N=14-13$ through $N=16-15$ and $N=24-23$ through $N=28-27$ \citep{Johansson+etal+1984, Kawaguchi+etal+1995, He+etal+2008, Gong+etal+2015}. A total of 15 transitions of C$_4$H have been detected in our work. There are four rotational transitions ($N=9-8$ through $N=12-11$) of C$_4$H\,(${\nu=0}$), each of these rotational transitions is split into spin-rotation doublets due to fine-structure, and these are presented in Figure~\ref{fig:C4H}. Additionally, seven transitions of C$_4$H\,(${\nu=1}$) are shown in Figure~\ref{fig:C2H+C4H+C2S}. However, the $\rm^{2}\Pi_{1/2}$\,C$_{4}$H\,$J=19/2-17/2$, ${\nu7=1f}$ line (93863.300 MHz) is blended with C$_2$S\,$N=8-7$, $J=7-6$ (93870.098 MHz), as shown in the second figure on the left panel of Figure~\ref{fig:C2H+C4H+C2S}.

{\bf CN.} The rotational emission from the cyanide radical CN was first detected by \cite{Jefferts+etal+1970} toward W\,51. Subsequently, the $N=1-0$ transition of this molecule was also detected in IRC\,+10216 \citep{Wilson+etal+1971}. Further transitions such as $N=2-1$ and $N=3-2$ of CN were also detected in this source by \cite{Avery+etal+1992}. We have detected the two spin-rotation fine structure lines of CN\,$N=1-0$ ($J=3/2-1/2$ and $J=1/2-1/2$), which are then further split into nine hyper-fine components due to the nuclear spin of nitrogen. These components include $J=3/2-1/2$  ($F=3/2-1/2$, $F=5/2-3/2$, $F=1/2-1/2$, $F=3/2-3/2$, $F=1/2-3/2$) and $J=1/2-1/2$ ($F=1/2-1/2$, $F=1/2-3/2$, $F=3/2-1/2$, $F=3/2-3/2$). The spectral line profiles are shown in Figure~\ref{fig:CN}, although we note that the five hyper-fine lines of CN\,$N=1-0$, $J=3/2-1/2$ are blended.

{\bf C$_3$N.} \cite{Guelin+etal+1977} detected a new molecule toward the CSE of IRC\,+10216, identifying it as the cyanoethynyl radical C$_3$N based on theoretical considerations. The $N=2-1$, $N=3-2$ through $N=5-4$, $N=8-7$ through $N=11-10$, $N=13-12$, $N=14-13$ through $N=17-16$, $N=23-22$, $N=24-23$, and $N=27-26$ transition lines of C$_3$N in this source have been reported \citep{Guelin+etal+1977, Johansson+etal+1984, Henkel+etal+1985, Kawaguchi+etal+1995, Cernicharo+etal+2000, He+etal+2008, Thaddeus+etal+2008, Gong+etal+2015}. A total of six fine-structure lines of C$_3$N have been detected in our work, and displayed in Figure~\ref{fig:C3N+c-C3H2}. These lines derive from three rotational transitions ($N=9-8$ through $N=11-10$) of C$_3$N. All of the detected rotational transition lines are split into spin-rotation doublets due to the interactions of fine structure.

{\bf $c$-C$_3$H$_2$.} A strong unidentified line at 85.339\,GHz was first detected by \cite{Thaddeus+etal+1981} toward Ori\,A, Sgr\,B2 (OH), and TMC-1. This line was subsequently identified as  cyclopropenylidene $c$-C$_3$H$_2$ by \cite{Thaddeus+etal+1985a}. The rotational transitions of $c$-C$_3$H$_2$ have also been detected in other spectral line surveys of this source \citep{Kawaguchi+etal+1995, Cernicharo+etal+2000, He+etal+2008, Gong+etal+2015, Zhang+etal+2017}. In our work, we have detected one transition of $c$-C$_3$H$_{2}$, shown at the bottom of the right panel in Figure~\ref{fig:C3N+c-C3H2}.

{\bf HC$_3$N.} Cyanoacetylene HC$_3$N was detected by \cite{Turner+etal+1970} and \cite{Turner+etal+1971} toward Sgr\,B2. \cite{Morris+etal+1975} also detected this molecule in IRC\,+10216 using the NRAO--11\,m telescope. Various transitions of HC$_3$N and its isotopologues have been reported in this source.  For instance, the $J=2-1$ transitions of HC$_3$N, H$^{13}$CCCN, HC$^{13}$CCN and HCC$^{13}$CN were detected by \cite{Zhang+etal+2017}, the $J=4-3$ and $J=5-4$ transitions of HC$_3$N and its isotope were detected by \cite{Kawaguchi+etal+1995} and the $J=9-8$ and $J=10-9$ transitions of HC$_3$N and its isotope were detected by \cite{Johansson+etal+1984}. \cite{Nyman+etal+1993} detected HC$_3$N $J=10-9$ and $J=12-11$ in this source,  as well as HC$^{13}$CCN and HCC$^{13}$CN. We detected three lines ($J=10-9$ through $J=12-11$) of HC$_3$N and two relatively weak isotopic lines (H$^{13}$CCCN\,$J=10-9$ and H$^{13}$CCCN\,$J=13-12$). We also detected three transitions of HC$^{13}$CCN ($J=10-9$ through $J=12-11$)  and three transitions of HCC$^{13}$CN ($J=10-9$ through $J=12-11$), but all detected HC$^{13}$CCN lines are blended with HCC$^{13}$CN transitions. These lines are shown in Figure~\ref{fig:HC3N}. To our knowledge, the $J=13-12$ transition of H$^{13}$CCCN is detected in IRC\,+10216 for the first time.

{\bf HC$_5$N}. Cyanodiacetylene HC$_5$N is a linear molecule, which was observed in Sgr\,B2 by \cite{Avery+etal+1976}. In IRC\,+10216, \cite{Winnewisser+etal+1978} confirmed the HC$_5$N transition reported by \cite{Churchwell+etal+1978}. The $J=2-1$, $J=5-4$ to $J=9-8$, $J=28-27$ to $J=40-39$ and $J=49-48$ to $J=54-53$ transitions of HC$_5$N have been detected in IRC\,+10216 \citep{Johansson+etal+1984, Cernicharo+etal+1986a, Cernicharo+etal+2000, Araya+etal+2003, Gong+etal+2015, Agundez+etal+2017, Zhang+etal+2017}. We have detected nine rotational transition lines of HC$_5$N from $J=32-31$ to $J=40-39$. These lines are shown in Figure~\ref{fig:S1-HC5N+SiO+12CO+13CO} and Figure~\ref{fig:HC5N}.

{\bf HCN.} Hydrogen cyanide HCN was first detected in W49, W51, W3\, (OH), Orion\,A, DR\,21 (OH) and Sgr\,A (NH$_{3}$A) \citep{Howard+etal+1970, Snyder+etal+1971}. HCN has been detected in IRC\,+10216 by \cite{Morris+etal+1971}. The transitions of $J=2-1$ \citep{Menten+etal+2018}, $J=3-2$ and $J=4-3$ \citep{Avery+etal+1992} of HCN, and the $J=1-0$, $J=3-2$, $J=4-3$, and $J=8-7$ transitions of H$^{13}$CN \citep{Schoier+etal+2007} have been detected as well. The $J=1-0$ transitions from HCN and H$^{13}$CN were detected in our line survey. The HCN\,$J=1-0$ transition contains hyperfine structure, the frequency intervals between the individual features are too small \citep{Olofsson+etal+1982} causing us to detect both absorption and emission in the line profile in Figure~\ref{fig:HCN+HNC+29SiO} \citep[see also][]{Dayal+etal+1995}.

%For the three hyperfine structures of HCN\,$J=1-0$, the radiation of the $F=1-1$ line is not affected by the emission of the main line emission ($F=2-1$) and $F=0-1$ line. The emission of the main line emission is absorbed by the transition of $F=1-1$, while the $F=0-1$ line is absorbed by both the transition of $F=1-1$ and $F=2-1$ \citep{Olofsson+etal+1982}. Due to the interaction of these hyper-fine structures, we can see emission and absorption characteristics in Figure~\ref{fig:HCN+HNC+29SiO} \citep[see also][]{Dayal+etal+1995}.

{\bf HNC.} An unidentified emission line at 89.190\,GHz was detected by \cite{Buhl+etal+1970} in W3\,(OH), Orion, L134, Sgr\,A (NH$_{3}$A) and W\,51, first attributed by \cite{Snyder+etal+1972} to the $J=1-0$ transition of hydrogen isocyanide HNC. \cite{Brown+etal+1976} reported that HNC had been detected in IRC\,+10216. Using the IRAM--30\,m telescope and the HIFI instrument onboard the Herschel Space Observatory, the transitions of HNC\,$J=3-2$, $J=1-0$ and $J=6-5$ through $J=12-11$ were detected in IRC\,+10216 \citep{Daniel+etal+2012}. In our work, we detected the $J=1-0$ transition from HNC, as shown in Figure~\ref{fig:HCN+HNC+29SiO}. Although HNC\,$J=1-0$ contains hyper-fine structures, the frequency interval between these structures is too small to detect in IRC\,+10216 \citep{Olofsson+etal+1982}.

{\bf CO.} Carbon monoxide CO at 115\,GHz was observed by \cite{Jefferts+etal+1970a} in at least five galactic sources, including Orion\,A, Sgr\,A, Sgr\,B2, W39, and W51. Subsequently, CO was also observed in IRC\,+10216 \citep{Solomon+etal+1971}.
Other transitions of CO and $^{13}$CO in IRC\,+10216 had also been reported \citep{Avery+etal+1992, Cernicharo+etal+2015}. In our work, we have detected the $J=1-0$ transitions from CO and $^{13}$CO, as shown in Figure~\ref{fig:S1-HC5N+SiO+12CO+13CO}.

{\bf SiO.} The $J=3-2$ transition of silicon monoxide SiO was detected in Sgr\,B2 by \cite{Wilson+etal+1971}. Subsequently, \cite{Morris+etal+1975} observed the $J=2-1$ transition of SiO in the CSE of IRC\,+10216. The transitions observed by previous research include $J=1-0$ through $J=8-7$ \citep{Avery+etal+1992, Kawaguchi+etal+1995, Cernicharo+etal+2000, He+etal+2008, Tenenbaum+etal+2010, Agundez+etal+2012}. In our study, we detected the $J=2-1$ rotational transitions of SiO and $^{29}$SiO, and their line profiles are illustrated in Figure~\ref{fig:S1-HC5N+SiO+12CO+13CO} and Figure~\ref{fig:HCN+HNC+29SiO}.

{\bf SiS.} Silicon monosulfide SiS was first detected by \cite{Morris+etal+1975} toward IRC\,+10216. The maser and quasi-thermal emission of the SiS\,$J=1-0$, and $J=2-1$ lines in the CSE of IRC\,+10216 have been studied by \cite{Gong+etal+2017}. The transitions of SiS and $^{29}$SiS observed by \cite{Agundez+etal+2012} include $J=5-4$, $J=6-5$ and $J=8-7$ through $J=19-18$. Among them, the $J=11-10$, $J=14-13$, and $J=15-14$ transitions are masers, which was also reported by \cite{Fonfria+etal+2006} and \cite{Fonfria+etal+2018}. The $J=5-4$, $J=6-5$ and $J=8-7$ through $J=20-19$ from Si$^{34}$S were also detected by \cite{Agundez+etal+2012}. We have detected the $J=5-4$ and $J=6-5$ lines from SiS, as well as the $J=5-4$ transitions from $^{29}$SiS and the $J=6-5$ line from Si$^{34}$S. The line profiles are presented in Figure~\ref{fig:SiS+SiC2}.

{\bf SiC$_2$.} Silacyclopropynylidene, SiC$_2$, was the first ring molecule identified in an astronomical source. Its millimeter- and centimeter-wave transitions were identified by \cite{Thaddeus+etal+1984} and \cite{Snyder+etal+1985}, respectively, in the CSE of IRC\,+10216. High resolution spectra (channel width 195\,kHz) of SiC$_2$ at 70--350\,GHz \citep{Cernicharo+etal+2018}, 554.5-636.5\,GHz \citep{Cernicharo+etal+2010} and the spatial distribution of SiC$_2$ in IRC\,+10216 \citep{Fonfria+etal+2014, Velilla+Prieto+etal+2015, Velilla-Prieto+etal+2019} have also been reported.
We have detected the
$J_{\rm K_a,K_c}$ = 4$_{0,4}$-3$_{0,3}$,
$J_{\rm K_a,K_c}$ = 4$_{2,3}$-3$_{2,2}$,
$J_{\rm K_a,K_c}$ = 4$_{2,2}$-3$_{2,1}$, and
$J_{\rm K_a,K_c}$ = 5$_{0,5}$-4$_{0,4}$ transitions of SiC$_2$, and the line profiles are shown in Figure~\ref{fig:SiS+SiC2}.

{\bf CS.} The $J=3-2$ transition of carbon monosulfide CS was detected in IRC\,+10216, Orion\,A, W51 and DR\,21 by \cite{Penzias+etal+1971}. The transitions of $J=1-0$ through $J=7-6$ from CS and the transitions of $J=1-0$ through $J=3-2$ and $J=5-4$ through $J=7-6$ from $^{13}$CS and C$^{34}$S in the CSE of IRC\,+10216 have also been extensively reported \citep{Johansson+etal+1984, Avery+etal+1992, Lucas+etal+1995, Kawaguchi+etal+1995, He+etal+2008, Agundez+etal+2012, Velilla-Prieto+etal+2019}. We detected strong CS\,$J=2-1$ emission and its isotopic lines, $^{13}$CS\,$J=2-1$ and C$^{34}$S\,$J=2-1$. The line profiles of CS, $^{13}$CS and C$^{34}$S are depicted in Figure~\ref{fig:CS+MgNC+CH3CN}.

{\bf C$_2$S.} An unknown interstellar line at 45.379\,GHz from Sgr\,B2, TMC-1, TMC-1  (NH$_{3}$), and TMC2 was detected by \cite{Suzuki+etal+1984}. \cite{Kaifu+etal+1987} detected seven unknown strong lines in the frequency regions of 22--24 and 36--50\,GHz from TMC-1. The four unknown lines at 45.379, 43.981, 38.866 and 22.344\,GHz were identified to trace the linear C$_2$S radical \citep{Saito+etal+1987}. Twelve spectral lines of C$_2$S in the frequency range of 81--157\, GHz have also been detected by \cite{Cernicharo+etal+1987} towards IRC\,+10216. In our work, we have detected two transitions of C$_2$S, $N=9-8$, $J=8-7$ and $N=8-7$, $J=7-6$, but the transition of C$_2$S\,$N=8-7$, $J=7-6$ is blended with the $\rm^{2}{\Pi}_{1/2}\,C_{4}H\,{\nu7=1f}$ line. The second figure in the left panel of Figure~\ref{fig:C2H+C4H+C2S} shows the C$_2$S spectral line blended with C$_4$H. Figure~\ref{fig:CS+MgNC+CH3CN} shows a single C$_2$S spectral line.

{\bf C$_3$S.} \cite{Kaifu+etal+1987} detected seven unknown strong lines in the frequency regions of 22--24 and 36--50\, GHz from TMC-1, of which three were identified as linear C$_3$S radicals at 3.123, 40.465, and 46.246\,GHz by \cite{Yamamoto+etal+1987}. Meanwhile, \cite{Cernicharo+etal+1987} used the IRAM 30\,m telescope to detect 11 transition lines of C$_3$S between 75 and 168\,GHz in IRC\,+10216. Only one transition line of C$_3$S\,$J=18-17$ was detected in our work, and the spectral line is shown in Figure~\ref{fig:CS+MgNC+CH3CN}.

{\bf MgNC.} \cite{Guelin+etal+1986} observed six unidentified spectral lines in IRC\,+10216 with the IRAM--30\,m telescope, and they considered that these molecules were likely to be HSiCC (or its isomer HCCSi) or HSCC. Subsequently, \cite{Kawaguchi+etal+1993} detected linear MgNC radicals for the first time using microwave spectroscopy in the laboratory, and they confirmed that these unrecognized lines were linear MgNC free radicals. The $N=2-1$ through $N=4-3$ rotational transitions, as well as the rotational transitions of $N=12-11$ and $N=13-12$ of MgNC in the CSE of IRC\,+10216 have been observed by \cite{Kawaguchi+etal+1995}, \cite{He+etal+2008}, and \cite{Gong+etal+2015}. In our work, we only detected the $N=8-7$\,$J=19/2-17/2$ rotational transition of MgNC, which is displayed in the second sub-picture of the right panel of Figure~\ref{fig:CS+MgNC+CH3CN}.

{\bf CH$_3$CN.} Methyl cyanide CH$_3$CN has been detected by \cite{Solomon+etal+1971} in Sgr\,B and Sgr\,A. \cite{Johansson+etal+1984} detected CH$_3$CN, the first species containing the methyl group, in IRC\,+10216. In this source, the $J=2-1$ transitions in the $K=0$ and $K=1$ states of CH$_3$CN were detected by \cite{Kawaguchi+etal+1995}. The $J=8-7$, $J=12-11$ and $J=13-12$ transitions in the $K=0, 1, 2, 3$ states and the $J=14-13$ transitions in the $K=0$ and  $K=1$ states of CH$_3$CN were detected by \cite{He+etal+2008}. Other higher J lines of CH$_3$CN in the frequency range $293.9-354.8$\, GHz were observed by \cite{Patel+etal+2011}. We only detected the $J=6-5$ transitions in the $K=0$, $K=1$, and $K=2$ states of CH$_3$CN and these blended emission lines are shown in the last sub-picture on the right panel of Figure~\ref{fig:CS+MgNC+CH3CN}.

%%%%%%%%%%%%%%%%%%%%%%%%%%%%%%%%%%%%%%%%%%%%%%%
%              Appendix  E
%%%%%%%%%%%%%%%%%%%%%%%%%%%%%%%%%%%%%%%%%%%%%%%
\section{Rotational diagrams for the identified species} \label{Rotational diagrams}
%This appendix includes the rotational diagrams for the identified species from this survey, as shown in Figs.~(\ref{fig:T1} $-$~\ref{fig:T5}).
The rotational diagrams for the identified species from this survey are shown in Figs.~\ref{fig:T1} $-$~\ref{fig:T5}.

%%%%%%%%%%%%%%%%%%%%%%%%%
\renewcommand\thefigure{\Alph{section}\arabic{figure}}
\setcounter{figure}{0}
\begin{figure*}
\gridline{\fig{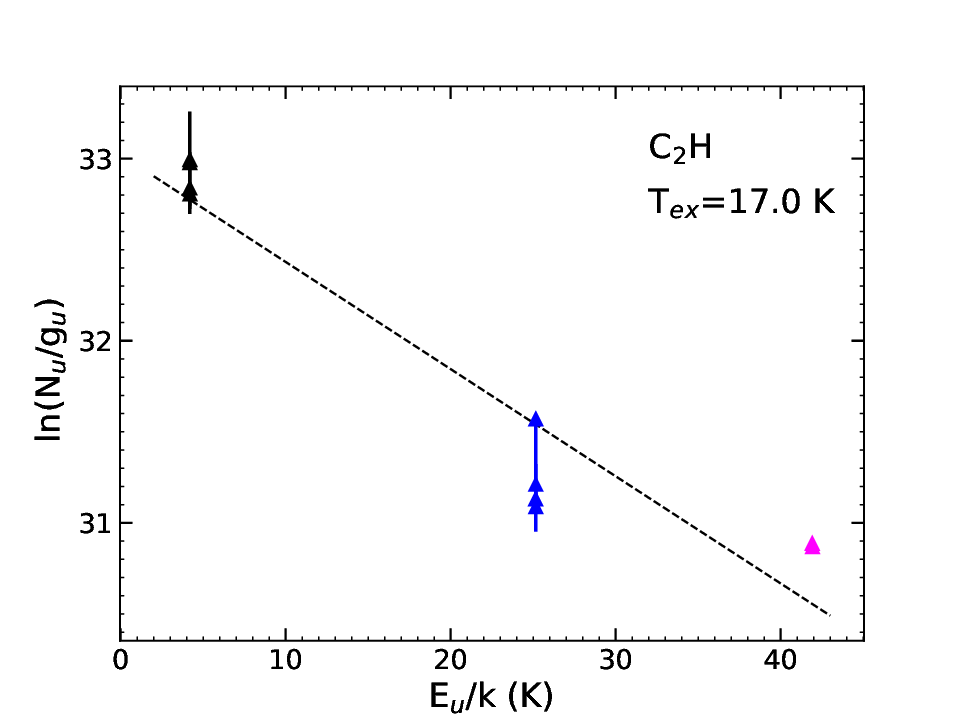}{0.45\textwidth}{ }
          \fig{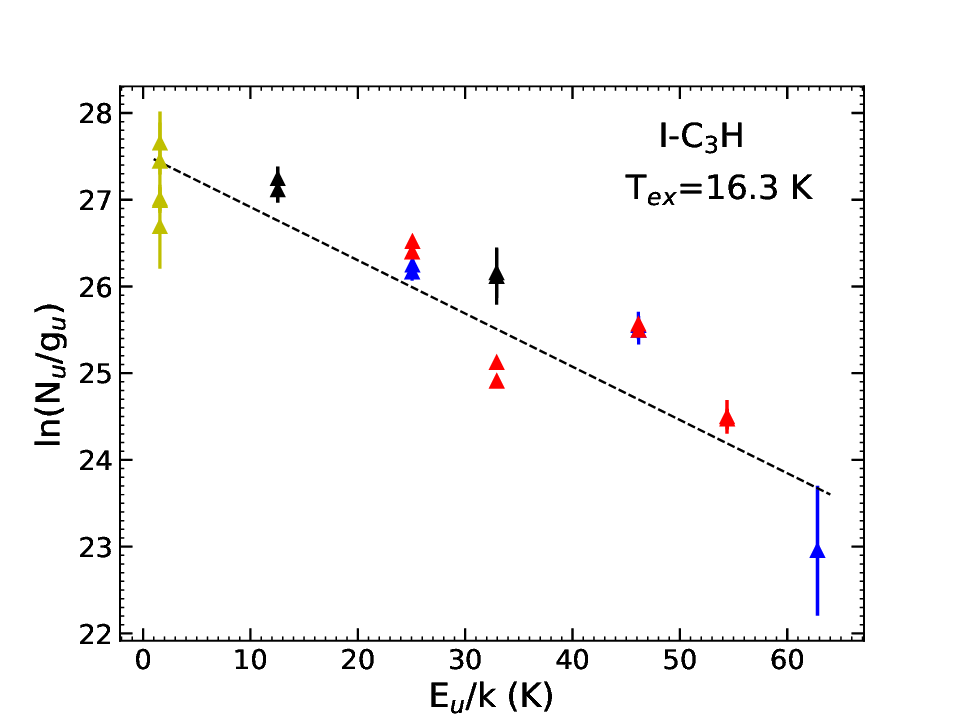}{0.45\textwidth}{ }
          }
\gridline{\fig{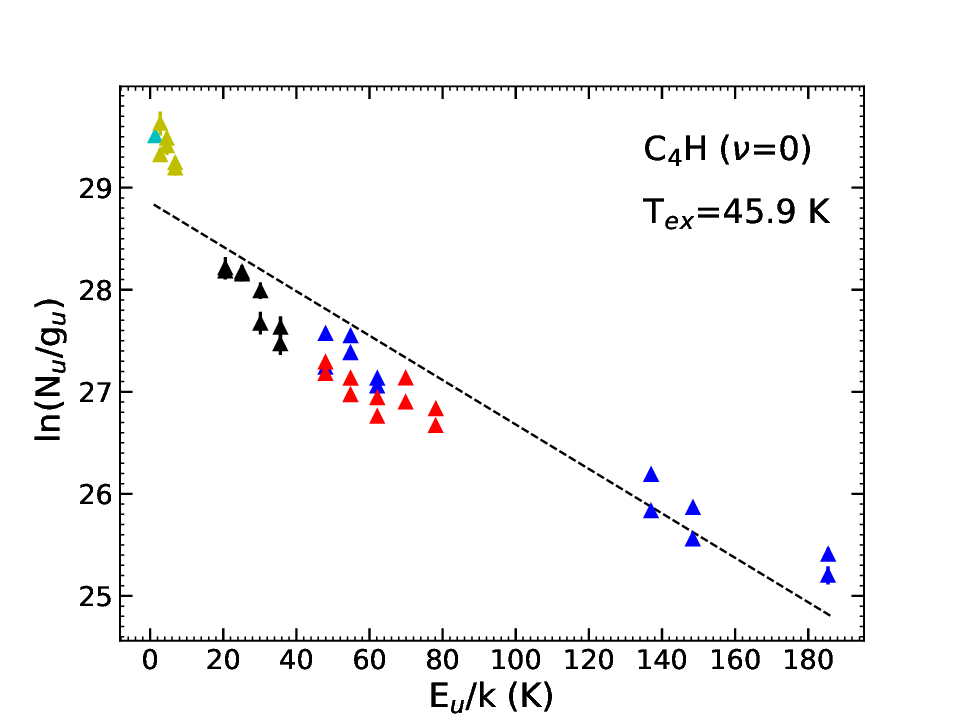}{0.45\textwidth}{ }
          \fig{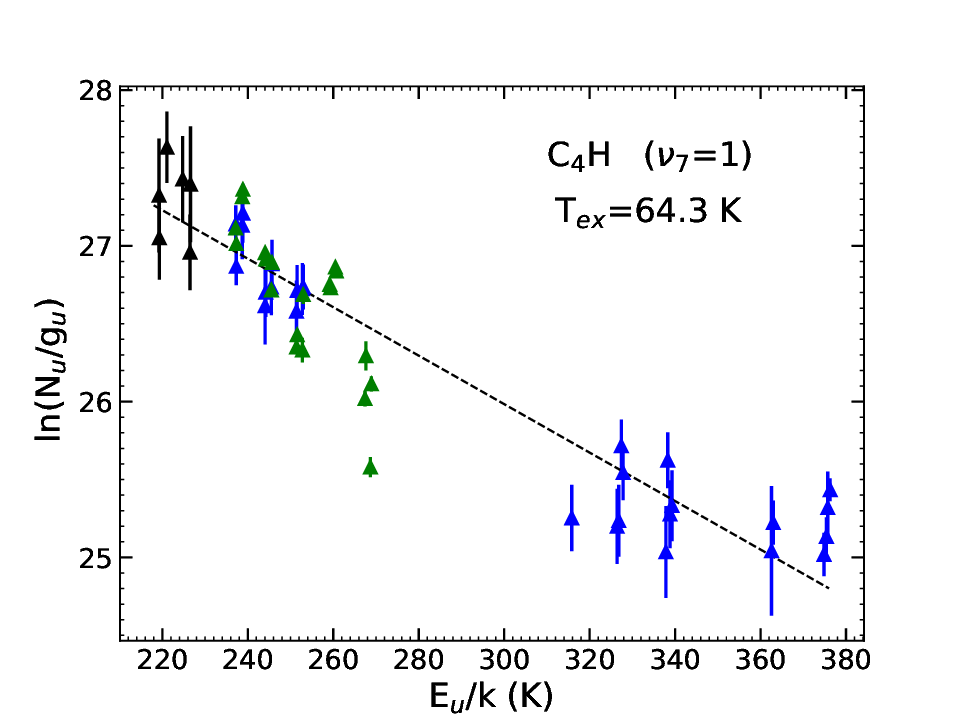}{0.45\textwidth}{ }
          }
\gridline{
          \fig{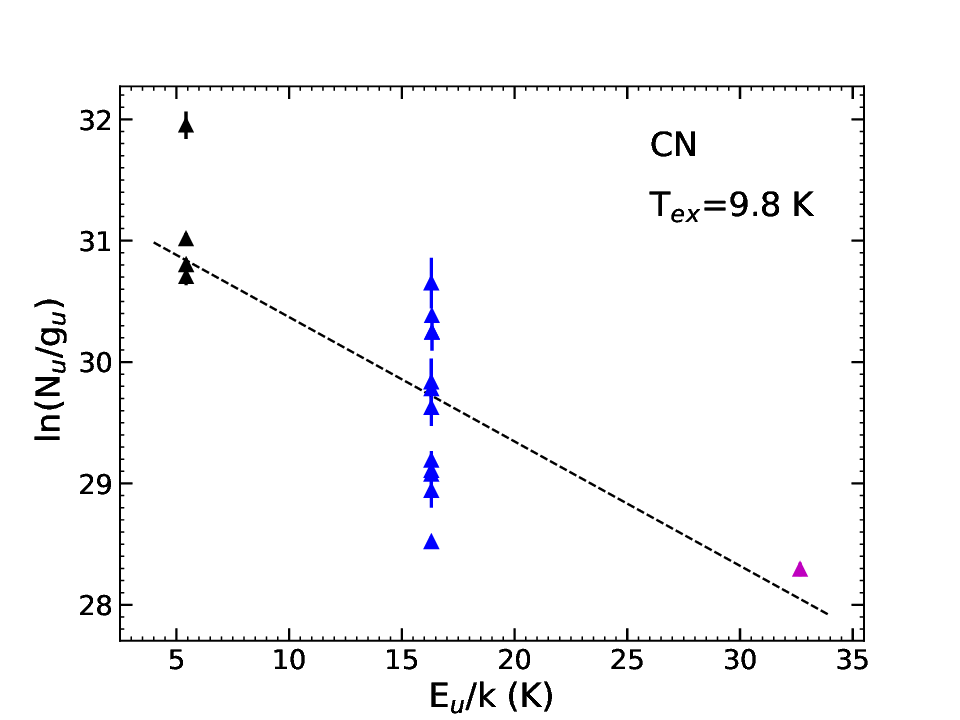}{0.45\textwidth}{ }
          \fig{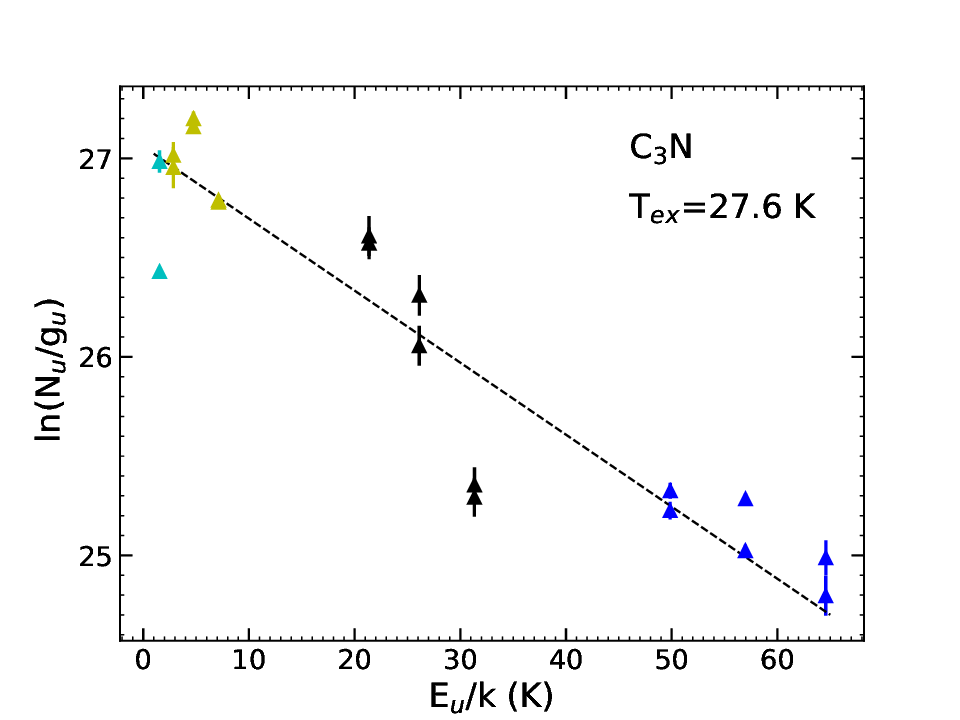}{0.45\textwidth}{ }
          }
\caption{Rotational diagrams for the identified molecules in IRC\,+10216 in the $\lambda$ 3\,mm wavelength band.
The results obtained from the linear least-squares fitting are plotted in black dashed lines.
The adopted data points with error bars are from our PMO--13.7\,m observations (black triangles),
 \protect\cite{He+etal+2008} (blue triangles),
 \protect\cite{Gong+etal+2015} (cyan triangles),
 \protect\cite{Zhang+etal+2017} (dark-orange triangles),
  \protect\cite{Kawaguchi+etal+1995} (yellow-green triangles),
  \protect\cite{Cernicharo+etal+2000} (red triangles),
  \protect\cite{Agundez+etal+2012} (green triangles), and
\protect\cite{Groesbeck+etal+1994} (magenta triangles), respectively.
Note that the error bars are invisible if their values are smaller than the adopted data points.
\label{fig:T1}}
\end{figure*}

\begin{figure*}
\gridline{\fig{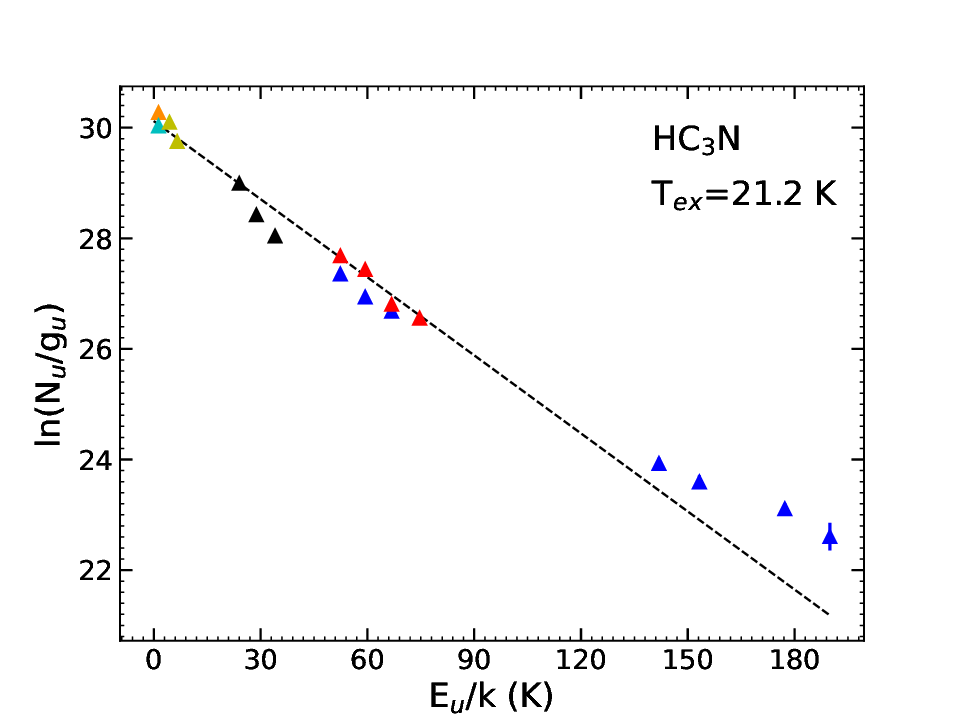}{0.45\textwidth}{ }
          \fig{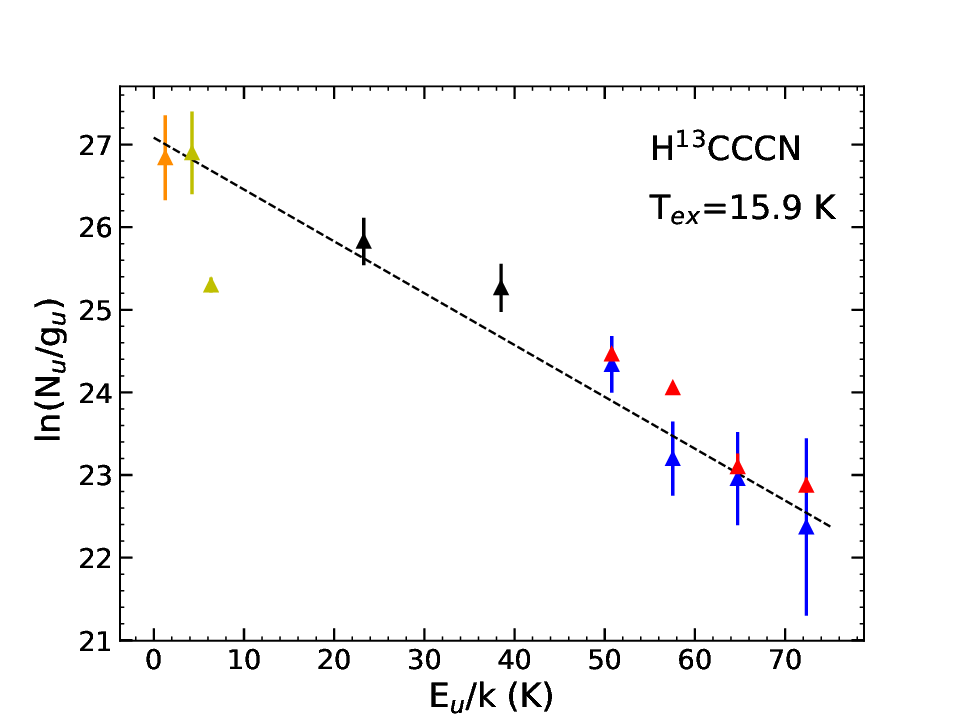}{0.45\textwidth}{ }
          }

\gridline{\fig{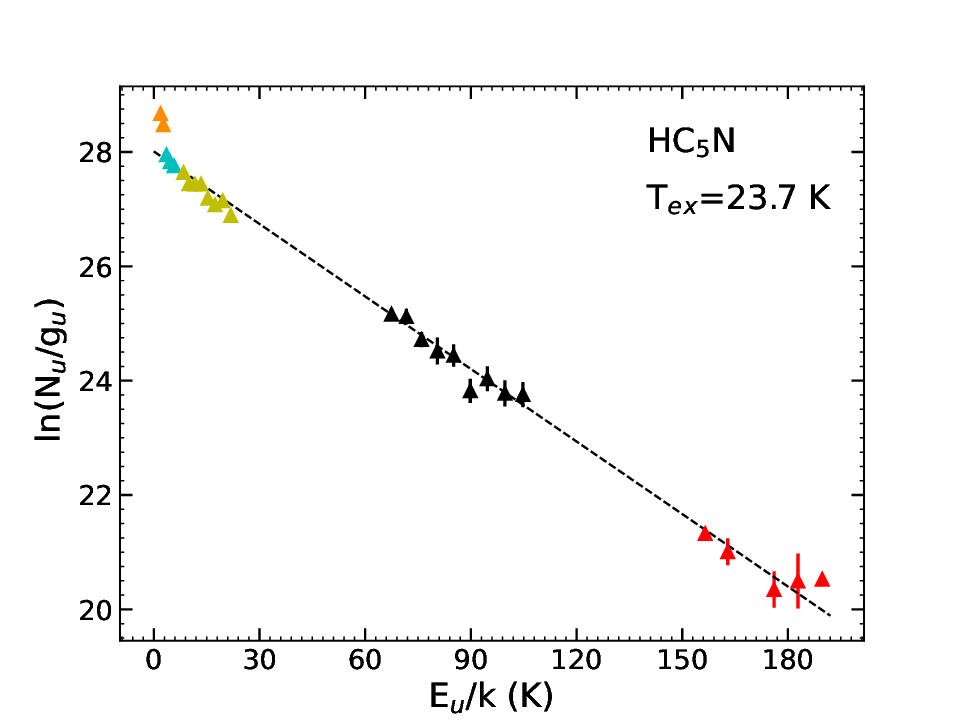}{0.45\textwidth}{ }
          \fig{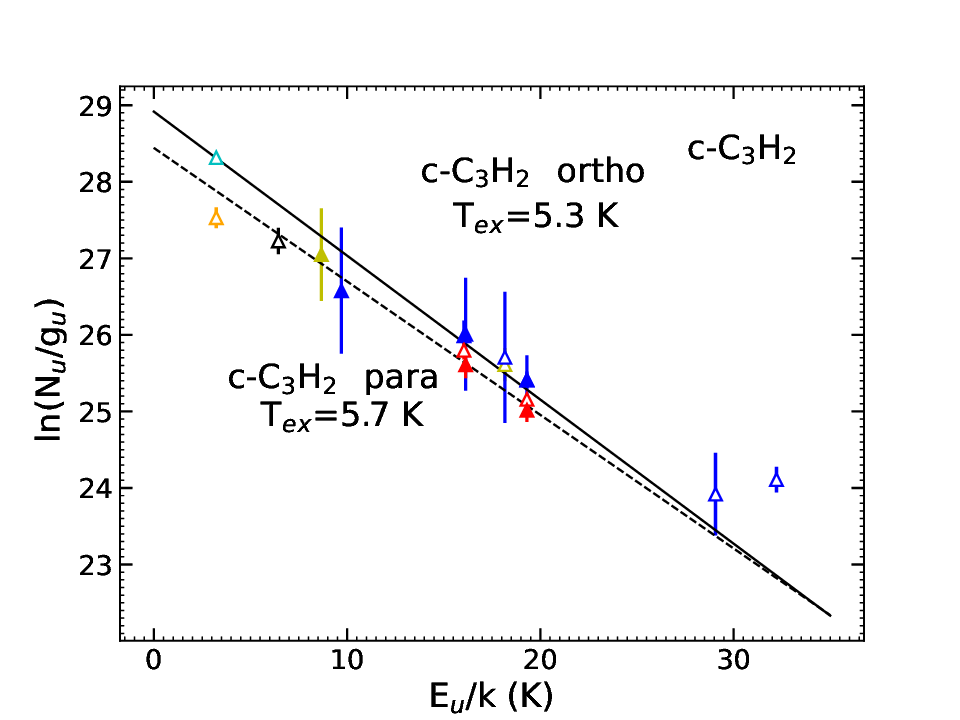}{0.45\textwidth}{ }
          }
\gridline{\fig{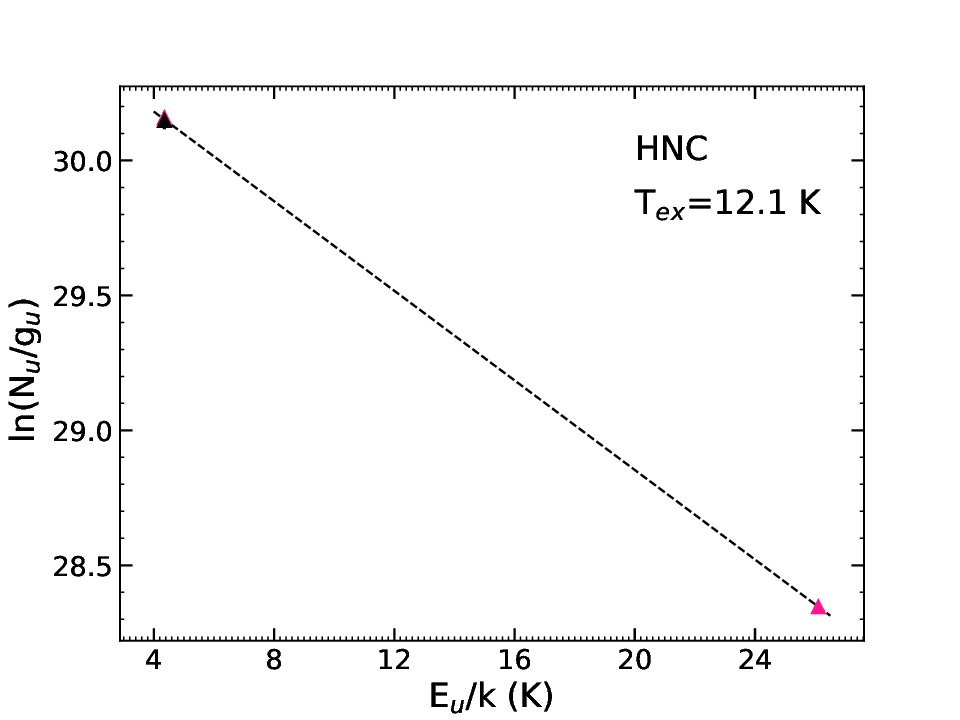}{0.45\textwidth}{ }
          \fig{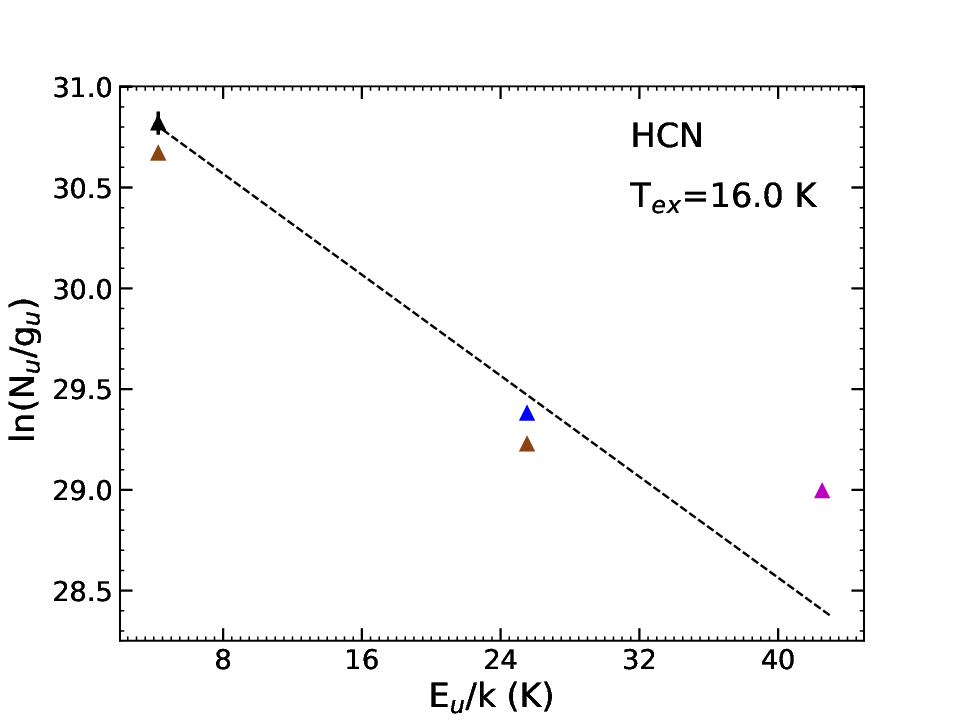}{0.45\textwidth}{ }
          }
\caption{Same as Figure~\ref{fig:T1}, but for different molecular species.
Note, molecular $c$-C$_{3}$H$_{2}$ is separated into ortho (open triangles) and para (filled triangles) states and fitted in solid (ortho) and dotted (para) lines. The data in saddle-brown triangles with error bars are obtained from \protect\cite{Nyman+etal+1993}.}
\label{fig:T2}
\end{figure*}

\begin{figure*}
\gridline{\fig{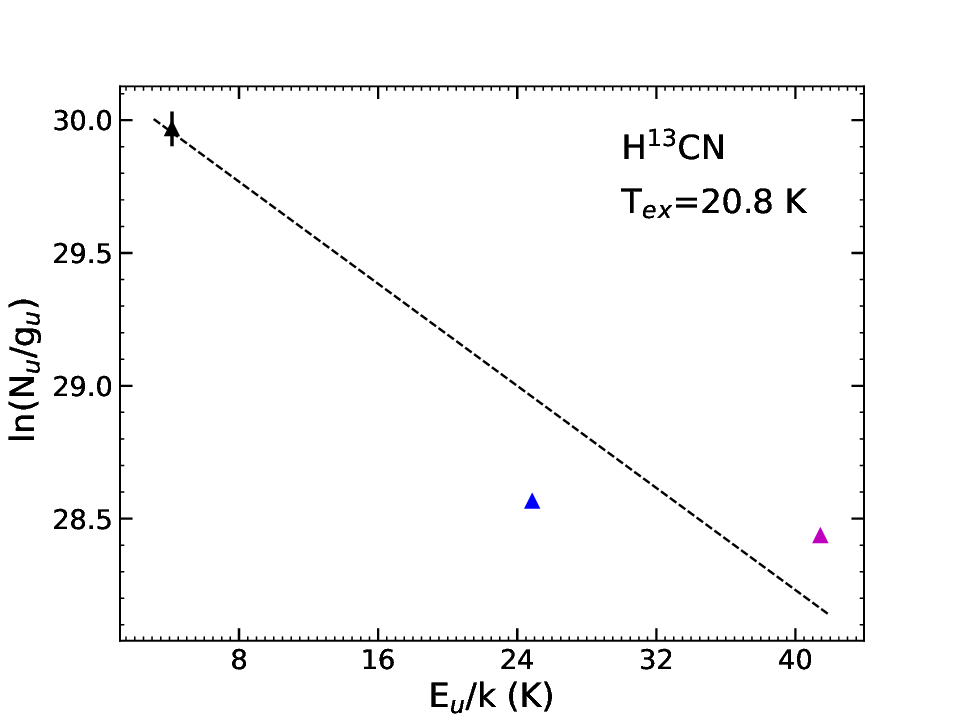}{0.45\textwidth}{ }
          \fig{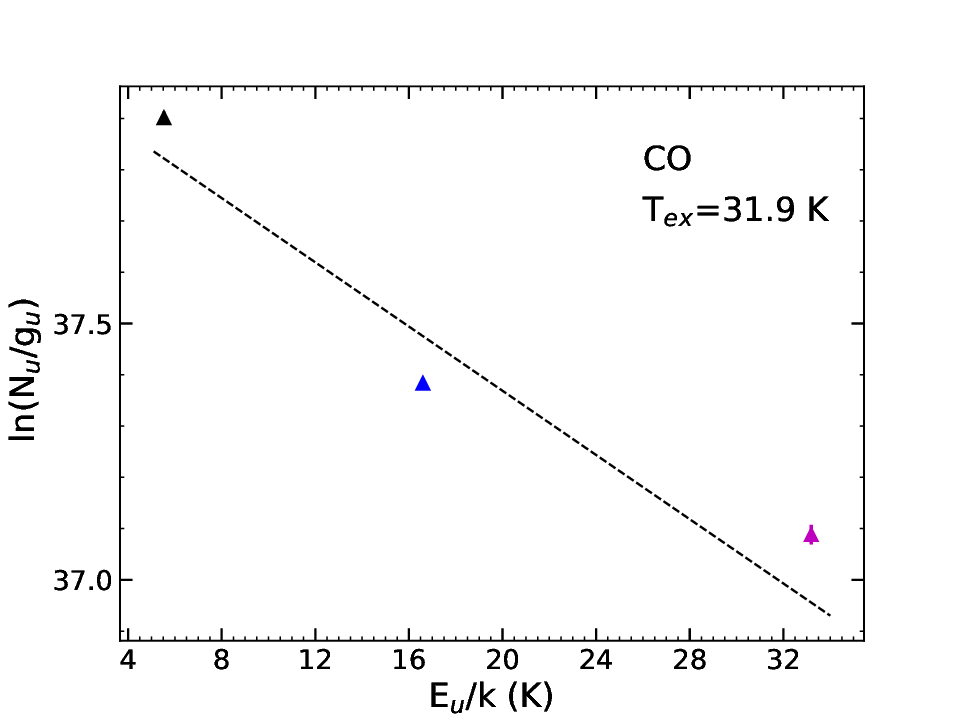}{0.45\textwidth}{ }
          }
\gridline{\fig{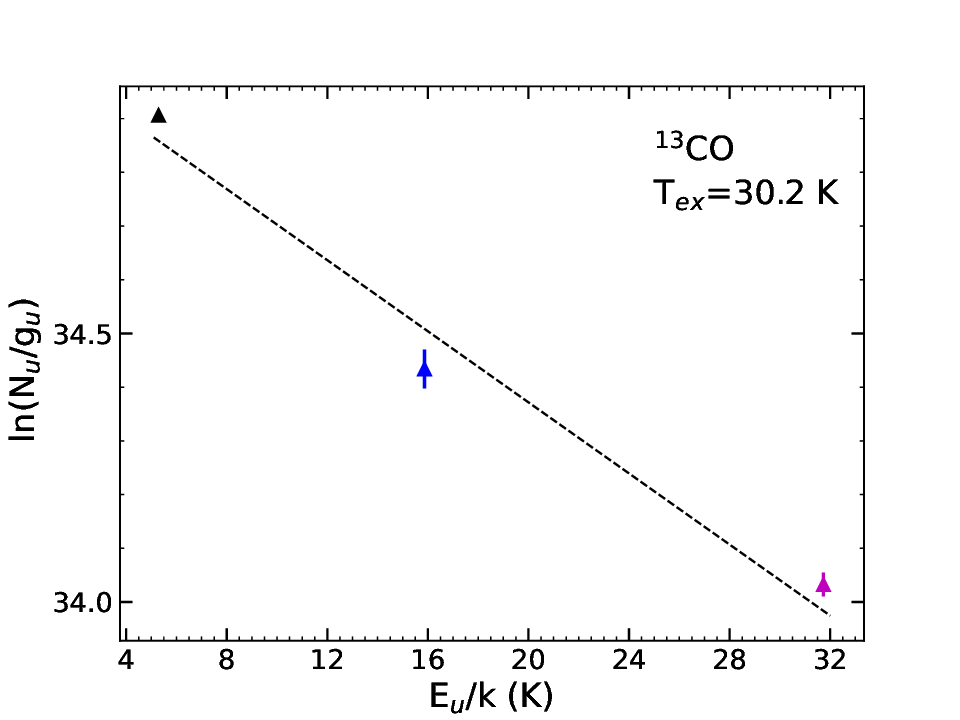}{0.45\textwidth}{ }
          \fig{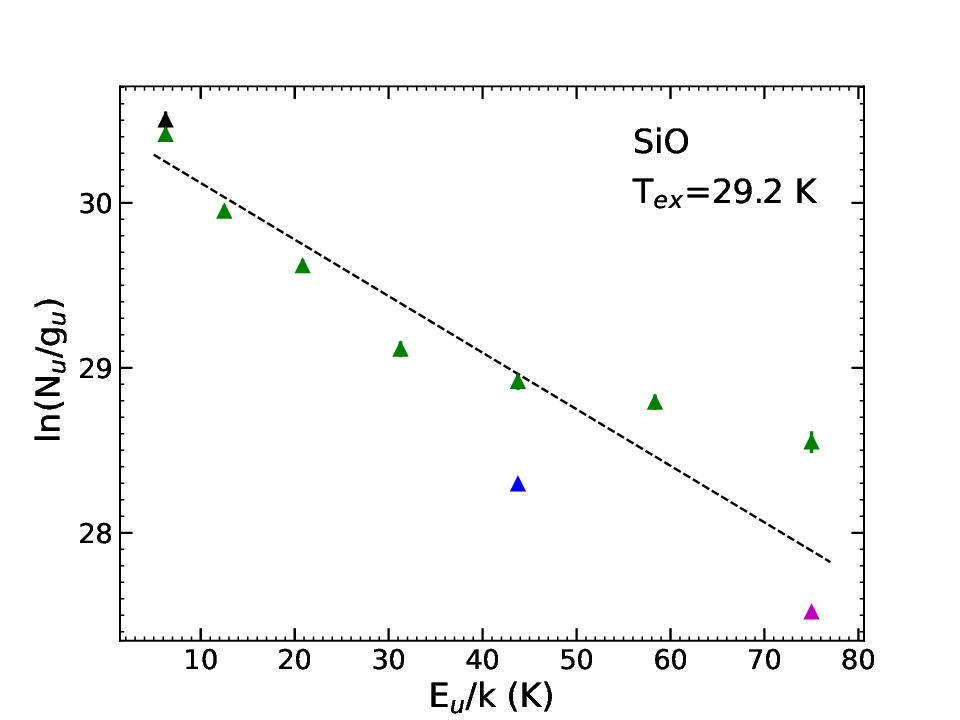}{0.45\textwidth}{ }
          }
\gridline{\fig{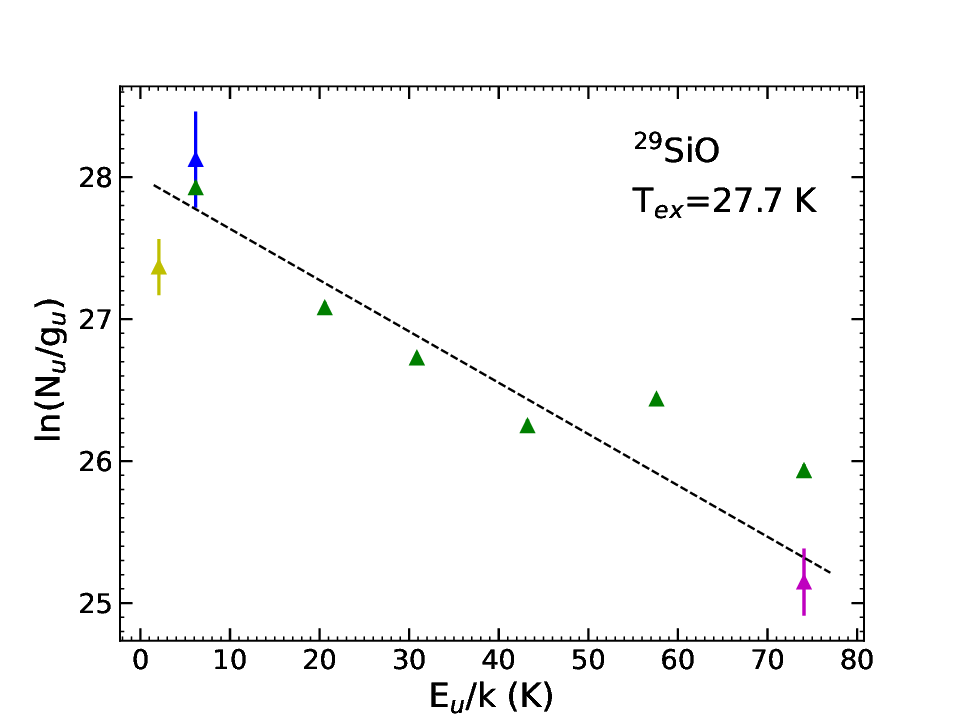}{0.45\textwidth}{ }
          \fig{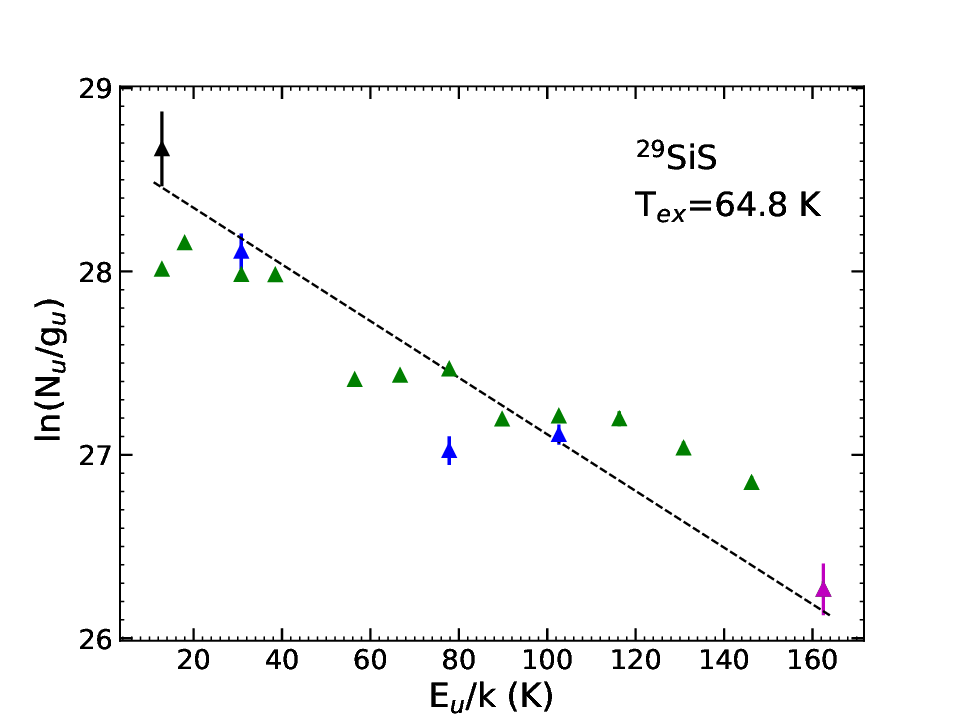}{0.45\textwidth}{ }
          }
\caption{Same as Figure~\ref{fig:T1}.
\label{fig:T3}}
\end{figure*}

\begin{figure*}
\gridline{\fig{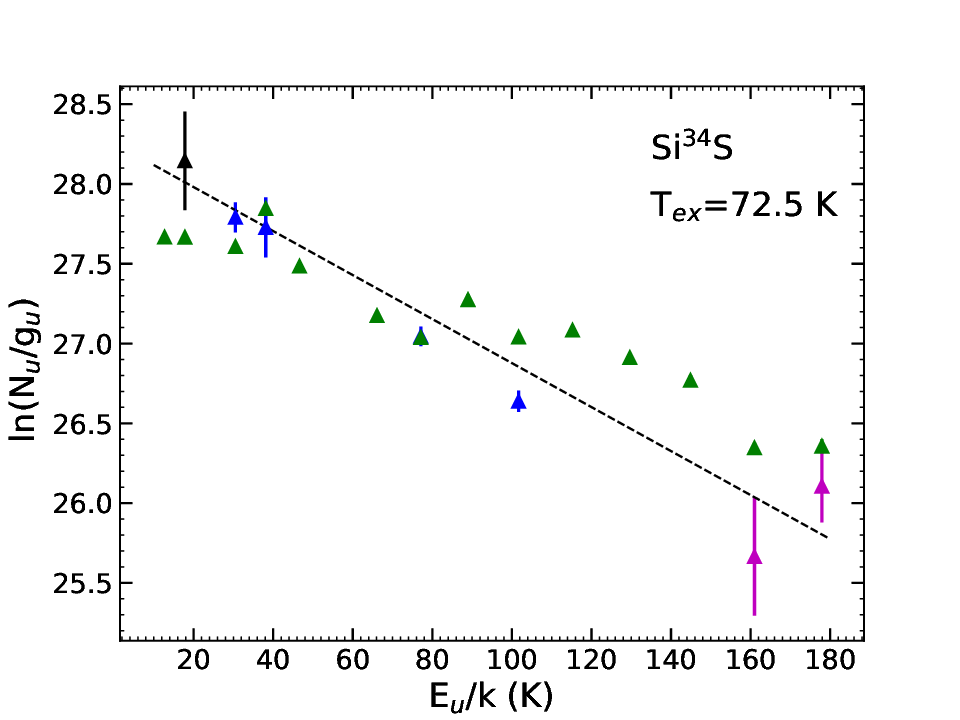}{0.45\textwidth}{ }
          \fig{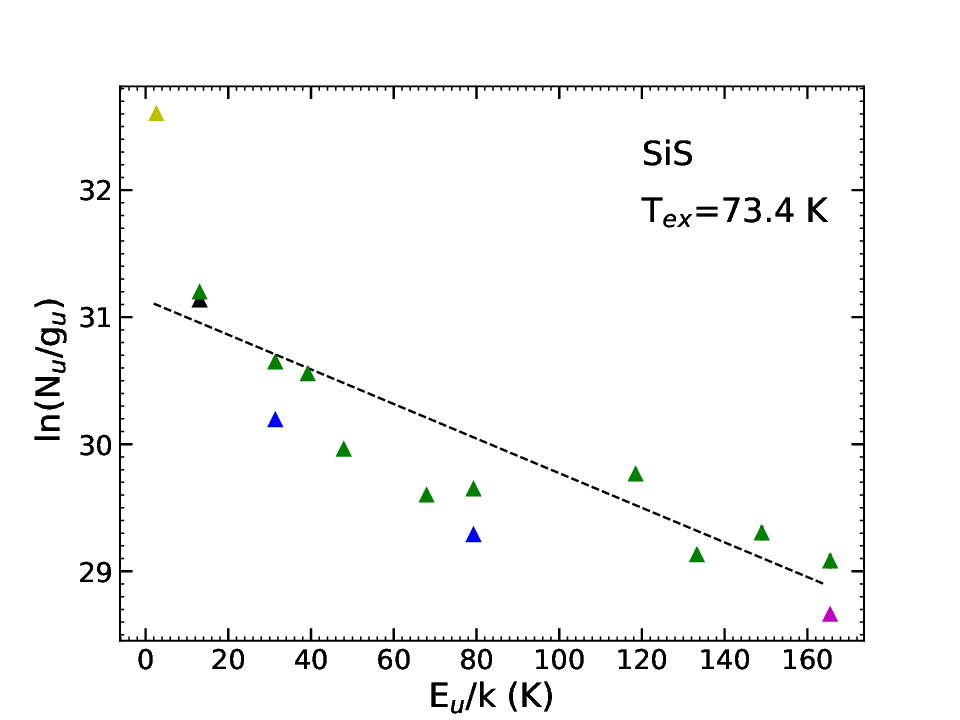}{0.45\textwidth}{ }
          }
\gridline{\fig{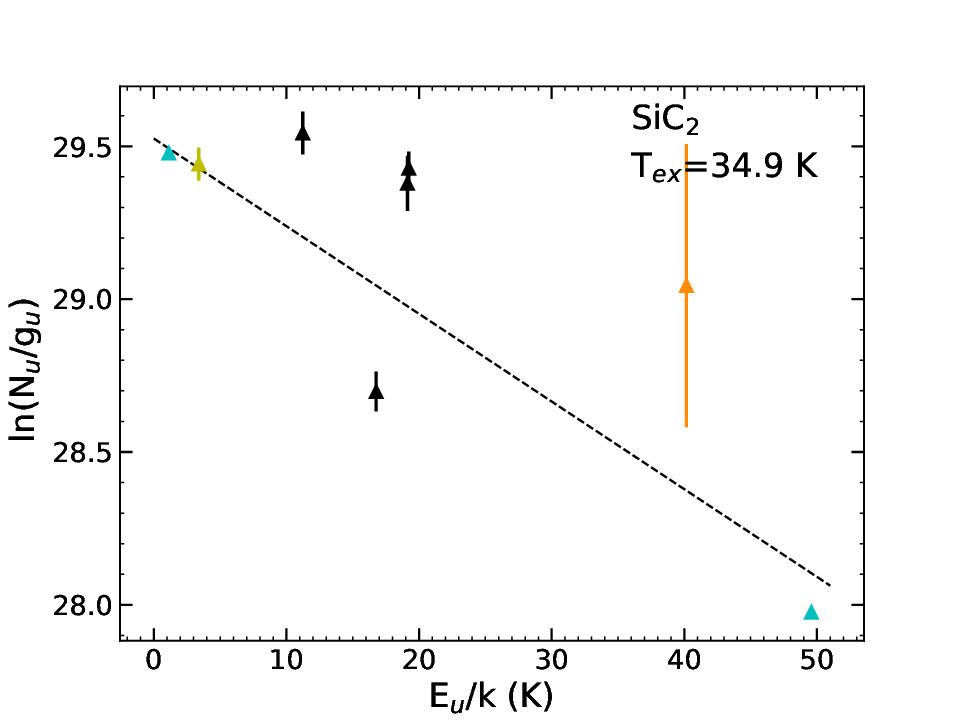}{0.45\textwidth}{ }
          \fig{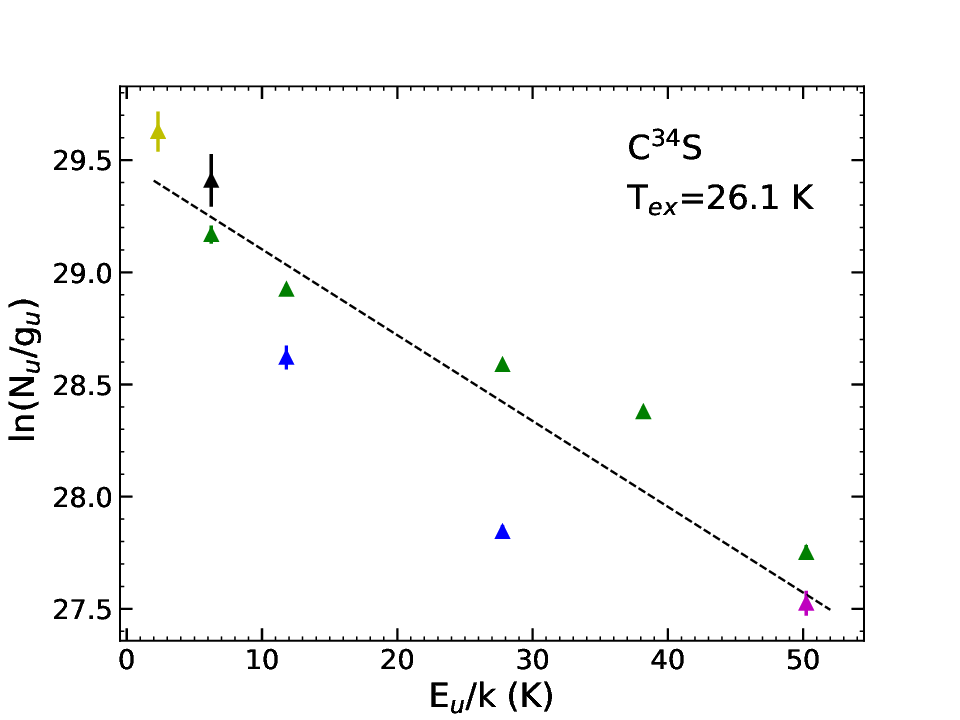}{0.45\textwidth}{ }
          }
\gridline{\fig{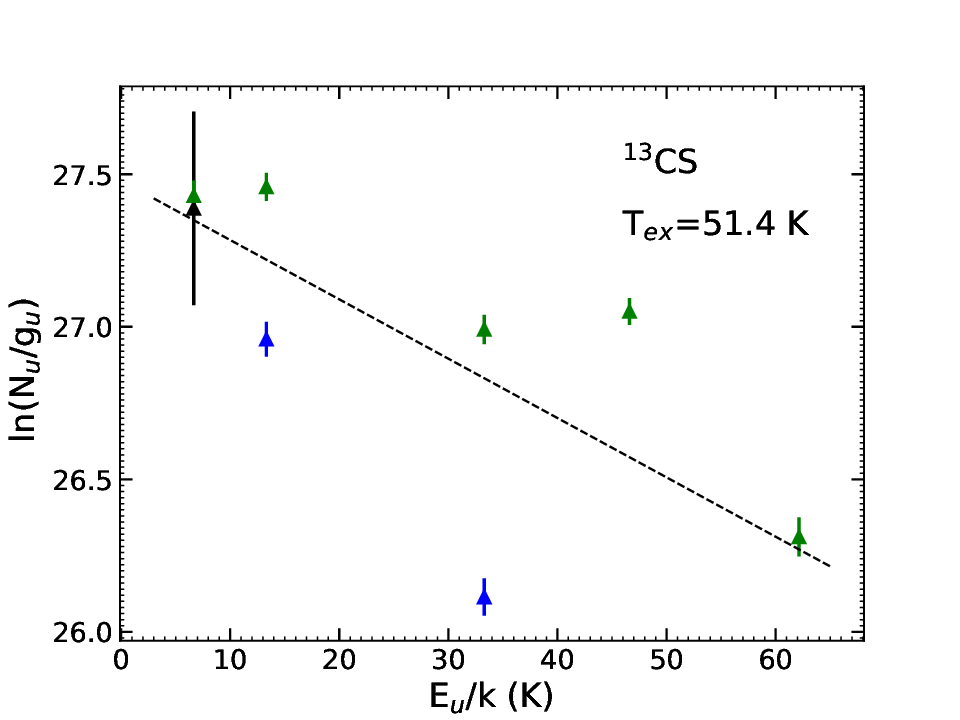}{0.45\textwidth}{ }
          \fig{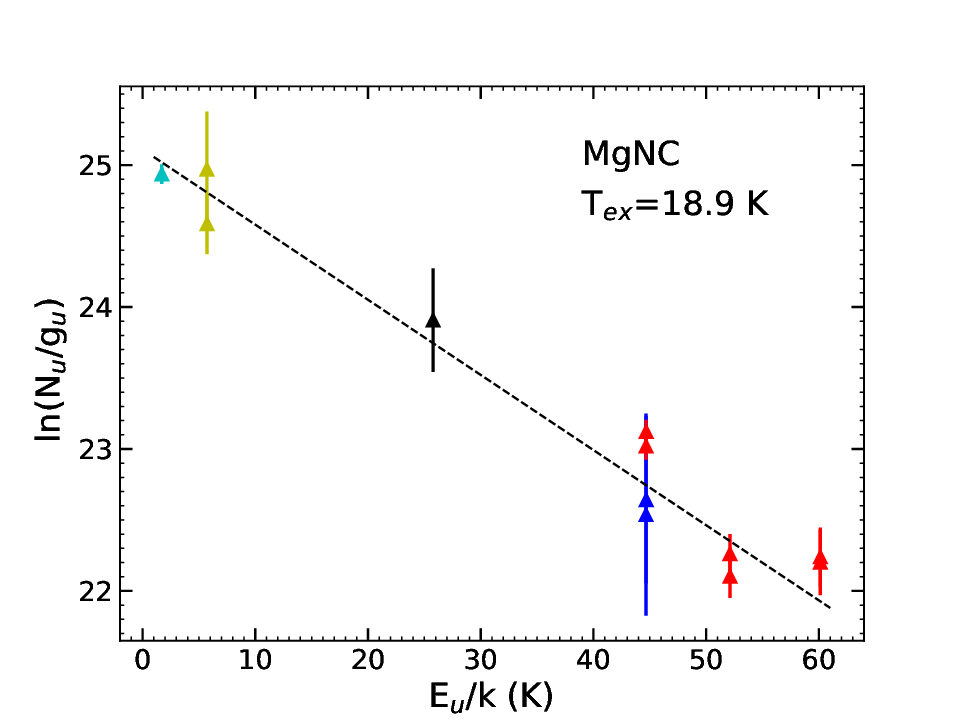}{0.45\textwidth}{ }
          }
\caption{Same as Figure~\ref{fig:T1}.
\label{fig:T4}}
\end{figure*}

\begin{figure*}

\gridline{\fig{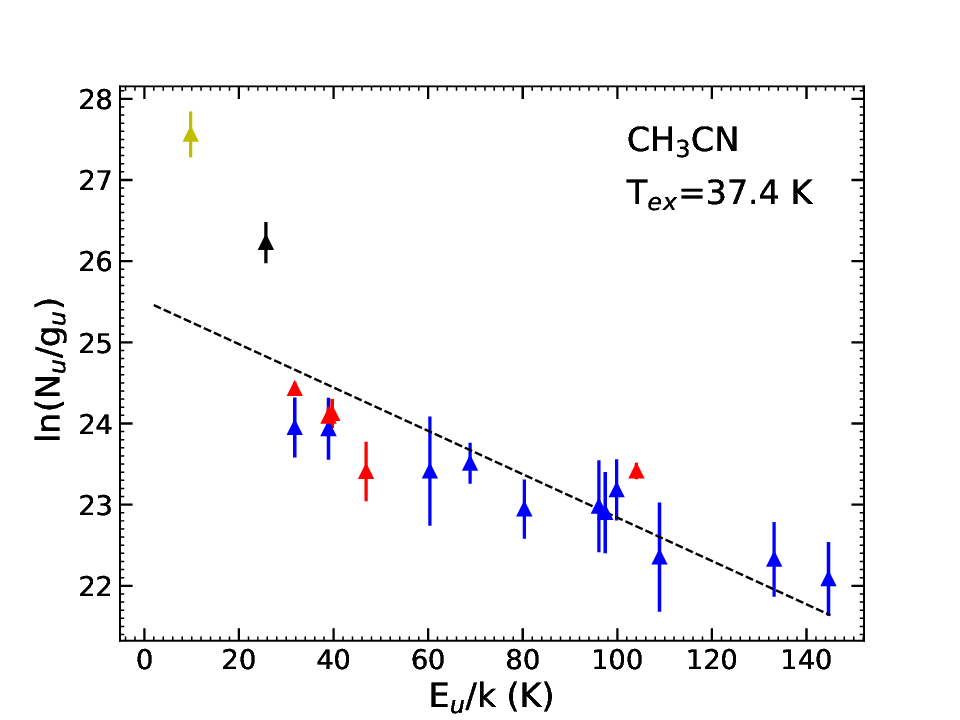}{0.45\textwidth}{ }
          \fig{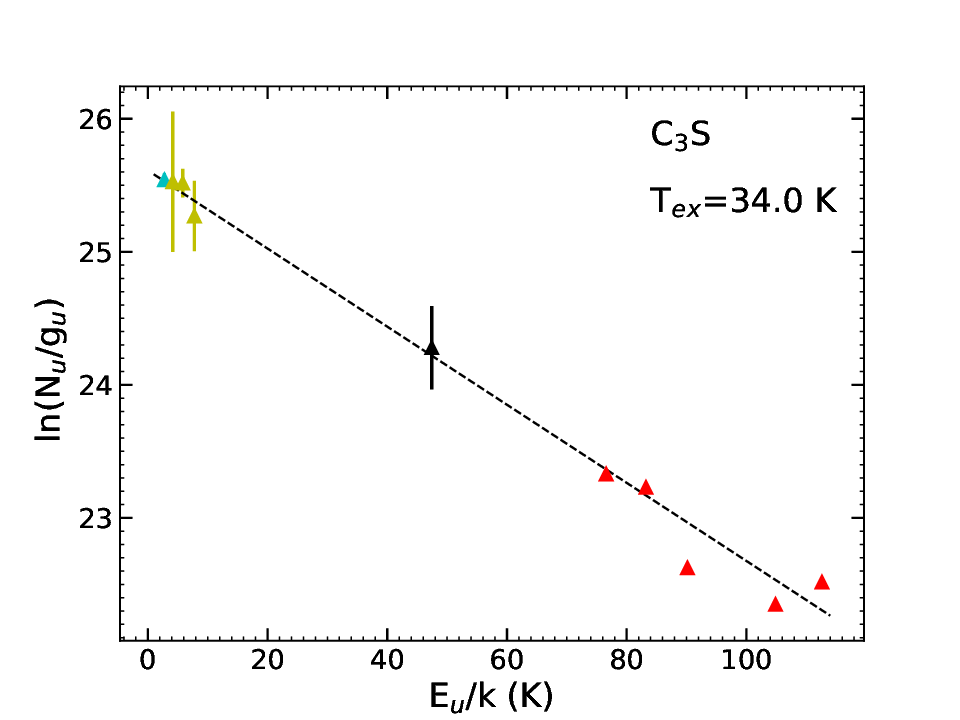}{0.45\textwidth}{ }
          }
\gridline{\fig{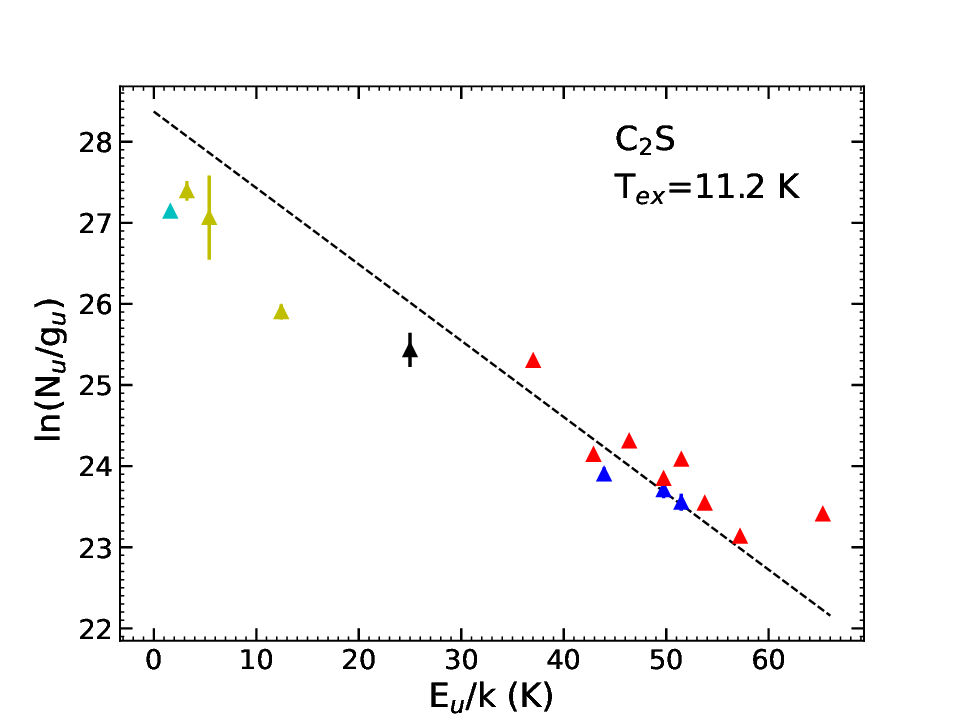}{0.45\textwidth}{ }
          \fig{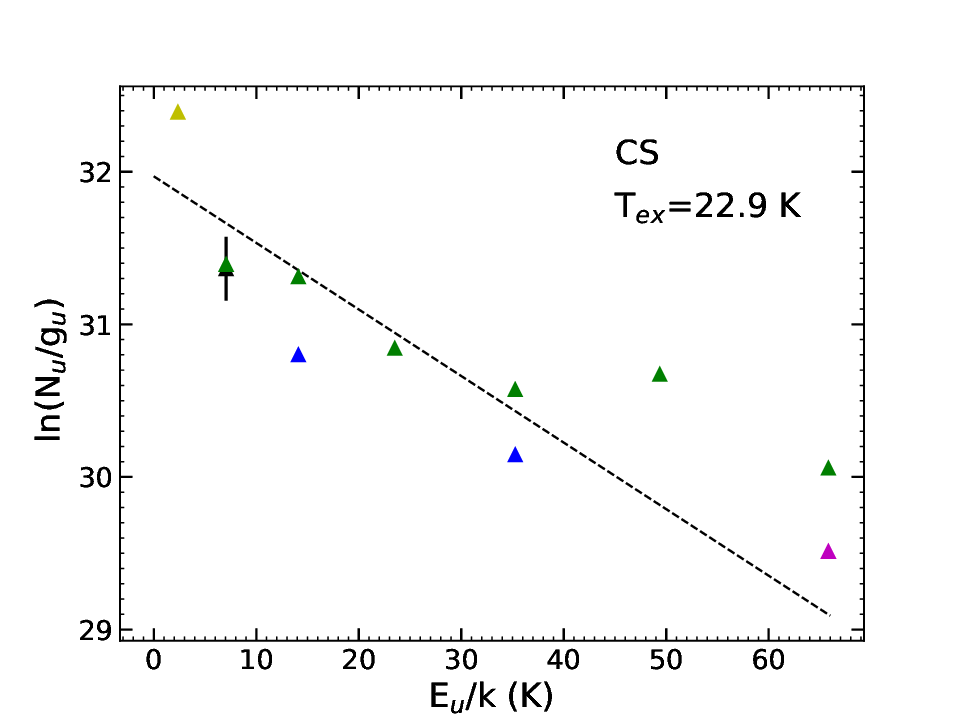}{0.45\textwidth}{ }
          }
\caption{Same as Figure~\ref{fig:T1}.
\label{fig:T5}}
\end{figure*}

%%%%%%%%%%%%%%%%%%%%%%%%

%%%%%%%%%%%%%%%%%%%%%%%%%%%%%%%%%%%%%%%%%%%%%%%
%              Appendix  F
%%%%%%%%%%%%%%%%%%%%%%%%%%%%%%%%%%%%%%%%%%%%%%%

\section{Multi-components of the CSE species \label{Appendix Comparison of rotational diagrams}}
\renewcommand\thefigure{\Alph{section}\arabic{figure}}
\setcounter{figure}{0}
\begin{figure*}
\gridline{\fig{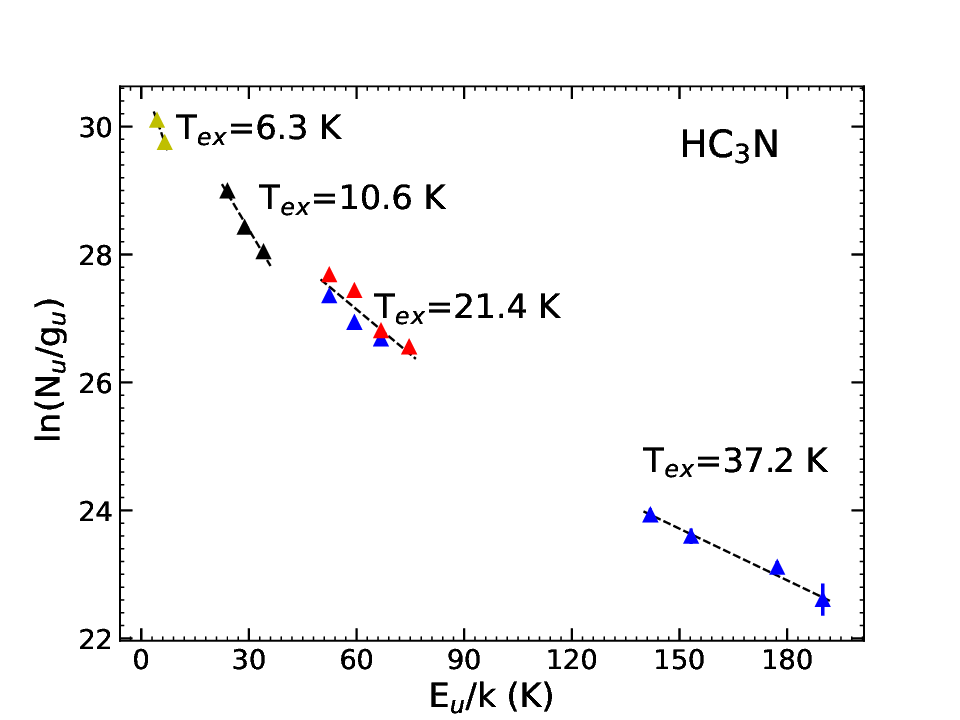}{0.45\textwidth}{ }
          \fig{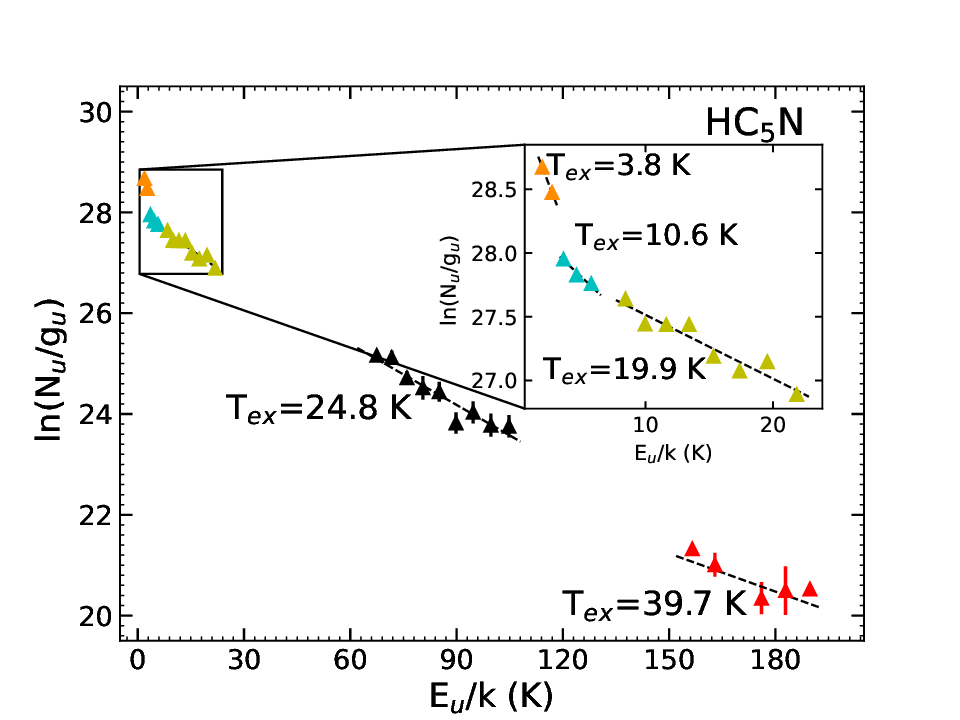}{0.45\textwidth}{ }
          }
\gridline{\fig{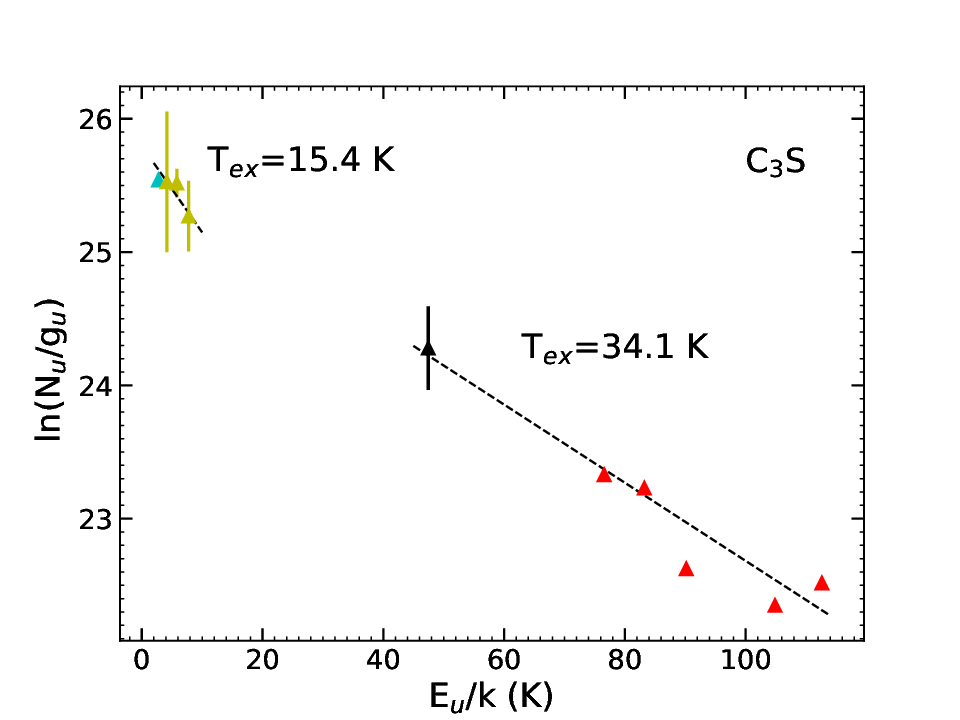}{0.45\textwidth}{ }
          \fig{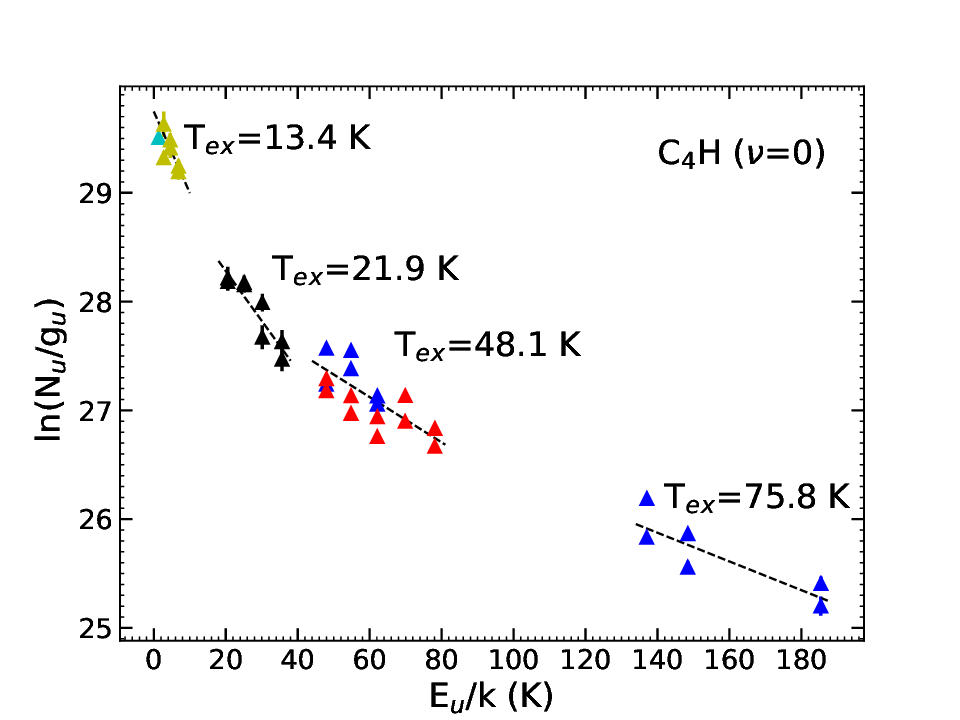}{0.45\textwidth}{ }
          }
\gridline{\fig{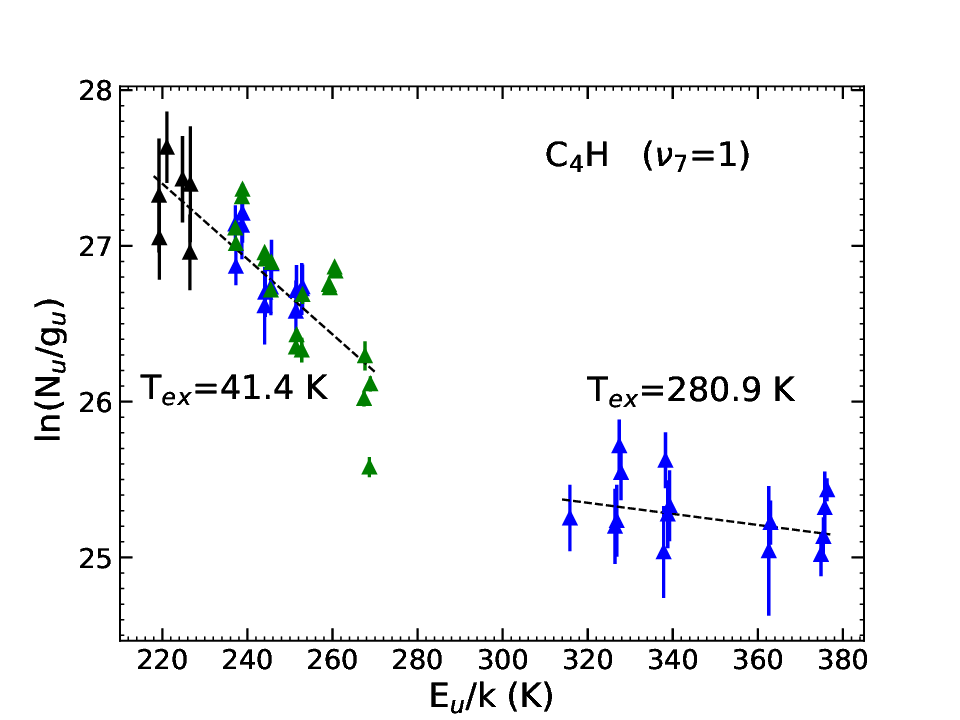}{0.45\textwidth}{ }
          \fig{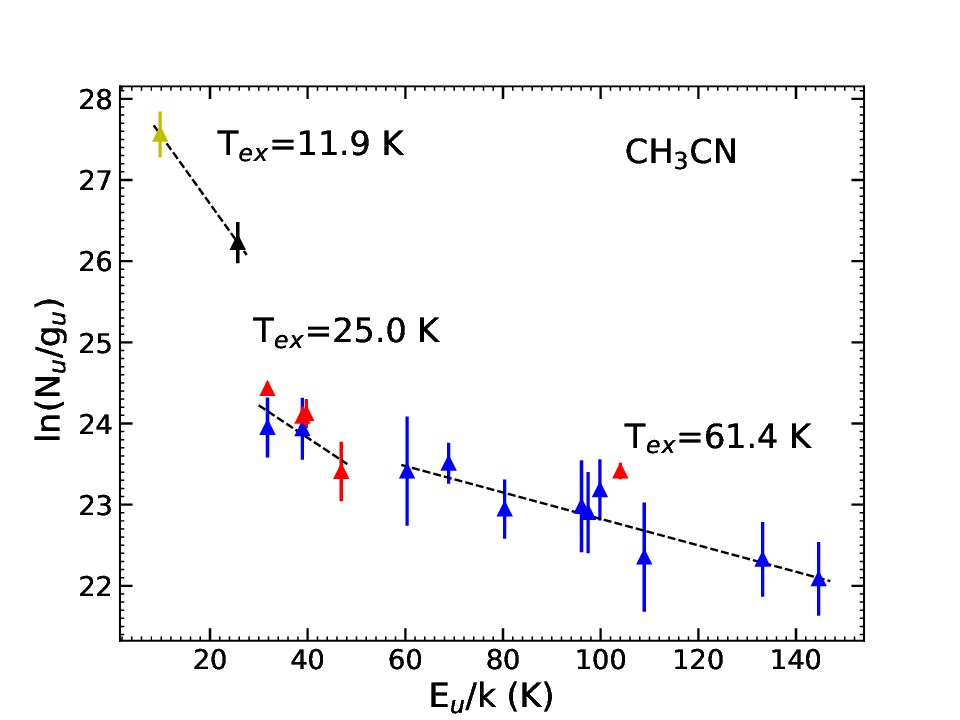}{0.45\textwidth}{ }
}

\caption{Multi-components revealed in the rotational diagrams of several species: HC$_3$N, HC$_5$N, C$_3$S, C$_4$H\, (${\nu}$  = 0), C$_4$H\, (${\nu}$  = 1)   and CH$_3$CN.
%The black dashed line is a linear least square fit for different upper-state energy levels and is shown in each box.
%The black triangles with error bars are from our PMO 13.7 m observations.
%The blue triangles with error bars are obtained from \protect\cite{He+etal+2008}.
%The cyan triangles with error bars are obtained from \protect\cite{Gong+etal+2015}.
%The dark orange triangles with error bars are obtained from \protect\cite{Zhang+etal+2017}.
%The yellow-green triangles with error bars are obtained from \protect\cite{Kawaguchi+etal+1995}.
%The red triangles with error bars are obtained from \protect\cite{Cernicharo+etal+2000}.
%The green triangles with error bars are obtained from \protect\cite{Agundez+etal+2012}.
%The magenta triangles with error bars are obtained from \protect\cite{Groesbeck+etal+1994}.
%The error bars of some data are smaller than the data points, so the error bars are invisible.
The results obtained from the linear least-squares fitting are plotted in black dashed lines.
The adopted data points with error bars are from our PMO--13.7\,m observations (black triangles),
 \protect\cite{He+etal+2008} (blue triangles),
 \protect\cite{Gong+etal+2015} (cyan triangles),
 \protect\cite{Zhang+etal+2017} (dark-orange triangles),
 \protect\cite{Kawaguchi+etal+1995} (yellow-green triangles),
 \protect\cite{Cernicharo+etal+2000} (red triangles),
\protect\cite{Agundez+etal+2012} (green triangles), and
\protect\cite{Groesbeck+etal+1994} (magenta triangles), respectively.
Note that the error bars are invisible if their values are smaller than the adopted data points.
\label{fig:T6}}
\end{figure*}

\begin{figure*}
\gridline{\fig{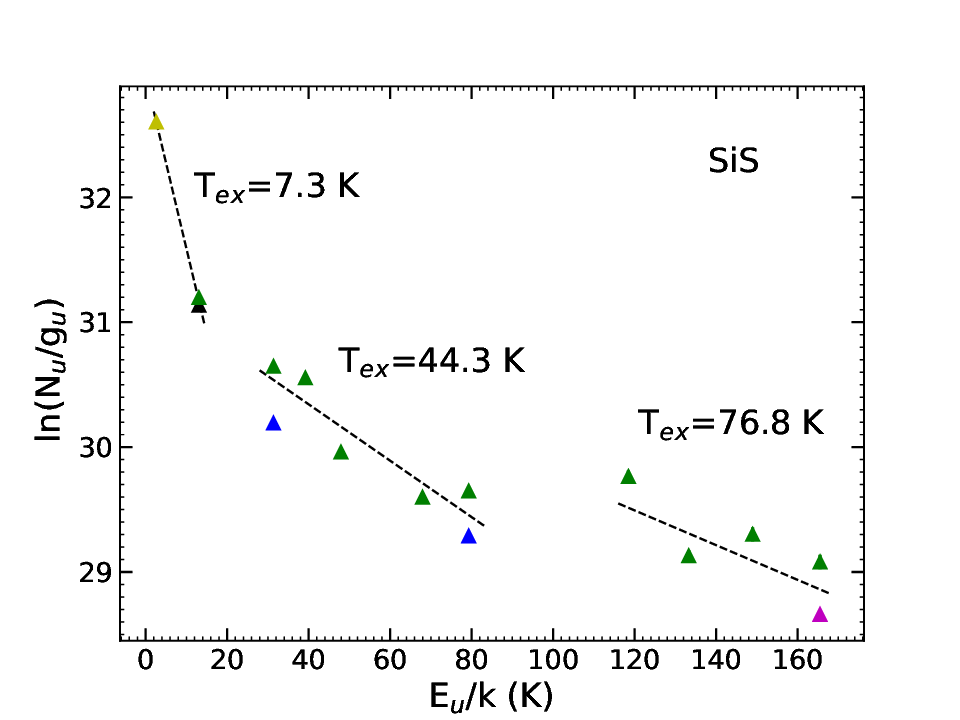}{0.45\textwidth}{ }
}
\caption{Same as Figure~\ref{fig:T6} but for SiS.
\label{fig:T7}}
\end{figure*}

%%%%%%%%%%%%%%%%%%%%%%%%%%
Both HC$_3$N and HC$_5$N in the CSE of IRC\,+10216 have two temperature components, tracing a warm region probed by high energy rotational lines and a cooler region probed by low energy lines. \cite{Bell+etal+1992} find that HC$_3$N has a cold component with $T_{\rm ex}=12.7\,{\pm}\,1$\,K, $N =(7.9\,{\pm}\,0.2) \times 10^{14}$\,cm$^{-2}$, and a warm component with $T_{\rm ex}=48\,{\pm}\,6$\,K, and $N=(2.1\,{\pm}\,1.4) \times 10^{14}$\,cm$^{-2}$. Meanwhile, they find that HC$_5$N has a cold component with $T_{\rm ex}{\sim}$13\,K and $N=3.2 \times 10^{14}$\,cm$^{-2}$ and a warm component with $T_{\rm ex}{\sim}$25\,K and $N=3.7 \times 10^{14}$\,cm$^{-2}$. \cite{Zhang+etal+2017} also compared the excitation temperatures of HC$_5$N, HC$_7$N, and HC$_{9}$N at different frequency ranges and find that the excitation temperature  obtained by fitting the high rotation transition of these molecules was correspondingly higher, indicating that the high rotational transitions of these molecules in the CSE of IRC\,+10216 trace the warm inner molecular region. We also compare the excitation temperatures of HC$_3$N and HC$_5$N at different upper-state energy levels.  Our results show that the excitation temperatures gradually increase with higher excited rotational transitions.

In Figures~\ref{fig:T6} and ~\ref{fig:T7}, we compare the excitation temperatures obtained from different upper-state energy levels. For HC$_3$N, the ratios of these are estimated to be scaled as: ${\rm HC_{3}N}\,(E_{\rm u}/k<10\,{\rm K}):{\rm HC_{3}N}\,( 20\,{\rm K}<E_{\rm u}/k<35\,{\rm K})  :{\rm HC_{3}N}\, (50\,{\rm K}<E_{\rm u}/k<80\,{\rm K}):{\rm HC_{3}N}\, (E_{\rm u}/k>140\,{\rm K}) = 1:1.7:3.4:5.9$. The excitation temperatures of HC$_5$N are estimated to be scaled as: ${\rm HC_{5}N}\, (E_{\rm u}/k<3\,{\rm K})  :{\rm HC_{5}N}\,(3\,{\rm K}<E_{\rm u}/k<6\,{\rm K}):{\rm HC_{5}N}\,(8\,{\rm K}<E_{\rm u}/k<25\,{\rm K}):{\rm HC_{5}N}\, (60\,{\rm K}<E_{\rm u}/k<105\,{\rm K}):{\rm HC_{5}N}\, (E_{\rm u}/k>150\,{\rm K}) = 1:2.8:5.2:6.5:10.4$. This indicates that the high-J transitions of the HC$_3$N molecules in the CSE of IRC\,+10216 also trace the warmer molecular regions.

\cite{Bell+etal+1993} find that C$_3$S also has at least two excitation temperature components, of which the excitation temperature of the low energy part is about 17\,K and that of the high energy part is about 47\,K. We also compare the excitation temperature of C$_3$S at different upper-state energies. When $E_{\rm u}/k$ is less than 10\,K, the excitation temperature is 15.4\,K, and when $E_{\rm u}/k$ is greater than 40\,K, the excitation temperature is 34.1\,K.

Our analysis using the rotational diagram method shows that C$_4$H\, (${\nu}=0$), C$_4$H\, (${\nu}=1$), CH$_3$CN and SiS also have different excitation temperature components. Therefore, we also compared the excitation temperatures of these molecules at different upper-state energies. For CH$_3$CN, the scaled ratios are estimated to be ${\rm CH_{3}CN}\,(E_{\rm u}/k<30\,{\rm K}) :{\rm CH_{3}CN}\,(30\,{\rm K}<E_{\rm u}/k<50\,{\rm K}) :{\rm CH_{3}CN}\,(E_{\rm u}/k>50\,{\rm K})=1:1.4:3.5$. For SiS, these are estimated to be ${\rm SiS}\,(E_{\rm u}/k<15\,{\rm K}): {\rm SiS}\,(20\,{\rm K}<E_{\rm u}/k<80\,{\rm K}): {\rm SiS}\, (E_{\rm u}/k>115\,{\rm K})=1:6.1:10.5$. For C$_4$H  (${\nu}=0$), these are estimated to be ${\rm C_{4}H}\,(E_{\rm u}/k<10\,{\rm K}) :{\rm C_{4}H}\,(20\,{\rm K}<E_{\rm u}/k<40\,{\rm K}) :{\rm C_{4}H}\, (40\,{\rm K}<E_{\rm u}/k<80\,\rm K) :{\rm C_{4}H}\,(E_{\rm u}/k>130\,{\rm K}) = 1:1.6:3.6:5.7$. The excitation temperatures of C$_4$H\,(${\nu}= 1$) are estimated to be ${\rm C_{4}H}\,(E_{\rm u}/k<260\,{\rm K}) :{\rm C_{4}H}\,(E_{\rm u}/k>315\,{\rm K})=1:6.8$.

Table~\ref{table:7} gives information about the excitation temperature, column density in specific levels, and fractional abundances of the molecules, relative to H$_{2}$, for C$_4$H\,(${\nu}$ = 0), C$_4$H\, (${\nu}$ = 1), HC$_3$N, HC$_5$N, CH$_3$CN, C$_3$S, and SiS, at different upper energy levels. Figure~\ref{fig:T6} and Figure~\ref{fig:T7} show that the thermal structures of HC$_3$N, HC$_5$N, C$_3$S, CH$_3$CN, SiS, C$_4$H\,(${\nu}$ = 0) and  C$_4$H\,(${\nu}$ = 1) are stratified \citep{Chau+etal+2012}. Similarly to HC$_3$N and HC$_5$N, their excitation temperatures increase with the increase of rotational energy level. Therefore, like HC$_5$N, HC$_7$N, and HC$_{9}$N, the high-J transitions of these molecules in the CSE of IRC\,+10216 can trace the warmer molecular regions. The column density and abundance derived from different transitions of the same molecule are different, which may be due to opacity effects \citep{Zhang+etal+2017}.

The low rotational lines observed in the CSE of AGB stars are seen to trace the low temperature region, while the information on warmer regions is determined by the higher energy lines. Some molecules with stratified excitation temperatures have already been mapped with high resolution, allowing us to determine the extended ranges of the cold and warm components based on the results. These findings have been summarized in Table~\ref{table:8}.

Thanks to previously obtained maps of C$_4$H (${\nu}$=0) by high resolution telescopes or interferometers, we  have discovered that the spatial distribution of low-rotational transitions ($N=9-8$ through $N=12-11$) form a hollow shell structure with a width of about 3$^{\prime\prime}$ and a radius of about 15$^{\prime\prime}$ \citep{Agundez+etal+2017}. In contrast, high rotational transitions ($N=36-35$) show a compact spatial distribution (radius $\leq$ 2.7 $^{\prime\prime}$) \citep{Patel+etal+2011}. The spatial distribution between low- and the high-rotational transitions ($N=27-26$) has both characteristics, that is, there is a compact distribution in the center and a bright shell structure outside \citep{Cooksy+etal+2015}.

The molecule at issue is CH$_3$CN. \cite{Agundez+etal+2015} observed the spatial distribution of CH$_3$CN ($J=14-13$) using the ALMA telescope, which showed a hollow shell with a radius of about 2$^{\prime\prime}$ and a width of about 1$^{\prime\prime}$. However, the CH$_3$CN observed by \cite{Patel+etal+2011} using SMA ($J=16-15$ to $J=19-18$) indicated a compact spatial distribution with a radius of about 3$^{\prime\prime}$. \cite{Agundez+etal+2015} points out that the higher rotational transition lines observed with the SMA may cause the emission to shift inward, and that the compact distribution in the center may be due to the insufficient resolution of the SMA. There are chemical models that predict that most CH$_3$CN should be  distributed at a large radius \cite[about 18$^{\prime\prime}$, see ][]{Agundez+etal+2008, Li+etal+2014}, and other chemical models predict that most CH$_3$CN is located within a radius of about 15$^{\prime\prime}$, but with a compact distribution of about 1$^{\prime\prime}$ in the center \citep{Agundez+etal+2015}, inconsistent with the ALMA observations. We speculate that the low-J transitions of CH$_3$CN are likely distributed over a large radius, and the emission may move gradually inwards as the rotational energy level increases, similar to C$_4$H (${\nu}$=0).

%%%%%%%%%%%%%%%%%%%%%%
\renewcommand\thetable{\Alph{section}\arabic{table}}
\setcounter{table}{0}
\begin{deluxetable*}{ccccccccccccccc}
%\tablenum{7}
%\tablecaption{Multi-components of molecular C$_4$H, HC$_3$N, HC$_5$N, CH$_3$CN, C$_3$S, and SiS, in addition to their excitation temperatures, column densities, and beam-averaged fractional abundances.
\tablecaption{Multi-components  fits for C$_4$H, HC$_3$N, HC$_5$N, CH$_3$CN, C$_3$S, and SiS.     \label{table:7}}
%%%%%%%%%%%%% Note:   column density and fractional abundance obtained here are in specific levels.
%%%
\tablewidth{0pt}
\tablehead{
\colhead{Species}       & \colhead{$E_{\rm u}/k$}    & \colhead{$T_{\rm ex}$}   & \colhead{$N$}            & \colhead{$f (N/N_{\rm H_{2}}$)  }\\
\colhead{ }             & \colhead{ (K)  }              & \colhead{  (K)   }          & \colhead{ (cm$^{-2}$)  }    & \colhead{  }
}
\startdata
HC$_3$N	                &  $E_{\rm u}/k<10$            & 6.3 (0.0)   	         & 6.88 (0.01)   $\times$ $10^{14}$	&	2.67 $\times$ $10^{-7}$  \\
		                &  $20<E_{\rm u}/k<35$         & 10.6 (1.4)   	         & 1.78 (0.70)   $\times$ $10^{15}$	&	6.92 $\times$ $10^{-7}$  \\
		                &  $50<E_{\rm u}/k<80$         & 21.4 (4.7)   	         & 1.00 (0.67)   $\times$ $10^{15}$	&	3.89 $\times$ $10^{-7}$  \\
		                &  $E_{\rm u}/k>140$           & 37.1 (3.0)   	         & 1.93 (0.74)   $\times$ $10^{14}$	&	7.51 $\times$ $10^{-8}$  \\
HC$_5$N	                &  $E_{\rm u}/k<3$             & 3.8 (0.0)  	         & 2.81 (0.00)   $\times$ $10^{14}$	&	1.09 $\times$ $10^{-7}$  \\
		                &  $3<E_{\rm u}/k<6$           & 10.6 (2.1)  	         & 3.19 (0.69)   $\times$ $10^{14}$	&	1.24 $\times$ $10^{-7}$  \\
		                &  $8<E_{\rm u}/k<25$          & 19.9 (2.8)  	         & 4.59 (0.82)   $\times$ $10^{14}$	&	1.78 $\times$ $10^{-7}$  \\
		                &  $60<E_{\rm u}/k<105$        & 24.8 (3.3)  	         & 4.65 (2.28)   $\times$ $10^{14}$	&	1.81 $\times$ $10^{-7}$  \\
		                &  $E_{\rm u}/k>150$           & 39.7 (17.4)  	         & 4.53 (9.00)   $\times$ $10^{13}$	&	1.76 $\times$ $10^{-8}$  \\
C$_3$S		            &  $E_{\rm u}/k<10$            & 15.4 (4.0)   	         & 1.79 (0.49)   $\times$ $10^{13}$	&	9.25 $\times$ $10^{-9}$  \\
		                &  $E_{\rm u}/k>40$ 	       & 34.1 (2.4)             & 3.29 (0.46)   $\times$ $10^{13}$	&	1.70 $\times$ $10^{-8}$  \\
C$_4$H (${\nu}$  = 0)  	    &  $E_{\rm u}/k<10$            & 13.4 (5.2)   	         & 1.96 (0.81)   $\times$ $10^{15}$	&	7.60 $\times$ $10^{-7}$  \\
		                &  $20<E_{\rm u}/k<40$         & 21.9 (3.7)   	         & 1.84 (0.51)   $\times$ $10^{15}$	&	7.14 $\times$ $10^{-7}$  \\
		                &  $40<E_{\rm u}/k<80$         & 48.1 (12.3)   	     & 1.77 (0.71)   $\times$ $10^{15}$	&	6.87 $\times$ $10^{-7}$  \\
		                &  $E_{\rm u}/k>130$           & 75.8 (20.1)   	     & 1.46 (0.92)   $\times$ $10^{15}$	&	5.67 $\times$ $10^{-7}$  \\
C$_4$H (${\nu}$  = 1)      	&  $E_{\rm u}/k<260$           & 41.4 (4.2)  	         & 1.67 (1.00)   $\times$ $10^{17}$	&	6.49 $\times$ $10^{-5}$  \\
		                &  $E_{\rm u}/k>315$           & 280.9 (209.1)          & 3.59 (4.68)   $\times$ $10^{15}$	&	1.39 $\times$ $10^{-6}$  \\
CH$_3$CN	            &  $E_{\rm u}/k<30$            & 11.9 (0.0)   	         & 2.07 (0.27)   $\times$ $10^{14}$	&	8.03 $\times$ $10^{-8}$  \\
		                &  $30<E_{\rm u}/k<50$         & 25.0 (10.4)            & 3.35 (3.10)   $\times$ $10^{13}$	&	1.30 $\times$ $10^{-8}$  \\
		                &  $E_{\rm u}/k>50$            & 61.4 (10.9)  	         & 5.06 (2.30)   $\times$ $10^{13}$	&	1.97 $\times$ $10^{-8}$  \\
SiS		                &  $E_{\rm u}/k<15$            & 7.3 (0.3)  	         & 3.17 (0.32)   $\times$ $10^{15}$	&	9.02 $\times$ $10^{-7}$  \\
		                &  $20<E_{\rm u}/k<80$         & 44.3 (8.5)  	         & 3.84 (1.28)   $\times$ $10^{15}$	&	1.09 $\times$ $10^{-6}$  \\
		                &  $E_{\rm u}/k>115$           & 76.8 (33.3)            & 6.32 (5.96)   $\times$ $10^{15}$	&	1.80 $\times$ $10^{-6}$  \\
\enddata
\end{deluxetable*}
%%%%%%%%%%%%%%%%%%%%%%%%

%%%%%%%%%%%%%%%%%%%%%%
\renewcommand\thetable{\Alph{section}\arabic{table}}
\setcounter{table}{1}
\begin{deluxetable*}{ccccccccccccccc}
%\tablenum{8}
\tablecaption{The cold and warm components of the extended radius of molecular C$_4$H, HC$_3$N, HC$_5$N, CH$_3$CN, and SiS.            \label{table:8}}
\tablewidth{0pt}
\tablehead{
\colhead{Species}         & \colhead{Radial extent (r)}    & \colhead{Telescope}   & \colhead{Comments}\\
\colhead{ }               & \colhead{ ($^{\prime\prime}$)  }     & \colhead{   }         & \colhead{ }      }
\startdata
HC$_3$N	(warm)            &  r $\leq$ 5 $^{(1)}$                & ALMA   	            & $J=28-27$, $J=30-29$, and $J=38-37$ \\
HC$_3$N	(cool)            &  7 $\leq$ r $\leq$ 18.5 $^{(2)}$    & ALMA   	            & $J=10-9$ to $J=12-11$  \\
HC$_5$N	(warm)	          &  13 $\leq$ r $\leq$ 18 $^{(2)}$     & ALMA  	            & $J=32-31$ to $J=43-42$  \\
C$_4$H (${\nu}$=0) (warm) &  r $\leq$ 2.7 $^{(3)}$   	        & SMA    	            & $N=36-35$  \\
C$_4$H (${\nu}$=0) (cool) &  12 $\leq$ r $\leq$ 20 $^{(2)}$   	& ALMA    	            & $N=9-8$ to $N=12-11$  \\
C$_4$H (${\nu}$=1) (cool) &  12 $\leq$ r $\leq$ 18 $^{(4,5)}$  	& IRAM, SMA	            & $N=27-26$  \\
CH$_3$CN (warm)	          &  r $\leq$ 3.2 $^{(3)}$    	        & SMA	                & $J=16-15$ to $J=19-18$  \\
SiS	(warm)	              &  r $\leq$ 2 $^{(6)}$   	            & CARMA	                & $J=14-13 $  \\
SiS	(cool)	              &  r $\leq$ 10 $^{(7, 8, 9)}$   	    & ALMA, BIMA, IRAM	    & $J=5-4$, $J=6-5$, $J=8-7$, $J=9-8$, and $J=12-11$\\
\enddata
\tablecomments{The radial extent was estimated to be the value at which the brightness distribution had fallen to half its peak value. References : (1) \cite{Siebert+etal+2022}; (2) \cite{Agundez+etal+2017}; (3) \cite{Patel+etal+2011}; (4) \cite{Guelin+etal+1999}; (5) \cite{Cooksy+etal+2015}; (6) \cite{Fonfria+etal+2014}; (7) \cite{Velilla-Prieto+etal+2019}; (8) \cite{Bieging+etal+1993}; (9) \cite{Lucas+etal+1995}. }
\end{deluxetable*}
%%%%%%%%%%%%%%%%%%%%%%%%

%%%%%%%%%%%%%%%%%%%%%%%%%%%%%%%%%%%%%%%%%%%%%%%
%              Appendix  G
%%%%%%%%%%%%%%%%%%%%%%%%%%%%%%%%%%%%%%%%%%%%%%%
\section{Detailed comparison of emission lines obtained with different angular resolutions} \label{comparisons of line shapes}
Here, we provide a comparison of the line intensities and line shapes of the same transition of a molecule obtained by telescopes with different angular resolutions, including IRAM--30\,m, Onsala--20\,m, FCRAO--14\,m, TRAO--14\,m, PMO--13.7\,m,  NRAO--11\,m,  the Number Two 10.4\,m telescope of the Owens Valley Radio Observatory, and the 7\,m Bell Laboratories Antenna at Crawford Hill.

(1) The impact of different angular resolutions on the integrated line intensities.
Comparison of emission lines observed with different angular resolutions is more uncertain if any of the observed lines are weak and we suggest that such comparisons be made only if the S/N of the lines are greater than or equal to 5.
For the strong lines (S/N $\geq$ 5), the integrated intensity of the same molecular line from different angular resolutions is approximately identical, i.e., $\int T_{\rm R}dv \approx Constant$, where $T_{\rm R}=\frac{T_{\rm mb}}{{\eta}_{\rm BD}}$, ${\eta}_{\rm BD}=\frac{\theta_{\rm s}^{2}}{\theta_{s}^{2}+\theta_{\rm beam}^{2}}$ (see Section ~\ref{sec:Excitation temperature and column density}), $T_{\rm mb}$=$T_{\rm A}^{*}/{\eta}$ (see Section ~\ref{sec:Observations and data reduction}). In other words, the ratio of the integrated line intensity ($\int T_{\rm R}dv$) between the same molecular transition from different angular resolutions is close to unity.
For instance, \cite{Zhang+etal+2017} found that the ratios of the integrated intensities observed with the TMRT--65\,m for HC$_7$N  ($v=0, J=16-15$), SiS ($v=0, J=1-0$), and HC$_3$N ($v=0, J=2-1$) to those values derived by  \cite{Gong+etal+2015} using the Efflesberg--100\,m telescope were 1 : (0.863 $\pm$ 0.121), 1: (0.928 $\pm$ 0.029), and 1 : (0.899 $\pm$ 0.020), respectively.
We compare the integrated intensities of SiS, HNC, and HCN in IRC +10216 observed by the PMO--13.7\,m, TRAO--14\,m, and IRAM--30\,m telescopes, and find that the ratios among SiS ($J=5-4$), HNC ($J=1-0$), and HCN ($J=1-0$) are 1 : 0.82 : 0.72, 1 : 0.88 : 0.86, and 1 : 0.83 : 1.08, respectively. The ratio of the integrated intensities of $^{13}$CO deduced by the observations from PMO--13.7\,m, TRAO--14\,m, and 7\,m Bell Laboratories Antenna at Crawford Hill telescopes is 1:0.88:0.96.

(2) The impact of different angular resolutions on the line shapes.
For the strong emission lines (S/N $\geq$ 5), the observed spectral line shape is the same for the same transition of the same molecule with different angular resolutions.
If the S/N is less than 5, the observed line profiles might differ.
Take molecular C$_4$H as an example. Using the PMO--13.7\,m telescope, we observed double-peaked structures of C$_4$H lines in all the $N=9-8, 10-9, 11-10, 12-11$ transitions with S/N $\geq$ 5,  as shown in Figure~\ref{fig:C4H}.
 \cite{Guelin+etal+1978} observed similar transitions of C$_4$H via the NRAO--11\,m telescope with the S/N less than 5, and obtained different conclusions:
the shapes of the $N=9-8$ and $N= 10-9$ lines were difficult to judge due to the poor S/N;
the $N=11-10$ line showed a flat-topped structure;
the $N=12-11$ line exhibited a double-peaked structure.
The C$_4$H ($N=9-8$) observed by the Onsala--20\,m telescope \citep{Johansson+etal+1984} and the C$_4$H ($N=10-9$) observed by the IRAM--30\,m telescope \citep{Cernicharo+etal+1986} both possess double-peaked line shapes.

%However, unlike the C$_4$H $^2{\Pi }_{1/2}$ $J=21/2-19/2$, ${\nu}7=1f$ spectral line profile observed by \cite{Yamamoto+etal+1987} with the IRAM--30\,m telescope,
%they have a double-peaked structure, while we observe a flat-topped structure. This result may be caused by the S/N of our lines being less than 5.
(3) The impact of different angular resolutions on the double peaked spectra.
 For the optically thin species that exhibit double-peaked line shapes, a more considerable intensity contrast can be found between the line wings and the line center while using a telescope with higher spatial resolution \citep{Olofsson+etal+1982}.
 For instance, the C$_2$H $N=1-0$, $J=3/2-1/2$, $F=2-1$ was observed by the IRAM--30\,m \citep{Beck+etal+2012}, Onsala--20\,m \citep{Truong-Bach+etal+1987}, TRAO--14\,m  \citep{Park+etal+2008}, and PMO--13.7\,m (this work) telescopes towards IRC\,+10216.
 All of these spectra showed double-peaked shapes while S/N $\geq$ 5.
 The calculated intensity contrasts between the line wings and the line center are  2.3, 1.4, 1.3, and 1.8. These values for HNC ($J=1-0$) are 1.8, 1.7, 1.2, and 1.4, respectively.
Similarly, for $^{13}$CO observed by the IRAM--30\,m \citep{Kahane+etal+1992}, Onsala--20\,m \citep{Olofsson+etal+1982}, TRAO--14\,m \citep{Park+etal+2008},  PMO--13.7\,m (this work), NRAO--11\,m  \citep{Kuiper+etal+1976, Wannier+etal+1978}, and the 7\,m Bell Laboratories Antenna at Crawford Hill \citep{knapp+etal+1985} telescopes, these values are obtained to be 3.4, 2.6, 2.2, 2.4, 1.9, and 1.8, respectively.

\bibliography{IRC+10216}{} \label{references}

\begin{thebibliography}{}
\expandafter\ifx\csname natexlab\endcsname\relax\def\natexlab#1{#1}\fi
\providecommand{\url}[1]{\href{#1}{#1}}
\providecommand{\dodoi}[1]{doi:~\href{http://doi.org/#1}{\nolinkurl{#1}}}
\providecommand{\doeprint}[1]{\href{http://ascl.net/#1}{\nolinkurl{http://ascl.net/#1}}}
\providecommand{\doarXiv}[1]{\href{https://arxiv.org/abs/#1}{\nolinkurl{https://arxiv.org/abs/#1}}}

\bibitem[{{Ag{\'u}ndez} \& {Cernicharo}(2006)}]{Agundez+2006}
{Ag{\'u}ndez}, M., \& {Cernicharo}, J. 2006, \apj, 650, 374,
  \dodoi{10.1086/506313}

\bibitem[{{Ag{\'u}ndez} {et~al.}(2007){Ag{\'u}ndez}, {Cernicharo}, \&
  {Gu{\'e}lin}}]{Agundez+etal+2007}
{Ag{\'u}ndez}, M., {Cernicharo}, J., \& {Gu{\'e}lin}, M. 2007, \apjl, 662, L91,
  \dodoi{10.1086/519561}

\bibitem[{{Ag{\'u}ndez} {et~al.}(2014){Ag{\'u}ndez}, {Cernicharo}, \&
  {Gu{\'e}lin}}]{Agundez+etal+2014}
---. 2014, \aap, 570, A45, \dodoi{10.1051/0004-6361/201424542}

\bibitem[{{Ag{\'u}ndez} {et~al.}(2008{\natexlab{a}}){Ag{\'u}ndez},
  {Cernicharo}, {Pardo}, {Gu{\'e}lin}, \& {Phillips}}]{Agundez+etal+2008a}
{Ag{\'u}ndez}, M., {Cernicharo}, J., {Pardo}, J.~R., {Gu{\'e}lin}, M., \&
  {Phillips}, T.~G. 2008{\natexlab{a}}, \aap, 485, L33,
  \dodoi{10.1051/0004-6361:200810193}

\bibitem[{{Ag{\'u}ndez} {et~al.}(2015){Ag{\'u}ndez}, {Cernicharo},
  {Quintana-Lacaci}, {Velilla Prieto}, {Castro-Carrizo}, {Marcelino}, \&
  {Gu{\'e}lin}}]{Agundez+etal+2015}
{Ag{\'u}ndez}, M., {Cernicharo}, J., {Quintana-Lacaci}, G., {et~al.} 2015,
  \apj, 814, 143, \dodoi{10.1088/0004-637X/814/2/143}

\bibitem[{{Ag{\'u}ndez} {et~al.}(2011){Ag{\'u}ndez}, {Cernicharo}, {Waters},
  {Decin}, {Encrenaz}, {Neufeld}, {Teyssier}, \& {Daniel}}]{Agundez+etal+2011}
{Ag{\'u}ndez}, M., {Cernicharo}, J., {Waters}, L.~B.~F.~M., {et~al.} 2011,
  \aap, 533, L6, \dodoi{10.1051/0004-6361/201117578}

\bibitem[{{Ag{\'u}ndez} {et~al.}(2012){Ag{\'u}ndez}, {Fonfr{\'\i}a},
  {Cernicharo}, {Kahane}, {Daniel}, \& {Gu{\'e}lin}}]{Agundez+etal+2012}
{Ag{\'u}ndez}, M., {Fonfr{\'\i}a}, J.~P., {Cernicharo}, J., {et~al.} 2012,
  \aap, 543, A48, \dodoi{10.1051/0004-6361/201218963}

\bibitem[{{Ag{\'u}ndez} {et~al.}(2008{\natexlab{b}}){Ag{\'u}ndez},
  {Fonfr{\'\i}a}, {Cernicharo}, {Pardo}, \& {Gu{\'e}lin}}]{Agundez+etal+2008}
{Ag{\'u}ndez}, M., {Fonfr{\'\i}a}, J.~P., {Cernicharo}, J., {Pardo}, J.~R., \&
  {Gu{\'e}lin}, M. 2008{\natexlab{b}}, \aap, 479, 493,
  \dodoi{10.1051/0004-6361:20078956}

\bibitem[{{Ag{\'u}ndez} {et~al.}(2010){Ag{\'u}ndez}, {Cernicharo},
  {Gu{\'e}lin}, {Kahane}, {Roueff}, {K{\l}os}, {Aoiz}, {Lique}, {Marcelino},
  {Goicoechea}, {Gonz{\'a}lez Garc{\'\i}a}, {Gottlieb}, {McCarthy}, \&
  {Thaddeus}}]{Agundez+etal+2010}
{Ag{\'u}ndez}, M., {Cernicharo}, J., {Gu{\'e}lin}, M., {et~al.} 2010, \aap,
  517, L2, \dodoi{10.1051/0004-6361/201015186}

\bibitem[{{Ag{\'u}ndez} {et~al.}(2017){Ag{\'u}ndez}, {Cernicharo},
  {Quintana-Lacaci}, {Castro-Carrizo}, {Velilla Prieto}, {Marcelino},
  {Gu{\'e}lin}, {Joblin}, {Mart{\'\i}n-Gago}, {Gottlieb}, {Patel}, \&
  {McCarthy}}]{Agundez+etal+2017}
{Ag{\'u}ndez}, M., {Cernicharo}, J., {Quintana-Lacaci}, G., {et~al.} 2017,
  \aap, 601, A4, \dodoi{10.1051/0004-6361/201630274}

\bibitem[{{Anderson} \& {Ziurys}(2014)}]{Anderson+etal+2014}
{Anderson}, J.~K., \& {Ziurys}, L.~M. 2014, \apjl, 795, L1,
  \dodoi{10.1088/2041-8205/795/1/L1}

\bibitem[{{Andrews} {et~al.}(2022){Andrews}, {De Beck}, \&
  {Hirvonen}}]{Andrews+etal+2022}
{Andrews}, H., {De Beck}, E., \& {Hirvonen}, P. 2022, \mnras, 510, 383,
  \dodoi{10.1093/mnras/stab3244}

\bibitem[{{Apponi} {et~al.}(1999){Apponi}, {McCarthy}, {Gottlieb}, \&
  {Thaddeus}}]{Apponi+etal+1999}
{Apponi}, A.~J., {McCarthy}, M.~C., {Gottlieb}, C.~A., \& {Thaddeus}, P. 1999,
  \apjl, 516, L103, \dodoi{10.1086/311998}

\bibitem[{{Araya} {et~al.}(2003){Araya}, {Hofner}, {Goldsmith}, {Slysh}, \&
  {Takano}}]{Araya+etal+2003}
{Araya}, E., {Hofner}, P., {Goldsmith}, P., {Slysh}, S., \& {Takano}, S. 2003,
  \apj, 596, 556, \dodoi{10.1086/377533}

\bibitem[{{Asplund} {et~al.}(2009){Asplund}, {Grevesse}, {Sauval}, \&
  {Scott}}]{Asplund+etal+2009}
{Asplund}, M., {Grevesse}, N., {Sauval}, A.~J., \& {Scott}, P. 2009, \araa, 47,
  481, \dodoi{10.1146/annurev.astro.46.060407.145222}

\bibitem[{{Avery} {et~al.}(1976){Avery}, {Broten}, {MacLeod}, {Oka}, \&
  {Kroto}}]{Avery+etal+1976}
{Avery}, L.~W., {Broten}, N.~W., {MacLeod}, J.~M., {Oka}, T., \& {Kroto}, H.~W.
  1976, \apjl, 205, L173, \dodoi{10.1086/182117}

\bibitem[{{Avery} {et~al.}(1992){Avery}, {Amano}, {Bell}, {Feldman}, {Johns},
  {MacLeod}, {Matthews}, {Morton}, {Watson}, {Turner}, {Hayashi}, {Watt}, \&
  {Webster}}]{Avery+etal+1992}
{Avery}, L.~W., {Amano}, T., {Bell}, M.~B., {et~al.} 1992, \apjs, 83, 363,
  \dodoi{10.1086/191742}

\bibitem[{{Bakker} {et~al.}(1997){Bakker}, {van Dishoeck}, {Waters}, \&
  {Schoenmaker}}]{Bakker+etal+1997}
{Bakker}, E.~J., {van Dishoeck}, E.~F., {Waters}, L.~B.~F.~M., \&
  {Schoenmaker}, T. 1997, \aap, 323, 469.
\newblock \doarXiv{astro-ph/9610063}

\bibitem[{{Battino} {et~al.}(2022){Battino}, {Pignatari}, {Tattersall},
  {Denissenkov}, \& {Herwig}}]{Battino+etal+2022}
{Battino}, U., {Pignatari}, M., {Tattersall}, A., {Denissenkov}, P., \&
  {Herwig}, F. 2022, Universe, 8, 170, \dodoi{10.3390/universe8030170}

\bibitem[{{Becklin} {et~al.}(1969){Becklin}, {Frogel}, {Hyland}, {Kristian}, \&
  {Neugebauer}}]{Becklin+etal+1969}
{Becklin}, E.~E., {Frogel}, J.~A., {Hyland}, A.~R., {Kristian}, J., \&
  {Neugebauer}, G. 1969, \apjl, 158, L133, \dodoi{10.1086/180450}

\bibitem[{{Bell}(1993)}]{Bell+etal+1993b}
{Bell}, M.~B. 1993, \apj, 417, 305, \dodoi{10.1086/173313}

\bibitem[{{Bell} {et~al.}(1993){Bell}, {Avery}, \& {Feldman}}]{Bell+etal+1993}
{Bell}, M.~B., {Avery}, L.~W., \& {Feldman}, P.~A. 1993, \apjl, 417, L37,
  \dodoi{10.1086/187088}

\bibitem[{{Bell} {et~al.}(1992{\natexlab{a}}){Bell}, {Avery}, {MacLeod}, \&
  {Matthews}}]{Bell+etal+1992a}
{Bell}, M.~B., {Avery}, L.~W., {MacLeod}, J.~M., \& {Matthews}, H.~E.
  1992{\natexlab{a}}, \apj, 400, 551, \dodoi{10.1086/172017}

\bibitem[{{Bell} {et~al.}(1992{\natexlab{b}}){Bell}, {Feldman}, \&
  {Avery}}]{Bell+etal+1992}
{Bell}, M.~B., {Feldman}, P.~A., \& {Avery}, L.~W. 1992{\natexlab{b}}, \apj,
  396, 643, \dodoi{10.1086/171745}

\bibitem[{{Bernath} {et~al.}(1989){Bernath}, {Hinkle}, \&
  {Keady}}]{Bernath+etal+1989}
{Bernath}, P.~F., {Hinkle}, K.~H., \& {Keady}, J.~J. 1989, Science, 244, 562,
  \dodoi{10.1126/science.244.4904.562}

\bibitem[{{Betz}(1981)}]{Betz+etal+1981}
{Betz}, A.~L. 1981, \apjl, 244, L103, \dodoi{10.1086/183490}

\bibitem[{{Betz} {et~al.}(1979){Betz}, {McLaren}, \& {Spears}}]{Betz+etal+1979}
{Betz}, A.~L., {McLaren}, R.~A., \& {Spears}, D.~L. 1979, \apjl, 229, L97,
  \dodoi{10.1086/182937}

\bibitem[{{Bieging} \& {Nguyen-Quang-Rieu}(1989)}]{Bieging+etal+1989}
{Bieging}, J.~H., \& {Nguyen-Quang-Rieu}. 1989, \apjl, 343, L25,
  \dodoi{10.1086/185502}

\bibitem[{{Bieging} \& {Rieu}(1988)}]{Bieging+etal+1988}
{Bieging}, J.~H., \& {Rieu}, N.-Q. 1988, \apjl, 329, L107,
  \dodoi{10.1086/185187}

\bibitem[{{Bieging} \& {Tafalla}(1993)}]{Bieging+etal+1993}
{Bieging}, J.~H., \& {Tafalla}, M. 1993, \aj, 105, 576, \dodoi{10.1086/116455}

\bibitem[{{Brown} {et~al.}(1976){Brown}, {Godfrey}, {Storey}, \&
  {Clark}}]{Brown+etal+1976}
{Brown}, R.~D., {Godfrey}, P.~D., {Storey}, J.~W.~V., \& {Clark}, F.~O. 1976,
  \nat, 262, 672, \dodoi{10.1038/262672a0}

\bibitem[{{Buhl} \& {Snyder}(1970)}]{Buhl+etal+1970}
{Buhl}, D., \& {Snyder}, L.~E. 1970, \nat, 228, 267, \dodoi{10.1038/228267a0}

\bibitem[{{Bujarrabal} {et~al.}(1981){Bujarrabal}, {Gu{\'e}lin}, {Morris}, \&
  {Thaddeus}}]{Bujarrabal+etal+1981}
{Bujarrabal}, V., {Gu{\'e}lin}, M., {Morris}, M., \& {Thaddeus}, P. 1981, \aap,
  99, 239

\bibitem[{{Busso} {et~al.}(1999){Busso}, {Gallino}, \&
  {Wasserburg}}]{Busso+etal+1999}
{Busso}, M., {Gallino}, R., \& {Wasserburg}, G.~J. 1999, \araa, 37, 239,
  \dodoi{10.1146/annurev.astro.37.1.239}

\bibitem[{{Cabezas} {et~al.}(2013){Cabezas}, {Cernicharo}, {Alonso},
  {Ag{\'u}ndez}, {Mata}, {Gu{\'e}lin}, \& {Pe{\~n}a}}]{Cabezas+etal+2013}
{Cabezas}, C., {Cernicharo}, J., {Alonso}, J.~L., {et~al.} 2013, \apj, 775,
  133, \dodoi{10.1088/0004-637X/775/2/133}

\bibitem[{{Cabezas} {et~al.}(2023){Cabezas}, {Pardo}, {Ag{\'u}ndez}, {Tercero},
  {Marcelino}, {Endo}, {de Vicente}, {Gu{\'e}lin}, \&
  {Cernicharo}}]{Cabezas+etal+2023}
{Cabezas}, C., {Pardo}, J.~R., {Ag{\'u}ndez}, M., {et~al.} 2023, \aap, 672,
  L12, \dodoi{10.1051/0004-6361/202346462}

\bibitem[{{Cernicharo} {et~al.}(2011){Cernicharo}, {Ag{\'u}ndez}, \&
  {Gu{\'e}lin}}]{Cernicharo+etal+2011}
{Cernicharo}, J., {Ag{\'u}ndez}, M., \& {Gu{\'e}lin}, M. 2011, in The Molecular
  Universe, ed. J.~{Cernicharo} \& R.~{Bachiller}, Vol. 280, 237--248,
  \dodoi{10.1017/S1743921311025014}

\bibitem[{{Cernicharo} {et~al.}(2013){Cernicharo}, {Daniel}, {Castro-Carrizo},
  {Agundez}, {Marcelino}, {Joblin}, {Goicoechea}, \&
  {Gu{\'e}lin}}]{Cernicharo+etal+2013}
{Cernicharo}, J., {Daniel}, F., {Castro-Carrizo}, A., {et~al.} 2013, \apjl,
  778, L25, \dodoi{10.1088/2041-8205/778/2/L25}

\bibitem[{{Cernicharo} {et~al.}(1991{\natexlab{a}}){Cernicharo}, {Gottlieb},
  {Gu{\'e}lin}, {Killian}, {Paubert}, {Thaddeus}, \&
  {Vrtilek}}]{Cernicharo+etal+1991b}
{Cernicharo}, J., {Gottlieb}, C.~A., {Gu{\'e}lin}, M., {et~al.}
  1991{\natexlab{a}}, \apjl, 368, L39, \dodoi{10.1086/185943}

\bibitem[{{Cernicharo} {et~al.}(1991{\natexlab{b}}){Cernicharo}, {Gottlieb},
  {Gu{\'e}lin}, {Killian}, {Thaddeus}, \& {Vrtilek}}]{Cernicharo+etal+1991a}
---. 1991{\natexlab{b}}, \apjl, 368, L43, \dodoi{10.1086/185944}

\bibitem[{{Cernicharo} {et~al.}(1989){Cernicharo}, {Gottlieb}, {Gu{\'e}lin},
  {Thaddeus}, \& {Vrtilek}}]{Cernicharo+etal+1989}
{Cernicharo}, J., {Gottlieb}, C.~A., {Gu{\'e}lin}, M., {Thaddeus}, P., \&
  {Vrtilek}, J.~M. 1989, \apjl, 341, L25, \dodoi{10.1086/185449}

\bibitem[{{Cernicharo} \& {Gu{\'e}lin}(1987)}]{Cernicharo+etal+1987a}
{Cernicharo}, J., \& {Gu{\'e}lin}, M. 1987, \aap, 183, L10

\bibitem[{{Cernicharo} \& {Gu{\'e}lin}(1996)}]{Cernicharo+etal+1996}
---. 1996, \aap, 309, L27

\bibitem[{{Cernicharo} {et~al.}(2007){Cernicharo}, {Gu{\'e}lin}, {Ag{\'u}ndez},
  {Kawaguchi}, {McCarthy}, \& {Thaddeus}}]{Cernicharo+etal+2007}
{Cernicharo}, J., {Gu{\'e}lin}, M., {Ag{\'u}ndez}, M., {et~al.} 2007, \aap,
  467, L37, \dodoi{10.1051/0004-6361:20077415}

\bibitem[{{Cernicharo} {et~al.}(2008){Cernicharo}, {Gu{\'e}lin}, {Ag{\'u}ndez},
  {McCarthy}, \& {Thaddeus}}]{Cernicharo+etal+2008}
{Cernicharo}, J., {Gu{\'e}lin}, M., {Ag{\'u}ndez}, M., {McCarthy}, M.~C., \&
  {Thaddeus}, P. 2008, \apjl, 688, L83, \dodoi{10.1086/595583}

\bibitem[{{Cernicharo} {et~al.}(1987{\natexlab{a}}){Cernicharo}, {Gu{\'e}lin},
  {Hein}, \& {Kahane}}]{Cernicharo+etal+1987}
{Cernicharo}, J., {Gu{\'e}lin}, M., {Hein}, H., \& {Kahane}, C.
  1987{\natexlab{a}}, \aap, 181, L9

\bibitem[{{Cernicharo} {et~al.}(2000){Cernicharo}, {Gu{\'e}lin}, \&
  {Kahane}}]{Cernicharo+etal+2000}
{Cernicharo}, J., {Gu{\'e}lin}, M., \& {Kahane}, C. 2000, \aaps, 142, 181,
  \dodoi{10.1051/aas:2000147}

\bibitem[{{Cernicharo} {et~al.}(1991{\natexlab{c}}){Cernicharo}, {Gu{\'e}lin},
  {Kahane}, {Bogey}, \& {Demuynck}}]{Cernicharo+etal+1991}
{Cernicharo}, J., {Gu{\'e}lin}, M., {Kahane}, C., {Bogey}, M., \& {Demuynck},
  C. 1991{\natexlab{c}}, \aap, 246, 213

\bibitem[{{Cernicharo} {et~al.}(1987{\natexlab{b}}){Cernicharo}, {Gu{\'e}lin},
  {Menten}, \& {Walmsley}}]{Cernicharo+etal+1987b}
{Cernicharo}, J., {Gu{\'e}lin}, M., {Menten}, K.~M., \& {Walmsley}, C.~M.
  1987{\natexlab{b}}, \aap, 181, L1

\bibitem[{{Cernicharo} {et~al.}(2004){Cernicharo}, {Gu{\'e}lin}, \&
  {Pardo}}]{Cernicharo+etal+2004}
{Cernicharo}, J., {Gu{\'e}lin}, M., \& {Pardo}, J.~R. 2004, \apjl, 615, L145,
  \dodoi{10.1086/426439}

\bibitem[{{Cernicharo} {et~al.}(1986{\natexlab{a}}){Cernicharo}, {Kahane},
  {Gomez-Gonzalez}, \& {Gu{\'e}lin}}]{Cernicharo+etal+1986}
{Cernicharo}, J., {Kahane}, C., {Gomez-Gonzalez}, J., \& {Gu{\'e}lin}, M.
  1986{\natexlab{a}}, \aap, 164, L1

\bibitem[{{Cernicharo} {et~al.}(1986{\natexlab{b}}){Cernicharo}, {Kahane},
  {Gomez-Gonzalez}, \& {Gu{\'e}lin}}]{Cernicharo+etal+1986a}
---. 1986{\natexlab{b}}, \aap, 167, L5

\bibitem[{{Cernicharo} {et~al.}(1986{\natexlab{c}}){Cernicharo}, {Kahane},
  {Gomez-Gonzalez}, \& {Gu{\'e}lin}}]{Cernicharo+etal+1986c}
---. 1986{\natexlab{c}}, \aap, 167, L9

\bibitem[{{Cernicharo} {et~al.}(2015{\natexlab{a}}){Cernicharo}, {Marcelino},
  {Ag{\'u}ndez}, \& {Gu{\'e}lin}}]{Cernicharo+etal+2015}
{Cernicharo}, J., {Marcelino}, N., {Ag{\'u}ndez}, M., \& {Gu{\'e}lin}, M.
  2015{\natexlab{a}}, \aap, 575, A91, \dodoi{10.1051/0004-6361/201424565}

\bibitem[{{Cernicharo} {et~al.}(2010{\natexlab{a}}){Cernicharo}, {Decin},
  {Barlow}, {Ag{\'u}ndez}, {Royer}, {Vandenbussche}, {Wesson}, {Polehampton},
  {De Beck}, {Blommaert}, {Daniel}, {De Meester}, {Exter}, {Feuchtgruber},
  {Gear}, {Goicoechea}, {Gomez}, {Groenewegen}, {Hargrave}, {Huygen}, {Imhof},
  {Ivison}, {Jean}, {Kerschbaum}, {Leeks}, {Lim}, {Matsuura}, {Olofsson},
  {Posch}, {Regibo}, {Savini}, {Sibthorpe}, {Swinyard}, {Vandenbussche}, \&
  {Waelkens}}]{Cernicharo+etal+2010a}
{Cernicharo}, J., {Decin}, L., {Barlow}, M.~J., {et~al.} 2010{\natexlab{a}},
  \aap, 518, L136, \dodoi{10.1051/0004-6361/201014553}

\bibitem[{{Cernicharo} {et~al.}(2010{\natexlab{b}}){Cernicharo}, {Waters},
  {Decin}, {Encrenaz}, {Tielens}, {Ag{\'u}ndez}, {De Beck}, {M{\"u}ller},
  {Goicoechea}, {Barlow}, {Benz}, {Crimier}, {Daniel}, {di Giorgio}, {Fich},
  {Gaier}, {Garc{\'\i}a-Lario}, {de Koter}, {Khouri}, {Liseau}, {Lombaert},
  {Erickson}, {Pardo}, {Pearson}, {Shipman}, {S{\'a}nchez Contreras}, \&
  {Teyssier}}]{Cernicharo+etal+2010}
{Cernicharo}, J., {Waters}, L.~B.~F.~M., {Decin}, L., {et~al.}
  2010{\natexlab{b}}, \aap, 521, L8, \dodoi{10.1051/0004-6361/201015150}

\bibitem[{{Cernicharo} {et~al.}(2015{\natexlab{b}}){Cernicharo}, {McCarthy},
  {Gottlieb}, {Ag{\'u}ndez}, {Velilla Prieto}, {Baraban}, {Changala},
  {Gu{\'e}lin}, {Kahane}, {Martin-Drumel}, {Patel}, {Reilly}, {Stanton},
  {Quintana-Lacaci}, {Thorwirth}, \& {Young}}]{Cernicharo+etal+2015a}
{Cernicharo}, J., {McCarthy}, M.~C., {Gottlieb}, C.~A., {et~al.}
  2015{\natexlab{b}}, \apjl, 806, L3, \dodoi{10.1088/2041-8205/806/1/L3}

\bibitem[{{Cernicharo} {et~al.}(2017){Cernicharo}, {Ag{\'u}ndez}, {Velilla
  Prieto}, {Gu{\'e}lin}, {Pardo}, {Kahane}, {Marka}, {Kramer}, {Navarro},
  {Quintana-Lacaci}, {Fonfr{\'\i}a}, {Marcelino}, {Tercero}, {Moreno},
  {Massalkhi}, {Santander-Garc{\'\i}a}, {McCarthy}, {Gottlieb}, \&
  {Alonso}}]{Cernicharo+etal+2017}
{Cernicharo}, J., {Ag{\'u}ndez}, M., {Velilla Prieto}, L., {et~al.} 2017, \aap,
  606, L5, \dodoi{10.1051/0004-6361/201731672}

\bibitem[{{Cernicharo} {et~al.}(2018){Cernicharo}, {Gu{\'e}lin}, {Ag{\'u}ndez},
  {Pardo}, {Massalkhi}, {Fonfr{\'\i}a}, {Velilla Prieto}, {Quintana-Lacaci},
  {Marcelino}, {Marka}, {Navarro}, \& {Kramer}}]{Cernicharo+etal+2018}
{Cernicharo}, J., {Gu{\'e}lin}, M., {Ag{\'u}ndez}, M., {et~al.} 2018, \aap,
  618, A4, \dodoi{10.1051/0004-6361/201833335}

\bibitem[{{Cernicharo} {et~al.}(2019{\natexlab{a}}){Cernicharo},
  {Velilla-Prieto}, {Ag{\'u}ndez}, {Pardo}, {Fonfr{\'\i}a}, {Quintana-Lacaci},
  {Cabezas}, {Berm{\'u}dez}, \& {Gu{\'e}lin}}]{Cernicharo+etal+2019}
{Cernicharo}, J., {Velilla-Prieto}, L., {Ag{\'u}ndez}, M., {et~al.}
  2019{\natexlab{a}}, \aap, 627, L4, \dodoi{10.1051/0004-6361/201936040}

\bibitem[{{Cernicharo} {et~al.}(2019{\natexlab{b}}){Cernicharo}, {Cabezas},
  {Pardo}, {Ag{\'u}ndez}, {Berm{\'u}dez}, {Velilla-Prieto}, {Tercero},
  {L{\'o}pez-P{\'e}rez}, {Gallego}, {Fonfr{\'\i}a}, {Quintana-Lacaci},
  {Gu{\'e}lin}, \& {Endo}}]{Cernicharo+etal+2019b}
{Cernicharo}, J., {Cabezas}, C., {Pardo}, J.~R., {et~al.} 2019{\natexlab{b}},
  \aap, 630, L2, \dodoi{10.1051/0004-6361/201936372}

\bibitem[{{Cernicharo} {et~al.}(2023{\natexlab{a}}){Cernicharo}, {Cabezas},
  {Pardo}, {Ag{\'u}ndez}, {Roncero}, {Tercero}, {Marcelino}, {Gu{\'e}lin},
  {Endo}, \& {de Vicente}}]{Cernicharo+etal+2023a}
---. 2023{\natexlab{a}}, \aap, 672, L13, \dodoi{10.1051/0004-6361/202346467}

\bibitem[{{Cernicharo} {et~al.}(2023{\natexlab{b}}){Cernicharo}, {Pardo},
  {Cabezas}, {Ag{\'u}ndez}, {Tercero}, {Marcelino}, {Fuentetaja}, {Gu{\'e}lin},
  \& {de Vicente}}]{Cernicharo+etal+2023}
{Cernicharo}, J., {Pardo}, J.~R., {Cabezas}, C., {et~al.} 2023{\natexlab{b}},
  \aap, 670, L19, \dodoi{10.1051/0004-6361/202245816}

\bibitem[{{Changala} {et~al.}(2022){Changala}, {Gupta}, {Cernicharo}, {Pardo},
  {Ag{\'u}ndez}, {Cabezas}, {Tercero}, {Gu{\'e}lin}, \&
  {McCarthy}}]{Changala+etal+2022}
{Changala}, P.~B., {Gupta}, H., {Cernicharo}, J., {et~al.} 2022, \apjl, 940,
  L42, \dodoi{10.3847/2041-8213/aca144}

\bibitem[{{Chapovsky}(2021)}]{Chapovsky+etal+2021}
{Chapovsky}, P.~L. 2021, \mnras, 503, 1773, \dodoi{10.1093/mnras/stab407}

\bibitem[{{Charbonnel}(1995)}]{Charbonnel+etal+1995}
{Charbonnel}, C. 1995, \apjl, 453, L41, \dodoi{10.1086/309744}

\bibitem[{{Chau} {et~al.}(2012){Chau}, {Zhang}, {Nakashima}, {Deguchi}, \&
  {Kwok}}]{Chau+etal+2012}
{Chau}, W., {Zhang}, Y., {Nakashima}, J.-i., {Deguchi}, S., \& {Kwok}, S. 2012,
  \apj, 760, 66, \dodoi{10.1088/0004-637X/760/1/66}

\bibitem[{{Churchwell} {et~al.}(1978){Churchwell}, {Winnewisser}, \&
  {Walmsley}}]{Churchwell+etal+1978}
{Churchwell}, E., {Winnewisser}, G., \& {Walmsley}, C.~M. 1978, \aap, 67, 139

\bibitem[{{Clegg} {et~al.}(1982){Clegg}, {Hinkle}, \&
  {Lambert}}]{Clegg+etal+1982}
{Clegg}, R.~E.~S., {Hinkle}, K.~H., \& {Lambert}, D.~L. 1982, \mnras, 201, 95,
  \dodoi{10.1093/mnras/201.1.95}

\bibitem[{{Cooksy} {et~al.}(2015){Cooksy}, {Gottlieb}, {Killian}, {Thaddeus},
  {Patel}, {Young}, \& {McCarthy}}]{Cooksy+etal+2015}
{Cooksy}, A.~L., {Gottlieb}, C.~A., {Killian}, T.~C., {et~al.} 2015, \apjs,
  216, 30, \dodoi{10.1088/0067-0049/216/2/30}

\bibitem[{{Cordiner} \& {Millar}(2009)}]{Cordiner+etal+2009}
{Cordiner}, M.~A., \& {Millar}, T.~J. 2009, \apj, 697, 68,
  \dodoi{10.1088/0004-637X/697/1/68}

\bibitem[{{Cristallo} {et~al.}(2009){Cristallo}, {Straniero}, {Gallino},
  {Piersanti}, {Dom{\'\i}nguez}, \& {Lederer}}]{Cristallo+etal+2009}
{Cristallo}, S., {Straniero}, O., {Gallino}, R., {et~al.} 2009, \apj, 696, 797,
  \dodoi{10.1088/0004-637X/696/1/797}

\bibitem[{{Crosas} \& {Menten}(1997)}]{Crosas+Menten+1997}
{Crosas}, M., \& {Menten}, K.~M. 1997, \apj, 483, 913, \dodoi{10.1086/304256}

\bibitem[{{Daniel} {et~al.}(2012){Daniel}, {Ag{\'u}ndez}, {Cernicharo}, {De
  Beck}, {Lombaert}, {Decin}, {Kahane}, {Gu{\'e}lin}, \&
  {M{\"u}ller}}]{Daniel+etal+2012}
{Daniel}, F., {Ag{\'u}ndez}, M., {Cernicharo}, J., {et~al.} 2012, \aap, 542,
  A37, \dodoi{10.1051/0004-6361/201118449}

\bibitem[{{Dayal} \& {Bieging}(1995)}]{Dayal+etal+1995}
{Dayal}, A., \& {Bieging}, J.~H. 1995, \apj, 439, 996, \dodoi{10.1086/175237}

\bibitem[{{De Beck} {et~al.}(2010){De Beck}, {Decin}, {de Koter}, {Justtanont},
  {Verhoelst}, {Kemper}, \& {Menten}}]{DeBeck+etal+2010}
{De Beck}, E., {Decin}, L., {de Koter}, A., {et~al.} 2010, \aap, 523, A18,
  \dodoi{10.1051/0004-6361/200913771}

\bibitem[{{De Beck} {et~al.}(2012){De Beck}, {Lombaert}, {Ag{\'u}ndez},
  {Daniel}, {Decin}, {Cernicharo}, {M{\"u}ller}, {Min}, {Royer},
  {Vandenbussche}, {de Koter}, {Waters}, {Groenewegen}, {Barlow}, {Gu{\'e}lin},
  {Kahane}, {Pearson}, {Encrenaz}, {Szczerba}, \& {Schmidt}}]{Beck+etal+2012}
{De Beck}, E., {Lombaert}, R., {Ag{\'u}ndez}, M., {et~al.} 2012, \aap, 539,
  A108, \dodoi{10.1051/0004-6361/201117635}

\bibitem[{{Decin} {et~al.}(2010){Decin}, {Justtanont}, {De Beck}, {Lombaert},
  {de Koter}, {Waters}, {Marston}, {Teyssier}, {Sch{\"o}ier}, {Bujarrabal},
  {Alcolea}, {Cernicharo}, {Dominik}, {Melnick}, {Menten}, {Neufeld},
  {Olofsson}, {Planesas}, {Schmidt}, {Szczerba}, {de Graauw}, {Helmich},
  {Roelfsema}, {Dieleman}, {Morris}, {Gallego}, {D{\'\i}ez-Gonz{\'a}lez}, \&
  {Caux}}]{Decin+etal+2010}
{Decin}, L., {Justtanont}, K., {De Beck}, E., {et~al.} 2010, \aap, 521, L4,
  \dodoi{10.1051/0004-6361/201015069}

\bibitem[{{Draine}(1978)}]{Draine78}
{Draine}, B.~T. 1978, \apjs, 36, 595, \dodoi{10.1086/190513}

\bibitem[{{Epchtein} {et~al.}(1987){Epchtein}, {Le Bertre}, {Lepine}, {Marques
  Dos Santos}, {Matsuura}, \& {Picazzio}}]{Epchtein+etal+1987}
{Epchtein}, N., {Le Bertre}, T., {Lepine}, J.~R.~D., {et~al.} 1987, \aaps, 71,
  39

\bibitem[{{Fonfr{\'\i}a} {et~al.}(2015){Fonfr{\'\i}a}, {Cernicharo}, {Richter},
  {Fern{\'a}ndez-L{\'o}pez}, {Velilla Prieto}, \& {Lacy}}]{Fonfria+etal+2015}
{Fonfr{\'\i}a}, J.~P., {Cernicharo}, J., {Richter}, M.~J., {et~al.} 2015,
  \mnras, 453, 439, \dodoi{10.1093/mnras/stv1634}

\bibitem[{{Fonfr{\'\i}a} {et~al.}(2014){Fonfr{\'\i}a},
  {Fern{\'a}ndez-L{\'o}pez}, {Ag{\'u}ndez}, {S{\'a}nchez-Contreras}, {Curiel},
  \& {Cernicharo}}]{Fonfria+etal+2014}
{Fonfr{\'\i}a}, J.~P., {Fern{\'a}ndez-L{\'o}pez}, M., {Ag{\'u}ndez}, M.,
  {et~al.} 2014, \mnras, 445, 3289, \dodoi{10.1093/mnras/stu1968}

\bibitem[{{Fonfr{\'\i}a} {et~al.}(2017){Fonfr{\'\i}a}, {Hinkle}, {Cernicharo},
  {Richter}, {Ag{\'u}ndez}, \& {Wallace}}]{Fonfria+etal+2017}
{Fonfr{\'\i}a}, J.~P., {Hinkle}, K.~H., {Cernicharo}, J., {et~al.} 2017, \apj,
  835, 196, \dodoi{10.3847/1538-4357/835/2/196}

\bibitem[{{Fonfr{\'\i}a} {et~al.}(2018){Fonfr{\'\i}a},
  {Fern{\'a}ndez-L{\'o}pez}, {Pardo}, {Ag{\'u}ndez}, {S{\'a}nchez Contreras},
  {Velilla Prieto}, {Cernicharo}, {Santander-Garc{\'\i}a}, {Quintana-Lacaci},
  {Castro-Carrizo}, \& {Curiel}}]{Fonfria+etal+2018}
{Fonfr{\'\i}a}, J.~P., {Fern{\'a}ndez-L{\'o}pez}, M., {Pardo}, J.~R., {et~al.}
  2018, \apj, 860, 162, \dodoi{10.3847/1538-4357/aac5e3}

\bibitem[{{Fonfr{\'\i}a Exp{\'o}sito} {et~al.}(2006){Fonfr{\'\i}a
  Exp{\'o}sito}, {Ag{\'u}ndez}, {Tercero}, {Pardo}, \&
  {Cernicharo}}]{Fonfria+etal+2006}
{Fonfr{\'\i}a Exp{\'o}sito}, J.~P., {Ag{\'u}ndez}, M., {Tercero}, B., {Pardo},
  J.~R., \& {Cernicharo}, J. 2006, \apjl, 646, L127, \dodoi{10.1086/507104}

\bibitem[{{Fong} {et~al.}(2003){Fong}, {Meixner}, \& {Shah}}]{Fong+etal+2003}
{Fong}, D., {Meixner}, M., \& {Shah}, R.~Y. 2003, \apjl, 582, L39,
  \dodoi{10.1086/346034}

\bibitem[{{Fong} {et~al.}(2006){Fong}, {Meixner}, {Sutton}, {Zalucha}, \&
  {Welch}}]{Fong+etal+2006}
{Fong}, D., {Meixner}, M., {Sutton}, E.~C., {Zalucha}, A., \& {Welch}, W.~J.
  2006, \apj, 652, 1626, \dodoi{10.1086/508127}

\bibitem[{{Ford} {et~al.}(2003){Ford}, {Neufeld}, {Goldsmith}, \&
  {Melnick}}]{Ford+etal+2003}
{Ford}, K.~E.~S., {Neufeld}, D.~A., {Goldsmith}, P.~F., \& {Melnick}, G.~J.
  2003, \apj, 589, 430, \dodoi{10.1086/374552}

\bibitem[{{Ford} {et~al.}(2004){Ford}, {Neufeld}, {Schilke}, \&
  {Melnick}}]{Ford+etal+2004}
{Ford}, K.~E.~S., {Neufeld}, D.~A., {Schilke}, P., \& {Melnick}, G.~J. 2004,
  \apj, 614, 990, \dodoi{10.1086/423886}

\bibitem[{{Gallardo Cava} {et~al.}(2022){Gallardo Cava}, {Bujarrabal},
  {Alcolea}, {G{\'o}mez-Garrido}, \&
  {Santander-Garc{\'\i}a}}]{Gallardo+etal+2022}
{Gallardo Cava}, I., {Bujarrabal}, V., {Alcolea}, J., {G{\'o}mez-Garrido}, M.,
  \& {Santander-Garc{\'\i}a}, M. 2022, \aap, 659, A134,
  \dodoi{10.1051/0004-6361/202142339}

\bibitem[{{Gensheimer}(1997{\natexlab{a}})}]{Gensheimer+etal+1997}
{Gensheimer}, P.~D. 1997{\natexlab{a}}, \apss, 251, 199,
  \dodoi{10.1023/A:1000744924767}

\bibitem[{{Gensheimer}(1997{\natexlab{b}})}]{Gensheimer+etal+1997a}
---. 1997{\natexlab{b}}, \apjl, 479, L75, \dodoi{10.1086/310576}

\bibitem[{{Goldhaber} \& {Betz}(1984)}]{Goldhaber+etal+1984}
{Goldhaber}, D.~M., \& {Betz}, A.~L. 1984, \apjl, 279, L55,
  \dodoi{10.1086/184255}

\bibitem[{{Goldhaber} {et~al.}(1987){Goldhaber}, {Betz}, \&
  {Ottusch}}]{Goldhaber+etal+1987}
{Goldhaber}, D.~M., {Betz}, A.~L., \& {Ottusch}, J.~J. 1987, \apj, 314, 356,
  \dodoi{10.1086/165066}

\bibitem[{{Gong} {et~al.}(2015){Gong}, {Henkel}, {Spezzano}, {Thorwirth},
  {Menten}, {Wyrowski}, {Mao}, \& {Klein}}]{Gong+etal+2015}
{Gong}, Y., {Henkel}, C., {Spezzano}, S., {et~al.} 2015, \aap, 574, A56,
  \dodoi{10.1051/0004-6361/201424819}

\bibitem[{{Gong} {et~al.}(2017){Gong}, {Henkel}, {Ott}, {Menten}, {Morris},
  {Keller}, {Claussen}, {Grasshoff}, \& {Mao}}]{Gong+etal+2017}
{Gong}, Y., {Henkel}, C., {Ott}, J., {et~al.} 2017, \apj, 843, 54,
  \dodoi{10.3847/1538-4357/aa7853}

\bibitem[{{Groesbeck} {et~al.}(1994){Groesbeck}, {Phillips}, \&
  {Blake}}]{Groesbeck+etal+1994}
{Groesbeck}, T.~D., {Phillips}, T.~G., \& {Blake}, G.~A. 1994, \apjs, 94, 147,
  \dodoi{10.1086/192076}

\bibitem[{{Gu{\'e}lin} \& {Cernicharo}(1991)}]{Guelin+etal+1991}
{Gu{\'e}lin}, M., \& {Cernicharo}, J. 1991, \aap, 244, L21

\bibitem[{{Gu{\'e}lin} {et~al.}(1986){Gu{\'e}lin}, {Cernicharo}, {Kahane}, \&
  {Gomez-Gonzales}}]{Guelin+etal+1986}
{Gu{\'e}lin}, M., {Cernicharo}, J., {Kahane}, C., \& {Gomez-Gonzales}, J. 1986,
  \aap, 157, L17

\bibitem[{{Gu{\'e}lin} {et~al.}(1990){Gu{\'e}lin}, {Cernicharo}, {Paubert}, \&
  {Turner}}]{Guelin+etal+1990}
{Gu{\'e}lin}, M., {Cernicharo}, J., {Paubert}, G., \& {Turner}, B.~E. 1990,
  \aap, 230, L9

\bibitem[{{Gu{\'e}lin} {et~al.}(1978){Gu{\'e}lin}, {Green}, \&
  {Thaddeus}}]{Guelin+etal+1978}
{Gu{\'e}lin}, M., {Green}, S., \& {Thaddeus}, P. 1978, \apjl, 224, L27,
  \dodoi{10.1086/182751}

\bibitem[{{Gu{\'e}lin} {et~al.}(1993){Gu{\'e}lin}, {Lucas}, \&
  {Cernicharo}}]{Guelin+etal+1993}
{Gu{\'e}lin}, M., {Lucas}, R., \& {Cernicharo}, J. 1993, \aap, 280, L19

\bibitem[{{Gu{\'e}lin} {et~al.}(1997){Gu{\'e}lin}, {Lucas}, \&
  {Neri}}]{Guelin+etal+1997}
{Gu{\'e}lin}, M., {Lucas}, R., \& {Neri}, R. 1997, IAU Symposium, 170, 359

\bibitem[{{Gu{\'e}lin} {et~al.}(2000){Gu{\'e}lin}, {Muller}, {Cernicharo},
  {Apponi}, {McCarthy}, {Gottlieb}, \& {Thaddeus}}]{Guelin+etal+2000}
{Gu{\'e}lin}, M., {Muller}, S., {Cernicharo}, J., {et~al.} 2000, \aap, 363, L9

\bibitem[{{Gu{\'e}lin} {et~al.}(2004){Gu{\'e}lin}, {Muller}, {Cernicharo},
  {McCarthy}, \& {Thaddeus}}]{Guelin+etal+2004}
{Gu{\'e}lin}, M., {Muller}, S., {Cernicharo}, J., {McCarthy}, M.~C., \&
  {Thaddeus}, P. 2004, \aap, 426, L49, \dodoi{10.1051/0004-6361:200400074}

\bibitem[{{Gu{\'e}lin} {et~al.}(1998){Gu{\'e}lin}, {Neininger}, \&
  {Cernicharo}}]{Guelin+etal+1998}
{Gu{\'e}lin}, M., {Neininger}, N., \& {Cernicharo}, J. 1998, \aap, 335, L1.
\newblock \doarXiv{astro-ph/9805105}

\bibitem[{{Gu{\'e}lin} {et~al.}(1999){Gu{\'e}lin}, {Neininger}, {Lucas}, \&
  {Cernicharo}}]{Guelin+etal+1999}
{Gu{\'e}lin}, M., {Neininger}, N., {Lucas}, R., \& {Cernicharo}, J. 1999, in
  The Physics and Chemistry of the Interstellar Medium, ed. V.~{Ossenkopf},
  J.~{Stutzki}, \& G.~{Winnewisser}, 326

\bibitem[{{Gu{\'e}lin} \& {Thaddeus}(1977)}]{Guelin+etal+1977}
{Gu{\'e}lin}, M., \& {Thaddeus}, P. 1977, \apjl, 212, L81,
  \dodoi{10.1086/182380}

\bibitem[{{Habing} \& {Olofsson}(2004)}]{Habing+etal+2004}
{Habing}, H.~J., \& {Olofsson}, H. 2004, {Asymptotic Giant Branch Stars},
  \dodoi{10.1007/978-1-4757-3876-6}

\bibitem[{{Halfen} {et~al.}(2008){Halfen}, {Clouthier}, \&
  {Ziurys}}]{Halfen+etal+2008}
{Halfen}, D.~T., {Clouthier}, D.~J., \& {Ziurys}, L.~M. 2008, \apjl, 677, L101,
  \dodoi{10.1086/588024}

\bibitem[{{Hall} \& {Ridgway}(1978)}]{Hall+etal+1978}
{Hall}, D.~N.~B., \& {Ridgway}, S.~T. 1978, \nat, 273, 281,
  \dodoi{10.1038/273281a0}

\bibitem[{{He} {et~al.}(2017){He}, {Dinh-V-Trung}, \&
  {Hasegawa}}]{He+etal+2017}
{He}, J.~H., {Dinh-V-Trung}, \& {Hasegawa}, T.~I. 2017, \apj, 845, 38,
  \dodoi{10.3847/1538-4357/aa7e76}

\bibitem[{{He} {et~al.}(2008){He}, {Dinh-V-Trung}, {Kwok}, {M{\"u}ller},
  {Zhang}, {Hasegawa}, {Peng}, \& {Huang}}]{He+etal+2008}
{He}, J.~H., {Dinh-V-Trung}, {Kwok}, S., {et~al.} 2008, \apjs, 177, 275,
  \dodoi{10.1086/587142}

\bibitem[{{He} {et~al.}(2019){He}, {Kami{\'n}ski}, {Mennickent}, {Shenavrin},
  {Mardones}, {Wang}, {Tang}, {Schmidt}, {Szczerba}, \& {Ge}}]{He+etal+2019}
{He}, J.~H., {Kami{\'n}ski}, T., {Mennickent}, R.~E., {et~al.} 2019, \apj, 883,
  165, \dodoi{10.3847/1538-4357/ab3d37}

\bibitem[{{Henkel} {et~al.}(1985){Henkel}, {Matthews}, {Morris}, {Terebey}, \&
  {Fich}}]{Henkel+etal+1985}
{Henkel}, C., {Matthews}, H.~E., {Morris}, M., {Terebey}, S., \& {Fich}, M.
  1985, \aap, 147, 143

\bibitem[{{Herwig}(2005)}]{Herwig+etal+2005}
{Herwig}, F. 2005, \araa, 43, 435,
  \dodoi{10.1146/annurev.astro.43.072103.150600}

\bibitem[{{Hinkle} {et~al.}(1988){Hinkle}, {Keady}, \&
  {Bernath}}]{Hinkle+etal+1988}
{Hinkle}, K.~W., {Keady}, J.~J., \& {Bernath}, P.~F. 1988, Science, 241, 1319,
  \dodoi{10.1126/science.241.4871.1319}

\bibitem[{{Howard} {et~al.}(1970){Howard}, {Buhl}, \&
  {Snyder}}]{Howard+etal+1970}
{Howard}, W.~E., {Buhl}, D., \& {Snyder}, L.~E. 1970, \iaucirc, 2251, 1

\bibitem[{{Jefferts} {et~al.}(1970{\natexlab{a}}){Jefferts}, {Penzias}, \&
  {Wilson}}]{Jefferts+etal+1970}
{Jefferts}, K.~B., {Penzias}, A.~A., \& {Wilson}, R.~W. 1970{\natexlab{a}},
  \apjl, 161, L87, \dodoi{10.1086/180576}

\bibitem[{{Jefferts} {et~al.}(1970{\natexlab{b}}){Jefferts}, {Penzias},
  {Penzias}, \& {Wilson}}]{Jefferts+etal+1970a}
{Jefferts}, K.~B., {Penzias}, N.~J., {Penzias}, A.~A., \& {Wilson}, R.~W.
  1970{\natexlab{b}}, \iaucirc, 2231, 1

\bibitem[{{Jeste} {et~al.}(2023){Jeste}, {Wiesemeyer}, {Menten}, \&
  {Wyrowski}}]{Jeste+etal+2023}
{Jeste}, M., {Wiesemeyer}, H., {Menten}, K.~M., \& {Wyrowski}, F. 2023, \aap,
  675, A139, \dodoi{10.1051/0004-6361/202346034}

\bibitem[{{Jewell} \& {Snyder}(1984)}]{Jewell+etal+1984}
{Jewell}, P.~R., \& {Snyder}, L.~E. 1984, \apj, 278, 176,
  \dodoi{10.1086/161779}

\bibitem[{{Johansson} {et~al.}(1984){Johansson}, {Andersson}, {Ellder},
  {Friberg}, {Hjalmarson}, {Hoglund}, {Irvine}, {Olofsson}, \&
  {Rydbeck}}]{Johansson+etal+1984}
{Johansson}, L.~E.~B., {Andersson}, C., {Ellder}, J., {et~al.} 1984, \aap, 130,
  227

\bibitem[{{Johansson} {et~al.}(1985){Johansson}, {Andersson}, {Elder},
  {Friberg}, {Hjalmarson}, {Hoglund}, {Olofsson}, {Rydbeck}, \&
  {Irvine}}]{Johansson+etal+1985}
{Johansson}, L.~E.~B., {Andersson}, C., {Elder}, J., {et~al.} 1985, \aaps, 60,
  135

\bibitem[{{Kahane} {et~al.}(1992){Kahane}, {Cernicharo}, {Gomez-Gonzalez}, \&
  {Gu{\'e}lin}}]{Kahane+etal+1992}
{Kahane}, C., {Cernicharo}, J., {Gomez-Gonzalez}, J., \& {Gu{\'e}lin}, M. 1992,
  \aap, 256, 235

\bibitem[{{Kahane} {et~al.}(1988){Kahane}, {Gomez-Gonzalez}, {Cernicharo}, \&
  {Gu{\'e}lin}}]{Kahane+etal+1988}
{Kahane}, C., {Gomez-Gonzalez}, J., {Cernicharo}, J., \& {Gu{\'e}lin}, M. 1988,
  \aap, 190, 167

\bibitem[{{Kaifu} {et~al.}(1987){Kaifu}, {Suzuki}, {Ohishi}, {Miyaji},
  {Ishikawa}, {Kasuga}, {Morimoto}, \& {Saito}}]{Kaifu+etal+1987}
{Kaifu}, N., {Suzuki}, H., {Ohishi}, M., {et~al.} 1987, \apjl, 317, L111,
  \dodoi{10.1086/184922}

\bibitem[{{Kasai} {et~al.}(2007){Kasai}, {Kagi}, \&
  {Kawaguchi}}]{Kasai+etal+2007}
{Kasai}, Y., {Kagi}, E., \& {Kawaguchi}, K. 2007, \apjl, 661, L61,
  \dodoi{10.1086/518555}

\bibitem[{{Kawaguchi} {et~al.}(1993){Kawaguchi}, {Kagi}, {Hirano}, {Takano}, \&
  {Saito}}]{Kawaguchi+etal+1993}
{Kawaguchi}, K., {Kagi}, E., {Hirano}, T., {Takano}, S., \& {Saito}, S. 1993,
  \apjl, 406, L39, \dodoi{10.1086/186781}

\bibitem[{{Kawaguchi} {et~al.}(1995){Kawaguchi}, {Kasai}, {Ishikawa}, \&
  {Kaifu}}]{Kawaguchi+etal+1995}
{Kawaguchi}, K., {Kasai}, Y., {Ishikawa}, S.-I., \& {Kaifu}, N. 1995, \pasj,
  47, 853

\bibitem[{{Kawaguchi} {et~al.}(2007){Kawaguchi}, {Fujimori}, {Aimi}, {Takano},
  {Okabayashi}, {Gupta}, {Br{\"u}nken}, {Gottlieb}, {McCarthy}, \&
  {Thaddeus}}]{Kawaguchi+etal+2007}
{Kawaguchi}, K., {Fujimori}, R., {Aimi}, S., {et~al.} 2007, \pasj, 59, L47,
  \dodoi{10.1093/pasj/59.5.L47}

\bibitem[{{Keene} {et~al.}(1993){Keene}, {Young}, {Phillips}, {Buettgenbach},
  \& {Carlstrom}}]{Keene+etal+1993}
{Keene}, J., {Young}, K., {Phillips}, T.~G., {Buettgenbach}, T.~H., \&
  {Carlstrom}, J.~E. 1993, \apjl, 415, L131, \dodoi{10.1086/187050}

\bibitem[{{Keller} {et~al.}(2015){Keller}, {Menten}, {Kami{\'n}ski}, \&
  {Claussen}}]{Keller+etal+2015}
{Keller}, D., {Menten}, K.~M., {Kami{\'n}ski}, T., \& {Claussen}, M.~J. 2015,
  in Astronomical Society of the Pacific Conference Series, Vol. 497, Why
  Galaxies Care about AGB Stars III: A Closer Look in Space and Time, ed.
  F.~{Kerschbaum}, R.~F. {Wing}, \& J.~{Hron}, 123

\bibitem[{{Khouri} {et~al.}(2014){Khouri}, {de Koter}, {Decin}, {Waters},
  {Maercker}, {Lombaert}, {Alcolea}, {Blommaert}, {Bujarrabal}, {Groenewegen},
  {Justtanont}, {Kerschbaum}, {Matsuura}, {Menten}, {Olofsson}, {Planesas},
  {Royer}, {Schmidt}, {Szczerba}, {Teyssier}, \& {Yates}}]{Khouri+etal+2014}
{Khouri}, T., {de Koter}, A., {Decin}, L., {et~al.} 2014, \aap, 570, A67,
  \dodoi{10.1051/0004-6361/201424298}

\bibitem[{{Knapp} \& {Chang}(1985)}]{knapp+etal+1985}
{Knapp}, G.~R., \& {Chang}, K.~M. 1985, \apj, 293, 281, \dodoi{10.1086/163235}

\bibitem[{{Koelemay} \& {Ziurys}(2023)}]{Koelemay+etal+2023ApJ}
{Koelemay}, L.~A., \& {Ziurys}, L.~M. 2023, \apjl, 958, L6,
  \dodoi{10.3847/2041-8213/ad0899}

\bibitem[{{Kuiper} {et~al.}(1976){Kuiper}, {Knapp}, {Knapp}, \&
  {Brown}}]{Kuiper+etal+1976}
{Kuiper}, T.~B.~H., {Knapp}, G.~R., {Knapp}, S.~L., \& {Brown}, R.~L. 1976,
  \apj, 204, 408, \dodoi{10.1086/154183}

\bibitem[{{Le Gal} {et~al.}(2017){Le Gal}, {Xie}, {Herbst}, {Talbi}, {Guo}, \&
  {Muller}}]{LeGal+etal+2017}
{Le Gal}, R., {Xie}, C., {Herbst}, E., {et~al.} 2017, \aap, 608, A96,
  \dodoi{10.1051/0004-6361/201731566}

\bibitem[{{Li} {et~al.}(2013){Li}, {Wang}, {Gu}, \& {Zheng}}]{Li+etal+2013}
{Li}, J., {Wang}, J.~Z., {Gu}, Q.~S., \& {Zheng}, X.~W. 2013, \aap, 555, A18,
  \dodoi{10.1051/0004-6361/201220943}

\bibitem[{{Li} {et~al.}(2014){Li}, {Millar}, {Walsh}, {Heays}, \& {van
  Dishoeck}}]{Li+etal+2014}
{Li}, X., {Millar}, T.~J., {Walsh}, C., {Heays}, A.~N., \& {van Dishoeck},
  E.~F. 2014, \aap, 568, A111, \dodoi{10.1051/0004-6361/201424076}

\bibitem[{{Linke} {et~al.}(1977){Linke}, {Goldsmith}, {Wannier}, {Wilson}, \&
  {Penzias}}]{Linke+etal+1977}
{Linke}, R.~A., {Goldsmith}, P.~F., {Wannier}, P.~G., {Wilson}, R.~W., \&
  {Penzias}, A.~A. 1977, \apj, 214, 50, \dodoi{10.1086/155229}

\bibitem[{{Lovas}(2004)}]{Lovas+2004}
{Lovas}, F.~J. 2004, Journal of Physical and Chemical Reference Data, 33, 177,
  \dodoi{10.1063/1.1633275}

\bibitem[{{Lucas} {et~al.}(1995){Lucas}, {Gu{\'e}lin}, {Kahane}, {Audinos}, \&
  {Cernicharo}}]{Lucas+etal+1995}
{Lucas}, R., {Gu{\'e}lin}, M., {Kahane}, C., {Audinos}, P., \& {Cernicharo}, J.
  1995, \apss, 224, 293, \dodoi{10.1007/BF00667860}

\bibitem[{{Massalkhi} {et~al.}(2018){Massalkhi}, {Ag{\'u}ndez}, {Cernicharo},
  {Velilla Prieto}, {Goicoechea}, {Quintana-Lacaci}, {Fonfr{\'\i}a}, {Alcolea},
  \& {Bujarrabal}}]{Massalkhi+etal+2018}
{Massalkhi}, S., {Ag{\'u}ndez}, M., {Cernicharo}, J., {et~al.} 2018, \aap, 611,
  A29, \dodoi{10.1051/0004-6361/201732038}

\bibitem[{{McCarthy} {et~al.}(2006){McCarthy}, {Gottlieb}, {Gupta}, \&
  {Thaddeus}}]{McCarthy+etal+2006}
{McCarthy}, M.~C., {Gottlieb}, C.~A., {Gupta}, H., \& {Thaddeus}, P. 2006,
  \apjl, 652, L141, \dodoi{10.1086/510238}

\bibitem[{{McElroy} {et~al.}(2013){McElroy}, {Walsh}, {Markwick}, {Cordiner},
  {Smith}, \& {Millar}}]{McElroy+etal+2013}
{McElroy}, D., {Walsh}, C., {Markwick}, A.~J., {et~al.} 2013, \aap, 550, A36,
  \dodoi{10.1051/0004-6361/201220465}

\bibitem[{{McGuire}(2018)}]{McGuire+etal+2018}
{McGuire}, B.~A. 2018, \apjs, 239, 17, \dodoi{10.3847/1538-4365/aae5d2}

\bibitem[{{McGuire}(2022)}]{McGuire+etal+2022}
---. 2022, \apjs, 259, 30, \dodoi{10.3847/1538-4365/ac2a48}

\bibitem[{{Melnick} {et~al.}(2001){Melnick}, {Neufeld}, {Ford}, {Hollenbach},
  \& {Ashby}}]{Melnick+etal+2001}
{Melnick}, G.~J., {Neufeld}, D.~A., {Ford}, K.~E.~S., {Hollenbach}, D.~J., \&
  {Ashby}, M. L.~N. 2001, \nat, 412, 160.
\newblock \doarXiv{astro-ph/0107212}

\bibitem[{{Menten} {et~al.}(2012){Menten}, {Reid}, {Kami{\'n}ski}, \&
  {Claussen}}]{Menten+etal+2012}
{Menten}, K.~M., {Reid}, M.~J., {Kami{\'n}ski}, T., \& {Claussen}, M.~J. 2012,
  \aap, 543, A73, \dodoi{10.1051/0004-6361/201219422}

\bibitem[{{Menten} {et~al.}(2018){Menten}, {Wyrowski}, {Keller}, \&
  {Kami{\'n}ski}}]{Menten+etal+2018}
{Menten}, K.~M., {Wyrowski}, F., {Keller}, D., \& {Kami{\'n}ski}, T. 2018,
  \aap, 613, A49, \dodoi{10.1051/0004-6361/201732296}

\bibitem[{{Milam} {et~al.}(2008){Milam}, {Halfen}, {Tenenbaum}, {Apponi},
  {Woolf}, \& {Ziurys}}]{Milam+etal+2008}
{Milam}, S.~N., {Halfen}, D.~T., {Tenenbaum}, E.~D., {et~al.} 2008, \apj, 684,
  618, \dodoi{10.1086/589135}

\bibitem[{{Milam} {et~al.}(2009){Milam}, {Woolf}, \&
  {Ziurys}}]{Milam+etal+2009}
{Milam}, S.~N., {Woolf}, N.~J., \& {Ziurys}, L.~M. 2009, \apj, 690, 837,
  \dodoi{10.1088/0004-637X/690/1/837}

\bibitem[{{Millar}(2020)}]{Millar+2020}
{Millar}, T.~J. 2020, Chinese Journal of Chemical Physics, 33, 668,
  \dodoi{10.1063/1674-0068/cjcp2008145}

\bibitem[{{Millar} {et~al.}(2000){Millar}, {Herbst}, \&
  {Bettens}}]{Millar+etal+2000}
{Millar}, T.~J., {Herbst}, E., \& {Bettens}, R.~P.~A. 2000, \mnras, 316, 195,
  \dodoi{10.1046/j.1365-8711.2000.03560.x}

\bibitem[{{Morisawa} {et~al.}(2006){Morisawa}, {Fushitani}, {Kato}, {Hoshina},
  {Simizu}, {Watanabe}, {Miyamoto}, {Kasai}, {Kawaguchi}, \&
  {Momose}}]{Morisawa+etal+2006}
{Morisawa}, Y., {Fushitani}, M., {Kato}, Y., {et~al.} 2006, \apj, 642, 954,
  \dodoi{10.1086/501041}

\bibitem[{{Morris} {et~al.}(1975){Morris}, {Gilmore}, {Palmer}, {Turner}, \&
  {Zuckerman}}]{Morris+etal+1975}
{Morris}, M., {Gilmore}, W., {Palmer}, P., {Turner}, B.~E., \& {Zuckerman}, B.
  1975, \apjl, 199, L47, \dodoi{10.1086/181846}

\bibitem[{{Morris} {et~al.}(1971){Morris}, {Zuckerman}, {Palmer}, \&
  {Turner}}]{Morris+etal+1971}
{Morris}, M., {Zuckerman}, B., {Palmer}, P., \& {Turner}, B.~E. 1971, \apjl,
  170, L109, \dodoi{10.1086/180850}

\bibitem[{{M{\"u}ller} {et~al.}(2005){M{\"u}ller}, {Schl{\"o}der}, {Stutzki},
  \& {Winnewisser}}]{Muller+etal+2005}
{M{\"u}ller}, H. S.~P., {Schl{\"o}der}, F., {Stutzki}, J., \& {Winnewisser}, G.
  2005, Journal of Molecular Structure, 742, 215,
  \dodoi{10.1016/j.molstruc.2005.01.027}

\bibitem[{{Nguyen-Q-Rieu} {et~al.}(1984){Nguyen-Q-Rieu}, {Graham}, \&
  {Bujarrabal}}]{Nguyen-Q-Rieu+etal+1984}
{Nguyen-Q-Rieu}, {Graham}, D., \& {Bujarrabal}, V. 1984, \aap, 138, L5

\bibitem[{{Nyman} {et~al.}(1993){Nyman}, {Olofsson}, {Johansson}, {Booth},
  {Carlstrom}, \& {Wolstencroft}}]{Nyman+etal+1993}
{Nyman}, L.~A., {Olofsson}, H., {Johansson}, L.~E.~B., {et~al.} 1993, \aap,
  269, 377

\bibitem[{{Ohishi} {et~al.}(1989){Ohishi}, {Kaifu}, {Kawaguchi}, {Murakami},
  {Saito}, {Yamamoto}, {Ishikawa}, {Fujita}, {Shiratori}, \&
  {Irvine}}]{Ohishi+etal+1989}
{Ohishi}, M., {Kaifu}, N., {Kawaguchi}, K., {et~al.} 1989, \apjl, 345, L83,
  \dodoi{10.1086/185558}

\bibitem[{{Olofsson}(1996)}]{Olofsson+etal+1996}
{Olofsson}, H. 1996, \apss, 245, 169, \dodoi{10.1007/BF00642225}

\bibitem[{{Olofsson} {et~al.}(1982){Olofsson}, {Johansson}, {Hjalmarson}, \&
  {Nguyen-Quang-Rieu}}]{Olofsson+etal+1982}
{Olofsson}, H., {Johansson}, L.~E.~B., {Hjalmarson}, A., \&
  {Nguyen-Quang-Rieu}. 1982, \aap, 107, 128

\bibitem[{{Pardo} {et~al.}(2021){Pardo}, {Cabezas}, {Fonfr{\'\i}a},
  {Ag{\'u}ndez}, {Tercero}, {de Vicente}, {Gu{\'e}lin}, \&
  {Cernicharo}}]{Pardo+etal+2021}
{Pardo}, J.~R., {Cabezas}, C., {Fonfr{\'\i}a}, J.~P., {et~al.} 2021, \aap, 652,
  L13, \dodoi{10.1051/0004-6361/202141671}

\bibitem[{{Pardo} {et~al.}(2007){Pardo}, {Cernicharo}, {Goicoechea},
  {Gu{\'e}lin}, \& {Asensio Ramos}}]{Pardo+etal+2007}
{Pardo}, J.~R., {Cernicharo}, J., {Goicoechea}, J.~R., {Gu{\'e}lin}, M., \&
  {Asensio Ramos}, A. 2007, \apj, 661, 250, \dodoi{10.1086/513734}

\bibitem[{{Pardo} {et~al.}(2018){Pardo}, {Cernicharo}, {Velilla Prieto},
  {Fonfr{\'\i}a}, {Ag{\'u}ndez}, {Quintana-Lacaci}, {Massalkhi}, {Tercero},
  {G{\'o}mez-Garrido}, {de Vicente}, {Gu{\'e}lin}, {Kramer}, {Marka},
  {Teyssier}, \& {Neufeld}}]{Pardo+etal+2018}
{Pardo}, J.~R., {Cernicharo}, J., {Velilla Prieto}, L., {et~al.} 2018, \aap,
  615, L4, \dodoi{10.1051/0004-6361/201833303}

\bibitem[{{Pardo} {et~al.}(2022){Pardo}, {Cernicharo}, {Tercero}, {Cabezas},
  {Berm{\'u}dez}, {Ag{\'u}ndez}, {Gallego}, {Tercero}, {G{\'o}mez-Garrido}, {de
  Vicente}, \& {L{\'o}pez-P{\'e}rez}}]{Pardo+etal+2022}
{Pardo}, J.~R., {Cernicharo}, J., {Tercero}, B., {et~al.} 2022, \aap, 658, A39,
  \dodoi{10.1051/0004-6361/202142263}

\bibitem[{{Park} {et~al.}(2006){Park}, {Wakelam}, \& {Herbst}}]{Park+etal+2006}
{Park}, I.~H., {Wakelam}, V., \& {Herbst}, E. 2006, \aap, 449, 631,
  \dodoi{10.1051/0004-6361:20054420}

\bibitem[{{Park} {et~al.}(2008){Park}, {Cho}, {Lee}, \&
  {Yang}}]{Park+etal+2008}
{Park}, J.~A., {Cho}, S.-H., {Lee}, C.~W., \& {Yang}, J. 2008, \aj, 136, 2350,
  \dodoi{10.1088/0004-6256/136/6/2350}

\bibitem[{{Patel} {et~al.}(2009){Patel}, {Young}, {Br{\"u}nken}, {Wilson},
  {Thaddeus}, {Menten}, {Reid}, {McCarthy}, {Dinh-V-Trung}, {Gottlieb}, \&
  {Hedden}}]{Patel+etal+2009}
{Patel}, N.~A., {Young}, K.~H., {Br{\"u}nken}, S., {et~al.} 2009, \apj, 692,
  1205, \dodoi{10.1088/0004-637X/692/2/1205}

\bibitem[{{Patel} {et~al.}(2011){Patel}, {Young}, {Gottlieb}, {Thaddeus},
  {Wilson}, {Menten}, {Reid}, {McCarthy}, {Cernicharo}, {He}, {Br{\"u}nken},
  {Trung}, \& {Keto}}]{Patel+etal+2011}
{Patel}, N.~A., {Young}, K.~H., {Gottlieb}, C.~A., {et~al.} 2011, \apjs, 193,
  17, \dodoi{10.1088/0067-0049/193/1/17}

\bibitem[{{Peng} {et~al.}(2013){Peng}, {Humphreys}, {Testi}, {Baudry},
  {Wittkowski}, {Rawlings}, {de Gregorio-Monsalvo}, {Vlemmings}, {Nyman},
  {Gray}, \& {de Breuck}}]{Peng+etal+2013}
{Peng}, T.~C., {Humphreys}, E.~M.~L., {Testi}, L., {et~al.} 2013, \aap, 559,
  L8, \dodoi{10.1051/0004-6361/201322466}

\bibitem[{{Penzias} {et~al.}(1971){Penzias}, {Solomon}, {Wilson}, \&
  {Jefferts}}]{Penzias+etal+1971}
{Penzias}, A.~A., {Solomon}, P.~M., {Wilson}, R.~W., \& {Jefferts}, K.~B. 1971,
  \apjl, 168, L53, \dodoi{10.1086/180784}

\bibitem[{{Pickett} {et~al.}(1998){Pickett}, {Poynter}, {Cohen}, {Delitsky},
  {Pearson}, \& {M{\"u}ller}}]{Pickett+etal+1998}
{Pickett}, H.~M., {Poynter}, R.~L., {Cohen}, E.~A., {et~al.} 1998, \jqsrt, 60,
  883, \dodoi{10.1016/S0022-4073(98)00091-0}

\bibitem[{{Portinari} {et~al.}(1998){Portinari}, {Chiosi}, \&
  {Bressan}}]{Portinari+etal+1998}
{Portinari}, L., {Chiosi}, C., \& {Bressan}, A. 1998, \aap, 334, 505,
  \dodoi{10.48550/arXiv.astro-ph/9711337}

\bibitem[{{Pulliam} {et~al.}(2011){Pulliam}, {Edwards}, \&
  {Ziurys}}]{Pulliam+etal+2011}
{Pulliam}, R.~L., {Edwards}, J.~L., \& {Ziurys}, L.~M. 2011, \apj, 743, 36,
  \dodoi{10.1088/0004-637X/743/1/36}

\bibitem[{{Pulliam} {et~al.}(2010){Pulliam}, {Savage}, {Ag{\'u}ndez},
  {Cernicharo}, {Gu{\'e}lin}, \& {Ziurys}}]{Pulliam+etal+2010}
{Pulliam}, R.~L., {Savage}, C., {Ag{\'u}ndez}, M., {et~al.} 2010, \apjl, 725,
  L181, \dodoi{10.1088/2041-8205/725/2/L181}

\bibitem[{{Qiu} {et~al.}(2022){Qiu}, {Zhang}, {Zhang}, \&
  {Nakashima}}]{Qiu+etal+2022}
{Qiu}, J.-J., {Zhang}, Y., {Zhang}, J.-S., \& {Nakashima}, J.-i. 2022, \apjs,
  259, 56, \dodoi{10.3847/1538-4365/ac5180}

\bibitem[{{Quintana-Lacaci} {et~al.}(2016){Quintana-Lacaci}, {Ag{\'u}ndez},
  {Cernicharo}, {Bujarrabal}, {S{\'a}nchez Contreras}, {Castro-Carrizo}, \&
  {Alcolea}}]{Quintana-Lacaci+etal+2016}
{Quintana-Lacaci}, G., {Ag{\'u}ndez}, M., {Cernicharo}, J., {et~al.} 2016,
  \aap, 592, A51, \dodoi{10.1051/0004-6361/201527688}

\bibitem[{{Reach} {et~al.}(2022){Reach}, {Ruaud}, {Wiesemeyer}, {Riquelme},
  {Tram}, {Cernicharo}, {Smith}, \& {Chambers}}]{Reach+etal+2022}
{Reach}, W.~T., {Ruaud}, M., {Wiesemeyer}, H., {et~al.} 2022, \apj, 926, 69,
  \dodoi{10.3847/1538-4357/ac4162}

\bibitem[{{Remijan} {et~al.}(2007){Remijan}, {Hollis}, {Lovas}, {Cordiner},
  {Millar}, {Markwick-Kemper}, \& {Jewell}}]{Remijan+etal+2007}
{Remijan}, A.~J., {Hollis}, J.~M., {Lovas}, F.~J., {et~al.} 2007, \apjl, 664,
  L47, \dodoi{10.1086/520704}

\bibitem[{{Ridgway} {et~al.}(1976){Ridgway}, {Hall}, {Wojslaw}, {Kleinmann}, \&
  {Weinberger}}]{Ridgway+etal+1976}
{Ridgway}, S.~T., {Hall}, D.~N.~B., {Wojslaw}, R.~S., {Kleinmann}, S.~G., \&
  {Weinberger}, D.~A. 1976, \nat, 264, 345, \dodoi{10.1038/264345a0}

\bibitem[{{Sahai} {et~al.}(1984){Sahai}, {Wootten}, \&
  {Clegg}}]{Sahai+etal+1984}
{Sahai}, R., {Wootten}, A., \& {Clegg}, R.~E.~S. 1984, \apj, 284, 144,
  \dodoi{10.1086/162394}

\bibitem[{{Saito} {et~al.}(1987){Saito}, {Kawaguchi}, {Yamamoto}, {Ohishi},
  {Suzuki}, \& {Kaifu}}]{Saito+etal+1987}
{Saito}, S., {Kawaguchi}, K., {Yamamoto}, S., {et~al.} 1987, \apjl, 317, L115,
  \dodoi{10.1086/184923}

\bibitem[{{Sch{\"o}ier} {et~al.}(2007){Sch{\"o}ier}, {Fong}, {Bieging},
  {Wilner}, {Young}, \& {Hunter}}]{Schoier+etal+2007}
{Sch{\"o}ier}, F.~L., {Fong}, D., {Bieging}, J.~H., {et~al.} 2007, \apj, 670,
  766, \dodoi{10.1086/522196}

\bibitem[{{Shan} {et~al.}(2012){Shan}, {Yang}, {Shi}, {Yao}, {Zuo}, {Lin},
  {Chen}, {Zhang}, {Duan}, {Cao}, {Li}, {Li}, {Liu}, \&
  {Zhong}}]{Shan+etal+2012}
{Shan}, W., {Yang}, J., {Shi}, S., {et~al.} 2012, IEEE Transactions on
  Terahertz Science and Technology, 2, 593, \dodoi{10.1109/TTHZ.2012.2213818}

\bibitem[{{Siebert} {et~al.}(2022){Siebert}, {Van de Sande}, {Millar}, \&
  {Remijan}}]{Siebert+etal+2022}
{Siebert}, M.~A., {Van de Sande}, M., {Millar}, T.~J., \& {Remijan}, A.~J.
  2022, \apj, 941, 90, \dodoi{10.3847/1538-4357/ac9e52}

\bibitem[{{Smith} {et~al.}(2015){Smith}, {Zijlstra}, \&
  {Fuller}}]{Smith+etal+2015}
{Smith}, C.~L., {Zijlstra}, A.~A., \& {Fuller}, G.~A. 2015, \mnras, 454, 177,
  \dodoi{10.1093/mnras/stv1934}

\bibitem[{{Snyder} \& {Buhl}(1971)}]{Snyder+etal+1971}
{Snyder}, L.~E., \& {Buhl}, D. 1971, \apjl, 163, L47, \dodoi{10.1086/180664}

\bibitem[{{Snyder} \& {Buhl}(1972)}]{Snyder+etal+1972}
---. 1972, Annals of the New York Academy of Sciences, 194, 17,
  \dodoi{10.1111/j.1749-6632.1972.tb12687.x}

\bibitem[{{Snyder} {et~al.}(1985){Snyder}, {Henkel}, {Hollis}, \&
  {Lovas}}]{Snyder+etal+1985}
{Snyder}, L.~E., {Henkel}, C., {Hollis}, J.~M., \& {Lovas}, F.~J. 1985, \apjl,
  290, L29, \dodoi{10.1086/184436}

\bibitem[{{Solomon} {et~al.}(1971){Solomon}, {Jefferts}, {Penzias}, \&
  {Wilson}}]{Solomon+etal+1971}
{Solomon}, P., {Jefferts}, K.~B., {Penzias}, A.~A., \& {Wilson}, R.~W. 1971,
  \apjl, 163, L53, \dodoi{10.1086/180665}

\bibitem[{{Suzuki} {et~al.}(1984){Suzuki}, {Kaifu}, {Miyaji}, {Morimoto},
  {Ohishi}, \& {Saito}}]{Suzuki+etal+1984}
{Suzuki}, H., {Kaifu}, N., {Miyaji}, T., {et~al.} 1984, \apj, 282, 197,
  \dodoi{10.1086/162191}

\bibitem[{{Tenenbaum} {et~al.}(2006){Tenenbaum}, {Apponi}, {Ziurys},
  {Ag{\'u}ndez}, {Cernicharo}, {Pardo}, \& {Gu{\'e}lin}}]{Tenenbaum+etal+2006}
{Tenenbaum}, E.~D., {Apponi}, A.~J., {Ziurys}, L.~M., {et~al.} 2006, \apjl,
  649, L17, \dodoi{10.1086/508166}

\bibitem[{{Tenenbaum} {et~al.}(2010{\natexlab{a}}){Tenenbaum}, {Dodd}, {Milam},
  {Woolf}, \& {Ziurys}}]{Tenenbaum+etal+2010a}
{Tenenbaum}, E.~D., {Dodd}, J.~L., {Milam}, S.~N., {Woolf}, N.~J., \& {Ziurys},
  L.~M. 2010{\natexlab{a}}, \apjl, 720, L102,
  \dodoi{10.1088/2041-8205/720/1/L102}

\bibitem[{{Tenenbaum} {et~al.}(2010{\natexlab{b}}){Tenenbaum}, {Dodd}, {Milam},
  {Woolf}, \& {Ziurys}}]{Tenenbaum+etal+2010}
---. 2010{\natexlab{b}}, \apjs, 190, 348, \dodoi{10.1088/0067-0049/190/2/348}

\bibitem[{{Tenenbaum} \& {Ziurys}(2008)}]{Tenenbaum+etal+2008}
{Tenenbaum}, E.~D., \& {Ziurys}, L.~M. 2008, \apjl, 680, L121,
  \dodoi{10.1086/589973}

\bibitem[{{Tercero} {et~al.}(2021){Tercero}, {L{\'o}pez-P{\'e}rez}, {Gallego},
  {Beltr{\'a}n}, {Garc{\'\i}a}, {Patino-Esteban}, {L{\'o}pez-Fern{\'a}ndez},
  {G{\'o}mez-Molina}, {Diez}, {Garc{\'\i}a-Carre{\~n}o}, {Malo}, {Amils},
  {Serna}, {Albo}, {Hern{\'a}ndez}, {Vaquero}, {Gonz{\'a}lez-Garc{\'\i}a},
  {Barbas}, {L{\'o}pez-Fern{\'a}ndez}, {Bujarrabal}, {G{\'o}mez-Garrido},
  {Pardo}, {Santander-Garc{\'\i}a}, {Tercero}, {Cernicharo}, \& {de
  Vicente}}]{Tercero+etal+2021}
{Tercero}, F., {L{\'o}pez-P{\'e}rez}, J.~A., {Gallego}, J.~D., {et~al.} 2021,
  \aap, 645, A37, \dodoi{10.1051/0004-6361/202038701}

\bibitem[{{Thaddeus} {et~al.}(1984){Thaddeus}, {Cummins}, \&
  {Linke}}]{Thaddeus+etal+1984}
{Thaddeus}, P., {Cummins}, S.~E., \& {Linke}, R.~A. 1984, \apjl, 283, L45,
  \dodoi{10.1086/184330}

\bibitem[{{Thaddeus} {et~al.}(2008){Thaddeus}, {Gottlieb}, {Gupta},
  {Br{\"u}nken}, {McCarthy}, {Ag{\'u}ndez}, {Gu{\'e}lin}, \&
  {Cernicharo}}]{Thaddeus+etal+2008}
{Thaddeus}, P., {Gottlieb}, C.~A., {Gupta}, H., {et~al.} 2008, \apj, 677, 1132,
  \dodoi{10.1086/528947}

\bibitem[{{Thaddeus} {et~al.}(1985{\natexlab{a}}){Thaddeus}, {Gottlieb},
  {Hjalmarson}, {Johansson}, {Irvine}, {Friberg}, \&
  {Linke}}]{Thaddeus+etal+1985}
{Thaddeus}, P., {Gottlieb}, C.~A., {Hjalmarson}, A., {et~al.}
  1985{\natexlab{a}}, \apjl, 294, L49, \dodoi{10.1086/184507}

\bibitem[{{Thaddeus} {et~al.}(1981){Thaddeus}, {Gu{\'e}lin}, \&
  {Linke}}]{Thaddeus+etal+1981}
{Thaddeus}, P., {Gu{\'e}lin}, M., \& {Linke}, R.~A. 1981, \apjl, 246, L41,
  \dodoi{10.1086/183549}

\bibitem[{{Thaddeus} {et~al.}(1985{\natexlab{b}}){Thaddeus}, {Vrtilek}, \&
  {Gottlieb}}]{Thaddeus+etal+1985a}
{Thaddeus}, P., {Vrtilek}, J.~M., \& {Gottlieb}, C.~A. 1985{\natexlab{b}},
  \apjl, 299, L63, \dodoi{10.1086/184581}

\bibitem[{{Truong-Bach} {et~al.}(1987){Truong-Bach}, {Nguyen-Q-Rieu}, {Omont},
  {Olofsson}, \& {Johansson}}]{Truong-Bach+etal+1987}
{Truong-Bach}, {Nguyen-Q-Rieu}, {Omont}, A., {Olofsson}, H., \& {Johansson},
  L.~E.~B. 1987, \aap, 176, 285

\bibitem[{{Tsuge} {et~al.}(2021){Tsuge}, {Kouchi}, \&
  {Watanabe}}]{Tsuge+etal+2021}
{Tsuge}, M., {Kouchi}, A., \& {Watanabe}, N. 2021, \apj, 923, 71,
  \dodoi{10.3847/1538-4357/ac2a33}

\bibitem[{{Tucker} {et~al.}(1974){Tucker}, {Kutner}, \&
  {Thaddeus}}]{Tucker+etal+1974}
{Tucker}, K.~D., {Kutner}, M.~L., \& {Thaddeus}, P. 1974, \apjl, 193, L115,
  \dodoi{10.1086/181646}

\bibitem[{{Turner}(1970)}]{Turner+etal+1970}
{Turner}, B.~E. 1970, \iaucirc, 2268, 2

\bibitem[{{Turner}(1971)}]{Turner+etal+1971}
---. 1971, \apjl, 163, L35, \dodoi{10.1086/180662}

\bibitem[{{Turner}(1992)}]{Turner+etal+1992}
---. 1992, \apjl, 388, L35, \dodoi{10.1086/186324}

\bibitem[{{Turner} {et~al.}(1994){Turner}, {Steimle}, \&
  {Meerts}}]{Turner+etal+1994}
{Turner}, B.~E., {Steimle}, T.~C., \& {Meerts}, L. 1994, \apjl, 426, L97,
  \dodoi{10.1086/174043}

\bibitem[{{Van de Sande} \& {Millar}(2019)}]{VandeSande+etal+2019}
{Van de Sande}, M., \& {Millar}, T.~J. 2019, \apj, 873, 36,
  \dodoi{10.3847/1538-4357/ab03d4}

\bibitem[{{Van de Sande} \& {Millar}(2022)}]{VandeSande+etal+2022}
---. 2022, \mnras, 510, 1204, \dodoi{10.1093/mnras/stab3282}

\bibitem[{{Van de Sande} {et~al.}(2021){Van de Sande}, {Walsh}, \&
  {Millar}}]{VandeSande+etal+2021}
{Van de Sande}, M., {Walsh}, C., \& {Millar}, T.~J. 2021, \mnras, 501, 491,
  \dodoi{10.1093/mnras/staa3689}

\bibitem[{{van der Veen} {et~al.}(1998){van der Veen}, {Huggins}, \&
  {Matthews}}]{van+etal+1998}
{van der Veen}, W.~E.~C.~J., {Huggins}, P.~J., \& {Matthews}, H.~E. 1998, \apj,
  505, 749, \dodoi{10.1086/306191}

\bibitem[{{Velilla-Prieto} {et~al.}(2019){Velilla-Prieto}, {Cernicharo},
  {Ag{\'u}ndez}, {Fonfr{\'\i}a}, {Quintana-Lacaci}, {Marcelino}, \&
  {Castro-Carrizo}}]{Velilla-Prieto+etal+2019}
{Velilla-Prieto}, L., {Cernicharo}, J., {Ag{\'u}ndez}, M., {et~al.} 2019, \aap,
  629, A146, \dodoi{10.1051/0004-6361/201834717}

\bibitem[{{Velilla Prieto} {et~al.}(2015){Velilla Prieto}, {Cernicharo},
  {Quintana-Lacaci}, {Ag{\'u}ndez}, {Castro-Carrizo}, {Fonfr{\'\i}a},
  {Marcelino}, {Z{\'u}{\~n}iga}, {Requena}, {Bastida}, {Lique}, \&
  {Gu{\'e}lin}}]{Velilla+Prieto+etal+2015}
{Velilla Prieto}, L., {Cernicharo}, J., {Quintana-Lacaci}, G., {et~al.} 2015,
  \apjl, 805, L13, \dodoi{10.1088/2041-8205/805/2/L13}

\bibitem[{{Velilla Prieto} {et~al.}(2017){Velilla Prieto}, {S{\'a}nchez
  Contreras}, {Cernicharo}, {Ag{\'u}ndez}, {Quintana-Lacaci}, {Bujarrabal},
  {Alcolea}, {Balan{\c{c}}a}, {Herpin}, {Menten}, \&
  {Wyrowski}}]{Prieto+etal+2017}
{Velilla Prieto}, L., {S{\'a}nchez Contreras}, C., {Cernicharo}, J., {et~al.}
  2017, \aap, 597, A25, \dodoi{10.1051/0004-6361/201628776}

\bibitem[{{Wannier} \& {Linke}(1978)}]{Wannier+etal+1978}
{Wannier}, P.~G., \& {Linke}, R.~A. 1978, \apj, 225, 130,
  \dodoi{10.1086/156474}

\bibitem[{{Wasserburg} {et~al.}(2017){Wasserburg}, {Karakas}, \&
  {Lugaro}}]{Wasserburg+etal+2017}
{Wasserburg}, G.~J., {Karakas}, A.~I., \& {Lugaro}, M. 2017, \apj, 836, 126,
  \dodoi{10.3847/1538-4357/836/1/126}

\bibitem[{{Wilson} {et~al.}(1971){Wilson}, {Solomon}, {Penzias}, \&
  {Jefferts}}]{Wilson+etal+1971}
{Wilson}, R.~W., {Solomon}, P.~M., {Penzias}, A.~A., \& {Jefferts}, K.~B. 1971,
  \apjl, 169, L35, \dodoi{10.1086/180809}

\bibitem[{{Wilson} \& {Rood}(1994)}]{Wilson+etal+1994}
{Wilson}, T.~L., \& {Rood}, R. 1994, \araa, 32, 191,
  \dodoi{10.1146/annurev.aa.32.090194.001203}

\bibitem[{{Winnewisser} \& {Walmsley}(1978)}]{Winnewisser+etal+1978}
{Winnewisser}, G., \& {Walmsley}, C.~M. 1978, \aap, 70, L37

\bibitem[{{Woods} {et~al.}(2003){Woods}, {Sch{\"o}ier}, {Nyman}, \&
  {Olofsson}}]{Woods+etal+2003}
{Woods}, P.~M., {Sch{\"o}ier}, F.~L., {Nyman}, L.~{\r{A}}., \& {Olofsson}, H.
  2003, \aap, 402, 617, \dodoi{10.1051/0004-6361:20030190}

\bibitem[{{Wootten} {et~al.}(1982){Wootten}, {Lichten}, {Sahai}, \&
  {Wannier}}]{Wootten+etal+1982}
{Wootten}, A., {Lichten}, S.~M., {Sahai}, R., \& {Wannier}, P.~G. 1982, \apj,
  257, 151, \dodoi{10.1086/159973}

\bibitem[{{Yamamoto} {et~al.}(1987){Yamamoto}, {Saito}, {Gu{\'e}lin},
  {Cernicharo}, {Suzuki}, \& {Ohishi}}]{Yamamoto+etal+1987}
{Yamamoto}, S., {Saito}, S., {Gu{\'e}lin}, M., {et~al.} 1987, \apjl, 323, L149,
  \dodoi{10.1086/185076}

\bibitem[{{Yan} {et~al.}(2023){Yan}, {Henkel}, {Kobayashi}, {Menten}, {Gong},
  {Zhang}, {Yu}, {Yang}, {Xie}, \& {Wang}}]{Yan+etal+2023}
{Yan}, Y.~T., {Henkel}, C., {Kobayashi}, C., {et~al.} 2023, \aap, 670, A98,
  \dodoi{10.1051/0004-6361/202244584}

\bibitem[{{Yang} {et~al.}(2023){Yang}, {Zhang}, {Qiu}, {Ao}, \&
  {Li}}]{Yang+etal+2023}
{Yang}, K., {Zhang}, Y., {Qiu}, J., {Ao}, Y., \& {Li}, X. 2023, \pasj, 75, 853,
  \dodoi{10.1093/pasj/psad043}

\bibitem[{{Zack} {et~al.}(2011){Zack}, {Halfen}, \& {Ziurys}}]{Zack+etal+2011}
{Zack}, L.~N., {Halfen}, D.~T., \& {Ziurys}, L.~M. 2011, \apjl, 733, L36,
  \dodoi{10.1088/2041-8205/733/2/L36}

\bibitem[{{Zhang} {et~al.}(2017){Zhang}, {Zhu}, {Li}, {Chen}, {Wang}, \&
  {Zhang}}]{Zhang+etal+2017}
{Zhang}, X.-Y., {Zhu}, Q.-F., {Li}, J., {et~al.} 2017, \aap, 606, A74,
  \dodoi{10.1051/0004-6361/201730791}

\bibitem[{{Zhang} {et~al.}(2020){Zhang}, {Chau}, {Nakashima}, \&
  {Kwok}}]{Zhang+etal+2020}
{Zhang}, Y., {Chau}, W., {Nakashima}, J.-i., \& {Kwok}, S. 2020, \pasj, 72, 46,
  \dodoi{10.1093/pasj/psaa032}

\bibitem[{{Zhang} {et~al.}(2008){Zhang}, {Kwok}, \&
  {Dinh-V-Trung}}]{Zhang+etal+2008}
{Zhang}, Y., {Kwok}, S., \& {Dinh-V-Trung}. 2008, \apj, 678, 328,
  \dodoi{10.1086/529428}

\bibitem[{{Zhang} {et~al.}(2009{\natexlab{a}}){Zhang}, {Kwok}, \&
  {Dinh-V-Trung}}]{Zhang+etal+2009a}
---. 2009{\natexlab{a}}, \apj, 691, 1660, \dodoi{10.1088/0004-637X/691/2/1660}

\bibitem[{{Zhang} {et~al.}(2009{\natexlab{b}}){Zhang}, {Kwok}, \&
  {Nakashima}}]{Zhang+etal+2009}
{Zhang}, Y., {Kwok}, S., \& {Nakashima}, J.-i. 2009{\natexlab{b}}, \apj, 700,
  1262, \dodoi{10.1088/0004-637X/700/2/1262}

\bibitem[{{Zhang} {et~al.}(2013){Zhang}, {Kwok}, {Nakashima}, {Chau}, \&
  {Dinh-V-Trung}}]{Zhang+etal+2013}
{Zhang}, Y., {Kwok}, S., {Nakashima}, J.-i., {Chau}, W., \& {Dinh-V-Trung}.
  2013, \apj, 773, 71, \dodoi{10.1088/0004-637X/773/1/71}

\bibitem[{{Ziurys} {et~al.}(1995){Ziurys}, {Apponi}, {Gu{\'e}lin}, \&
  {Cernicharo}}]{Ziurys+etal+1995}
{Ziurys}, L.~M., {Apponi}, A.~J., {Gu{\'e}lin}, M., \& {Cernicharo}, J. 1995,
  \apjl, 445, L47, \dodoi{10.1086/187886}

\bibitem[{{Ziurys} {et~al.}(1994){Ziurys}, {Apponi}, \&
  {Phillips}}]{Ziurys+etal+1994}
{Ziurys}, L.~M., {Apponi}, A.~J., \& {Phillips}, T.~G. 1994, \apj, 433, 729,
  \dodoi{10.1086/174682}

\bibitem[{{Ziurys} {et~al.}(2002){Ziurys}, {Savage}, {Highberger}, {Apponi},
  {Gu{\'e}lin}, \& {Cernicharo}}]{Ziurys+etal+2002}
{Ziurys}, L.~M., {Savage}, C., {Highberger}, J.~L., {et~al.} 2002, \apjl, 564,
  L45, \dodoi{10.1086/338775}

\bibitem[{{Zuckerman}(1987)}]{Zuckerman+etal+1987}
{Zuckerman}, B. 1987, in Astrochemistry, ed. M.~S. {Vardya} \& S.~P.
  {Tarafdar}, Vol. 120, 345--355

\end{thebibliography}
\bibliographystyle{aasjournal}

\end{document}